\pdfoutput=1
\documentclass[11pt,a4paper]{article}
\usepackage{jheppub}
\usepackage{amsfonts,amssymb,amsmath,mathtools}
\usepackage{booktabs,array}
\usepackage{tikz}
\usetikzlibrary{arrows.meta,calc}
\usetikzlibrary{intersections}
\usepackage{subcaption}
\usepackage{float}
\usepackage{needspace}

\definecolor{planback}{gray}{0.92}
\definecolor{planframe}{gray}{0.55}
\definecolor{NGblue}{RGB}{18,55,110}

\newcommand{\algsl}{\mathfrak{sl}}

\newcommand{\ii}{i}
\newcommand{\dd}{\mathrm{d}}
\renewcommand{\Re}{{\rm Re}\,}
\renewcommand{\Im}{{\rm Im}\,}
\newcommand{\svx}{\mathsf{x}}
\newcommand{\svxd}{\dot{\svx}}
\newcommand{\svy}{\mathsf{y}}
\newcommand{\svyd}{\dot{\svy}}
\newcommand{\Qp}{\mathbb Q_{+}}
\newcommand{\cD}{\mathcal D}
\newcommand{\bfS}{\mathbf S}
\newcommand{\bz}{\boldsymbol z}
\newcommand{\bI}{\widehat{\mathsf I}}
\newcommand{\LRpair}[2]{\bigl(\!\bigl(#1\,\mathord{\mid}\,#2\bigr)\!\bigr)}
\newcommand{\towerbracket}[2]{\mathopen{\boldsymbol{\{}}#1,#2\mathclose{\boldsymbol{\}}}}
\newcommand{\Bgood}{\mathbf{B}}
\newcommand{\zerosov}{\mathtt{0}}
\newcommand{\beq}{\begin{equation}}
\newcommand{\eeq}{\end{equation}}

\graphicspath{{./}{../}{figures/}}

\usepackage{listings}
\lstdefinestyle{bsolver}{
  language=Mathematica, basicstyle=\ttfamily\footnotesize,
  keywordstyle=\color{NGblue}, commentstyle=\color{gray},
  stringstyle=\color{black}, showstringspaces=false,
  columns=fullflexible, keepspaces=true, breaklines=true,
  frame=single, framerule=0.3pt, rulecolor=\color{planframe},
  backgroundcolor=\color{planback}, xleftmargin=0.5em,
  xrightmargin=0.5em, aboveskip=1em, belowskip=1em}

\title{Correlation functions from integrability: 2D fishnets}

\author[a]{Carlos Bercini,}
\author[a]{Simon Ekhammar,}
\author[a]{Nikolay Gromov,}
\author[c]{Paul Ryan}

\affiliation[a]{Department of Mathematics, King's College London,
  The Strand, London WC2R 2LS, UK}
\affiliation[c]{Deutsches Elektronen-Synchrotron DESY,
  Notkestr.~85, 22607 Hamburg, Germany}

\emailAdd{carlos.bercini@kcl.ac.uk}
\emailAdd{simon.ekhammar@kcl.ac.uk}
\emailAdd{nikolay.gromov@kcl.ac.uk}
\emailAdd{paul.ryan@desy.de}

\makeatletter
\gdef\@fpheader{Prepared for submission to JHEP\hfill{\normalfont\texttt{DESY-26-172}}}
\makeatother

\abstract{%
We derive three-point functions of the two-dimensional bi-scalar fishnet
CFT at finite coupling, directly from the Q-functions that determine its
spectrum. Using separation of variables for the underlying twisted spin
chain, we show that the separated wave function is a product of some special bilinear combinations of
Q-functions corresponding to two copies of the $\mathfrak{sl}(2)$ symmetry algebra. At finite coupling the structure constants with one
or two unprotected operators of arbitrary length and spin reduce to finite determinants of single
integrals of Q-functions. The norm, in turn, reduces to a finite variation of the gluing conditions with respect to the conserved charges, similar to the Gaudin determinant for Bethe-state norms. Our construction is expected to provide further structural hints for the computation of non-perturbative structure constants in planar $\mathcal N=4$ SYM.}

\begin{document}
\maketitle

\newpage
\section{Introduction and overview}
\label{sec:intro}

In integrable planar gauge theories the spectrum of local operators is
encoded in Q-functions. In planar $\mathcal N=4$ supersymmetric Yang--Mills
(SYM) theory and in ABJM theory they are
governed by the Quantum Spectral Curve (QSC), a closed set of functional
relations supplemented by analyticity and gluing
conditions~\cite{Gromov:2013pga,Gromov:2014caa,Cavaglia:2014exa}. A QSC
has also been proposed for AdS$_3$/CFT$_2$~\cite{Ekhammar:2021pys,Cavaglia:2021eqr},
and in fishnet theories the same structure reduces to Baxter equations
with a quantisation condition, both in four~\cite{Gromov:2017cja} and in
two~\cite{Ekhammar:2025fvn} dimensions. In all these cases finding
the Q-functions is usually considered an intermediate step: one reads off
the scaling dimension from their large-$u$ asymptotics, and the rest of the
information they carry is discarded.

This has recently started to change. The first hints came from Wilson
lines~\cite{Cavaglia:2018lxi,Giombi:2018qox}, weak-coupling three-point functions and orthogonality of
Q-functions~\cite{Bercini:2022jxo,Bargheer:2025kli,Bargheer:2026kon}, and the
four-dimensional
fishnet~\cite{Cavaglia:2021mft}. In planar $\mathcal N=4$ SYM, the hexagon
representation of structure constants~\cite{Basso:2015zoa} was then completed for short
operators, with wrapping corrections resummed using QSC
data~\cite{Basso:2022nny}, and at small spin it was written directly in terms
of the QSC fully non-perturbatively~\cite{Basso:2025mca}. At finite spin,
however, the conjecture of~\cite{Basso:2022nny} expresses the result as an
infinite sum of multiple integrals of unbounded dimension, which is not
suitable for evaluation at finite coupling. Nevertheless,
in~\cite{Ekhammar:2026structure} it was rewritten in a form resembling separation of variables (SoV), which packs all integrations into
determinants of single integrals of Q-functions.

For two half-BPS operators
and any operator with parity symmetry, this form allows both analytic
computations at weak coupling and high-precision numerics from weak to strong
coupling.
Could the Q-functions then determine the most general correlation functions completely,
 at finite coupling? In this paper we {derive}
exactly how to arrange them to obtain three-point coefficients in the
two-dimensional bi-scalar fishnet theory. We show from first principles how
determinants of this kind arise, and we expect them to serve as a guide for
generalisations to structure constants with two unprotected operators and
beyond.

Fishnet theories are the natural laboratory for this question. While in
theories like ${\cal N}=4$ SYM integrability is still mysterious, despite a
huge number of highly non-trivial tests, in fishnet theories one is able to
{\it derive} it from first principles. In particular, one can prove that the
graph-building operators, which generate the planar Feynman diagrams, belong
to a commuting family of spin-chain transfer
matrices~\cite{Kazakov:2018qbr}. The eigenfunctions of this spin chain have a
direct field-theory meaning: they are correlators of an unprotected operator
with $J$ elementary scalars~\cite{Cavaglia:2020hdb}. In 2D, SoV for the
$SL(2,\mathbb C)$ spin chain gives an explicit basis in which these wave
functions factorise into products of
Q-functions~\cite{Derkachov:2001yn,Derkachov:2018rot}.\footnote{Fishnet theories are non-unitary, and their
dilatation operator is in general not diagonalisable: some operators
organise into Jordan blocks and form logarithmic
multiplets~\cite{Ipsen:2018fmu,Ahn:2020zly,NietoGarcia:2021kgh,Ahn:2021emp,Ahn:2022snr}. Throughout
this paper we restrict to the non-logarithmic sector, in which the
wave functions are true eigenvectors of the commuting Hamiltonians. For
generalised eigenvectors the separated wave function no longer factorises;
SoV for such non-diagonalisable chains was recently developed
in~\cite{NietoGarcia:2026jva}.} This is the starting
point for a first-principles construction of their overlaps, and hence of
correlation functions, at arbitrary chain length.

As AI makes numerical and symbolic experiments easier to run at scale,
a structure found in one theory can quickly guide calculations across a
family of models. With this in mind, we set out to identify the key elements
of the non-perturbative structure in a setting where every step can be
derived and checked. Our main findings are the following. First, the norm,
which in $\mathcal N=4$ SYM still requires TBA Fredholm determinants, is
expressed entirely through Q-functions. Furthermore, when two copies of the same state are glued, its two-point normalisation collapses to a finite determinant of charge
variations of the connection matrix of the Baxter equation, much like the
Gaudin formula for Bethe-state norms~\cite{Gaudin:1981,Korepin:1982gg}. Second, we obtain three-point
coefficients with one or with two unprotected operators, for any lengths and
spins and without assuming any $\mathbb Z_2$ symmetry of the states. Third,
the shadow representation of the spin chain, invisible at tree level,
contributes at finite coupling and is needed to complete the result. Finally,
our correlators are formulated naturally in the presence of a colour twist,
a topological line defect compatible with integrability, and the untwisted
results follow in a controlled limit. In $\mathcal N=4$ SYM, twisted
structure constants and their untwisting limit have so far been constructed
only at weak coupling~\cite{Bargheer:2026kon}; our finite-coupling
construction suggests a way to include integrable topological defects there
as well.

While the model studied here is two-dimensional, structurally it is very
close to its four-dimensional counterparts. Its Q-functions obey Baxter
equations and gluing conditions closely analogous to those of the
four-dimensional fishnet~\cite{Cavaglia:2021mft} and of the QSC, and three- and four-dimensional
fishnets themselves arise in double-scaling limits of deformed ABJM and
$\mathcal N=4$ SYM, respectively~\cite{Caetano:2016ydc}. The determinants of
Q-function moments that we derive are similar in structure to those
of~\cite{Ekhammar:2026structure}, and indeed the fishnet formulas developed
here helped motivate their interpretation as SoV overlaps. Since in our paper the
norm is also expressed through Q-functions alone, we hope that our results
will help to extend those constructions further, ultimately leading to a
solution of a fully interacting gauge theory for the first time.

\paragraph{The theory and what we compute.}
The bi-scalar fishnet theory contains two complex $N\times N$ matrix
scalars $\phi_1$ and $\phi_2$, coupled by a single chiral quartic vertex
with coupling $\xi$, whose Lagrangian is recalled in~\eqref{eq:2d-fishnet-lagrangian}. The vertex has
no Hermitian conjugate, so the theory is not unitary, but in exchange its
planar graphs are extremely rigid: they are regular square lattices, or
fishnets. We study the simplest single-trace operators,
$\operatorname{tr}[\phi_1^J]$ dressed with derivatives. In the planar limit
the only graphs that correct them are wheels of $\phi_2$ propagators wrapped
around the $J$ lines of $\phi_1$. The resulting scaling dimensions
$\Delta(\xi)$ were found in~\cite{Ekhammar:2025fvn} from Baxter
equations and a gluing condition, in close analogy with the QSC of
$\mathcal N=4$ SYM and of AdS$_3$. In this paper we ask what the same
Q-functions say about the remaining conformal data.

\begin{figure}[t]
\centering
\includegraphics[width=0.5\linewidth]{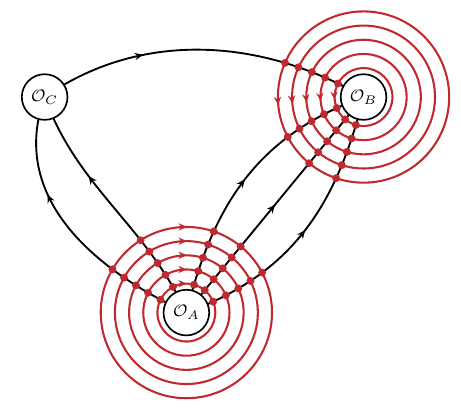}
\caption{A planar heavy--heavy--light (HHL) three-point function.
The black lines are $\phi_1$ propagators and the red lines are the
$\phi_2$ fishnet wheels. The heavy operators $\mathcal O_A$ and
$\mathcal O_B$ are dressed by wheels and carry the coupling dependence. At
the light operator $\mathcal O_C$ the neighbouring $\phi_1$ and
$\phi_1^\dagger$ lines have opposite orientation, so no wheel can pass
through it, and its dimension keeps its free value. Replacing
$\mathcal O_A$ by a free operator gives the HLL configuration.}
\label{fig:twisted-fishnet-chain}
\end{figure}

The first quantity we compute is the two-point normalisation
$\mathcal N_2$ of such an operator. This is the analogue of a Bethe-state
norm, and it is needed to normalise any three-point coefficient. The
second consists of three-point coefficients of these operators with
operators of mixed field content, such as
$\operatorname{tr}[\phi_1^{M}\phi_1^{\dagger\,M'}]$. No wheel can wrap
around a mixed operator in the planar limit, so it keeps its free
dimension~\cite{Basso:2018cvy}. We therefore call operators of the first
kind \emph{heavy} ($\bullet$) and the coupling-independent ones
\emph{light} ($\circ$). We compute two families of structure constants: the
HLL coefficient $\mathcal C_\ell^{\bullet\circ\circ}$, with one heavy
operator, and the HHL coefficient $\mathcal C_\ell^{\bullet\bullet\circ}$,
with two heavy operators sharing $\ell$ propagators, see
Figure~\ref{fig:twisted-fishnet-chain}. The HLL coefficients control the
operator product expansion of four-point functions such as
$\langle\operatorname{tr}[\phi_1^{J-M}(x_1)\phi_1^{M}(x_2)]\,
\operatorname{tr}[\phi_1^{\dagger\,J-M}(x_3)\phi_1^{\dagger\,M}(x_4)]\rangle$,
which so far have been studied at finite coupling only for the simplest case of
$J=2$~\cite{Kazakov:2018qbr,Gromov:2018hut}.
Equivalently, an HLL
coefficient is the overlap of the heavy operator with the single-trace
operator $\operatorname{tr}[(\phi_1^\dagger(x_1))^{\ell}(\phi_1^\dagger(x_2))^{J-\ell}]$,
whose fields sit at two different points. In $\mathcal N=4$ super
Yang--Mills such an operator is not gauge invariant, so this overlap has no
analogue there, while in fishnet theories it is perfectly fine and often
considered for its simplicity. As far as we know, HHL
coefficients have only been obtained at $J=2$ in four
dimensions~\cite{Gromov:2018hut}.
Similar HHL correlators were also obtained via the hexagon approach~\cite{Basso:2018cvy}.

\paragraph{The difficulty.}
All these quantities are encoded in the CFT wave function of the heavy
operator~\cite{Gromov:2019aku,Gromov:2019bsj,Ekhammar:2025fvn},
\begin{equation}\label{eq:mathcalAInter}
  \psi(z_1,z_2,\dots,z_J) \propto \langle \mathcal{O}_{\Delta,S}(0)\, \operatorname{tr}\left[ \phi^{\dagger}_1(z_1) \dots \phi^{\dagger}_1(z_{J}) \right] \rangle\,,
\end{equation}
which is an eigenfunction of the integrable $SL(2,\mathbb C)$ spin chain
that builds the fishnet graphs. Gluing two wave functions together gives
the three-point functions above, while gluing a wave function to its
conjugate gives the norm. In practice, however, $\psi$ is known in closed
form only for the shortest chains, and the gluing integrals over $2J$ real
coordinates become intractable already at $J>3$. The norm is even more
subtle: the self-overlap of a wave function diverges logarithmically, and its logarithmic coefficient is proportional to the two-point normalisation $\mathcal N_2$, as given in~\eqref{eq:coordinate-general-norm-relation}. What we need is a
representation in which these integrals reduce to the same one-variable
Q-functions that fix the spectrum.

\paragraph{Twist and separation of variables.}
Our strategy has two steps. First, we deform the theory by a colour
twist~\cite{Cavaglia:2020hdb}. This preserves planar integrability, breaks
the conformal group down to the transformations that fix two points, lifts
spectral degeneracies, and makes the SoV construction regular. Second, in the separated basis the
wave function of every state becomes a product of a single function
$\Qp$,
\begin{equation}
 \Phi(\svx,\dot\svx)\propto\prod_{\alpha=1}^{J}\Qp(\svx_\alpha,\dot\svx_\alpha),
 \qquad
 \Qp(u,\dot u)=\sum_{i,j=1}^2 q_i^{\downarrow}(u)\,\Gamma^{ij}\,\dot q_j^{\uparrow}(\dot u),
\end{equation}
which glues the two Baxter solutions with the same constant matrix
$\Gamma$ that quantises the spectrum~\cite{Ekhammar:2025fvn}. The
glued Q-function thus does double duty: its gluing fixes $\Delta(\xi)$, and
its values form the wave function. Scalar products then become $J\times J$
determinants whose entries are sums over sites of one-dimensional integrals
of q-bilinears. Closing the contours turns these integrals into residue sums
over towers of poles. At zero coupling only the ordinary towers contribute,
and the determinant reduces to the product of two $\algsl_2$ SoV
determinants~\cite{Gromov:2022waj}. At finite coupling the shadow
towers, attached to the representation with site spin $1-s$, contribute as
well and complete the result.

\paragraph{Three-point functions.}
To reach three-point functions we need to cut a chain into two pieces. We
show that this cut is very simple in the separated basis: freezing $J-\ell$
separated variables at their vacuum values factorises the SoV basis into
two shorter chains~\cite{Bargheer:2026kon}. As a result, the HHL overlap is
an $\ell\times\ell$ determinant. Its measure involves only the $\ell$
glued sites, while the Q-functions in the integrand belong to the full
length-$J_A$ and length-$J_B$ states.

\paragraph{The norm as a Jacobian.}
The two-point normalisation needs more work. When the twists of two copies
of the same state approach each other, their overlap diverges as
$\log\varepsilon$, where $\varepsilon$ is the difference between their twists. In coordinate
space, as for the four-dimensional
fishnet~\cite{Gromov:2019bsj,Cavaglia:2021mft}, the coefficient of this logarithm is
$\mathcal N_2$ multiplied by the spectral slope
$\mathcal D_\xi=\xi^2\partial\Delta/\partial\xi^2$. In the SoV determinant
the divergence is more severe: individual entries contain double
logarithms. We find, however, that on shell all double logarithms cancel,
and all single logarithms cancel except the one in the corner entry. The
norm is then the remaining $(J-1)\times(J-1)$ minor,
see~\eqref{eq:norms-two-point-normalisation}. Rewriting this minor in terms
of the connection matrix $\Omega$, which relates two bases of the Q-functions $q^\uparrow=\Omega q^\downarrow$, the
infinite tower sums collapse to a finite residue
formula~\eqref{eq:norms-connection-matrix}. It pairs the derivatives of
$\Omega$ with respect to the conserved charges in the two chiral sectors,
evaluated at $2J$ points in one period strip. This is the fishnet analogue
of the Gaudin formula~\cite{Gaudin:1981,Korepin:1982gg}, which expresses the norm of a Bethe state as the
Jacobian of the Bethe equations. Here the role of the Bethe equations is
played by the gluing condition.

\paragraph{Removing the twist.}
The twist can be considered as a regulator. As it is removed, the overlaps and the
norms both diverge as powers of the twist angle, and only their ratio, the
normalised structure constant, stays finite. We carry out this limit for
the Q-functions and for the determinants. It reduces the result to Q-values
of the untwisted theory together with explicit Gamma-function factors. For
$J=2$ all ingredients are explicit. The HLL coefficient reproduces the
result of~\cite{Kazakov:2018qbr}, and we find a new
closed form for the HHL coefficient of two interacting operators,
\begin{equation}
 \mathcal C_1^{\bullet\bullet\circ}
 =2e^{-\frac{\pi\ii}{2}(S_A-S_B)}
 \frac{\Gamma(h_A+h_B-\tfrac12)}
 {\Gamma(\tfrac32-\dot h_A-\dot h_B)}
 \left[\frac{\Gamma(2-2\dot h_A)\Gamma(2-2\dot h_B)}
 {(\Delta_A-1)(\Delta_B-1)
 \Gamma(2h_A-1)\Gamma(2h_B-1)}\right]^{1/2},
\end{equation}
where $h$ and $\dot h$ are the conformal weights, $\Delta=h+\dot h$ and
$S=\dot h-h$. Its coupling dependence enters only through the exactly known
dimensions~\eqref{eq:J2-spectrum}.

\begin{figure}[t!]
\centering
\begin{subfigure}[b]{0.40\linewidth}
\centering
\includegraphics[width=\linewidth]{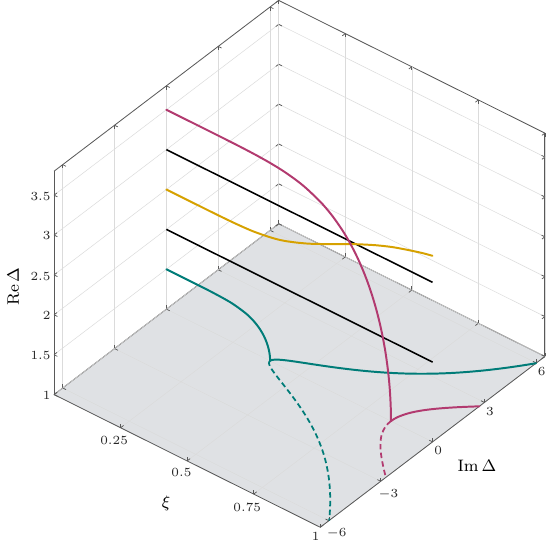}
\caption{}
\label{fig:intro-J3-dimensions}
\end{subfigure}\hfill
\begin{subfigure}[b]{0.57\linewidth}
\centering
\includegraphics[width=\linewidth]{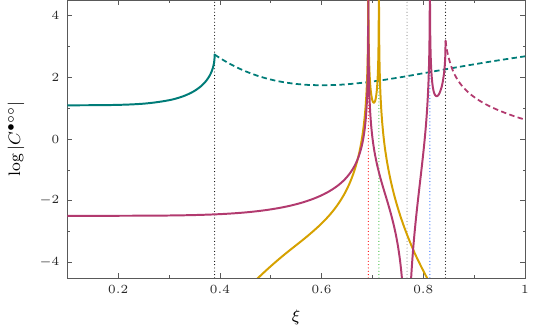}
\caption{}
\label{fig:intro-J3-logabsC}
\end{subfigure}
\caption{Conformal data for three scalar trajectories at length $J=3$ as
functions of the coupling $\xi$. (a)~The coloured curves show the first
three scaling-dimension trajectories of~\cite{Ekhammar:2025fvn},
which start at $\Delta=\tfrac32,\tfrac52,\tfrac72$ at zero coupling.
The black lines show coupling-independent states with $\Delta=2,3$.
The plane at $\Delta=1$ marks where a trajectory can meet its shadow
and continue into the complex plane.
(b)~The same colours show the logarithm of the absolute value of the
corresponding HLL structure constant with two light, coupling-independent
operators. Dashed segments indicate complex coefficients. Dotted lines mark
collisions between states.}
\label{fig:intro-J3}
\end{figure}

\paragraph{New finite-coupling data.}
From $J=3$ onwards no structure constants of these operators were known.
Our formulas, implemented in the \texttt{BSolver2D} package (available on GitHub as part of this paper), compute these structure constants at any coupling. Figure~\ref{fig:intro-J3} shows three scalar trajectories of length three. Across these trajectories, we see crossings with other unprotected trajectories and with protected states. Trajectories~1 and~3 also meet their shadows at $\Delta=1$ and become complex. At a shadow collision the dimension has a square-root branch point and $\mathcal D_\xi$ diverges. At the crossing of trajectories~2 and~3 near $\xi=0.692$, both structure constants develop a pole. The spectral slope stays finite and the pole comes from the SoV determinant as the state becomes orthogonal to itself.

\paragraph{Checks.}
Wherever possible we test each step independently. At $J=1$ every
overlap can be computed directly in coordinate space. At $J=2$ the
single-valuedness of the coordinate wave function reproduces the QSC gluing
condition. The untwisted $J=2$ coefficients are rederived directly from
conformal symmetry. At $J=3$ the leading weak-coupling corrections agree
with an independent perturbative computation from the conformal block
expansion of a four-point function. The finite-twist overlap formulas are
compared with direct coordinate-space integration numerically in
Appendix~\ref{app:bsolver}.

\paragraph{Outline.}
We start in Section~\ref{sec:setup} by reviewing the 2D bi-scalar fishnet
theory, its colour twist and its integrable spin chain, and we set up the
Baxter equations, the gluing matrix and the connection matrices.
Section~\ref{sec:gluing} explains how CFT wave functions are glued into
correlators. There we introduce the planar bracket and relate the
logarithmic divergence of the self-overlap to the two-point
normalisation. In Section~\ref{sec:solvable} we work out the one-site
chain in full and follow the passage from coordinate integrals to
Q-functions and residue sums. Section~\ref{sec:sov} constructs the
separated basis, identifies the separated wave function with the glued
$\Qp$, writes the scalar product as a moment determinant and establishes
the splitting of the SoV basis. Section~\ref{sec:twisted-overlaps} uses
this splitting to derive the HHL and HLL overlaps at finite twist.
In Section~\ref{sec:norms} we extract the norm from the equal-twist
logarithm and reduce it to the connection-matrix residue formula. In
Section~\ref{sec:untwisted} we remove the twist and assemble the
normalised OPE coefficients, with the two-site chain as a fully
explicit example. Section~\ref{sec:CFTData} turns these formulas into
finite-coupling CFT data at lengths two and three.

The appendices collect derivations and checks. Appendix~\ref{app:bsolver} describes the
\texttt{BSolver2D} package with runnable examples and numerical tests of our
formulas. Appendix~\ref{sec:coordinate-J2-ode}
computes the twisted two-site wave function in coordinate space and
recovers the quantisation condition from single-valuedness.
Appendix~\ref{app:J1} collects the closed-form one-site
results, and Appendix~\ref{app:omega} derives the connection-matrix
formulas behind the norm. Appendix~\ref{sec:J2-untwisted} derives the untwisted two-site
results directly in coordinate space.
Appendix~\ref{app:4pt} gives the perturbative four-point computation used as
a weak-coupling check. Finally, Appendix~\ref{app:untwisting-descendants}
treats twisted states that become conformal descendants once the twist is
removed.

\paragraph{Notation.} The paper uses many decorated symbols, so we collect the main ones in the table below together with the equations that define them. A dot denotes the antiholomorphic counterpart of an undotted quantity, e.g.\ $\dot h$, $\dot\lambda$, $\dot u$, $\dot q$, $\dot\Omega$. A tilde denotes the shadow counterpart, obtained by $s\to1-s$. Finally, boldface denotes quantities at zero twist.

\newpage
\begingroup
\renewcommand{\arraystretch}{1.20}
\small

\noindent
\begin{tabular*}{\textwidth}{@{\extracolsep{\fill}}lll@{}}
\toprule
\multicolumn{3}{@{}l}{\textbf{Theory and kinematics}}\\
\midrule
$\xi$ & fishnet coupling & \eqref{eq:2d-fishnet-lagrangian}\\
$[z]^{\{\alpha,\dot\alpha\}}$ & $z^\alpha\bar z^{\dot\alpha}$, single-valued for $\alpha-\dot\alpha\in\mathbb Z$ & \eqref{eq:complex-propagator}\\
$J$ & length of the operator, i.e.\ the number of sites & \eqref{eq:CFT-wavefunction}\\
$h,\dot h$; $\Delta$, $S$ & conformal weights; $\Delta=h+\dot h$, $S=\dot h-h$ & \eqref{eq:weights-dimension-spin}\\
$s$, $\theta_\alpha$ & site spin and inhomogeneities & \eqref{eq:fishnet-point}\\
$\lambda=e^{\varrho+\ii\phi}$ & twist eigenvalue, with $\dot\lambda=\bar\lambda$ on the Euclidean slice & \eqref{eq:euclidean-colour-twist}\\
$G_{\rm D}$, $G_{\rm C}$ & diagonal and companion frames of the colour twist & \eqref{eq:diagonal-colour-twist}, \eqref{eq:companion-colour-twist}\\
$\psi$; $\psi^{\rm D}$, $\psi^{\rm C}$ & CFT wave function in different frames & \eqref{eq:CFT-wavefunction}, \eqref{eq:diagonal-companion-wavefunction}\\
$\Omega_J$ & reference-state wave function at zero coupling & \eqref{eq:reference-state-frames}\\
\midrule
\multicolumn{3}{@{}l}{\textbf{Spin chain and Q-functions}}\\
\midrule
$t(u)$, $a_\beta$ & transfer-matrix eigenvalue and its coefficients & \eqref{eq:transfer-matrix-expansion}\\
$q_a^\downarrow$, $q_a^\uparrow$ & Baxter solutions with poles below ($\downarrow$) or above ($\uparrow$) & \eqref{eq:baxter-asymptotics}\\
$\widetilde q_a$ & shadow Q-functions, Baxter solutions with $s\to1-s$ & \eqref{eq:shadow-q-map}\\
$\mu^\downarrow,\mu^\uparrow$; $\mu(u,\dot u)$ & measures relating $q$ and $\widetilde q$ & \eqref{eq:shadow-chain-mu-factors}, \eqref{eq:overlaps-one-pair-measure}\\
$\Omega$, $\widetilde\Omega$ & $\ii$-periodic connection matrices, $q^\uparrow=\Omega q^\downarrow$ & \eqref{eq:baxter-connection-matrix}\\
$\Gamma=\operatorname{diag}(1,c)$ & gluing matrix and gluing constant $c$ & \eqref{eq:baxter-gluing-condition}\\
$\Qp$, $\widetilde{\mathbb Q}_+$ & glued Q-function and its shadow & \eqref{eq:glued-Qplus}, \eqref{eq:shadow-glued-Qplus}\\
$\svx_{\alpha,n}$, $\widetilde\svx_{\alpha,n}$ & ordinary points $\theta_\alpha-\ii(s+n)$ and its shadow & \eqref{eq:J1-overlap-lower-pole-families}, \eqref{eq:J1-glued-Q-fixed-m-poles}\\
$\mathsf R_\alpha$, $\widetilde{\mathsf R}_\alpha$ & residue matrices of $\Omega^{-1}$ and $\widetilde\Omega^{-1}$ & \eqref{eq:setup-omega-residue-matrices}\\
\midrule
\multicolumn{3}{@{}l}{\textbf{Pairings and overlaps}}\\
\midrule
$\Phi$, $\widetilde\Phi$ & separated wave function and its shadow & \eqref{eq:sov-separated-wavefunction-definition}, \eqref{eq:overlaps-shadow-wavefunction}\\
$\bfS$, $\cD$, $\mathcal R$ & shadow transform, inversion, site reversal & \eqref{eq:shadow-operator}, \eqref{eq:full-inversion}, \eqref{eq:chain-reversal-R}\\
$\langle f|g\rangle$ & companion pairing & \eqref{eq:companion-pairing}\\
$\LRpair{f}{g}$ & planar bracket & \eqref{eq:left-right-bracket}\\
$\LRpair{\psi_A}{\psi_B}_\ell$ & planar overlap, where $A$ and $B$ share $\ell$ lines & \eqref{eq:HHL-target-correlator}\\
$(f)_\alpha$ & regularised chiral residue sum at site $\alpha$ & \eqref{eq:lower-tower-contour}\\
$\towerbracket{f}{g}_\alpha$ & paired site bracket & \eqref{eq:paired-tower-bracket}\\
$k_\alpha$, $k_\alpha^{\rm c}$ & residue-to-value weight and its phase-adjusted form & \eqref{eq:J1-measure-conversion-coefficients}, \eqref{eq:contour-conversion-coefficients}\\
$\mathcal M_{nm}$ & SoV moment matrix & \eqref{eq:overlaps-moment-residues}\\
$\mathcal N_J^{\rm SoV}$, $\mathcal A$ & SoV normalisation and state normalisation & \eqref{eq:overlaps-moment-determinant}, \eqref{eq:sov-state-normalisation}\\
\midrule
\multicolumn{3}{@{}l}{\textbf{Norms and structure constants}}\\
\midrule
$\mathcal C_{\log}$ & coefficient of the logarithm in the self-overlap & \eqref{eq:coordinate-logarithmic-overlap}\\
$\mathcal N_2$, $\mathcal D_\xi$ & two-point normalisation; $\mathcal D_\xi=\xi^2\partial_{\xi^2}\Delta$ & \eqref{eq:coordinate-general-norm-relation}\\
$\mathcal N_2^{\theta=0}$ & norm at $\theta_\alpha=0$, $J\lim_{\theta\to0}\mathcal N_2$ & \eqref{eq:norm-theta-zero}\\
$\bullet$, $\circ$ & heavy (wheel-dressed) and light (zero-coupling) & \eqref{eq:untwisting-HHL-OPE-definition}\\
$\mathcal C_\ell^{\bullet\circ\circ}$, $\mathcal C_\ell^{\bullet\bullet\circ}$ & HLL and HHL structure constants with $\ell$ bridges & \eqref{eq:untwisting-HHL-OPE-definition}\\
\bottomrule
\end{tabular*}
\endgroup

\newpage
\section{The twisted two-dimensional fishnet chain}
\label{sec:setup}

This section reviews the two-dimensional bi-scalar fishnet theory and its integrable structure. Its purpose is to set the stage and fix the conventions and notations used throughout this work. Readers familiar with this material can safely skip this section.

\subsection{Fishnet CFT and CFT wave function}
\label{sec:fishnet-dictionary}

\paragraph{Fields and Lagrangian.}
We follow the conventions of~\cite{Kazakov:2018qbr,Ekhammar:2025fvn}, and define the bi-scalar fishnet theory by the following Lagrangian\footnote{Strictly speaking, this Lagrangian must be supplemented by double-trace interactions, which are required to preserve conformality in both four~\cite{Sieg:2016vap,Grabner:2017pgm} and
two~\cite{Kazakov:2018qbr} dimensions. We omit them here.}
\begin{equation}
 \mathcal L
 =N\operatorname{tr}\!\left[
   \phi_1^\dagger\Box^{1/2}\phi_1
  +\phi_2^\dagger\Box^{1/2}\phi_2
  +4\pi\xi^2\phi_1^\dagger\phi_2^\dagger\phi_1\phi_2
 \right],
 \qquad \Box=-\partial_\mu\partial^\mu .
 \label{eq:2d-fishnet-lagrangian}
\end{equation}
Here $\phi_1$ and $\phi_2$ are two complex $N\times N$ matrix scalar fields of bare dimension $1/2$. We work in the planar limit $N\to\infty$ with the coupling $\xi$ playing the role of a 't~Hooft coupling. This model is the two-dimensional member of the family of bi-scalar fishnet CFTs of~\cite{Kazakov:2018qbr}. The family generalises the four-dimensional bi-scalar fishnet, which arises from a double-scaling limit of the $\gamma$-deformed $\mathcal N=4$ SYM~\cite{Gurdogan:2015csr}. Unlike in four dimensions, no parent theory of the two-dimensional model is known~\cite{Ekhammar:2025fvn}.

Importantly, the quartic vertex in~\eqref{eq:2d-fishnet-lagrangian} appears without its Hermitian conjugate. The action is therefore not real and the theory is not unitary. In exchange, perturbation theory is extremely rigid. At each vertex a $\phi_1$ line crosses a $\phi_2$ line with a fixed orientation, so every planar Feynman graph is a fragment of a regular square lattice: the ``fishnet'' graphs, shown to be integrable by Zamolodchikov almost fifty years ago~\cite{Zamolodchikov:1980mb}.

In two dimensions, the fishnet theory has a non-local kinetic term involving a fractional power of the Laplacian, $(\partial_\mu\partial^\mu)^\beta$. We define it by analytic continuation in $\beta$ of the integral representation
\begin{equation}
 (\partial_\mu\partial^\mu)^\beta f(x)
 =\frac{(-4)^\beta\Gamma(1+\beta)}{\pi\Gamma(-\beta)}
   \int\!\dd^2y\,\frac{f(y)}{|x-y|^{2+2\beta}},
 \label{eq:fractional-laplacian}
\end{equation}
with the usual distributional prescription. Correspondingly, the Green function of this analytically continued operator is
\begin{equation}
 D_\beta(x-y)
 =\frac{\Gamma(1-\beta)}{4^\beta\pi\Gamma(\beta)}
  \frac{1}{|x-y|^{2-2\beta}} .
 \label{eq:fractional-propagator}
\end{equation}

Even though general correlation functions are neither holomorphic nor antiholomorphic, in two-dimensional Euclidean space it is convenient to use the complex coordinates,
\begin{equation}
 z=x^1+\ii x^2,\quad
 \quad \bar z=x^1-\ii x^2,
\end{equation}
and to define the generalised propagator
\begin{equation}
 [z]^{\{\alpha,\dot\alpha\}} \equiv z^\alpha\bar z^{\dot\alpha}
 =|z|^{\alpha+\dot\alpha}
   e^{\ii(\alpha-\dot\alpha)\arg z}\,.
 \label{eq:complex-propagator}
\end{equation}
We see that single-valuedness on the ${
\mathbb R}^2$ plane requires that \(\alpha-\dot\alpha\in\mathbb Z\).

When the two exponents of the complex propagator coincide and there is no risk of confusion we abbreviate it as
\([z]^\alpha\equiv[z]^{\{\alpha,\alpha\}}\) and in particular, \([z-w]^{1/2}=|z-w|\) is the norm.

\paragraph{Local operators.} Up to permutations of the fields, we consider single-trace operators represented schematically as
\begin{equation}
 \mathcal O
 = \operatorname{tr}\!\left[
\partial^K\bar\partial^{\dot K}
   \phi_1^J\phi_2^M
   (\phi_1\phi_1^\dagger)^n
   (\phi_2^\dagger\phi_2)^m
  \right].
 \label{eq:fishnet-local-operators}
\end{equation}
Even though there is no supersymmetry, we borrow the language of
$\mathcal{N}=4$ super Yang--Mills and call an operator \textit{protected} or \textit{unprotected} depending on whether its scaling dimension coincides with its bare dimension for all values of the coupling in the planar limit. For example, $\operatorname{tr}[\phi_{1}^{J}]$ is unprotected while $\operatorname{tr}[\phi_{1}^{M} (\phi^{\dagger}_1)^{J-M}]$ with $0<M<J$ is protected.

The theory has a $\mathbb Z_4$ symmetry that cyclically permutes $\{\phi_1,\phi_2,\phi_1^\dagger,\phi_2^\dagger\}$
and leaves the spectrum invariant. We can therefore take $J\geq M$ and regard $\operatorname{tr}[\phi_1^J]$ as the spin chain vacuum and the $\phi_2$ fields as magnons.
Here we restrict to the zero-magnon sector $M=0$.

\paragraph{Conformal algebra.}
The complexification of the Euclidean conformal algebra $\mathfrak{so}(1,3)$ factorises into two commuting $\mathfrak{sl}(2)$ algebras, which we refer to as the undotted and dotted. In our conventions, their generators are realised as
\begin{equation}
 \begin{aligned}
  \hat S^z&=-z\partial-s,
  &\hat S^+&=\partial,
  &\hat S^-&=-z^2\partial-2sz,\\
  \hat{\dot S}^{z}&=-\bar z\bar\partial-\dot s,
  &\hat{\dot S}^{+}&=\bar\partial,
  &\hat{\dot S}^{-}&=-\bar z^2\bar\partial-2\dot s\,\bar z.
 \end{aligned}
 \label{eq:Sl2Generators}
\end{equation}
Thus the dotted realisation is obtained from the undotted one by
\((z,\partial,s)\mapsto(\bar z,\bar\partial,\dot s)\). Nonetheless, it is an independent copy of the algebra rather than the complex conjugate of the undotted sector. When $\dot s=1-s^*$ and $s-\dot s\in\mathbb Z$, the two sectors form a non-spinorial unitary principal-series representation of
$SL(2,\mathbb C)$~\cite{Derkachov:2001yn}. Here we do not impose these conditions; in particular, the physical fishnet weights in the zero-magnon sector are
$s=\dot s=1/4$, which lie outside this series. From here on we set $s=\dot{s}$.

We also need global $\mathfrak{sl}(2)$ generators $ \widehat{\mathbb S}^{a}$ acting on $J$ variables $z_\alpha$, $\alpha=1,\dots,J$, built from the local ones by
\begin{equation}
   \widehat{\mathbb S}^{a}=\sum_{\alpha=1}^{J}\hat S^a_\alpha, \qquad a\in\{z,+,-\},
   \label{eq:global-sl2-generators}
\end{equation}
where $\hat S^a_\alpha$ is obtained from $\hat S^a$ in \eqref{eq:Sl2Generators} by $(z,\partial,s)\mapsto (z_\alpha,\partial_\alpha,s)$. Note that we have assumed the spin $s$ is the same in each copy, and the same construction holds for the dotted sector.

\begin{figure}[H]
\centering
\includegraphics[width=0.35\linewidth]{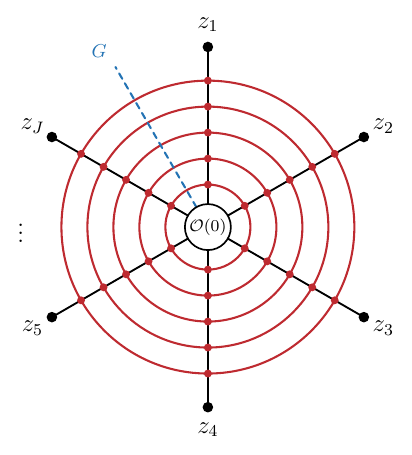}
\caption{The CFT wave function of a local operator is a $(J+1)$-point correlation function, whose typical Feynman diagram is shown above. The operator $\mathcal{O}$ sits at the origin and is probed by a cyclic chain of $J$ identical scalars $\phi_1^\dagger(z_\alpha)$. The black lines are the external $\phi_1$ propagators and each red circuit represents an additional $\phi_2$ fishnet wheel. The dashed blue line is the colour-twist cut introduced below.}
\label{fig:zero-magnon-wavefunction}
\end{figure}

\paragraph{CFT wave function.}
A key ingredient in the integrability description of the fishnet theory is the so-called \textit{CFT wave function} $\psi$ of the operator $\mathcal O$. In the zero-magnon sector, this wave function is obtained by probing $\mathcal O$ with $J$ scalar $\phi_1^\dagger$ fields,
\begin{equation}
 \psi(z_1,\ldots,z_J)
 =\left\langle
   \mathcal O(0)\,
   \operatorname{tr}\!\left[
    \phi_1^\dagger(z_1)\cdots\phi_1^\dagger(z_J)
   \right]
  \right\rangle ,
 \label{eq:CFT-wavefunction}
\end{equation}
with its typical diagram shown in Figure~\ref{fig:zero-magnon-wavefunction}. Here, the dependence on $\bar{z}_i$ is kept implicit.

The CFT wave function was introduced for the four-dimensional fishnet
in~\cite{Gromov:2019aku,Gromov:2019bsj}, building on the $J=2$ analysis of~\cite{Grabner:2017pgm}. Knowing $\psi$ is equivalent to knowing the operator $\mathcal{O}$, and it gives a scheme-independent definition of $\mathcal O$ at finite coupling, in contrast with the scheme-dependent mixing of the single traces \eqref{eq:fishnet-local-operators} in perturbation theory. In particular,
\begin{equation}
 \widehat{\mathbb S}^{z}\psi=h\psi,
 \qquad
 \hat{\dot{\mathbb S}}^{z}\psi
     =\dot h\psi,
 \qquad
 \Delta=h+\dot h,
 \qquad S=\dot h-h\,.
 \label{eq:weights-dimension-spin}
\end{equation}
Note that although $\psi$ depends on the precise operator $\mathcal{O}$ we are probing, we leave this dependence implicit and so in what follows the CFT wave function will always refer to some fixed $\mathcal{O}$ unless otherwise stated.

\subsection{Twisted fishnets and colour-twist operators}
\label{sec:colour-twist}

In this paper, it will prove very convenient to deform the theory by a colour twist~\cite{Cavaglia:2020hdb}. Twisting does two things for us: it breaks part of the conformal symmetry, lifting spectral degeneracies, and it makes the separation of variables construction more regular.

\paragraph{Twisted Feynman diagrams.}
Any theory with a planar limit and a global symmetry admits a twist construction, implemented by a symmetry transformation $G$. In the planar limit, a correlator of single-trace operators is a sum of graphs drawn on a plane, with the operators as marked points. The twist is defined by drawing an oriented line, the \textit{cut}, which starts at the twisted operator; the blue dashed line in Figure~\ref{fig:zero-magnon-wavefunction} is an example. Every propagator crossing the cut is replaced by its transform under $G$: one of its endpoints $x$ is replaced by $G(x)$ and the propagator gains a Jacobian factor that depends on the weight of the field it carries. Since $G$ is a symmetry of the interaction vertex, moving the cut across a vertex leaves the integral unchanged. Hence the path of the cut is a matter of convention and only its endpoints matter: a field carried once around an endpoint comes back transformed by $G$. In the fishchain description the twist becomes a target-space monodromy of the discretised worldsheet~\cite{Gromov:2019jfh}.

\paragraph{Colour-twist operators.}
The endpoint of the cut is a twist field. Placing it inside the colour trace defines a \textit{colour-twist operator}~\cite{Cavaglia:2020hdb},
\begin{equation}
 \mathcal O_G(0)
 =\operatorname{tr}\!\left[
   \mathcal T_G\,\phi_1^J
  \right](0)\,,
 \label{eq:colour-twist-operator}
\end{equation}
and the twisted CFT wave function $\psi$ is obtained by replacing $\mathcal O$ with $\mathcal O_G$ in~\eqref{eq:CFT-wavefunction}. Because a field carried around $\mathcal O_G$ comes back transformed by $G$, the operator is consistent only if it is left invariant by $G$, so it must sit at a fixed point of the spacetime transformation~\cite{Cavaglia:2020hdb}. We restrict to twists with two distinct fixed points and choose the conformal frame in which they are $0$ and $\infty$; any other choice conjugates $G$ and moves the operator together with the endpoints of its cut. In this frame the undotted and dotted twists are diagonal,
\begin{equation}
 G_{\rm D}(\lambda)=
  \begin{pmatrix}\lambda&0\\0&\lambda^{-1}\end{pmatrix},
 \qquad
 \dot G_{\rm D}(\dot\lambda)=
  \begin{pmatrix}\dot\lambda&0\\0&\dot\lambda^{-1}\end{pmatrix},
 \label{eq:diagonal-colour-twist}
\end{equation}
acting as \((x,\bar x)\mapsto(\lambda^2x,\dot\lambda^2\bar x)\), with the operator~\eqref{eq:colour-twist-operator} at $0$ and its cut running to $\infty$. The twist preserves planar integrability, but only the conformal transformations commuting with $G$ survive: dilations and rotations, which keep $0$ and $\infty$ fixed individually. States are therefore still labelled by $\Delta$ and $S$, but the broken generators no longer organise them into conformal multiplets~\cite{Cavaglia:2020hdb}. The twist eigenvalues need not lie on the unit circle. On the Euclidean plane the dotted twist is the complex conjugate of the undotted one, and we write
\begin{equation}
 \lambda=e^{\varrho+\ii\phi},
 \qquad
 \dot\lambda=e^{\varrho-\ii\phi}=\bar\lambda,
 \qquad \varrho,\phi\in\mathbb R.
 \label{eq:euclidean-colour-twist}
\end{equation}
In most of what follows we take a pure rotation, $\varrho=0$.

\paragraph{Graph-building operator.}
The planar diagrams contributing to \eqref{eq:CFT-wavefunction} are generated recursively by adding fishnet wheels to the tree-level correlator, as depicted in Figure~\ref{fig:zero-magnon-wavefunction}. With the diagonal twist \eqref{eq:diagonal-colour-twist} adding one such wheel is implemented
by the graph-building operator~\cite{Kazakov:2018qbr,Derkachov:2018rot,Ekhammar:2025fvn}
\begin{equation}
 [\widehat{\mathcal B}f](z_1,\ldots,z_J)
 =\frac{\xi^{2J}}{\pi^J}
   \int\prod_{\alpha=1}^{J}
   \frac{\dd^2w_\alpha}
   {[z_\alpha-w_\alpha]^{1/2}
    [w_\alpha-w_{\alpha-1}]^{1/2}}
   f(w_1,\ldots,w_J),
 \label{eq:graph-building-operator}
\end{equation}
where \(\dd^2w=\dd w\,\dd\bar w/2\), \(w_0= \lambda^2 w_J\), $\bar{w}_0 = \dot{\lambda}^2 \bar{w}_J$. The CFT wave function $\psi$ is left invariant by the action of the graph-building operator, that is
\begin{equation}
 \widehat{\mathcal B}\circ\psi
 =\psi.
 \label{eq:graph-building-stationarity}
\end{equation}
Integrability means that \(\widehat{\mathcal B}\) is not an isolated operator: it belongs to a large family of mutually commuting operators, the transfer matrices and Q-operators of a spin chain, which we explicitly construct in Section~\ref{sec:lax}.

\paragraph{Companion twist.}
In addition to the diagonal twist matrix $G_D$ \eqref{eq:diagonal-colour-twist} which has fixed points at $0$ and $\infty$, it is also very convenient (from a separation of variables perspective) to consider the \textit{companion twist matrix} $G_C$, in which both fixed points are finite. For generic $\lambda^2\ne1$, the companion twist is conjugate to the diagonal one:
\begin{equation}
 G_{\rm C}(\lambda)=
 \begin{pmatrix}\lambda+\lambda^{-1}&-1\\1&0\end{pmatrix},
 \qquad
 G_{\rm C}(\lambda)
 =H_\lambda G_{\rm D}(\lambda)H_\lambda^{-1},
 \qquad
 H_\lambda=\begin{pmatrix}\lambda&1\\1&\lambda\end{pmatrix}.
 \label{eq:companion-colour-twist}
\end{equation}

\begin{figure}[t]
\centering
\includegraphics[width=0.38\linewidth]{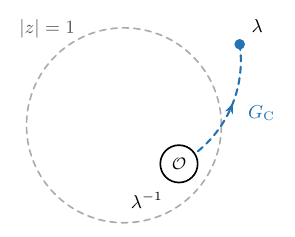}
\caption{The colour-twist cut in the companion frame. Under
$m_\lambda^{-1}$, the ray from $0$ to $\infty$ in the diagonal frame becomes an
oriented arc from $\lambda^{-1}=m_\lambda^{-1}(0)$ to
$\lambda=m_\lambda^{-1}(\infty)$. These endpoints are the two fixed points of
$G_{\rm C}$, and the colour-twist operator $\mathcal{O}$ remains attached to the first
one. The dashed grey circle is the unit circle; the case $|\lambda|>1$
is shown, with $\lambda^{-1}$ inside and $\lambda$ outside.}
\label{fig:companion-twist-fixed-points}
\end{figure}

The dotted twist is obtained by replacing $\lambda$ with $\dot\lambda$. In terms of coordinates, the conjugation to the companion frame is implemented by the M\"obius map
\begin{equation}
 z=m_\lambda^{-1}(x)=\frac{\lambda x+1}{x+\lambda},
 \qquad
 x=m_\lambda(z)=\frac{1-\lambda z}{z-\lambda}.
 \label{eq:diagonal-companion-map}
\end{equation}
Thus $G_{\rm C}$ acts as $z\mapsto\lambda+\lambda^{-1}-z^{-1}$, with fixed points $\lambda^{-1}=m_\lambda^{-1}(0)$ and $\lambda=m_\lambda^{-1}(\infty)$, and similarly in the dotted sector. The operator and the start of the twist cut move from $x=0$ to $z=\lambda^{-1}$, while the other endpoint moves from $x=\infty$ to $z=\lambda$, as shown in Figure~\ref{fig:companion-twist-fixed-points}.

The two frames describe the same twisted correlator. Hence, the twisted wave functions in the two frames differ only by Jacobian factors of the scalar fields $\phi_1^\dagger$ in~\eqref{eq:CFT-wavefunction}. More explicitly, the companion $\psi^{\rm C}$ and diagonal $\psi^{\rm D}$ twisted wave functions are related by
\begin{equation}
  \psi^{\rm C}(\bz,\bar{\bz})
 = \mathfrak{n}(\bz,\bar{\bz}) \times \psi^{\rm D}
 \bigl(m_\lambda(\bz),m_{\dot\lambda}(\bar{\bz})\bigr) \label{eq:diagonal-companion-wavefunction}
\end{equation}
where the Jacobian factors are
\begin{equation}
\mathfrak{n}(\bz,\bar{\bz}) = \mathfrak{n}_\lambda \prod_{\alpha=1}^{J}\frac{1}{(z_\alpha-\lambda)^{2 s}(\bar{z}_\alpha-\dot{\lambda})^{2 s}}, \quad \mathfrak{n}_\lambda = (\lambda^2-1)^{Js}(\dot\lambda^2-1)^{ Js}.
 \label{eq:diagonal-companion-wavefunction-norm}
\end{equation}
Here the constant $\mathfrak n_\lambda$ is kept explicitly, since it enters the normalisation of states below.

At zero coupling, the twisted trace $\mathcal O=\operatorname{tr}[\mathcal T\,\phi_1^J]$ defines the reference state $\Omega_J$. Contracting each probe $\phi_1^\dagger(z_\alpha)$ with a free field at the origin gives the diagonal-frame wave function. The frame transformation~\eqref{eq:diagonal-companion-wavefunction} then gives the companion-frame expression,
\begin{equation}
 \Omega_J^{\rm D}(\bz,\bar{\bz})
 =\prod_{\alpha=1}^{J}[z_\alpha]^{-2s},
 \qquad
 \Omega_J^{\rm C}(\bz,\bar{\bz})
 =\mathfrak n_\lambda\prod_{\alpha=1}^{J}[1-\lambda z_\alpha]^{-2s}.
 \label{eq:reference-state-frames}
\end{equation}

\paragraph{Normalisation.} We now pick a convenient normalisation of the CFT wave function. Our default normalisation, in the diagonal frame, is $\psi^{\rm D}(1,\dots,1)=1$. The value of the wave function when all $(z_\alpha,\bar{z}_\alpha)=(z,\bar{z})$ is then completely fixed by conformal symmetry
\begin{equation}
    \psi^{\rm D}(z,\dots,z) = [z]^{-\{h,\dot{h}\}-Js}\,.
\end{equation}
Equation~\eqref{eq:diagonal-companion-wavefunction} fixes the companion wave function $\psi^{\rm C}$ at $z_\alpha=1$. Since $m_\lambda(1)=1$, this results in
\begin{equation}
 \psi^{\rm C}(1,\dots,1)
 =\frac{(\lambda^2-1)^{Js}
 (\dot\lambda^2-1)^{Js}}
 {(1-\lambda)^{2Js}
 (1-\dot\lambda)^{2Js}}.
\end{equation}

As we will see, a useful object in the separation of variables construction is the value of the companion frame CFT wave function $\psi^{\rm C}$ when all $z_\alpha=0$. Since $m_\lambda(0)=-\lambda^{-1}$, the frame transformation
\eqref{eq:diagonal-companion-wavefunction} with the Jacobian
\eqref{eq:diagonal-companion-wavefunction-norm} gives
\begin{equation}
 \psi^{\rm C}(0,\ldots,0)
 =\frac{(\lambda^2-1)^{Js}
        (\dot\lambda^2-1)^{Js}}
       {(-\lambda)^{Js-h}
        (-\dot\lambda)^{Js-\dot h}}.
 \label{eq:companion-origin-normalisation}
\end{equation}

For scalar states at the fishnet
point, $h=\dot h$, $s=\dot s=1/4$ and the twist being a pure rotation
$\lambda=e^{\ii\phi}$, $\dot\lambda=e^{-\ii\phi}$, this reduces to
$\psi^{\rm C}(0,\ldots,0)=(2|\sin\phi|)^{J/2}$,
independently of the scaling dimension.

\subsection{Integrability of 2D bi-scalar fishnet theory}
\label{sec:lax}

Integrability embeds the graph-building operator $\widehat{\mathcal B}$ in a commuting family of transfer matrices and Q-operators. Following the non-compact spin chain framework~\cite{Lipatov:1993qn,Faddeev:1994zg,Derkachov:2001yn,Derkachov:2002wz,Derkachov:2002tf}, we review the construction of this family and explain how the graph-building operator and cyclic shift operators arise from Q-operators at special values of the spectral parameter~\cite{Kazakov:2018qbr,Ekhammar:2025fvn}.

\paragraph{Lax operator and transfer matrix.}
At each site $\alpha=1,\ldots,J$ we use the generators~\eqref{eq:Sl2Generators}, acting on $z_\alpha$ with spin $s$, to define the Lax operator
\begin{equation}
 \hat{L}_{a\alpha}(u-\theta_\alpha)
 =\begin{pmatrix}
  u-\theta_\alpha-\ii\hat S^z_\alpha
    &-\ii\hat S^+_\alpha\\
  -\ii\hat S^-_\alpha
    &u-\theta_\alpha+\ii\hat S^z_\alpha
 \end{pmatrix}.
 \label{eq:lax-operator}
\end{equation}
It is a $2\times2$ matrix in the auxiliary space, whose entries are differential operators acting on site $\alpha$. Here $u$ is the spectral parameter and $\theta_\alpha$ is an inhomogeneity. An analogous Lax operator $\hat{\dot{L}}_{a\alpha}$ is constructed in the dotted sector by replacing the differential operators $\hat{S}^z$ etc.\ with their dotted counterparts, and replacing $(u,\theta)\rightarrow(\dot{u},\dot{\theta})$. In principle $\dot{\theta}$ and $\theta$ can be independent, but for simplicity we always assume $\dot{\theta}_{\alpha} = \theta_{\alpha}$. The fishnet theory, for operators without magnons, corresponds to the homogeneous point
\begin{equation}
  s =\frac{1}{4}, \qquad \theta_\alpha = 0,\quad \alpha=1,\dots,J\,.
  \label{eq:fishnet-point}
\end{equation}
In what follows we mostly write relations only for the undotted sector, with the dotted sector satisfying an analogous set of relations.

The ordered product of local Lax operators is the so-called monodromy matrix,
\begin{equation}
 \widehat T_a(u)
 \equiv \hat{L}_{a1}(u-\theta_1)\cdots \hat{L}_{aJ}(u-\theta_J)
 =\begin{pmatrix}
   \widehat A(u)&\widehat B(u)\\
   \widehat C(u)&\widehat D(u)
  \end{pmatrix},
 \label{eq:spin-chain-monodromy}
\end{equation}
where the sites are ordered as the scalar probes in~\eqref{eq:CFT-wavefunction}.

The colour twist enters when we close the chain: before taking the trace over the auxiliary space, we multiply the monodromy matrix by the twist $G\in SL(2,\mathbb C)$ of Section~\ref{sec:colour-twist}. This defines the (twisted) transfer matrix
\begin{equation}
 \widehat t_G(u)
 \equiv \operatorname{tr}_a\!\left[\widehat T_a(u)G_a\right].
 \label{eq:twisted-transfer-matrix}
\end{equation}
We will often drop the $G$ subscript when it is clear from the context.

\paragraph{Commuting charges.} The transfer matrices commute at different values of the spectral parameters. At the homogeneous fishnet point \eqref{eq:fishnet-point} they also commute with the graph-building operator~\cite{Ekhammar:2025fvn}:
\begin{equation}
 [\widehat t(u),\widehat t(v)]=0,
 \qquad
 [\widehat t(u),\widehat{\mathcal B}]=0
 \qquad\text{for all }u,v.
 \label{eq:transfer-graph-builder-commutativity}
\end{equation}
Each Lax operator is linear in $u$, so the transfer matrix is a polynomial of degree $J$. We denote its coefficients by $\bI_\alpha$, and its eigenvalue on a common eigenstate by $t(u)$:
\begin{equation}
 \widehat t(u)
   =\sum_{\beta=0}^{J}u^\beta\,\bI_\beta,
 \qquad
 \widehat t(u)\psi
   =t(u)\psi,
 \qquad
 t(u)
   =\sum_{\beta=0}^{J}u^\beta a_\beta.
 \label{eq:transfer-matrix-expansion}
\end{equation}
The two leading coefficients are fixed by the twist and the total Cartan charge. In the diagonal frame, with $\Theta\equiv \sum_{\alpha=1}^{J}\theta_\alpha$, they can be read off directly from~\eqref{eq:spin-chain-monodromy}:
\begin{align}
 \bI_J&=\lambda+\lambda^{-1},
 &
 \bI_{J-1}
 &=-\bigl(\lambda+\lambda^{-1}\bigr)\Theta
   -\ii\bigl(\lambda-\lambda^{-1}\bigr)
       \widehat{\mathbb S}^{z},
 \label{eq:leading-transfer-charges}\\
 a_J&=\lambda+\lambda^{-1},
 &
 a_{J-1}
 &=-\bigl(\lambda+\lambda^{-1}\bigr)\Theta
   -\ii\bigl(\lambda-\lambda^{-1}\bigr)h.
 \label{eq:leading-transfer-eigenvalues}
\end{align}
The remaining charges $a_0,\ldots,a_{J-2}$ are fixed by the quantisation condition, as explained below in Section~\ref{sec:setup-baxter}. The dotted transfer matrix has the same structure, but with an independent set of dotted charges $\dot{\bI}_\beta$ with eigenvalues $\dot{a}_\beta$ which satisfy relations analogous to \eqref{eq:leading-transfer-eigenvalues} in terms of $\dot\lambda$ and $\dot h$. The dotted and undotted charges commute, since the two act on different variables $z$ and $\bar{z}$.

\paragraph{The Q-operator.}
An important object is the Baxter Q-operator $\widehat{\mathbb Q}_{+}(u,\dot u)$. It acts on both the undotted and dotted sectors and commutes with both families of transfer matrices~\cite{Derkachov:2001yn,DerkachovManashov:2006,Ekhammar:2025fvn}. In coordinate space it is an integral operator
\begin{equation}
 \bigl[\widehat{\mathbb Q}_{+}(u,\dot u)f\bigr]
       (\bz,\bar{\bz})
 =\int\prod_{\alpha=1}^{J}\dd^2w_\alpha\,
   \mathcal Q_{+}
   (u,\dot u\mid\bz,\bar{\bz};
                    \boldsymbol w,\bar{\boldsymbol w})
   f(\boldsymbol w,\bar{\boldsymbol w}).
 \label{eq:integral-Q-operator}
\end{equation}
We will not spell out the precise form of the kernel $\mathcal{Q}_+$ here, and instead focus on some key properties of the Q-operator we use later. The eigenvalues of the Q-operator play a key role below. On a given eigenstate $\psi$, we write
\begin{equation}
 \widehat{\mathbb Q}_{+}(u,\dot u)\psi
 =\Qp(u,\dot u)\psi\,.
 \label{eq:Q-operator-eigenvalue}
\end{equation}

\paragraph{Relation to the graph-building operator.}
At the fishnet point \eqref{eq:fishnet-point} the Q-operator contains the graph-building operator, as well as the identity operator, at special values of $(u,\dot{u})$ \cite{Ekhammar:2025fvn}. For our purposes we only need the combined relation which states
\begin{equation}
   \lim_{\epsilon\to0}\epsilon^J \frac{
 \widehat{\mathbb Q}_{+}\!\left(
   \frac{3\ii}{4}-\ii\epsilon,
   \frac{3\ii}{4}-\ii\epsilon\right)}{\widehat{\mathbb Q}_{+}\!\left(\frac{\ii}{4},\frac{\ii}{4}\right)} = \xi^{2J}\widehat{\mathcal{B}}^{-1}\,.
\label{eq:graph-building-from-Q}
\end{equation}

\paragraph{Cyclic shift and momentum.}
A third special point of the Q-operator gives the twisted cyclic shift operator $\widehat{U}$~\cite{Ekhammar:2025fvn}. In the diagonal twist frame, for a pure rotation twist $\lambda=e^{\ii\phi}$, it acts as
\begin{equation}
 [\widehat U f](z_1,\dots,z_J,\bar{z}_1,\dots,\bar{z}_J)=f
 (z_2,\ldots,z_J,e^{-2\ii\phi}z_1;
  \bar z_2,\ldots,\bar z_J,e^{2\ii\phi}\bar z_1)\,.
 \label{eq:twisted-shift-operator}
\end{equation}
In the $\phi\rightarrow 0$ limit this amounts to the usual cyclic shift of coordinates.

 The shift operator is obtained from the Q-operator via the relation \cite{Ekhammar:2025fvn}
\begin{equation}
 \widehat{U}=(-1)^J\lim_{\epsilon\to0}
 \frac{\widehat{\mathbb{Q}}_+\!\left(-\frac{3\ii}{4}-\ii\epsilon,
                  -\frac{3\ii}{4}-\ii\epsilon\right)}
      {\widehat{\mathbb{Q}}_+\!\left(\frac{3\ii}{4}-\ii\epsilon,
                   \frac{3\ii}{4}-\ii\epsilon\right)}\,.
 \label{eq:shift-eigenvalue-Qplus}
\end{equation}
Thus, $\widehat{U}$ commutes with the transfer matrices of the homogeneous chain and hence acts diagonally on the CFT wave function
\begin{equation}
    \widehat{U}\psi = U \psi\,.
\end{equation}

\paragraph{Cyclicity condition.} The CFT wave functions, without twist, describe single-trace operators, which are cyclic by definition, while generic spin chain eigenstates are not. Imposing cyclicity therefore selects a subset of the spin chain states. In the presence of the twist, cyclicity is promoted to quasi-cyclicity: as the $\phi_1^\dagger$ field in the CFT wave function is moved cyclically from the first to the last position the wave function picks up a phase factor
\begin{equation}
 [\widehat U \psi](\bz,\bar{\bz})
 =U\,\psi(\bz,\bar{\bz}),
 \qquad
 U^J=e^{-2\ii\phi S}.
 \label{eq:twisted-wavefunction-cyclicity}
\end{equation}
For states connected continuously to untwisted cyclic
single traces, one has \(U=e^{-2\ii\phi S/J}\), which reduces to the usual cyclicity condition $U=1$ as $\phi\to0$~\cite{Ekhammar:2025fvn}. For
\(\varrho\neq0\), the same statement holds once the primary Jacobian factors of the dilatation are included.

In the next subsection we turn these operator statements into a functional problem for $\Qp$. The TQ relations and the analyticity of $\Qp$ quantise the charges~\eqref{eq:transfer-matrix-expansion}, the graph-building relation~\eqref{eq:graph-building-from-Q} fixes the coupling, and cyclicity~\eqref{eq:shift-eigenvalue-Qplus} and~\eqref{eq:twisted-wavefunction-cyclicity} selects the single-trace states.

\subsection{Baxter equation, the gluing matrix and quantisation condition}
\label{sec:setup-baxter}

The Q-operator eigenvalue $\Qp(u,\dot{u})$ satisfies two Baxter equations: one in $u$ and one in $\dot{u}$. These equations shift $u$ or $\dot u$ by $\pm\ii$, and therefore preserve the lattice $u-\dot u\in\ii\mathbb Z$. The undotted equation is explicitly
\begin{equation}
 P_+(u)q(u+\ii)-t(u)q(u)+P_-(u)q(u-\ii)=0,
 \qquad
 P_\pm(u)\equiv \prod_{\alpha=1}^{J}
 \bigl(u-\theta_\alpha\pm\ii s\bigr).
 \label{eq:baxter-equation}
\end{equation}
The dotted equation is the same, with $(u,t,q)$ replaced by $(\dot u,\dot t,\dot q)$. Since $\Qp$ solves both the dotted and undotted Baxter equations, we can write $\Qp(u,\dot u)=\sum_{i,j=1}^2 C^{ij}q_i(u)\dot q_j(\dot u)$, where $q_i$ and $\dot q_j$ are two independent solutions of the undotted and dotted equations, and $C^{ij}$ is a $2\times 2$ matrix, $\ii$-periodic in both arguments. Below we choose a convenient basis of solutions and explain how the matrix elements $C^{ij}$ are fixed.

\paragraph{UHPA and LHPA solutions.}
For a generic twist the undotted equation has two solutions, which we distinguish by their large-$u$ asymptotics,
\begin{equation}
 q_1(u)\simeq\lambda^{\ii u}u^{h-Js},
 \qquad
 q_2(u)\simeq\lambda^{-\ii u}u^{-h-Js},
 \label{eq:baxter-asymptotics}
\end{equation}
with the total weight $Js$ as in~\eqref{eq:diagonal-companion-wavefunction-norm}. The asymptotics alone do not specify the solutions uniquely, and there are two natural choices: analytic in the upper or in the lower half-plane. We denote the first basis by $q_a^\downarrow$ and the second by $q_a^\uparrow$. The arrow indicates the direction in which the solution may have poles. More precisely, $q_a^\downarrow$ can have poles at $u=\theta_\alpha-\ii(1-s+n)$. Likewise $q_a^\uparrow$ may have poles at $u=\theta_\alpha+\ii(1-s+n)$, with $n\in\mathbb Z_{\geq0}$.

\paragraph{Connection matrices.}
The two bases solve the same difference equation and are therefore related by connection matrices,
\begin{equation}
 q^\uparrow_i(u)=\Omega_{i}{}^{j}(u)q_j^\downarrow(u), \qquad \dot q_i^\uparrow(\dot u)=\dot{\Omega}_{i}{}^{j}(\dot u)\dot{q}_{j}^\downarrow(\dot u).
 \label{eq:baxter-connection-matrix}
\end{equation}
These connection matrices must be $\ii$-periodic, since multiplying a solution by an $\ii$-periodic function leaves the Baxter equation~\eqref{eq:baxter-equation} unchanged.

\paragraph{Quantisation condition.}
 Let us now describe the quantisation condition in the form derived in~\cite{Ekhammar:2025fvn} that fixes the remaining free variables in $t(u)$ and $\dot{t}(\dot{u})$. We require that there exists a constant $c$ such that the two connection matrices satisfy
\begin{equation}
 \Gamma=\operatorname{diag}(1,c),
 \qquad
 \Gamma\Omega(u)=\dot\Omega(u)^{\mathsf T}\Gamma,
 \label{eq:baxter-gluing-condition}
\end{equation}
or, in components,
$\Gamma^{ik}\Omega_k{}^j=\Gamma^{jk}\dot\Omega_k{}^i$.
For fixed twist and coupling this condition quantises the undotted and dotted
conserved charges $a_\alpha$ and $\dot a_\alpha$ together and fixes the state-dependent gluing constant
$c$. This is the functional counterpart of coordinate-space single-valuedness of the CFT wave function, as we explain in
Appendix~\ref{sec:coordinate-J2-quantisation}.

The eigenvalue of the two-dimensional Q-operator is then the glued bilinear\footnote{Setting $\Gamma^{11}=1$ loses the original normalisation, so strictly speaking this bilinear gives the eigenvalue of $\widehat{\mathbb Q}_+$ only up to a state-dependent constant. In the rest of the paper we use the normalisation defined by~\eqref{eq:glued-Qplus}.}
\begin{equation}
 \Qp(u,\dot u)
 \equiv q_i^\downarrow(u)\Gamma^{ij}\dot q_j^\uparrow(\dot u)=
q_1^\downarrow(u)\dot q_1^\uparrow(\dot u)+
c\;q_2^\downarrow(u)\dot q_2^\uparrow(\dot u)
.
 \label{eq:glued-Qplus}
\end{equation}
On the physical lattice $u-\dot u\in\ii\mathbb Z$, the periodicity of the connection matrices and~\eqref{eq:baxter-gluing-condition} give an equivalent representation with the arrows switched,
\begin{equation}
 \Qp(u,\dot u)
 =q_i^\uparrow(u)\Gamma^{ij}\dot q_j^\downarrow(\dot u)=
q_1^\uparrow(u)\dot q_1^\downarrow(\dot u)+
c\;q_2^\uparrow(u)\dot q_2^\downarrow(\dot u)
,
 \qquad u-\dot u\in\ii\mathbb Z.
 \label{eq:glued-Qplus-arrow-switch}
\end{equation}
The two representations make the analyticity of $\Qp$ transparent.
In~\eqref{eq:glued-Qplus}, both factors are regular when $u$ lies above
the down-arrow poles and $\dot u$ below the up-arrow poles;
\eqref{eq:glued-Qplus-arrow-switch} covers the opposite region.
For real inhomogeneities with spins
$0<s<1$ their union contains the strip
\begin{equation}
 \left|\Im\frac{u+\dot u}{2}\right|<1-s+\frac{|n|}{2},
 \qquad u-\dot u=\ii n,\quad n\in\mathbb Z.
 \label{eq:setup-Qplus-analyticity-strip}
\end{equation}
At the fishnet point, the bound is $\frac34+\frac{|n|}{2}$.

For $0<s<1$, we will use the following regularity at the first ordinary and shadow poles.
At $u=\theta_\alpha-\ii s$, the arrow-switched representation gives a regular $\Qp(u,u+\ii)$ since $q^\uparrow(u)$ is below its poles and $\dot q^\downarrow(u+\ii)$ is above its poles.
The original representation gives the same regularity for $\Qp(u+\ii,u)$.

Hence
\begin{equation}
 \operatorname*{Res}_{u=\theta_\alpha-\ii s}\Qp(u,u+\ii)
 =\operatorname*{Res}_{u=\theta_\alpha-\ii s}\Qp(u+\ii,u)=0.
 \label{eq:setup-Qplus-first-pole-regularity}
\end{equation}
The same argument applies at $u=\theta_\alpha-\ii(1-s)$.
These pole-cancellation identities extend to generic chain data by
meromorphic continuation.

\paragraph{Injecting the coupling constant.}
We can now close the spectral problem. For given $h$ and $\dot h$, the quantisation condition fixes the charges $a_r$, $\dot a_r$ and the gluing constant $c$, and hence determines $\Qp$. Applying the identity~\eqref{eq:graph-building-from-Q} to an eigenstate and using the graph-building equation~\eqref{eq:graph-building-stationarity}, one finds at the fishnet point~\eqref{eq:fishnet-point}
\begin{equation}
 \xi^{2J}
 =\lim_{\epsilon\to0}\epsilon^J
 \frac{\Qp\!\left(\frac{3\ii}{4}-\ii\epsilon,
                         \frac{3\ii}{4}-\ii\epsilon\right)}
      {\Qp\!\left(\frac{\ii}{4},\frac{\ii}{4}\right)}.
 \label{eq:graph-builder-spectral-condition}
\end{equation}

The right-hand side is a normalisation-independent function of the quantised solution. It
defines $\xi(h,\dot h)$, or, upon locally inverting this relation,
$h(\xi)$ and $\dot h(\xi)$.
Together with
$\Delta=h+\dot h$ and $S=\dot h-h$, this gives the spectrum of local
operators in the fishnet CFT. Finally, the solutions corresponding to
single-trace operators must obey the twisted cyclicity condition
\eqref{eq:twisted-wavefunction-cyclicity}, with $U$ computed from~\eqref{eq:shift-eigenvalue-Qplus}.

\paragraph{The shadow Baxter equation.}
In addition to the Baxter equation and Q-functions described above, in this paper we also make use of the \textit{shadow} Baxter equation, corresponding to replacing $s$ by $1-s$ in \eqref{eq:baxter-equation}, yielding
\begin{equation}
 \widetilde P_+(u)\widetilde q(u+\ii)-t(u)\widetilde q(u)
 +\widetilde P_-(u)\widetilde q(u-\ii)=0,
 \qquad
 \widetilde P_\pm(u)\equiv \prod_{\alpha=1}^J
 [u-\theta_\alpha\pm\ii(1-s)].
 \label{eq:shadow-baxter-equation}
\end{equation}
The transfer matrix eigenvalue $t(u)$ is unchanged. The corresponding $\mathfrak{sl}(2)$ representations are equivalent~\cite{Derkachov:2001yn}, with the shadow transform \eqref{eq:shadow-operator} intertwining
them.

We now introduce the following measures
\begin{equation}
 \begin{aligned}
 \mu^\downarrow(u)
 &\equiv \prod_{\alpha=1}^J
 \frac{\Gamma(s-\ii(u-\theta_\alpha))}
      {\Gamma(1-s-\ii(u-\theta_\alpha))},\quad
 \mu^\uparrow(u)
 &\equiv \prod_{\alpha=1}^J
 \frac{\Gamma(s+\ii(u-\theta_\alpha))}
      {\Gamma(1-s+\ii(u-\theta_\alpha))}.
 \end{aligned}
 \label{eq:shadow-chain-mu-factors}
\end{equation}
These allow us to easily move between solutions of the Baxter equation and solutions of the shadow Baxter equation.

For either arrow, the measure satisfies
\begin{equation}
\mu(u\pm\ii)/
\mu(u)=P_\pm(u)/\widetilde P_\pm(u)\,,
\label{eq:mu-finite-diff}
\end{equation}
so solutions \eqref{eq:baxter-asymptotics} with pure asymptotics obey
\begin{equation}
 \begin{aligned}
 \widetilde q_a^\downarrow(u)
 &=e^{+\frac{\ii\pi}{2}(2Js-J)}
   \mu^\downarrow(u)q_a^\downarrow(u)\,,
   \quad
 \widetilde q_a^\uparrow(u)
 &=e^{-\frac{\ii\pi}{2}(2Js-J)}
   \mu^\uparrow(u)q_a^\uparrow(u).
 \end{aligned}
 \label{eq:shadow-q-map}
\end{equation}
Thus $\widetilde q$ is the canonical Q-function of the chain
with $s\to1-s$, and the leading asymptotics are
\begin{equation}
 \widetilde q_1(u)\simeq\lambda^{\ii u}u^{h+J(s-1)},
 \qquad
 \widetilde q_2(u)\simeq\lambda^{-\ii u}u^{-h+J(s-1)}.
 \label{eq:shadow-baxter-asymptotics}
\end{equation}
It is useful to
combine the two measure factors in \eqref{eq:shadow-chain-mu-factors} into the combination
\begin{equation}
 \mu(u,\dot u)
 \equiv \mu^\downarrow(u)\mu^\uparrow(\dot u)
 =\mu^\uparrow(u)\mu^\downarrow(\dot u),
 \qquad u-\dot u\in\ii\mathbb Z.
 \label{eq:overlaps-one-pair-measure}
\end{equation}
The glued shadow Q-function is then
\begin{equation}
 \widetilde{\mathbb Q}_+(u,\dot u)
 =\mu(u,\dot u)\Qp(u,\dot u).
 \label{eq:shadow-glued-Qplus}
\end{equation}

\subsection{Connection matrices}
\label{sec:setup-omega}

The connection matrices $\Omega$ and $\dot\Omega$ will play a crucial role in Section \ref{sec:norms}: they will be shown to encode the norm of CFT wave functions. To finish this section, we recall some useful properties of the connection matrices. We usually focus on $\Omega$; the dotted
matrix behaves in the same way unless we explicitly note so.

\paragraph{Wronskians and connection coefficients.}

Let us recall how the connection matrix $\Omega$ can be written directly in terms of the Q-functions \cite{Ekhammar:2025fvn}. First, we introduce the shifted Wronskian
\begin{equation}
 \mathcal W[f,g](u)\equiv f(u)g(u+\ii)-f(u+\ii)g(u)
 \label{eq:omega-casoratian-definition}
\end{equation}
of two functions $f$ and $g$. Plugging in UHPA and LHPA solutions of the Baxter equation we obtain
\begin{equation}
  \mathcal{W}[q_i^\uparrow,q_k^\downarrow]=\Omega_{i}{}^j\mathcal{W}[q_j^\downarrow,q_k^\downarrow]
  \label{eq:wronskian-soln}
\end{equation}
since $\Omega$ is $\ii$-periodic. Clearly the Wronskian is antisymmetric in $f$ and $g$. As a consequence, we have
\begin{equation}
  \mathcal{W}[q_j^\downarrow,q_k^\downarrow] = \epsilon_{jk}\, \mathcal{W}[q_1^\downarrow,q_2^\downarrow],\quad \epsilon=\begin{pmatrix}0&1\\-1&0\end{pmatrix}\,.
  \label{eq:wronskian-anti-sym}
\end{equation}
Equation \eqref{eq:wronskian-soln} together with \eqref{eq:wronskian-anti-sym} provides an inhomogeneous linear system of equations for the components of $\Omega$, easily solvable by Cramer's rule, and similarly for $\Omega^{-1}$:
\begin{equation}
 \Omega_{i}{}^j(u)=\epsilon^{jk}
 \frac{\mathcal W[q_i^\uparrow,q_k^\downarrow](u)}
      {\mathcal W[q_1^\downarrow,q_2^\downarrow](u)},
 \qquad (\Omega^{-1})_{i}{}^{j}(u)= \epsilon^{jk}
 \frac{\mathcal W[q_i^\downarrow,q_k^\uparrow](u)}
      {\mathcal W[q_1^\uparrow,q_2^\uparrow](u)}\,.
 \label{eq:omega-cramer}
\end{equation}
The same formulas with $q\mapsto\widetilde q$ give the shadow connection
matrices $\widetilde{\Omega}$.

\paragraph{Explicit form of denominator.}

The denominators $\mathcal{W}[q_1^\downarrow,q_2^\downarrow]$ and $\mathcal{W}[q_1^\uparrow,q_2^\uparrow]$ can be written completely explicitly in a state-independent way. For two solutions $f,g$ of the same Baxter
equation~\eqref{eq:baxter-equation}, subtracting their equations after
multiplication by the other solution gives
\begin{equation}
 P_+(u)\mathcal W[f,g](u)=P_-(u)\mathcal W[f,g](u-\ii).
 \label{eq:setup-wronskian-recurrence}
\end{equation}
Now consider the combination $P_+(u)\mu^\uparrow(u)\mathcal{W}[q_1^\uparrow,q_2^\uparrow](u)$.
Shifting $u\to u-\ii$ and using the recurrence~\eqref{eq:setup-wronskian-recurrence} for the Wronskian together with the finite-difference equation~\eqref{eq:mu-finite-diff} for $\mu^\uparrow$, we find that this combination is $\ii$-periodic. Since all factors are analytic in the lower half-plane, periodicity makes it analytic everywhere, and the Q-function asymptotics~\eqref{eq:baxter-asymptotics} then fix it to a constant,
\begin{equation}
 P_+(u)\mu^\uparrow(u)
 \mathcal W[q_1^\uparrow,q_2^\uparrow](u)
 =(\lambda-\lambda^{-1})e^{+\frac{\ii\pi}{2}J(2s-1)}.
 \label{eq:setup-wronskian-normalisation}
\end{equation}
For the down-arrow basis, we simply replace $\uparrow$ by $\downarrow$ and reverse the sign of the phase.

\paragraph{Asymptotics of $\Omega$.}
An $\ii$-periodic matrix depends on $u$ only through $e^{2\pi u}$. The
canonical normalisation~\eqref{eq:baxter-asymptotics} makes its limits
at $\Re u\to\pm\infty$ finite, so the entries are rational functions of
$e^{2\pi u}$. The limits are moreover triangular. A solution that is
subdominant at one end cannot contain the dominant one with a finite
coefficient. For $0<\phi<\pi$, $q_1$ is subdominant at
$+\infty$ and $q_2$ at $-\infty$; this gives
\begin{equation}
 \Omega(+\infty)=
 \begin{pmatrix}1&0\\ *&1\end{pmatrix},
 \qquad
 \Omega(-\infty)=
 \begin{pmatrix}
 e^{-2\pi\ii(h-Js)}&*\\
 0&e^{+2\pi\ii(h+Js)}
 \end{pmatrix}.
 \label{eq:setup-omega-limits}
\end{equation}
Here $*$ denotes a finite coefficient. The diagonal phases at $-\infty$
come from the two branches of $u^{\pm h-Js}$ on either side of the
cuts. The dotted twist is opposite, so the two solutions exchange roles.
Hence $\dot\Omega(\pm\infty)$ has the transposed triangular shape, with
$h\to\dot h$ in the phases.

\paragraph{Determinant.}
Taking the ratio of $\mathcal{W}[q_1^\uparrow,q_2^\uparrow]$ and $\mathcal{W}[q_1^\downarrow,q_2^\downarrow]$ allows us to extract the determinant of $\Omega$, which is independent of the quantum numbers of any particular state
\begin{equation}
 \det\Omega(u)
 =e^{\ii\pi J(2s-1)}
 \frac{\mu^\downarrow(u)}
      {\mu^\uparrow(u)}
 =\det\dot\Omega(u),
 \label{eq:setup-omega-determinant}
\end{equation}
with $\mu^{\downarrow}$ and $\mu^\uparrow$ given in~\eqref{eq:shadow-chain-mu-factors}.
The inverse can also be written using the antisymmetric tensor $\epsilon$
of~\eqref{eq:wronskian-anti-sym}:
\begin{equation}
 \Omega^{-1}=-\frac{\epsilon\Omega^{\mathsf T}\epsilon}{\det\Omega}.
 \label{eq:omega-two-by-two-inverse}
\end{equation}

\paragraph{Poles.}
There are two classes of points that will play a key role throughout the paper. The first, at $u\equiv \theta_\alpha-\ii s$ modulo $\ii$, we call \emph{ordinary points}. The second, at $u\equiv \theta_\alpha-\ii (1-s)$ modulo $\ii$, we call \emph{shadow points}. As a result of the defining relation $q^\uparrow = \Omega q^\downarrow$ in its Wronskian form~\eqref{eq:omega-cramer}, together with the pole positions of $q^\uparrow$ and $q^\downarrow$ given in Section~\ref{sec:setup-baxter} and the explicit denominators~\eqref{eq:setup-wronskian-normalisation},
$\Omega$ has simple poles only at the ordinary points, and $\Omega^{-1}$ only at the shadow points.

By~\eqref{eq:setup-omega-determinant}, $\det\Omega$ has simple poles at the ordinary points and simple zeros at the shadow points. At an ordinary point, the residue of $\Omega$ must therefore have vanishing determinant, while by~\eqref{eq:omega-two-by-two-inverse} $\Omega^{-1}$ is regular there. At a shadow point the roles of $\Omega$ and $\Omega^{-1}$ are exchanged. In both cases the residue of the singular matrix and the value of the regular one have rank one:
\begin{equation}
 \begin{array}{c|ccc}
 u\pmod{\ii}&\det\Omega&\Omega&\Omega^{-1}\\ \hline
 \theta_\alpha-\ii s
 &\text{simple pole}&\text{simple pole}&\text{regular, rank one}\\
 \theta_\alpha-\ii(1-s)
 &\text{simple zero}&\text{regular, rank one}&\text{simple pole}
 \end{array}
 \label{eq:setup-omega-pole-table}
\end{equation}
There are no other singularities.

\paragraph{Shadow map.}
The shadow map~\eqref{eq:shadow-q-map} multiplies both bases by scalar
factors. Hence the shadow connection matrix is
\begin{equation}
 \widetilde\Omega=\frac{\Omega}{\det\Omega},
 \qquad
 \widetilde\Omega^{-1}=(\det\Omega)\,\Omega^{-1}.
 \label{eq:setup-omega-shadow}
\end{equation}
By~\eqref{eq:setup-omega-pole-table}, the map exchanges the roles of the
ordinary and shadow points. For instance, $\widetilde\Omega^{-1}$ has its
simple poles at the ordinary points.

\paragraph{Residue matrices.}
By~\eqref{eq:setup-omega-pole-table} and~\eqref{eq:setup-omega-shadow},
$\Omega^{-1}$ has a simple pole at each shadow point
$u=\theta_\alpha-\ii(1-s)$, and $\widetilde\Omega^{-1}$ at each ordinary
point $u=\theta_\alpha-\ii s$. We label their residues, for each $\alpha$, by
\begin{equation}
 \begin{aligned}
 \mathsf R_\alpha&\equiv
 \ii e^{+\frac{\ii\pi}{2}J(2s-1)}
 \operatorname*{Res}_{u=\theta_\alpha-\ii(1-s)}\Omega^{-1}(u),\\
 \widetilde{\mathsf R}_\alpha&\equiv
 \ii e^{-\frac{\ii\pi}{2}J(2s-1)}
 \operatorname*{Res}_{u=\theta_\alpha-\ii s}\widetilde\Omega^{-1}(u).
 \end{aligned}
 \label{eq:setup-omega-residue-matrices}
\end{equation}
The shadow map $s\mapsto1-s$ exchanges the two definitions, and the dotted
counterparts are defined in the same way from $\dot\Omega$. By periodicity,
the residues at $\theta_\alpha-\ii(1-s+n)$ and $\theta_\alpha-\ii(s+n)$ are
the same for every $n\in\mathbb Z$. As we explain later, the $2J$ constant
$2\times2$ matrices~\eqref{eq:setup-omega-residue-matrices} and their $2J$ dotted counterparts, along with their derivatives, are the only state-dependent objects that enter the CFT wave function norm in Section~\ref{sec:norms}.

Since $\Omega$ is regular at the shadow points and $\widetilde\Omega$ at the
ordinary points, the inverse formula~\eqref{eq:omega-two-by-two-inverse}
shows that the poles of $\Omega^{-1}$ and $\widetilde\Omega^{-1}$ come from
$\det\Omega$ alone, that is, from $\mu^\downarrow$
in~\eqref{eq:setup-omega-determinant}. We define the site-dependent coefficients
\begin{equation}
 k_\alpha(s)
 \equiv \frac{\mu^{\uparrow}(\theta_\alpha-\ii s)}
         {\operatorname*{Res}_{u=\theta_\alpha-\ii s}\mu^{\downarrow}(u)}\,,
 \qquad
 k_\alpha(1-s)
 =\frac{(\mu^{\downarrow})'(\theta_\alpha-\ii(1-s))}
         {\mu^{\uparrow}(\theta_\alpha-\ii(1-s))}.
 \label{eq:J1-measure-conversion-coefficients}
\end{equation}
The second coefficient follows from the first by $s\mapsto1-s$, under which
$\mu^{\uparrow\downarrow}\mapsto1/\mu^{\uparrow\downarrow}$.

For later use with the contour brackets of Section~\ref{sec:J1-overlap-q-bilinears}, eq.~\eqref{eq:lower-tower-contour}, built from the canonical
shadow Q-functions~\eqref{eq:shadow-q-map}, we include the shadow-map phase in the conversion
coefficient
\begin{equation}
 k_\alpha^{\mathrm{c}}(s)\equiv e^{-\ii\pi J(2s-1)}k_\alpha(s).
 \label{eq:contour-conversion-coefficients}
\end{equation}
The substitution $s\mapsto1-s$ reverses the sign of the exponent.

The coefficients $k_\alpha$ relate the residue matrices to values of
the regular matrices:
\begin{equation}
 \begin{aligned}
 k_\alpha(s)\widetilde{\mathsf R}_\alpha
 &=\ii e^{+\frac{\ii\pi}{2}J(2s-1)}
 \Omega^{-1}(\theta_\alpha-\ii s),\\
 k_\alpha(1-s)\mathsf R_\alpha
 &=\ii e^{-\frac{\ii\pi}{2}J(2s-1)}
 \widetilde\Omega^{-1}(\theta_\alpha-\ii(1-s)).
 \end{aligned}
 \label{eq:setup-residue-value-conversion}
\end{equation}
The same holds for the dotted matrices.

\paragraph{Gluing.}
In terms of inverse matrices, the quantisation
condition~\eqref{eq:baxter-gluing-condition} reads
\begin{equation}
 \dot\Omega^{-1}=\Gamma^{-1}(\Omega^{-1})^{\mathsf T}\,\Gamma,
 \qquad \Gamma=\operatorname{diag}(1,c).
 \label{eq:setup-omega-gluing}
\end{equation}
Its local content at the singular points is the pole
cancellation~\eqref{eq:setup-Qplus-first-pole-regularity}: the physical
$\Qp$ is regular even where one arrow representation has poles term by term.

\section{Overlaps of the CFT wave functions and correlators}
\label{sec:gluing}

Having reviewed the integrability structure of the 2D fishnet theory, we now move on to describing how to glue CFT wave functions together to compute three-point structure constants of the theory.

\subsection{The coordinate-space scalar product and the planar bracket}
\label{sec:setup-bracket}
To compute norms and overlaps, we need a scalar product on the space of wave functions. Here we introduce two such pairings: the companion pairing $\langle f|g\rangle$ and the planar bracket $\LRpair{f}{g}$. We review these bilinear pairings and show that the planar bracket makes eigenstates with different transfer matrix eigenvalues orthogonal.

\paragraph{Shadow transform and companion pairing.} A pairing of two wave functions should be conformally invariant. Since both wave functions carry the same weight $s$ at every site, we must first map one of them to the shadow weight $1-s$. For equal chiral weights $s=\dot s$, this is realised by the shadow transform
\begin{equation}
 [\bfS^{(z)} f](z)
 \equiv -\frac{(2s-1)^2}{\pi^2}\int\dd^2w\,
   \frac{f(w)}{[z-w]^{2(1-s)}},
 \qquad
 \bfS\equiv \prod_{\alpha=1}^{J}\bfS^{(\alpha)},
 \label{eq:shadow-operator}
\end{equation}
normalised so that it inverts the complex propagator with no additional factors
\begin{equation}
 \bfS^{(z)}[z-w]^{-2s}=\delta^{(2)}(z-w).
 \label{eq:shadow-inverse-propagator}
\end{equation}
To glue the two operators, we place them at opposite fixed points of the twist. In the companion frame, these fixed points are exchanged by the inversion
\begin{equation}
 (\cD f)(\bz)
 \equiv \prod_{\alpha=1}^{J}[z_\alpha]^{-2s}\,
   f\bigl(z_1^{-1},\ldots,z_J^{-1}\bigr),
 \qquad \cD^2=1,
 \label{eq:full-inversion}
\end{equation}
which allows us to define the companion pairing as
\begin{equation}
 \langle f|g\rangle
 \equiv \int\prod_{\alpha=1}^{J}\dd^2z_\alpha\;
   f(\bz)\,[\bfS\cD g](\bz).
 \label{eq:companion-pairing}
\end{equation}
The companion kernel is symmetric in its two arguments, so inversion can equivalently act on
the first argument. Inversion exchanges the fixed points $\lambda$ and $\lambda^{-1}$ and
conjugates $G_{\rm C}$ into $G_{\rm C}^{-1}$. The joined twist cuts therefore carry inverse
holonomies around the two endpoints.
Our normalisation is the same as in~\cite{Gromov:2020fwh} and it gives
\begin{equation}
 \langle1|1\rangle=1
 \label{eq:one-site-vacuum-norm}
\end{equation}
where $|1\rangle$ is the constant identity function.

\begin{figure}[t]
\centering
\includegraphics[width=0.75\linewidth]{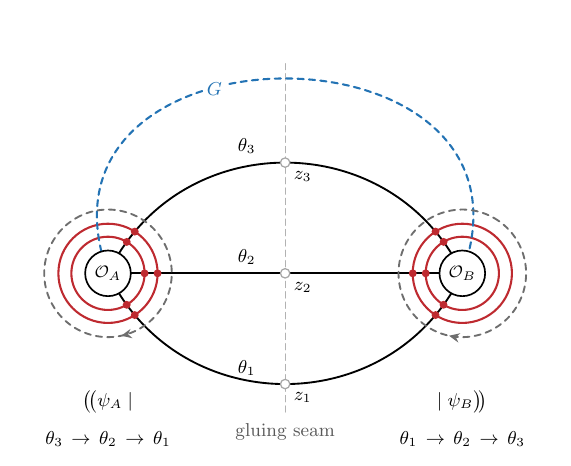}
\caption{The planar bracket $\LRpair{\psi_A}{\psi_B}$ at finite fixed
points $x_A,x_B$ ($x_A=\lambda^{-1}$, $x_B=\lambda$ in the companion frame).
Black lines are $\phi_1$ propagators, red circles are $\phi_2$ wheels,
and the blue dashed line is the joined twist cut. Open marks denote the
seam coordinates $z_\alpha$ integrated in the gluing. Grey dashed arrows
indicate clockwise site order: starting from the cut, it is $1,2,3$ at
$B$ and $3,2,1$ at $A$. Each line retains its $\theta_\alpha$ across the
seam; the first argument thus belongs to the reversed chain.}
\label{fig:bracket-gluing}
\end{figure}

\paragraph{Planar gluing and site reversal.}
The companion pairing glues site $\alpha$ of one wave function to site $\alpha$ of the other. The planar contraction of two cyclic traces, however, pairs site $\alpha$ with site $J+1-\alpha$, because the two traces are read in opposite orders, see Figure~\ref{fig:bracket-gluing}. To account for this, we introduce the reversal operator
\begin{equation}
 (\mathcal Rf)(z_1,\ldots,z_J)
 \equiv f(z_J,\ldots,z_1),
 \qquad \mathcal R^2=1.
 \label{eq:chain-reversal-R}
\end{equation}

Since the traces are cyclic, the pairing $\alpha\leftrightarrow J+1-\alpha$ is defined only up to a cyclic relabelling of the sites. The twist cut removes this ambiguity: on both sides the cut passes between sites $J$ and $1$, as in Figure~\ref{fig:bracket-gluing}. Any other labelling amounts to acting with the twisted cyclic shift~\eqref{eq:twisted-shift-operator} on one of the two wave functions.

Each propagator keeps its inhomogeneity $\theta_\alpha$ across the seam. Seen from the first wave function, the sites therefore appear in reversed order, so the two local site orders carry
\begin{equation}
 \boldsymbol\theta=(\theta_1,\ldots,\theta_J),\qquad
 \boldsymbol\theta^{\rm R}=(\theta_J,\ldots,\theta_1). \label{eq:reversed-list}
\end{equation}
Hence, we define the planar bracket as\footnote{Reversing the first argument is a convention: since $\langle\mathcal Rf|g\rangle=\langle f|\mathcal Rg\rangle$, where $\mathcal Rg$ is a state of the chain with inhomogeneities $\boldsymbol\theta^{\rm R}$, one could equally reverse $g$.}
\begin{equation}
 \LRpair{f}{g}_{\boldsymbol\theta}
 \equiv \bigl\langle
     \mathcal R f_{\boldsymbol\theta^{\rm R}}
     \bigm|
     g_{\boldsymbol\theta}
   \bigr\rangle,
 \label{eq:left-right-bracket}
\end{equation}
where the function $g$ is on the original chain and $f$ is a function on the reversed chain. The superscript $\rm R$ reverses the list of inhomogeneities as in~\eqref{eq:reversed-list}, while $\mathcal R$ reverses the coordinate arguments~\eqref{eq:chain-reversal-R} with the parameters held fixed:
\begin{equation}
 (\mathcal R f_{\boldsymbol\theta^{\rm R}})
 (z_1,\ldots,z_J)
 =f_{(\theta_J,\ldots,\theta_1)}
 (z_J,\ldots,z_1).
 \label{eq:reversal-parameters-and-coordinates}
\end{equation}
This simply means that each glued line keeps its $\theta_\alpha$ across the seam, i.e.\ the parameter list is not reversed a second time, as depicted in Figure~\ref{fig:bracket-gluing}. We write $((\psi_A|$ and $|\psi_B))$ for the two sides of the planar bracket, and keep angle brackets for the companion pairing~\eqref{eq:companion-pairing}.

\paragraph{Transpose and orthogonality.}
We now show that eigenstates with different transfer matrix eigenvalues are orthogonal with respect to the planar bracket. We define the transpose of an operator $O$ with respect to the companion pairing by $\langle Of|g\rangle=\langle f|O^{\mathsf T_{\langle\rangle}}g\rangle$. Conformal invariance gives $\langle \hat S^af|\cD{g}\rangle=-\langle f|\cD{\hat S^ag}\rangle$ and taking into account $\cD\hat S^z\cD=-\hat S^z$ and $\cD\hat S^\pm\cD=\hat S^\mp$ gives
\begin{equation}
 (\hat S^z_\alpha)^{\mathsf T_{\langle\rangle}}=\hat S^z_\alpha,
 \qquad
 (\hat S^\pm_\alpha)^{\mathsf T_{\langle\rangle}}=-\hat S^\mp_\alpha,
 \qquad
 (O_1O_2)^{\mathsf T_{\langle\rangle}}=O_2^{\mathsf T_{\langle\rangle}}O_1^{\mathsf T_{\langle\rangle}}.
 \label{eq:transpose-rules}
\end{equation}
These are transposition rules for a bilinear pairing, not Hermitian adjoints~\cite{Gromov:2022waj}.

 Applied to the Lax operator, they give
\begin{equation}
 L_{a\alpha}(u-\theta_\alpha)^{\mathsf T_{\langle\rangle}}
 =\sigma_3\,L_{a\alpha}(u-\theta_\alpha)^{t_{\rm a}}\,\sigma_3,
 \qquad
 \sigma_3= \begin{pmatrix}
     1 & 0 \\ 0 & -1
 \end{pmatrix}\,,
 \label{eq:lax-transpose}
\end{equation}
where $\mathsf T_{\langle\rangle}$ and $t_{\rm a}$ denote transposition in the quantum and auxiliary spaces. The transpose of the transfer matrix~\eqref{eq:twisted-transfer-matrix} is obtained by applying~\eqref{eq:lax-transpose} to each Lax operator. The inner factors of $\sigma_3$ cancel pairwise. The outer ones are absorbed by the twist, since $\sigma_3G_{\rm C}^{t_{\rm a}}\sigma_3=G_{\rm C}$, and likewise for $G_{\rm D}$. Thus, transposing reverses the order of the Lax operators:
\begin{equation}
 \widehat t_{\boldsymbol\theta}(u)^{\mathsf T_{\langle\rangle}}
 =\operatorname{tr}_a\!\bigl[L_{aJ}(u-\theta_J)\cdots
    L_{a1}(u-\theta_1)\,G_a\bigr]
 =\mathcal R\,\widehat t_{\boldsymbol\theta^{\rm R}}(u)\,
  \mathcal R.
 \label{eq:transfer-transpose-reversal}
\end{equation}
In other words, transposition maps the transfer matrix of the original chain to that of the reversed chain, $\widehat t_{\boldsymbol\theta}^{\mathsf T_{\langle\rangle}}\mathcal R=\mathcal R\widehat t_{\boldsymbol\theta^{\rm R}}$, and the planar bracket reads
\begin{equation}
 \LRpair{f}{\widehat t_{\boldsymbol\theta}\,g}
 =\LRpair{\widehat t_{\boldsymbol\theta^{\rm R}}f}
         {g} .
 \label{eq:bracket-adjointness}
\end{equation}
Hence, if $f$ and $g$ are eigenfunctions of the transfer matrices of the reversed and the original chain, with eigenvalues $t_A(u)$ and $t_B(u)$, then~\eqref{eq:bracket-adjointness} gives $(t_A(u)-t_B(u))\LRpair{f}{g}=0$, so states with different transfer matrix eigenvalues are orthogonal. On a reversal-invariant chain, such as the homogeneous one, \eqref{eq:bracket-adjointness} reduces to $\LRpair{f}{\widehat t g}=\LRpair{\widehat t f}{g}$.

\paragraph{One-site check.}
For $J=1$ we have $\mathcal R=1$, so the companion and planar brackets coincide, and the monomials $z^n$ give a simple illustration of the rules above. They are eigenfunctions of $\hat S^z$, with $\hat S^zz^n=-(n+s)z^n$. Since $\hat S^z$ is symmetric, $\langle\hat S^zz^n|z^m\rangle=\langle z^n|\hat S^zz^m\rangle$, which gives $(n-m)\langle z^n|z^m\rangle=0$: distinct monomials are orthogonal. Their norms follow from $(\hat S^\pm)^{\mathsf T_{\langle\rangle}}=-\hat S^\mp$, and their pairing gives
\begin{equation}
 \langle z^n\bar z^{\dot n}|z^m\bar z^{\dot m}\rangle
 =\delta_{nm}\delta_{\dot n\dot m}\,
  \frac{\Gamma(n+1)\Gamma(2s)}{\Gamma(n+2s)}\,
  \frac{\Gamma(\dot n+1)\Gamma(2s)}{\Gamma(\dot n+2s)} ,
 \label{eq:monomial-norms}
\end{equation}
which is the standard rank-one SoV pairing in each chirality~\cite{Gromov:2020fwh}.

\subsection{The logarithmic norm and the two-point function}
\label{sec:coordinate-log-norm}
We now return to the homogeneous fishnet chain,
$\theta_\alpha=\dot\theta_\alpha=0$ and $s=\dot s=\tfrac14$,
but keep the length $J$ arbitrary. Gluing a wave function to its dual with the
planar pairing~\eqref{eq:left-right-bracket} produces a two-point graph,
as in~\cite{Gromov:2019bsj,Cavaglia:2021mft}. The overlap nevertheless
has a logarithmic divergence. To see this, let us for simplicity place the two operators at $0$ and $\infty$, parametrise $z_1 = \rho e^{\ii \varphi}, z_{\alpha\geq 2} = z_1 \omega_{\alpha}$ and change integration variables to $\rho,\varphi$ and $\omega_\alpha$. Since wave functions in the diagonal twist frame have well-defined scaling behaviour, i.e.\ $\psi^{\rm D}(\rho \mathbf{z}) = \rho^{-\Delta-2 J s}\psi^{\rm D}(\mathbf{z})$, we can factor out all $\rho$ dependence. Combining the scaling factors of the two wave functions, the shadow transform and the Jacobian results in an overall remaining divergent scale integral $\int \frac{\dd\rho}{\rho}$. We regularise this integral by cutting out neighbourhoods of the two operators, restricting the radial scale to $\delta <\rho<R$, leading to the expression
$\int_\delta^R \mathrm d\rho/\rho=\log(R/\delta)$.

With the standard two-point coordinate factor removed, we define
\begin{equation}
 \LRpair{\psi}{\psi}_{\rm reg}
 =\mathcal C_{\log}\log(R/\delta)+O(1).
 \label{eq:coordinate-logarithmic-overlap}
\end{equation}

The operators must be renormalised in some scheme. For a
point-split operator $\mathcal O_\delta$, we write
\begin{equation}
 \mathcal O(x)=\lim_{\delta\to0}Z_\delta\mathcal O_\delta(x),
 \qquad Z_\delta=\delta^{-\Delta},
 \label{eq:coordinate-operator-renormalisation}
\end{equation}
and similarly for the dual operator.

\begin{figure}[H]
 \centering
 \includegraphics[width=.95\linewidth]{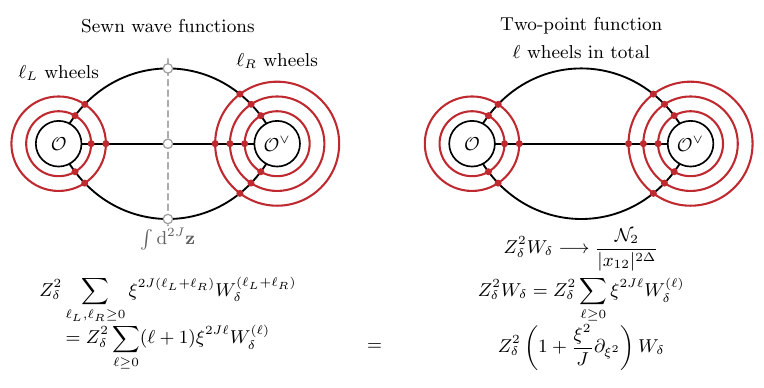}
 \caption{Sewing versus the two-point sum for renormalised operators.
 Red circles are fishnet wheels; the open marks denote the integrated
 seam, with the shadow kernels implicit.
 For $\ell=\ell_L+\ell_R$ there are $\ell+1$ possible seams, whereas
 the two-point graph is counted once.
 $W_\delta^{(\ell)}$ is the $\ell$-wheel coefficient.
 The wheel-counting derivative acts before the factors $Z_\delta^2$.}
 \label{fig:norm-wheel-counting}
\end{figure}

The finite normalisation is fixed by our wave function convention.
The regularised two-point function $W_\delta$ is given by the sum of its Feynman diagrams. The scalar two-point
normalisation ${\cal N}_2$ is defined by
\begin{equation}
 \big\langle\mathcal O(x_1)\mathcal O^\vee(x_2)\big\rangle
 =\lim_{\delta\to0}Z_\delta^2W_\delta
 =\frac{\mathcal N_2}{|x_{12}|^{2\Delta}},
 \label{eq:coordinate-point-split-two-point}
\end{equation}
where ${\cal O}^\vee$ is the operator conjugated in a suitable way to complete the diagram, as in Figure~\ref{fig:norm-wheel-counting}.
Now let us point out that the diagrams contributing to ${\cal N}_2$ and to the overlap \eqref{eq:coordinate-logarithmic-overlap} are the same. Their multiplicities differ, as illustrated in Figure~\ref{fig:norm-wheel-counting}. The counting below follows the one given for the four-dimensional fishnet in~\cite{Gromov:2019bsj,Cavaglia:2021mft}.

Indeed, a graph with $\ell$ wheels can be split between the two wave functions in
$\ell+1$ ways. Since each wheel carries $\xi^{2J}$, the sewn sum is
$Z_\delta^2(1+\xi^2\partial_{\xi^2}/J)W_\delta$:
the derivative counts wheels before operator renormalisation.
Both operators are point-split by $\delta$. In the frame where they sit
at $0$ and $\infty$ the conformally invariant radial range is therefore
$\rho_{\max}/\rho_{\min}=|x_{12}|^2/\delta^2$: the scale integral
in~\eqref{eq:coordinate-logarithmic-overlap} equals $2\log(R/\delta)$
with $R=|x_{12}|$, and the overlap is $2\,\mathcal C_{\log}\log(R/\delta)$.
Equation~\eqref{eq:coordinate-point-split-two-point}
gives $W_\delta=\delta^{2\Delta}\mathcal N_2/R^{2\Delta}$.
After applying the $Z$-factors and stripping the two-point coordinate
factor, we obtain
\begin{equation}
 \mathcal C_{\log}
 =-\frac{1}{J}\mathcal D_\xi\,\mathcal N_2,
 \qquad
 \mathcal D_\xi\equiv \xi^2\frac{\partial\Delta}{\partial\xi^2}.
 \label{eq:coordinate-general-norm-relation}
\end{equation}
The derivative follows the on-shell branch at fixed twist and quantum
numbers. The relation holds at generic inhomogeneities, where the wave
functions are not cyclic: the planar pairing and $\mathcal N_2$ both
refer to one fixed ordering of the legs. When $\theta_\alpha\to0$ the
$J$ cyclic images of the trace all contribute to the physical two-point
function, so the CFT norm of the cyclic operator is
\begin{equation}
 \mathcal N_2^{\theta=0}\equiv J\lim_{\theta_\alpha\to0}\mathcal N_2 .
 \label{eq:norm-theta-zero}
\end{equation}
For the free trace $\operatorname{tr}\phi_1^J$, whose wave function
$\Omega_J$ has single-ordering norm $1$, this gives
$\mathcal N_2^{\theta=0}=J$, the planar Wick count of
$\langle\operatorname{tr}\phi_1^J\,\operatorname{tr}\phi_1^{\dagger J}\rangle$.
At $J=1$ the two norms coincide.
We regularise the overlap by choosing slightly different twist values for the two operators. The twist difference effectively introduces a point-splitting regularisation, as the first link has a finite size controlled precisely by the difference of the twists. We work this out explicitly for $J=1$ in Section~\ref{sec:J1}.

\subsection{From wave functions to open three-point coefficients}
\label{sec:coordinate-norms}

\begin{figure}[t]
 \centering
 \includegraphics[width=.62\linewidth]{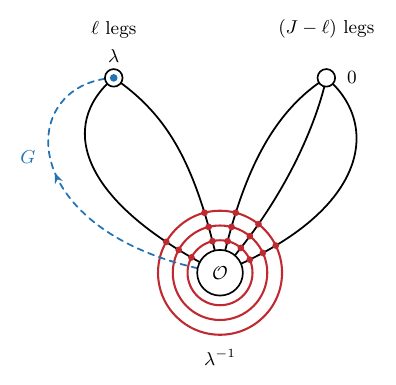}
 \caption{The OPE coefficient $\mathcal C_\ell^{\bullet\circ\circ}$
 in~\eqref{eq:coordinate-open-three-point}, shown in the companion frame.
 The operator sits at $\lambda^{-1}$ and the first $\ell$ probes at
 $\lambda$, the images of $0$ and $\infty$ in the diagonal frame;
 the other $J-\ell$ probes sit at $0$. Red circles are fishnet wheels,
 black lines are scalar propagators, and the dashed blue line is the
 twist cut. The two open circles denote groups of legs in the same
 ordered trace.}
 \label{fig:open-three-point}
\end{figure}

The wave function~\eqref{eq:mathcalAInter} also contains three-point data.
Split the ordered scalar probes into two open groups: the first $\ell$
legs at one point and the remaining $J-\ell$ at another, with
$1\leq\ell<J$, see Figure~\ref{fig:open-three-point}. We denote the resulting normalised open coefficient by
$\mathcal C_\ell^{\bullet\circ\circ}$: the bullet is the wheel-dressed
operator and the two circles are the open groups of scalar legs.

For the homogeneous scalar fishnet chain, put the first group at
infinity and the second at $1$. Each scalar propagator is $1/|z|$, so
stripping the $\ell$ external legs gives the numerator
$\lim_{z\to+\infty}|z|^\ell\psi_\lambda^{\rm D}(z,\ldots,z,1,\ldots,1)$.
Its pointwise normalisation is arbitrary and the two-point norm removes it.
The twisted structure constant in~\eqref{eq:coordinate-open-three-point}
uses the single-ordering norm $\mathcal N_2$ of~\eqref{eq:coordinate-general-norm-relation}, since the open groups of probes have no cyclic images. Thus, one can write the twisted structure constant as
\begin{equation}
 \mathcal C_\ell^{\bullet\circ\circ}(\lambda)
 =\frac{|\lambda^2-1|^{\ell-J/2}|\lambda|^{J/2-\Delta}
  \psi_\lambda^{\rm C}(\lambda^\ell,0^{J-\ell})}
 {\sqrt{\mathcal N_2(\lambda)}}.
 \label{eq:coordinate-open-three-point}
\end{equation}
This remains a non-trivial structure constant in the untwisting $\lambda\to 1$ limit. Here we take a scalar state, $h=\dot h=\Delta/2$, and powers in the
arguments denote repeated coordinates.

The wave function ratio needed for~\eqref{eq:coordinate-open-three-point}
is available directly in \texttt{BSolver2D} through \texttt{EllSplitterData}.

Coordinate wave functions make the physical meaning of these quantities
clear, but their construction quickly becomes very complicated as $J$ grows.
Below, in Section~\ref{sec:sov}, we turn to the SoV basis, which expresses overlaps and three-point data through
one-variable Q-functions. To motivate it, next we consider the simplest length-one case.

\section{Explicitly solvable \texorpdfstring{$J=1$}{J=1} case}
\label{sec:solvable}

As a warm-up for the general case we consider the length-one chain. In this case, all calculations can be done explicitly. In particular, the one-site overlap lets us follow the passage from coordinate wave functions to Q-functions and residue sums.

\subsection{Twisted \texorpdfstring{$J=1$}{J=1} chain}
\label{sec:J1}

In this subsection, we compute the overlap $\LRpair{\psi_A}{\psi_B}$ for states with different twists and conformal weights using the explicit coordinate wave functions.
For $J=1$, residual conformal symmetry fixes the wave function up to an overall constant. On the Euclidean slice $\dot\lambda=\bar\lambda$, its diagonal~\eqref{eq:diagonal-colour-twist} and companion~\eqref{eq:companion-colour-twist} frame expressions are
\begin{align}
 \psi^{\rm D}_{h,\dot h}(z,\bar z)
 &=\frac{1}{[z]^{\{h,\dot h\}+s}},
 \label{eq:reference-J1-diagonal}\\
 \psi^{\rm C}_{h,\dot h}(z,\bar z)
 &=\frac{[\lambda^2-1]^{s}}{
   [z-\lambda]^{\{-h,-\dot h\}+s}
   [1-\lambda z]^{\{h,\dot h\}+s}},
 \label{eq:reference-J1-companion}
\end{align}
where their relative normalisation follows from the frame change~\eqref{eq:diagonal-companion-wavefunction}. Up to the factor $[1-\lambda^{-2}]^{\{h,\dot h\}}$, the companion frame wave function~\eqref{eq:reference-J1-companion} is a three-point conformal structure with weights $h$ at $x=\lambda^{-1}$, $0$ at $y=\lambda$, and $s$ at the scalar probe $z$.

The spectrum of $J=1$ for $s=\tfrac{1}{4}$ is found from the condition~\eqref{eq:graph-building-stationarity}. Acting with the twisted graph-building operator, see~\eqref{eq:graph-building-operator}, on~\eqref{eq:reference-J1-diagonal} and using the chain relation (see Appendix \ref{app:J1}) gives~\cite{Ekhammar:2025fvn}
\begin{equation}
 \xi^2=(-1)^S\left[\frac{1-\lambda^2}{\lambda}\right]^{\{1/2,1/2\}}
 \frac{\Gamma(\tfrac34+h)\Gamma(\tfrac34-h)}
      {\Gamma(\tfrac14+\dot h)\Gamma(\tfrac14-\dot h)},
 \qquad S=\dot h-h\in\mathbb Z.
 \label{eq:J1-exact-coupling}
\end{equation}

\paragraph{The overlap from coordinate wave functions.}

Let us now consider the overlap $\LRpair{\psi^{\rm C}_{A}}{\psi^{\rm C}_{B}}$ of two states with possibly different twists and conformal weights, with the shorthand notation $\psi_A^{\rm C} = \psi^{\rm C}_{h_{A},\dot{h}_A}$ and the same for $\psi^{\rm C}_B$. Unlabelled quantities below refer to either state.

The overlap $\LRpair{\psi^{\rm C}_{A}}{\psi^{\rm C}_{B}}$ is defined in~\eqref{eq:left-right-bracket}. It depends on the reversal~\eqref{eq:chain-reversal-R}, which is trivial for a one-site chain, and on the inversion~\eqref{eq:full-inversion} that acts component-wise by $\boldsymbol\lambda\mapsto1/\boldsymbol\lambda$, so that the overlap between the two states reads
\begin{equation}\label{eq:J1-overlap-shadow-bracket}
  \LRpair{\psi^{\rm C}_{A}}{\psi^{\rm C}_{B}}
  = -\frac{(2s-1)^2}{\pi^2}\int \dd^2 z \,\dd^2 w \,
  \frac{\psi^{\rm C}_{A}(z)\,
  \left.\psi^{\rm C}_{B}(w)\right|_{\boldsymbol\lambda_B\to1/\boldsymbol\lambda_B}}
  {[z-w]^{2(1-s)}}\,.
\end{equation}

Before attacking this integral in full generality, let us consider the reference state $h_B=\dot h_B=s=\tfrac14$, corresponding to $\mathcal{O}_{B}\propto\operatorname{tr}\left[\mathcal{T}_{G_{\text{C}}(\lambda_B)}\phi\right]$ at zero coupling~\eqref{eq:J1-exact-coupling}. Integration over the scalar-insertion point of operator $B$ in the full overlap results in a delta function, see~\eqref{eq:shadow-inverse-propagator}, leading to
\begin{equation}
 \begin{aligned}
 \LRpair{\psi_A^{\rm C}}{\psi_B^{\rm C}}\bigg|_{h_{B}=\dot{h}_{B}=s=\frac{1}{4}}
 &=[\lambda_B^2-1]^{1/4}\psi_A^{\rm C}(\lambda_B)\,.
 \end{aligned}
 \label{eq:J1-overlap-zero-coupling}
\end{equation}
That is, the overlap is the CFT wave function of the first operator evaluated at the twist of the second operator.

\begin{figure}[htbp]
\centering
\includegraphics[width=0.4\linewidth]{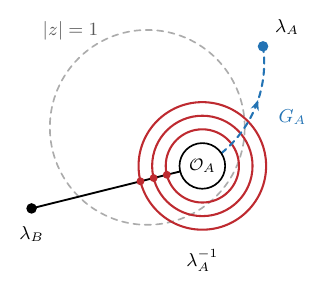}
\caption{The overlap \eqref{eq:J1-overlap-zero-coupling}. State $B$ sits at $\lambda_B$ and has zero coupling. The state $A$ sits at $\lambda_A^{-1}$ and it is surrounded by red wheels; its blue twist cut ends at $\lambda_A$.}
\label{fig:J1-zero-coupling-overlap}
\end{figure}

\begin{figure}[htbp]
\centering
\includegraphics[width=0.55\linewidth]{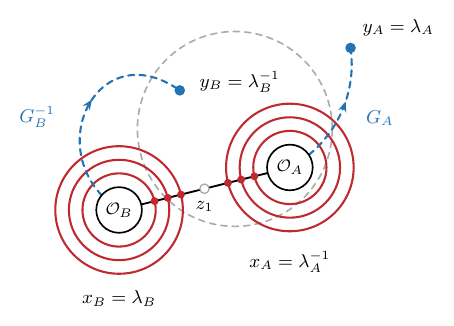}
\caption{A schematic depiction of the $J=1$ gluing for generic unequal twists.
The operators $\mathcal O_A,\mathcal O_B$ are joined by a single black
$\phi_1$ particle and surrounded by a cloud of red wheels. The dashed blue cuts end
at $\lambda_A$ and $\lambda_B^{-1}$.
Here inversion acts on $B$, as in
\eqref{eq:J1-overlap-shadow-bracket}, so the operator positions are
$x_A=\lambda_A^{-1}$ and $x_B=\lambda_B$, and the remaining endpoints are
$y_A=\lambda_A$ and $y_B=\lambda_B^{-1}$.
The open mark $z_1$ denotes the integrated gluing coordinate.}
\label{fig:J1-unequal-twist-gluing}
\end{figure}

For the general case, we solve the full integral \eqref{eq:J1-overlap-shadow-bracket} explicitly
in Appendix~\ref{app:J1-coordinate-overlap}; here we briefly explain the resulting structure. As each CFT wave function is a three-point correlator of operators of dimensions $\Delta$, $2s$ and $0$, their pairing through a weight-$s$ leg, see Figure~\ref{fig:J1-unequal-twist-gluing}, gives
a global conformal partial wave. This means that the solution is given as a combination
of a block of weights $(s,s)$ and its shadow block of weights
$(1-s,1-s)$. Writing the external points as
$(x_A,y_A)=(\lambda_A^{-1},\lambda_A)$ and
$(x_B,y_B)=(\lambda_B,\lambda_B^{-1})$, we introduce the cross-ratios
\begin{equation}
  \zeta
   =\frac{(x_A-y_A)(x_B-y_B)}{(x_A-x_B)(y_A-y_B)}
   =\frac{(1-\lambda_A^2)(1-\lambda_B^2)}
          {(1-\lambda_A\lambda_B)^2}
   =1-\chi,\qquad
  \chi=\left(\frac{\lambda_A-\lambda_B}{1-\lambda_A\lambda_B}\right)^2, \label{eq:cross-ratios-J1}
\end{equation}
with the dotted cross-ratios defined analogously.\footnote{On the Euclidean
slice, $\dot\lambda=\bar\lambda$ implies $\dot\chi=\bar\chi$ and
$\dot\zeta=\bar\zeta$, without requiring $|\lambda|=1$.}
Using these cross-ratios we can build the aforementioned blocks as $k_s(h_A,h_B;\zeta)$ and $k_{1-s}(h_A,h_B;\zeta)$, where explicitly
\begin{equation}
  k_a(p,q;\zeta)\equiv \zeta^a\,
  {}_2F_1\!\left(a-p,a+q;2a;\zeta\right)
  \label{eq:J1-global-block}
\end{equation}
is the undotted block with exchanged weight $a$ and external weight
differences $p,q$. The full solution is built by combining the dotted and undotted blocks, but single-valuedness on the Euclidean plane
allows only the diagonal products $k_s\dot k_s$ and
$k_{1-s}\dot k_{1-s}$ and fixes their relative coefficient. Restoring the external kinematic prefactor and the overall normalisation fixed in Appendix~\ref{app:J1-coordinate-overlap} gives
\begin{equation}\label{eq:J1CompanionOverlapClosed}
  \begin{aligned}
  \LRpair{\psi_A^{\rm C}}{\psi_B^{\rm C}}
  ={}
  \chi^{-\frac{h_A-h_B}{2}}
      \dot\chi^{-\frac{\dot h_A-\dot h_B}{2}}&\Bigl[
      (-1)^{S_A}\,k_s(h_A,h_B;\zeta)\,
                 k_s(\dot h_A,\dot h_B;\dot\zeta)
      \\
  &\quad
      +(-1)^{S_B}\beta_s\,k_{1-s}(h_A,h_B;\zeta)\,
                  k_{1-s}(\dot h_A,\dot h_B;\dot\zeta)
    \Bigr]\,.
  \end{aligned}
\end{equation}
Near $\zeta=1$, each function $k_a(h_A,h_B;\zeta)$, with $a=s,1-s$, is a mix of regular and irregular $(1-\zeta)^{h_A-h_B}$ series. The mixed terms proportional to
$(1-\zeta)^{h_A-h_B}$ and $(1-\dot\zeta)^{\dot h_A-\dot h_B}$
are generically multivalued. Their coefficients must therefore vanish
separately; the hypergeometric connection formula then fixes
\begin{equation}
 \beta_s
 =\frac{\Gamma(2s)\Gamma(2s-1)}{\Gamma(1-2s)\Gamma(2-2s)}
 \frac{\Gamma(1-s+\dot h_A)\Gamma(1-s-\dot h_A)
       \Gamma(1-s+\dot h_B)\Gamma(1-s-\dot h_B)}
      {\Gamma(s+h_A)\Gamma(s-h_A)\Gamma(s+h_B)\Gamma(s-h_B)}.
\end{equation}

\paragraph{The logarithm and the norm.}
The overlap~\eqref{eq:J1CompanionOverlapClosed} shows the logarithmic divergence of Section~\ref{sec:coordinate-log-norm} explicitly. Take
$h_A=h_B=\dot h_A=\dot h_B=h$ and $s=\tfrac14$, and let
$\boldsymbol\lambda_B\to\boldsymbol\lambda_A$, so that
$\chi,\dot\chi\to0$. Each block is logarithmic near $\zeta=1$;
the coefficient $\beta_s$ cancels the double logarithm term
$\log\chi\log\dot\chi$, leaving
\begin{equation}
 \LRpair{\psi_A^{\rm C}}{\psi_B^{\rm C}}
 =-\frac{2\pi^2\,\log(\chi\dot\chi)}
 {\cos(2\pi h)\,\Gamma(\tfrac14-h)^2\Gamma(\tfrac14+h)^2}
 +O(1).
 \label{eq:coordinate-J1-logarithmic-overlap}
\end{equation}
On the Euclidean slice, $\dot\chi=\bar\chi$ and
$\log(\chi\dot\chi)=4\log\left|
(\lambda_A-\lambda_B)/(1-\lambda_A\lambda_B)\right|$.

Comparing this divergence with
\eqref{eq:coordinate-general-norm-relation} gives the two-point normalisation for $J=1$ states\footnote{The twist separation is linear in the point-splitting distance.
However, both endpoint pairs collide. With
$\varepsilon=|(\lambda_A-\lambda_B)/(1-\lambda_A\lambda_B)|$, the diagonal-frame
radial cutoffs scale as $\rho_{\min}\sim\varepsilon$ and
$\rho_{\max}\sim\varepsilon^{-1}$, so
\[
 \int_{\varepsilon}^{1/\varepsilon}\frac{\mathrm d\rho}{\rho}
 =2\log(1/\varepsilon).
\]
The coefficient of $\log(1/\varepsilon)$ is therefore $2\mathcal C_{\log}$;
$\mathcal C_{\log}$ counts the UV logarithm at fixed $R$.}
\begin{equation}
 \mathcal N_2
 =-\partial_h
       \frac{\Gamma(\tfrac34-h)\Gamma(\tfrac34+h)}
            {\Gamma(\tfrac14-h)\Gamma(\tfrac14+h)}.
 \label{eq:coordinate-J1-two-point-normalisation}
\end{equation}
Here we used \eqref{eq:J1-exact-coupling}: for scalar states
$\Delta=2h$, differentiation at fixed twist gives
$
 \mathcal D_\xi=\frac{2\xi^2}{\partial_h\xi^2}
$.

\subsection{Overlaps using Q-functions}
\label{sec:J1-overlap-q-bilinears}
Let us now re-express \eqref{eq:J1-overlap-shadow-bracket} using $\mathbb{Q}_{+}$, setting $\theta=0$ without loss of generality. For $J=1$, the diagonal-frame kernel in
\eqref{eq:integral-Q-operator} is explicitly given by
\begin{equation}
 \begin{aligned}
 \mathcal Q_{+}(u,\dot u\mid z,\bar z;w,\bar w)
 ={}&\lambda^{\ii u}\dot\lambda^{\ii\dot u}
 [w-z]^{\{\ii u-s,\ii\dot u-s\}}\\
 &\times[\lambda^2w-z]^{\{-\ii u-s,-\ii\dot u-s\}}
 [(1-\lambda^2)w]^{2s-1}.
 \end{aligned}
 \label{eq:J1-integral-Q-kernel}
\end{equation}
Acting with $\hat{\mathbb{Q}}_+$ on the diagonal-frame wave function and changing
to companion coordinates gives the following Mellin-transform representation of $\mathbb{Q}_+$:
\begin{equation}
 \begin{aligned}
 \mathbb Q_{+}(\svx,\dot\svx)
 =\frac{M_{\psi^{\rm C}}(\svx,\dot\svx)}
 {2\pi^2\mathcal N_1^{\mathrm{SoV}}\mathcal A\;\mu_s(\svx,\dot\svx)}\,,
 \quad
 M_{\psi^{\rm C}}(\svx,\dot\svx)
 \equiv \int\dd^2z\,[z]^{-\ii\{\svx,\dot\svx\}+s-1}\psi^{\rm C}(z,\bar z)\,,
 \end{aligned}
 \label{eq:J1-Q-mellin}
\end{equation}
see
Appendix~\ref{app:J1-Q-mellin} for additional details. Here
the SoV normalisation factor is
\begin{equation}
 \mathcal N_1^{\mathrm{SoV}}\equiv \frac{\Gamma(1-2s)}{2\pi\Gamma(2s)}\,,
 \label{eq:J1-sov-resolution-normalisation}
\end{equation}
and the constant $\mathcal A$ is introduced to enforce the normalisation $\Gamma^{11}=1$, see~\eqref{eq:J1-Qplus-closed} for $\Gamma^{22}$. It is fixed by the value~\eqref{eq:companion-origin-normalisation} of $\psi^{\rm C}$ at the origin, see Appendix~\ref{app:J1-Q-mellin}, which gives
\begin{equation}
 \mathcal A
 =\frac{\psi^{\rm C}(0)}{\mathbb Q_{+}(-\ii s,-\ii s)}
 =\frac{[\lambda^2-1]^s\,[-\lambda]^{\{h-s,\dot h-s\}}}
 {\mathbb Q_{+}(-\ii s,-\ii s)}\,.
 \label{eq:J1-A-explicit}
\end{equation}
Finally, the single-site measure $\mu_s(u,\dot u)$ is the $J=1$, $\theta_1=\dot\theta_1=0$ specialisation of the chain measure $\mu(u,\dot u)$ in~\eqref{eq:overlaps-one-pair-measure}.

To express $\psi^{\rm C}$ in terms of $\mathbb{Q}_{+}$ we need to invert the Mellin transform in \eqref{eq:J1-Q-mellin}. To do so we introduce the
sum--integral notation
\begin{equation}
 \svx=\nu+\frac{\ii m}{2},\qquad \dot \svx=\nu-\frac{\ii m}{2},\qquad
 \int\dd^2\svx\equiv \sum_{m\in\mathbb Z}\int_{\mathbb R}\dd\nu,
 \label{eq:sov-sum-integral-notation}
\end{equation}
where the use of $\svx$ foreshadows the interpretation of $\svx$ as an SoV variable.

Using \eqref{eq:shadow-glued-Qplus} we find
\begin{equation}
 \psi^{\rm C}(z,\bar z)=\mathcal N_1^{\mathrm{SoV}}\mathcal{A}
 \int\dd^2\svx\,[z]^{\ii\{\svx,\dot\svx\}-s}\widetilde{\mathbb{Q}}_{+}(\svx,\dot\svx)\,,
 \label{eq:J1-psi-from-Q}
\end{equation}
defined outside its domain of convergence by analytic continuation in the weights.

We now apply this transform to the planar overlap
\eqref{eq:J1-overlap-shadow-bracket}. The shadow transform
maps each inverted Mellin mode to its dual, with multiplier
$\Gamma(2s)/[\pi\Gamma(1-2s)\mu_s]$. Simplifying the resulting expression gives
\begin{equation}
 \LRpair{\psi_A^{\rm C}}{\psi_B^{\rm C}}
 = \mathcal{N}_{1}^{\text{SoV}}\mathcal{A}_{A} \mathcal{A}_{B}
 \int\dd^2\svx\,\mathbb{Q}_{+,A}(\svx,\dot\svx)\,\widetilde{\mathbb{Q}}_{+,B}(\svx,\dot\svx).
 \label{eq:J1_Overlap_SoV}
\end{equation}
We will now evaluate \eqref{eq:J1_Overlap_SoV} by first fixing $m$ and then
moving the $\nu$-contour downwards. The twist exponentials in
\eqref{eq:baxter-asymptotics} cancel between the two chiral factors at
fixed $m$, leaving only powers of $\nu$.\footnote{For this deformation
we assume unimodular twists,
$\dot\lambda=\bar\lambda=\lambda^{-1}$; general twists are
understood by analytic continuation.}
We first work with weights $h,\dot h$ for which the contribution
at infinity vanishes, and understand the result elsewhere by analytic
continuation. We will consider $0<s<\tfrac12$ to ensure that $s\neq 1-s$.

\begin{figure}[htbp]
\centering
\includegraphics[width=0.95\linewidth]{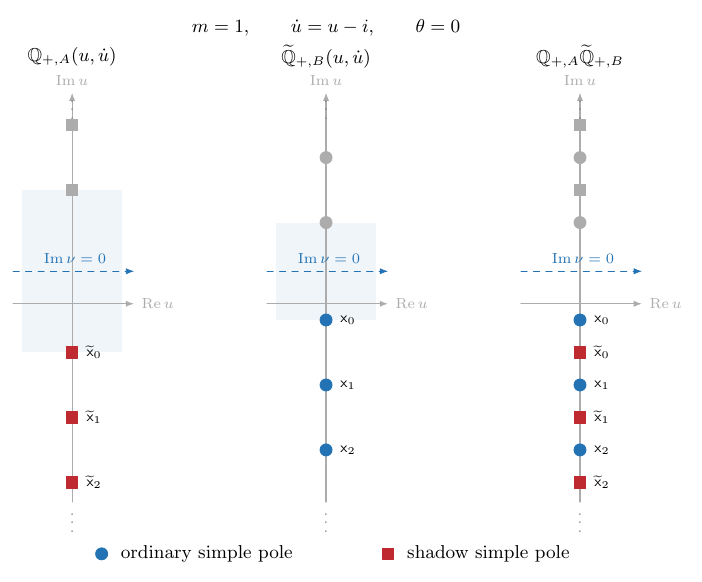}
\caption{Complementary pole towers at $m=1$ ($\dot u=u-\ii$) and
$\theta=0$. Labels are for general $s$; positions are drawn at
$s=\tfrac14$. Left: $\mathbb Q_{+,A}$ has possible simple shadow
poles (red squares). Centre: $\widetilde{\mathbb Q}_{+,B}$ has possible
simple ordinary poles (blue circles). Right: their product has both
towers. The dashed line is the integration contour
$\Im \nu=0$ ($\Im u=\tfrac12$); shading marks
the pole-free strip of each Q-function. Upper poles are grey and
unlabelled. The towers have unit spacing and are separated by $1-2s$.}
\label{fig:J1-nu-poles-m1}
\end{figure}

\paragraph{Two complementary pole towers.}
The two factors $\mathbb Q_{+,A}$ and $\widetilde{\mathbb Q}_{+,B}$
in the integrand have complementary poles, as shown in
Figure~\ref{fig:J1-nu-poles-m1}. To label these poles, we restore the inhomogeneities
$\theta_\alpha$ and the site index $\alpha=1,\ldots,J$, setting the stage for $J\geq 2$.

The shadow Q-function has possible poles on the ordinary tower,
\begin{equation}
 \begin{aligned}
 u=\svx_{\alpha,n}&\equiv \theta_\alpha-\ii(s+n),\\
 \dot u=\dot \svx_{\alpha,\dot n}&\equiv \dot\theta_\alpha-\ii(s+\dot n),
 \end{aligned}
 \qquad n,\dot n\geq0\,.
 \label{eq:J1-overlap-lower-pole-families}
\end{equation}

The original Q-function instead has possible poles on the shadow tower,
\begin{equation}
 \begin{aligned}
 u=\widetilde \svx_{\alpha,n}&\equiv \theta_\alpha-\ii(1-s+n),\\
 \dot u=\dot{\widetilde\svx}_{\alpha,\dot n}&\equiv \dot\theta_\alpha-\ii(1-s+\dot n),
 \end{aligned}
 \qquad n,\dot n\geq0\,.
 \label{eq:J1-glued-Q-fixed-m-poles}
\end{equation}

These are exchanged by the shadow map \eqref{eq:shadow-glued-Qplus}.
The two arrow representations
\eqref{eq:glued-Qplus} and \eqref{eq:glued-Qplus-arrow-switch}
require both tower indices to be non-negative, with $\dot n=n+m$.
For $J=1$, the pole-free strips around the real $\nu$-contour are
$|\Im \nu|<1-s+\tfrac{|m|}{2}$ for $\mathbb Q_+$ and
$|\Im \nu|<s+\tfrac{|m|}{2}$ for
$\widetilde{\mathbb Q}_+$.
At either tower one factor has a possible simple pole and the other
is regular. Moving the contour downwards therefore picks up simple
poles at $\svx_n$ and $\widetilde\svx_n$, with
$n\geq\max(0,-m)$ and $\dot n=n+m$.

\paragraph{Chiral residue brackets.}
To factorise the residues, we use the q-bilinear expressions~\eqref{eq:glued-Qplus} and~\eqref{eq:glued-Qplus-arrow-switch} to write
\begin{equation}
 \begin{aligned}
 \mathbb Q_{+,A}(u,\dot u)
 =\Gamma_A^{ik}q_i^{A\downarrow}(u)\dot q_k^{A\uparrow}(\dot u),
 \quad
 \widetilde{\mathbb Q}_{+,B}(u,\dot u)
 =\Gamma_B^{jl}\widetilde q_j^{B\uparrow}(u)
                    \dot{\widetilde q}_l^{B\downarrow}(\dot u).
 \end{aligned}
 \label{eq:J1-overlap-mixed-q-pairs}
\end{equation}

At an ordinary point the pole comes from
$\dot{\widetilde q}_l^{B\downarrow}$, while the undotted pair is
regular; at a shadow point it comes from $q_i^{A\downarrow}$,
while the dotted pair is regular. To sum the lower poles at site $\alpha$,
we define
\begin{equation}
 (f)_\alpha\equiv
 \frac{1}{2\pi\ii}
 \int_{\cap_{\alpha}} f(u)\,\dd u .
 \label{eq:lower-tower-contour}
\end{equation}
For distinct real inhomogeneities, $\cap_{\alpha}$ is the
counterclockwise boundary of the strip
$|\Re(u-\theta_\alpha)|<\varepsilon_\alpha$ with $\varepsilon_\alpha$ small enough
to exclude other sites.
The top segment joins the two descending sides, centred at
$\Re(u-\theta_\alpha)=0$. Hence \eqref{eq:lower-tower-contour} is nothing but the sum of residues of poles with real part $\theta_{\alpha}$ in the lower half-plane. As written, \eqref{eq:lower-tower-contour} is generically divergent, and we define it
by $\zeta$-regularisation, following the 4D fishnet prescription
of \cite{Cavaglia:2021mft}: subtract the
large-$n$ asymptotic tail, sum the convergent remainder, and restore
the tail by analytic continuation using the Lerch transcendent
(or the Hurwitz zeta function when the twist factor is unity).
This is also the prescription used by \texttt{BSolver2D}.
As concrete examples relevant for $J=1$, we have for $f$ an analytic function
\begin{equation}
  (\widetilde q_i^\downarrow f q_j^\uparrow)_\alpha
 =\sum_{n\geq0}
 \operatorname*{Res}_{u=\svx_{\alpha,n}}
 [f \widetilde q_i^\downarrow q_j^\uparrow ]\,,
 \quad
 (q_i^\downarrow f \widetilde q_j^\uparrow)_{\alpha} = \sum_{n\geq 0}\operatorname*{Res}_{u=\widetilde\svx_{\alpha,n}} [f q_i^\downarrow \widetilde q_j^\uparrow]\,.
 \label{eq:J1-ordinary-residue-bilinear}
\end{equation}
The dotted brackets are
defined in the same way. For $J=1$ we suppress $\alpha$ when convenient.

To factorise the contour residues, we use the conversion coefficients
$k_\alpha(s)$ and $k_\alpha(1-s)$ defined
in~\eqref{eq:J1-measure-conversion-coefficients}. The ordinary coefficient
uses the residue of $\mu^\downarrow$ at its pole, and the shadow
coefficient uses the derivative of $\mu^\downarrow$ at its zero.
For one site we have the relation $k(1-s)=-k(s)$.

At an ordinary pole $u=\svx_n$, the undotted factor is regular.
At a shadow pole $\dot u=\dot{\widetilde\svx}_{\dot n}$, the dotted
factor is regular. The shadow map~\eqref{eq:shadow-q-map} converts
these values into residues,
\begin{align*}
 \left.q_i^{A\downarrow}\widetilde q_j^{B\uparrow}\right|_{u=\svx_n}
 &=k^{\mathrm{c}}(s)
 \operatorname*{Res}_{u=\svx_n}
 \bigl[\widetilde q_i^{A\downarrow}q_j^{B\uparrow}\bigr],\\
 \left.\dot q_k^{A\uparrow}\dot{\widetilde q}_l^{B\downarrow}
 \right|_{\dot u=\dot{\widetilde\svx}_{\dot n}}
 &=k^{\mathrm{c}}(1-s)
 \operatorname*{Res}_{\dot u=\dot{\widetilde\svx}_{\dot n}}
 \bigl[\dot{\widetilde q}_k^{A\uparrow}\dot q_l^{B\downarrow}\bigr].
\end{align*}
The coefficients $k^{\mathrm{c}}$ are defined in~\eqref{eq:contour-conversion-coefficients}. The tilde thus moves from $B$ to $A$ in whichever chiral factor
is regular. See Appendix~\ref{app:J1-overlap-residues} for both residue
products.
\paragraph{The overlap in terms of bilinears.}
Summing over $m$ makes the two tower indices independent:
\begin{equation*}
 \sum_{m\in\mathbb Z}\ \sum_{n=\max(0,-m)}^\infty
 \quad\longrightarrow\quad
 \sum_{n=0}^\infty\ \sum_{\dot n=0}^\infty,
 \qquad m=\dot n-n.
\end{equation*}
After factorising each tower contribution, the independent sums give
the contour brackets~\eqref{eq:lower-tower-contour}. Including the
clockwise contour factor $-2\pi\ii$ gives
\begin{equation}
 \begin{aligned}
 &\LRpair{\psi_A^{\rm C}}{\psi_B^{\rm C}}
 =-2\pi\ii\,\mathcal N_1^{\mathrm{SoV}}\mathcal A_A\mathcal A_B
 \,\Gamma_A^{ik}\Gamma_B^{jl}\\
 &\quad\times\Bigl[
 k^{\mathrm{c}}(1-s)\,
 \bigl(q_i^{A\downarrow}\widetilde q_j^{B\uparrow}\bigr)\,
 \bigl(\dot q_l^{B\downarrow}\dot{\widetilde q}_k^{A\uparrow}\bigr)
 +k^{\mathrm{c}}(s)\,
 \bigl(\widetilde q_i^{A\downarrow}q_j^{B\uparrow}\bigr)\,
 \bigl(\dot{\widetilde q}_l^{B\downarrow}\dot q_k^{A\uparrow}\bigr)
 \Bigr].
 \end{aligned}
 \label{eq:J1-overlap-q-bilinears-skeleton}
\end{equation}
The first product comes from the shadow tower and the second from the
ordinary tower.

We use bold braces for the site bracket that combines the two tower
contributions from the contour brackets~\eqref{eq:lower-tower-contour}:
\begin{equation}
 \begin{aligned}
 \towerbracket{F}{G}_\alpha
 &\equiv k_\alpha(s)
   \bigl(\mu^\downarrow(u)F(u)\bigr)_\alpha
   \bigl(\mu^\downarrow(\dot u)G(\dot u)\bigr)_\alpha\\
 &\quad+k_\alpha(1-s)
   \bigl(\mu^\uparrow(u)F(u)\bigr)_\alpha
   \bigl(\mu^\uparrow(\dot u)G(\dot u)\bigr)_\alpha .
 \end{aligned}
 \label{eq:paired-tower-bracket}
\end{equation}
The functions $F$ and $G$ contain the undotted and dotted factors,
respectively. The contour variables $u$ and $\dot u$ are independent.
For example, $\towerbracket{u^{n-1}F}{\dot u^{m-1}G}_\alpha$ inserts
$u^{n-1}$ in the undotted integrands and $\dot u^{m-1}$ in the dotted ones.
The first term pairs the ordinary towers, and the second pairs the
shadow towers. The explicit
measures give the conversion coefficients $k_\alpha$ of
\eqref{eq:J1-measure-conversion-coefficients}. Applying the shadow
map~\eqref{eq:shadow-q-map} to the integrands gives the factors
$k_\alpha^{\mathrm{c}}$ used in~\eqref{eq:J1-overlap-q-bilinears-skeleton}.
At $J=1$ we suppress the site index, so the same overlap is
\begin{equation*}
 \LRpair{\psi_A^{\rm C}}{\psi_B^{\rm C}}
 =-2\pi\ii\,\mathcal N_1^{\mathrm{SoV}}\mathcal A_A\mathcal A_B
 \,\Gamma_A^{ik}\Gamma_B^{jl}\,
 \towerbracket{q_i^{A\downarrow}q_j^{B\uparrow}}
              {\dot q_k^{A\uparrow}\dot q_l^{B\downarrow}}.
\end{equation*}

Appendix~\ref{app:bsolver} gives a runnable
\texttt{BSolver2D} check of the complete normalised overlap.

This completes our analysis of the $J=1$ case. Over the next few sections we will generalise the construction to any $J$.

\section{Separation of variables}
\label{sec:sov}
Away from $J=1$, coordinate-space wave functions quickly become complicated. In this section we introduce a separation-of-variables basis in which the wave function factorises into a product of ${\mathbb Q}_+$ functions. These functions can be computed for any state using QSC methods and, in practice, for essentially any finite length $J$ with our package \texttt{BSolver2D}.

\subsection{Separation of variables basis in the companion frame}

Separation of variables for quantum spin chains was pioneered by Sklyanin~\cite{Sklyanin:1991ss,Sklyanin:1995bm} and applied to non-compact $SL(2,\mathbb C)$ chains in~\cite{Derkachov:2001yn}. We will employ a recent formulation~\cite{Cavaglia:2019pow,Gromov:2019wmz} originally envisioned to deal with higher-rank generalisations.

\paragraph{Bras and kets.}
Throughout this section, bra--ket notation refers to the companion pairing~\eqref{eq:companion-pairing}. A ket $|g\rangle$ is a function $g$, a bra $\langle f|$ is the functional $g\mapsto\langle f|g\rangle$, and operators act on bras through the transpose of Section~\ref{sec:setup-bracket}:
\begin{equation}
 \langle f|\hat{O}\equiv\langle \hat{O}^{\mathsf T_{\langle\rangle}}f|,
 \qquad
 \langle f|\hat{O}|g\rangle=\langle f|\hat{O}g\rangle .
 \label{eq:sov-bra-convention}
\end{equation}

A left eigenvector $\langle f|\hat{O}=O\,\langle f|$ is therefore an eigenfunction of $\hat{O}^{\mathsf T_{\langle\rangle}}$. Every SoV bra below is defined, and contracted, with this same pairing.

\paragraph{$\hat{\Bgood}$-operator.}
The starting point of our analysis is the separated variable operator $\hat{\Bgood}$ defined, in the companion frame, as
\beq
 \hat \Bgood (u)\equiv \hat{T}_{11}(u).
\eeq
The RTT relations imply $[\hat{\Bgood}(u),\hat{\Bgood}(v)]=0$.
Since $\hat{\Bgood}(u)$ is a polynomial in $u$ of degree $J$, so are its eigenvalues, and we label its common left eigenvectors by the roots $\svx_\alpha$, and its right eigenvectors by $\svy$:
\beq
 \langle\svx,\dot\svx|\hat{\Bgood}(u)
 =\prod_{\alpha=1}^{J}(u-\svx_\alpha)\langle\svx,\dot\svx|\;,
 \quad
\hat{\Bgood}(u)| \svy,\dot\svy \rangle
 =\prod_{\alpha=1}^{J}(u-\svy_\alpha)|\svy,\dot\svy \rangle\,,
 \label{eq:sov-basis-eigen}
\eeq
where we have collected $\svx=(\svx_1,\ldots,\svx_J),\,\dot{\svx}=(\dot{\svx}_1,\ldots,\dot{\svx}_J)$ and similarly for $\svy,\svyd$.
 By~\eqref{eq:sov-bra-convention} and the transposition rule~\eqref{eq:lax-transpose}, which reverses the order of the Lax operators as in~\eqref{eq:transfer-transpose-reversal}, the bra $\langle\svx,\dot\svx|$ is the function $\mathcal R\,X$, where $X$ is a right eigenvector of $\hat\Bgood$ on the reversed chain: $\hat\Bgood_{\boldsymbol\theta}^{\,\mathsf T_{\langle\rangle}}=\mathcal R\,\hat\Bgood_{\boldsymbol\theta^{\rm R}}\mathcal R$. Unless stated otherwise, all relations below also hold in the dotted sector.

\paragraph{SoV vacuum state.}
Following the convention of~\cite{Gromov:2020fwh,Gromov:2022waj},
we call the constant function $1$ the SoV vacuum $\langle\zerosov|$. Contracting the SoV vacuum with any function regular at the origin corresponds to the evaluation $\langle \zerosov|f\rangle = f(0,\dots,0)$.
 Using the monodromy matrix~\eqref{eq:spin-chain-monodromy}, we have
    \begin{equation}
 \langle\zerosov|\hat{\Bgood}(u)=P_+(u)\,\langle\zerosov|\,,
 \quad
 \langle\zerosov|\hat{T}_{21}(u)=0\,,
 \label{eq:sov-vacuum}
\end{equation}
with $P_\pm(u)=\prod_{\beta=1}^{J}(u-\theta_\beta\pm\ii s)$, see~\eqref{eq:baxter-equation} and hence
\begin{equation}
  \langle\zerosov| = \mathop{\bigotimes}\limits_{\alpha=1}^{J}\langle \svx_{\alpha,0},\dot{\svx}_{\alpha,0} |
  \quad
  \text{with}
  \quad
  \svx_{\alpha,0} = \dot{\svx}_{\alpha,0} = \theta_\alpha -\ii s\,.
\end{equation}

Finally, we fix the remaining normalisation of each SoV vector by its overlap with the reference state~\eqref{eq:reference-state-frames}, as in~\cite{Gromov:2020fwh,Gromov:2022waj}:
\beq
 \begin{aligned}
 \langle\svx,\dot\svx|\Omega_J^{\rm C}\rangle
 &=\mathfrak n_\lambda
 \prod_{\alpha=1}^{J}
 \lambda^{\ii(\svx_\alpha-\svx_{\alpha,0})}
 \dot\lambda^{\ii(\dot\svx_\alpha-\dot\svx_{\alpha,0})},\\
 \langle\Omega_J^{\rm C}|\svy,\dot\svy\rangle
 &=\mathfrak n_\lambda
 \prod_{\alpha=1}^{J}
 \lambda^{\ii(\svy_\alpha-\svx_{\alpha,0})}
 \dot\lambda^{\ii(\dot\svy_\alpha-\dot\svx_{\alpha,0})},
 \end{aligned}
 \label{eq:sov-reference-normalisation}
\eeq
with the normalisation $\mathfrak n_\lambda$ as
defined in~\eqref{eq:diagonal-companion-wavefunction-norm}, so that
$\langle\zerosov|\Omega_J^{\rm C}\rangle=\langle \Omega_J^{\rm C}|\zerosov\rangle = \mathfrak n_\lambda$.

\paragraph{Transfer matrix in separated variables.}
The operators $\hat{T}_{ij}$ satisfy the RTT relations. They imply that $\hat{T}_{12}(\svx_\alpha)$
lowers $\svx_\alpha$ by $\ii$, while $\hat{T}_{21}(\svx_\alpha)$ raises
it~\cite{Sklyanin:1995bm}. With the normalisation in~\eqref{eq:sov-reference-normalisation}, their action on the SoV basis is
\begin{equation}
 \begin{aligned}
 \langle \svx |\hat{T}_{12}(\svx_\alpha)
 =+P_-(\svx_\alpha)\langle \svx-\ii\mathtt{ e}_\alpha|\,,
 \quad
 \langle \svx |\hat{T}_{21}(\svx_\alpha)
 =-P_+(\svx_\alpha)\langle \svx+\ii \mathtt{e}_\alpha|\,.
 \end{aligned}
 \label{eq:sov-covector-shifts}
\end{equation}
 Here $\mathtt{e}_\alpha$ is a unit vector, shifting only the $\alpha$th component.
Since $P_+(\svx_{\alpha,0})=0$ we have
\begin{equation}
 \langle \svx |\hat{T}_{21}(\svx_{\alpha,0})=0,
 \qquad \svx_\alpha=\svx_{\alpha,0}\,.
 \label{eq:sov-lowest-root-annihilation}
\end{equation}
Let us now study the overlap between the SoV basis and the wave function. We start by defining the wave function in separated coordinates
\begin{equation}
 \;\Phi(\svx,\dot{\svx})
 \equiv \langle \svx,\dot{\svx}|\psi^{\rm C}\rangle,\;
 \label{eq:sov-separated-wavefunction-definition}
\end{equation}
whose essential property is that it satisfies a Baxter equation in each component of $\svx$ and $\svxd$ separately. To derive this, we start from the companion frame transfer matrix~\eqref{eq:twisted-transfer-matrix}
\begin{equation}
 \widehat t^{C}(u)
 =(\lambda+\lambda^{-1})\hat{T}_{11}(u)-\hat{T}_{21}(u)+\hat{T}_{12}(u),
 \qquad
 \widehat{t}^C(u)|\psi^{\rm C}\rangle
 =t^C(u)|\psi^{\rm C}\rangle,
 \label{eq:sov-companion-transfer}
\end{equation}
and contract with $\langle \svx|$ to find
\begin{equation}
 t^C(\svx_\alpha)\Phi(\svx)
 =P_+(\svx_\alpha)\Phi(\svx+\ii\mathtt{e}_\alpha)
 +P_-(\svx_\alpha)\Phi(\svx-\ii\mathtt{e}_\alpha).
 \label{eq:sov-wavefunction-baxter}
\end{equation}

Thus $\Phi$ obeys the Baxter equation~\eqref{eq:baxter-equation}
in every undotted and dotted variable.

Equation~\eqref{eq:sov-wavefunction-baxter} specifies the transfer matrix action
at $J$ distinct points $\svx_\alpha$.
Together with the known leading coefficient $\lambda+\lambda^{-1}$, these
values reconstruct the full degree-$J$ polynomial by Lagrange interpolation, specifying all integrals of motion
$\bI_r$ in~\eqref{eq:transfer-matrix-expansion} in the SoV basis. This is Sklyanin's SoV construction~\cite{Sklyanin:1991ss}.

The Baxter equation also strongly constrains the SoV wave function~\eqref{eq:sov-separated-wavefunction-definition}. Starting from the constant vacuum vector, successive actions of $\hat T_{12}$ generate the lower ordinary towers
\begin{equation}
 \svx_{\alpha,n_\alpha}
 =\theta_\alpha-\ii(s+n_\alpha),\qquad
 \dot \svx_{\alpha,\dot n_\alpha}
 =\dot\theta_\alpha-\ii(s+\dot n_\alpha),\qquad
 n_\alpha,\dot n_\alpha\in\mathbb Z_{\ge0}.
 \label{eq:sov-lower-towers}
\end{equation}
Equation~\eqref{eq:sov-lowest-root-annihilation} thus tells us that the Baxter equation becomes a recursion relation with the initial value
\begin{equation}
 \Phi(\svx_{\alpha,0},
\svxd_{\alpha,0})
 =\psi^{\rm C}(0,\ldots,0),
 \label{eq:sov-lowest-projection}
\end{equation}
where we used that $\langle\zerosov|f \rangle = f(0,\dots,0)$. Hence we can reconstruct $\Phi$ uniquely at every point on the ordinary tower, in terms of this initial value. However, we have not yet fixed $\Phi$ away from these points. This is what we turn to next.

\subsection{SoV wave function is the glued \texorpdfstring{$\Qp$}{Q+}}
\label{sec:sov-Qplus}

Since the wave function in separated coordinates~\eqref{eq:sov-separated-wavefunction-definition} satisfies the Baxter
equation~\eqref{eq:sov-wavefunction-baxter} in every separated variable,
it can be expanded in products of undotted and dotted
Baxter solutions with $\ii$-periodic coefficients. For $J=1$, \eqref{eq:J1-Q-mellin} shows that $\Phi(\svx,\svxd)\propto\mathbb{Q}_{+}(\svx,\svxd)$; we now argue that this persists for arbitrary lengths.

\paragraph{Regularity.}
The SoV wave function~\eqref{eq:sov-separated-wavefunction-definition} should be regular at $\svx_{\alpha,0}+\ii$, so that the factor $P_+(u)$ removes one term from the recursion. This boundary condition suggests how to pair
the undotted and dotted Baxter solutions.
The down-arrow functions are regular at these starting points,
so one might naively take any pair $q_a^\downarrow\dot q_a^\downarrow$
(with equal indices to ensure power-law asymptotics).
However, the inverted vacuum $\langle\infty|$, obtained from
$\langle\zerosov|$ by the inversion $\mathcal D$ of~\eqref{eq:full-inversion}
(see Section~\ref{sec:sov-projected-wavefunction}), is annihilated by
$\hat T_{12}$; the recursion then proceeds upwards in the complex plane and
requires regularity at $\theta_\alpha-\ii(1-s)$.
Here $q^\downarrow$ generally has a pole, as does $\dot q^\downarrow$ in the dotted
sector. Thus a generic down--down product cannot satisfy both boundary conditions.

The compromise is to take a mixed bilinear,
$\sum_{a,b=1}^{2} q_a^\downarrow C^{ab}\dot q_b^\uparrow$, holding the other pairs of separated variables fixed.
The undotted analyticity condition is satisfied, but the dotted recursion
may encounter a pole unless the matrix $C^{ab}$ is chosen carefully. We therefore require, for all non-negative $n$,
\begin{equation}
 \sum_{a,b}q^\downarrow_a(\svx_{\alpha,n})\,C^{ab}\,
 \operatorname*{Res}_{\dot\svx=\dot\svx_{\alpha,0}+\ii}
 \dot q^\uparrow_b(\dot\svx)=0,\qquad n\ge0 .
 \label{eq:sov-exit-pole-cancellation}
\end{equation}
For $C=\Gamma$, this is the pole cancellation in
\eqref{eq:setup-Qplus-first-pole-regularity} of Section~\ref{sec:setup-baxter}.
The two regular regions
of $\Qp$ cover all four exit points of the recursions.
Since~\eqref{eq:sov-exit-pole-cancellation} must hold for every $n$, together with the gluing condition~\eqref{eq:setup-omega-gluing} it fixes $C\propto\Gamma$, so this combination is unique up to an overall normalisation. We therefore obtain
\begin{equation}
\boxed{\Phi (\svx,\svxd) =\mathcal{A}\prod_{\alpha=1}^J\Qp(\svx_\alpha,\dot\svx_\alpha)\,.}
 \label{eq:sov-Qplus-statement}
\end{equation}
Using~\eqref{eq:companion-origin-normalisation}, the lowest-vector
condition fixes the proportionality factor
\begin{equation}
 \mathcal A
 =\frac{(\lambda^2-1)^{Js}
          (\dot\lambda^2-1)^{Js}}
         {(-\lambda)^{Js-h}
          (-\dot\lambda)^{Js-\dot h}}
 \prod_{\alpha=1}^{J}
 \Qp(\svx_{\alpha,0},\dot\svx_{\alpha,0})^{-1}.
 \label{eq:sov-state-normalisation}
\end{equation}
This motivates the factorisation on the towers. The argument is not
fully rigorous, but it explains why the special combination
${\mathbb Q}_+$ appears. The factorisation itself was proven in~\cite{Derkachov:2001yn}, where the SoV
states were constructed from commuting Q-operators.

\subsection{Determinant form of the SoV scalar product}
\label{sec:overlaps-resolution}
The factorisation~\eqref{eq:sov-Qplus-statement} reduces scalar products
to one-pair moments once the SoV measure is specified.
For the noncompact $SL(2,\mathbb C)$ chain, the SoV measure and
eigenfunctions were constructed in~\cite{Derkachov:2001yn,DerkachovManashov:2014},
and completeness in the unitary principal series was proved
in~\cite{Manashov:2023SoV}. Here we assume their continuation to the
bilinear ordinary--shadow pairing at our spins and twists.
At $J=1$ this pairing follows from the Mellin--Parseval
identity~\eqref{eq:J1_Overlap_SoV}.

To express the planar bracket in separated variables, we also need the shadow SoV wave function. Using the shadow convention~\eqref{eq:shadow-glued-Qplus}, we define it as
\begin{equation}
 \widetilde\Phi(\svx,\svxd)
 \equiv \mathcal A\prod_{\alpha=1}^{J}
 \widetilde{\mathbb Q}_{+}(\svx_\alpha,\dot\svx_\alpha),
 \label{eq:overlaps-shadow-wavefunction}
\end{equation}
where $\mathcal A$ is the same factor as in~\eqref{eq:sov-state-normalisation}.

As at $J=1$, we express the overlap as a sum--integral over the separated variables by inserting the SoV resolution of the identity. With the continuation assumed above and the notation of~\eqref{eq:sov-sum-integral-notation}, we obtain
\begin{equation}
 \LRpair{\psi_A}{\psi_B}
 =\mathcal N_J^{\mathrm{SoV}}
 \int\prod_{\alpha=1}^{J}\dd^2\svx_\alpha\,
 \Delta(\svx)\Delta(\svxd)
 \Phi_A\widetilde\Phi_B.
 \label{eq:overlaps-sov-sum-integral}
\end{equation}
Here $\Delta(\svx)=
\det \svx_{\alpha}^{\beta-1}=\prod_{\alpha<\beta}(\svx_\beta-\svx_\alpha)$ is the Vandermonde determinant, and the first state belongs to the reversed chain, as
in~\eqref{eq:left-right-bracket}.
As in the one-site case, we absorb the one-pair spectral measure~\eqref{eq:overlaps-one-pair-measure},
including contributions from all sites, into the wave function of the second state through
$\widetilde{\mathbb Q}_{+,B}$.
The overall normalisation $\mathcal N_J^{\mathrm{SoV}}$ is fixed by requiring
$\LRpair{1}{1}=1$~\eqref{eq:one-site-vacuum-norm}, resulting in
\begin{equation}
 \mathcal N_J^{\mathrm{SoV}}
 =\frac{1}{J!}\left[\frac{\Gamma(1-2s)}{2\pi\Gamma(2s)}\right]^J
 \prod_{\alpha<\beta}
 \frac{\Gamma(1-2s+\ii\theta_{\alpha\beta})
       \Gamma(1-2s-\ii\theta_{\alpha\beta})}
      {\Gamma(2s+\ii\theta_{\alpha\beta})
       \Gamma(2s-\ii\theta_{\alpha\beta})}\,, \quad \theta_{\alpha\beta}\equiv \theta_\alpha-\theta_\beta.
 \label{eq:overlaps-resolution-normalisation}
\end{equation}

To express \eqref{eq:overlaps-sov-sum-integral} in terms of q-bilinears, it is convenient to first write it as a determinant. We therefore define the moments
\begin{equation}
 \mathcal M^{A,B}_{nm}
 \equiv \int\dd^2\svx\,
 \svx^{n-1}\dot\svx^{m-1}
 \mathbb Q_{+,A}(\svx,\dot\svx)
 \widetilde{\mathbb Q}_{+,B}(\svx,\dot\svx),
 \qquad n,m=1,\ldots,J.
 \label{eq:overlap-moments}
\end{equation}
Substituting \eqref{eq:sov-Qplus-statement} and~\eqref{eq:overlaps-shadow-wavefunction} into \eqref{eq:overlaps-sov-sum-integral} and expanding both Vandermonde determinants, we can perform the pair integrals independently to obtain
\begin{equation*}
 \begin{aligned}
 &\int\prod_{\alpha=1}^{J}\dd^2\svx_\alpha\,
 \Delta(\svx)\Delta(\svxd)
 \prod_{\alpha=1}^{J}\mathbb Q_{+,A}(\svx_\alpha,\dot\svx_\alpha)
 \widetilde{\mathbb Q}_{+,B}(\svx_\alpha,\dot\svx_\alpha)\\
 &\qquad=\sum_{\sigma,\tau\in S_J}\operatorname{sgn}(\sigma)
 \operatorname{sgn}(\tau)
 \prod_{\alpha=1}^{J}\mathcal M^{A,B}_{\sigma(\alpha)\tau(\alpha)}
 =J!\det\mathcal M^{A,B}.
 \end{aligned}
\end{equation*}
For each fixed $\sigma$ we set
$\rho=\tau\circ\sigma^{-1}$. The sum over $\tau$ then gives $\det\mathcal M^{A,B}$, and the
remaining sum over $\sigma$ supplies $J!$. We therefore obtain
\begin{equation}
 \LRpair{\psi_A}{\psi_B}
 =\mathcal N_J^{\mathrm{SoV}}\mathcal A_A\mathcal A_B
 J!\det\limits_{1 \leq n,m \leq J}\mathcal M^{A,B}_{nm}.
 \label{eq:overlaps-moment-determinant}
\end{equation}
At $J=1$ this reduces to~\eqref{eq:J1_Overlap_SoV}, and at $s=\tfrac{1}{4}$ its
prefactor is simply $\mathcal A_A\mathcal A_B/(2\pi)^J$.
The identities hold where
the moments converge and extend elsewhere by a common analytic continuation.

We now apply the lower-contour calculation of
Section~\ref{sec:J1-overlap-q-bilinears} to each moment, with matching
dotted and undotted site data $\dot\theta_\alpha=\theta_\alpha$ and
$\dot s=s$. The contour bracket~\eqref{eq:lower-tower-contour} selects
the residues at site $\alpha$, and $k_\alpha^{\mathrm{c}}$ carries the
phase for the product of the two chiral brackets. Summing over the ordinary and shadow
towers gives
\begin{equation}
 \begin{aligned}
 \mathcal M^{A,B}_{nm}
 ={}&-2\pi\ii\sum_{\alpha=1}^{J}\Gamma_A^{ik}\Gamma_B^{jl}\,\Bigl[
 k_\alpha^{\mathrm{c}}(s)\,
 \bigl(\widetilde q_i^{A\downarrow}u^{n-1}q_j^{B\uparrow}\bigr)_\alpha\,
 \bigl(\dot{\widetilde q}_l^{B\downarrow}\dot u^{m-1}\dot q_k^{A\uparrow}\bigr)_\alpha\\
 &\qquad\qquad
 +k_\alpha^{\mathrm{c}}(1-s)\,
 \bigl(q_i^{A\downarrow}u^{n-1}\widetilde q_j^{B\uparrow}\bigr)_\alpha\,
 \bigl(\dot q_l^{B\downarrow}\dot u^{m-1}\dot{\widetilde q}_k^{A\uparrow}\bigr)_\alpha
 \Bigr].
 \end{aligned}
 \label{eq:overlaps-moment-residues}
\end{equation}
At $J=1$, substituting this moment into~\eqref{eq:overlaps-moment-determinant}
reproduces~\eqref{eq:J1-overlap-q-bilinears-skeleton}.

\Needspace{6\baselineskip}

Using the site bracket~\eqref{eq:paired-tower-bracket} to write the
moments~\eqref{eq:overlaps-moment-residues}, the planar overlap
\eqref{eq:overlaps-moment-determinant} becomes
\begin{equation}
 \boxed{
 \LRpair{\psi_A}{\psi_B}
 =(-2\pi\ii)^J\mathcal N_J^{\mathrm{SoV}}
  \mathcal A_A\mathcal A_B J!\,
 \det\limits_{1\le n,m\le J}\!\left[
 \sum_{\alpha=1}^{J}\Gamma_A^{ik}\Gamma_B^{jl}\,
 \towerbracket{u^{n-1}q_i^{A\downarrow}q_j^{B\uparrow}}
              {\dot u^{m-1}\dot q_k^{A\uparrow}\dot q_l^{B\downarrow}}_\alpha
 \right].}
 \label{eq:overlaps-bilinear-determinant}
\end{equation}\par

\paragraph{Tree-level polynomial states.}
At zero coupling and at the homogeneous fishnet point, the regular Baxter functions of the tree-level states with derivatives are $q_1(u)=\lambda^{\ii u}P(u)$ and $\dot q_1(\dot u)=\dot\lambda^{\ii\dot u}\dot P(\dot u)$, with $P$ and $\dot P$ polynomials.

For these states, $\Gamma^{22}=0$ and $\mathbb Q_+=q_1\dot q_1$. Due to the regularity of $\mathbb Q_+$, only the ordinary tower contributes and we find
\begin{equation}
\det\limits_{1 \leq n,m \leq J}\mathcal M^{A,B}_{nm}
 =(-2\pi\ii)^J\prod_{\alpha=1}^{J}k_\alpha^{\mathrm{c}}(s)
 \det_{\alpha,n}
 (\widetilde q_1^{A\downarrow}u^{n-1}q_1^{B\uparrow})_\alpha\,
 \det_{\alpha,m}
 (\dot{\widetilde q}_1^{B\downarrow}
  \dot u^{m-1}\dot q_1^{A\uparrow})_\alpha.
 \label{eq:overlaps-tree-chiral-determinants}
\end{equation}
Here the $\theta_\alpha$ are kept distinct, and the homogeneous limit is taken after evaluating the brackets. This is precisely the product of two SoV determinants for non-compact $\algsl_2$ spin chains~\cite{Gromov:2022waj}.

\subsection{SoV localisation and factorisation}
\label{sec:endpoints-collapse}
\label{sec:sov-projected-wavefunction}
\label{sec:HHL-localisation}

The unequal-twist overlap~\eqref{eq:overlaps-moment-determinant}
has the kinematics of a four-point correlator with two local operators
and two twist defect end points. To obtain general three-point
structure constants, we need an additional
operation that splits a chain into shorter factors.
This is supplied by SoV localisation in the companion twist
frame~\cite{Bargheer:2026kon}: fixing a block of separated variables
at their vacuum values makes the SoV state factorise into states
of the two subchains\footnote{A similar splitting of the SoV states in $\mathfrak{su}(2)$ spin chains was carried out in~\cite{Jiang:2015lda}. }. We establish this property in this section before
applying it to three-point functions.

\begin{figure}[t]
 \centering
 \includegraphics[width=0.72\textwidth]{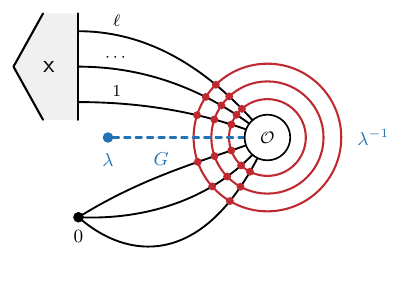}
 \caption{The localised SoV projection of a length-$J$ wave function.
 The first $\ell$ legs are paired jointly with the short-chain bra
 $\langle\svx^{(\ell)},\dot\svx^{(\ell)}|$, labelled simply
 $\langle\svx|$ in the diagram. The leg ordering follows the ket side of
 Figure~\ref{fig:bracket-gluing}.
 The remaining legs are paired with $\langle\zerosov|$, which evaluates their
 coordinates at the same point $0$. The operator is at $\lambda^{-1}$,
 and the twist cut ends at $\lambda$. Black spokes, red wheels and the blue twist cut
 follow the conventions of Figure~\ref{fig:zero-magnon-wavefunction}.}
 \label{fig:sov-localisation}
\end{figure}

Fix the last $J-\ell$ separated variables to their vacuum values,
$\svx_{\alpha,0}=\theta_\alpha-\ii s$ and
$\dot\svx_{\alpha,0}=\dot\theta_\alpha-\ii s$,
while keeping the first $\ell$ pairs generic. Writing
$\svx^{(\ell)}=(\svx_1,\ldots,\svx_\ell)$, and similarly for the dotted
variables, SoV localisation~\cite{Bargheer:2026kon} states that
\begin{equation}\label{eq:sov-localization}
 \langle\svx^{(\ell)},\dot\svx^{(\ell)},\zerosov^{(J-\ell)}|
 =
 \langle\svx^{(\ell)},\dot\svx^{(\ell)}|
 \otimes\langle\zerosov^{(J-\ell)}|,
\end{equation}
where the first factor is the SoV bra vector for the subchain with
inhomogeneities $\theta_1,\ldots,\theta_\ell$; the second is the vacuum
on sites $\ell+1,\ldots,J$. Both subchains use the
normalisation~\eqref{eq:sov-reference-normalisation}.
The resulting projection is illustrated in Figure~\ref{fig:sov-localisation}.

\paragraph{$\hat\Bgood$-operator factorisation.}
The mechanism behind the SoV localisation property is simple. Although $\hat\Bgood$
acts on the whole chain, its action on the tensor product \eqref{eq:sov-localization} reduces to the first $\ell$ sites, multiplied by the vacuum
eigenvalue of the remaining subchain. To see this, note that the ordering in~\eqref{eq:spin-chain-monodromy} gives
$\hat T(u)=\hat T^{(\ell)}(u)\hat T^{(J-\ell)}(u)$, where the factors
contain sites $1,\ldots,\ell$ and $\ell+1,\ldots,J$, respectively.
Taking the $11$ entry, since $\hat\Bgood=\hat T_{11}$, gives
\begin{equation}
 \hat\Bgood(u)=
 \hat\Bgood^{(\ell)}(u)\hat\Bgood^{(J-\ell)}(u)
 +\hat T_{12}^{(\ell)}(u)\hat T_{21}^{(J-\ell)}(u).
\end{equation}
By~\eqref{eq:sov-vacuum},
$\langle\zerosov^{(J-\ell)}|\hat T_{21}^{(J-\ell)}(u)=0$.
The cross term vanishes, and the two diagonal actions give
\begin{equation}\label{eq:B-factor-eigenvalue}
 \begin{aligned}
 &\bigl(\langle\svx^{(\ell)},\dot\svx^{(\ell)}|
       \otimes\langle\zerosov^{(J-\ell)}|\bigr)\hat\Bgood(u)\\
 &\quad=
 \prod_{\alpha=1}^{\ell}(u-\svx_\alpha)
 \prod_{\alpha=\ell+1}^{J}(u-\theta_\alpha+\ii s)
 \bigl(\langle\svx^{(\ell)},\dot\svx^{(\ell)}|
       \otimes\langle\zerosov^{(J-\ell)}|\bigr).
 \end{aligned}
\end{equation}
The dotted action follows in the same way. These are exactly the
eigenvalues of the full-chain bra on the left of
\eqref{eq:sov-localization}, by~\eqref{eq:sov-basis-eigen}.

\paragraph{Fixing the normalisation.}
For generic inhomogeneities, the joint SoV spectrum is nondegenerate,
so the two bra vectors are proportional.
To fix the proportionality constant, we pair both bras with the reference
state. This state factorises over the two subchains, and its normalisation
obeys $\mathfrak n_\lambda^{(\ell)}
\mathfrak n_\lambda^{(J-\ell)}=\mathfrak n_\lambda^{(J)}$.
The frozen variables contribute unit factors in
\eqref{eq:sov-reference-normalisation}, so both overlaps coincide and
the proportionality constant is $1$.

\paragraph{Dual SoV states.}
Similar statements hold for the corresponding ket SoV states $|\svy,\dot{\svy}\rangle$, but now with the SoV vacuum on the first
$J-\ell$ sites:
\begin{equation}\label{eq:sov-localization-ket}
 |\zerosov^{(J-\ell)},\svy^{(\ell)},\dot\svy^{(\ell)}\rangle
 =
 |\zerosov^{(J-\ell)}\rangle
 \otimes|\svy^{(\ell)},\dot\svy^{(\ell)}\rangle.
\end{equation}
Here $\svy^{(\ell)}=(\svy_{J-\ell+1},\ldots,\svy_J)$, and similarly
for the dotted variables. The short ket now belongs to the subchain
with inhomogeneities $\theta_{J-\ell+1},\ldots,\theta_J$.
Indeed, in
$\hat T(u)=\hat T^{(J-\ell)}(u)\hat T^{(\ell)}(u)$ the cross term
now vanishes because
$\hat T_{12}^{(J-\ell)}(u)|\zerosov^{(J-\ell)}\rangle=0$.
The eigenvalue comparison and reference normalisation proceed as
for the bra. Note that the reversal in the planar
pairing~\eqref{eq:left-right-bracket} exchanges the first and last
blocks, making these complementary bra and ket orderings consistent.

\paragraph{The projected wave function.} Figure~\ref{fig:sov-localisation} represents the overlap
$\langle\svx^{(\ell)},\dot\svx^{(\ell)},\zerosov^{(J-\ell)}
|\psi_J^{\rm C}\rangle$.
To express it as a short-chain overlap, define the rectangular vacuum projection
\begin{equation}
 \Pi_\ell\equiv
 1^{\otimes\ell}\otimes\langle\zerosov^{(J-\ell)}|
 :\mathcal H_J\longrightarrow\mathcal H_\ell.
 \label{eq:sov-vacuum-projector}
\end{equation}
Here $\mathcal H_J$ and $\mathcal H_\ell$ are the state spaces of
the full chain and its first $\ell$ sites.
Since $\langle\zerosov|f\rangle=f(0,\dots,0)$, $\Pi_\ell$ removes the last $J-\ell$
legs by evaluating their coordinates at zero:
\begin{equation}
(\Pi_\ell\psi_J^{\rm C})(z_1,\ldots,z_\ell)
=\psi_J^{\rm C}(z_1,\ldots,z_\ell,0,\ldots,0).
\end{equation}
The dotted coordinates are restricted in the same way.
 SoV localisation~\eqref{eq:sov-localization} then gives
\begin{equation}
 {}_\ell\langle\svx^{(\ell)},\dot\svx^{(\ell)}
 |\Pi_\ell\psi_J^{\rm C}\rangle
 =\psi_J^{\rm C}(0,\dots,0)
 \prod_{\alpha=1}^{\ell}
 \frac{\Qp(\svx_\alpha,\dot\svx_\alpha)}
      {\Qp(\svx_{\alpha,0},\dot\svx_{\alpha,0})}.
 \label{eq:sov-projected-Qplus}
\end{equation}
Indeed, fixing the last separated variables to their vacuum values
in~\eqref{eq:sov-Qplus-statement} and inserting
\eqref{eq:sov-state-normalisation} cancels their $J-\ell$ factors.
The $\Qp$ functions here belong to the original length-$J$
eigenstate; the subscript $\ell$ on the bra specifies only the
short-chain SoV basis.
The overall factor $\psi_J^{\rm C}(0,\dots,0)$ is the
companion-origin value given in~\eqref{eq:companion-origin-normalisation}.

Thus the restricted coordinate space wave function has a short-chain
SoV transform given by a product of the original $\Qp$ functions.
Before applying this result to structure constants, we also need the complementary projection
for a state on the reversed chain.

\paragraph{Localisation on the upper SoV vacuum.} The companion pairing also includes the inversion $z\mapsto1/z$. For the frozen legs to end at the origin after gluing, their coordinates must therefore be sent to infinity before pairing. Thus, one also needs to consider the following limit of wave functions
\begin{equation}
  \lim_{R_1,\ldots,R_{J-\ell}\to\infty}
  \prod_{\alpha=1}^{J-\ell}[R_\alpha]^{2s}\,
  \psi_J^{\rm C}(R_1,\ldots,R_{J-\ell},z_{J-\ell+1},\ldots,z_J),
  \label{eq:endpoints-infinity-projection}
 \end{equation}
where each $R_\alpha$ tends to infinity along a fixed ray. The factors $[R_\alpha]^{2s}$
supply the conformal weights required by
inversion~\eqref{eq:full-inversion}.

At one site this restriction is the inverted vacuum of Section~\ref{sec:sov-Qplus},
$\langle\infty|\equiv\lim_{R\to\infty}[R]^{2s}\langle R|$.
It obeys $\langle\infty|\hat T_{12}=0$, since differentiation by
$\hat T_{12}\propto\partial_z$ lowers the large-$R$ power. On the
leading term selected by this limit, $z\partial_z\to-2s$, so that
\begin{equation}
 \langle\infty|\hat T_{11}(u)=(u-\svx^\uparrow_{\alpha,0})\langle\infty|,
 \qquad \svx^\uparrow_{\alpha,0}\equiv\theta_\alpha+\ii s .
 \label{eq:HHL-vacuum-infinity}
\end{equation}

With the frozen block first, the monodromy splits as
$\hat T(u)=\hat T^{(J-\ell)}(u)\hat T^{(\ell)}(u)$.
The annihilation condition removes the cross term
$\hat T_{12}^{(J-\ell)}\hat T_{21}^{(\ell)}$ from its $11$ entry.
The eigenvalue comparison and reference-state normalisation used
above then identify the tensor product of the frozen infinity vacua and a
short-chain SoV bra with the full-chain SoV bra whose frozen variables
are $\svx^\uparrow_{\alpha,0}$. Contracting this tensor-product bra with $\psi_J^{\rm C}$ gives
\begin{equation}
 \bigl(\langle\infty|^{\otimes(J-\ell)}
   \otimes {}_\ell\langle\svx^{(\ell)},\dot\svx^{(\ell)}|\bigr)
   |\psi_J^{\rm C}\rangle
 =\psi_J^{\rm C}(0,\dots,0)
 \prod_{\alpha=\ell+1}^{J}\kappa_{\alpha}
 \prod_{\alpha=1}^{\ell}
 \frac{\Qp(\svx_\alpha,\dot\svx_\alpha)}
      {\Qp(\svx_{\alpha,0},\dot\svx_{\alpha,0})},
 \label{eq:HHL-bra-transform}
\end{equation}
where
\begin{equation}
 \kappa_{\alpha}\equiv
 \frac{\Qp(\svx^\uparrow_{\alpha,0},\dot\svx^\uparrow_{\alpha,0})}
      {\Qp(\svx_{\alpha,0},\dot\svx_{\alpha,0})}.
\end{equation}

The short-chain SoV bra acts on the remaining $\ell$ sites
in their inherited order on the reversed chain. Since the bra lives on
the reversed chain, its first $J-\ell$ coordinates carry the
inhomogeneities $\theta_J,\ldots,\theta_{\ell+1}$, which is the
labelling used in~\eqref{eq:HHL-bra-transform}.

Here $\Qp$ belongs to the length-$J$ state on that chain,
and $\psi_J^{\rm C}(0,\dots,0)$ is its origin value. Compared with
the zero projection~\eqref{eq:sov-projected-Qplus}, each frozen site
contributes the ratio $\kappa_\alpha$ of the upper- and lower-vacuum
Q-values.

For the reference state,
$\mathbb Q_{+,\Omega}=\lambda^{\ii\svx}\dot\lambda^{\ii\dot\svx}$,
these ratios are $1$ when $\dot\lambda=\lambda^{-1}$.
The upper-vacuum choice is essential: restricting the leading block
to zero leaves the monodromy cross term in place.
Numerical checks are given in Appendix~\ref{app:HHL-tests}.
Both projections are now available for the three-point functions
in the following section.

\section{Twisted overlaps}
\label{sec:twisted-overlaps}
\begin{figure}[t]
 \centering
 \includegraphics[width=0.6\textwidth]{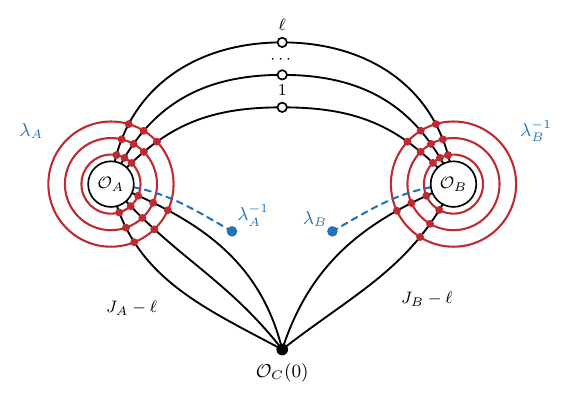}
 \caption{The HHL correlator~\eqref{eq:HHL-target-correlator}.
 Both upper insertions are dressed by wheels and share $\ell$
 propagators; their spectator legs meet at $\mathcal O_C(0)$.
 The diagram is drawn in the coordinates of the ket, with $B$ at
 $\lambda_B^{-1}$ and $A$ at $\lambda_A$ after the bra inversion.
 Open marks are the seam coordinates integrated when the two wave
 functions are glued. Spokes, wheels and twist cuts follow
 Figure~\ref{fig:zero-magnon-wavefunction}.}
 \label{fig:HHL-three-point}
\end{figure}

In Section~\ref{sec:sov} we developed the SoV framework for the fishnet CFT. In this section we put it to work on the planar three-point contraction shown in Figure~\ref{fig:HHL-three-point}.

\paragraph{The set-up.}
Throughout this section we work in the companion twist frame~\eqref{eq:companion-colour-twist}, and consider three single-trace operators labelled $A$, $B$ and $C$.

The first two are colour-twist operators in the zero-magnon sector of the theory. Operator $B$ is made of $J_B$ fields $\phi_1$ and operator $A$ of $J_A$ conjugate fields $\phi_1^\dagger$, in both cases with any number of derivatives, as in~\eqref{eq:fishnet-local-operators}. Each is specified by its CFT wave function~\eqref{eq:CFT-wavefunction}. Their twists are
\begin{equation}
 G_A\equiv G_{\rm C}(\lambda_A),
 \qquad
 G_B\equiv G_{\rm C}(\lambda_B),
 \label{eq:HHL-twisted-operators}
\end{equation}
together with their dotted counterparts. Operator $B$ sits at the fixed point $\lambda_B^{-1}$ of its twist. Operator $A$ carries the inverse twist, as required by the companion pairing~\eqref{eq:companion-pairing}, and after the inversion it sits at $\lambda_A$.

The third operator $\mathcal{O}_C$ sits at the origin and is untwisted,
\begin{equation}\label{eq:OC-at-origin}
 \mathcal O_C=\operatorname{tr}\!\left[
 \phi_1^{J_A-\ell}(\phi_1^\dagger)^{J_B-\ell}\right].
\end{equation}
Here $\ell=0,\dots,\min(J_A,J_B)$ is the number of propagators shared by $A$ and $B$; their remaining legs end on $\mathcal O_C$. The correlator we want to compute is then
\begin{equation}
 \left\langle
 \mathcal O_{A,G_A^{-1}}(\lambda_A)\,
 \mathcal O_{B,G_B}(\lambda_B^{-1})\,
 \mathcal O_C(0)
 \right\rangle\,.
 \label{eq:HHL-target-correlator}
\end{equation}

Note that operators $A$ and $B$ are dressed by fishnet wheels and carry the full coupling dependence, see Figure~\ref{fig:HHL-three-point}. Operator $C$ is not: because of its mixed field content, no wheels can be drawn around it in the planar theory, and it stays at its free value~\cite{Basso:2018cvy}. We call operators of the first kind \textit{heavy} and those of the second kind \textit{light}, and accordingly refer to~\eqref{eq:HHL-target-correlator} as the heavy--heavy--light (HHL) correlator.

\subsection{From correlators to Q-functions}
\label{sec:HHL-correlator}

In order to compute the HHL correlator~\eqref{eq:HHL-target-correlator} using the technology of the previous section we need to write it in terms of the CFT wave functions of $A$ and $B$ and the planar bracket.

\paragraph{Wave function coordinates and HHL overlap.}
First we need to identify which coordinates the CFT wave function should be put at. Generically, the CFT wave function for an operator of length $J$ is $\psi(z_1,\dots,z_J)$.
In the correlator set-up of Figure~\ref{fig:HHL-three-point} the wave function of $B$ has its last $J_B-\ell$ coordinates set to zero, as they contract with the operator \eqref{eq:OC-at-origin}. This leaves the $\ell$ shared coordinates $\boldsymbol z_\ell=(z_1,\ldots,z_\ell)$ to be glued with the CFT wave function of $A$.

For $A$ the situation is reversed. The companion pairing~\eqref{eq:companion-pairing} includes an inversion, which moves $A$ from $\lambda_A^{-1}$ to $\lambda_A$. Its $J_A-\ell$ spectator coordinates (i.e. those not contracted with the operator $B$) must therefore first be sent to infinity with the weights in~\eqref{eq:endpoints-infinity-projection}, so that after the inversion they end at the origin and contract with $C$. This is the reversed chain introduced in~\eqref{eq:left-right-bracket}, and we write its coordinates as $\boldsymbol R=(R_1,\ldots,R_{J_A-\ell})$.

Putting the two together, we define the HHL overlap $\LRpair{\psi_A^{\rm C}}{\psi_B^{\rm C}}_\ell$ by the planar bracket
\begin{equation}
 \LRpair{\psi_A^{\rm C}}{\psi_B^{\rm C}}_\ell=
 \lim_{\boldsymbol R\to\infty}
 \prod_{a=1}^{J_A-\ell}[R_a]^{2s}\,
 \LRpair{\psi_A^{\rm C}(\boldsymbol R,\boldsymbol z_\ell)}
        {\psi_B^{\rm C}(\boldsymbol z_\ell,0^{J_B-\ell})}.
 \label{eq:HHL-definition}
\end{equation}
By construction it is equal to the three-point function~\eqref{eq:HHL-target-correlator}
\begin{equation}
  \LRpair{\psi_A^{\rm C}}{\psi_B^{\rm C}}_\ell=
 \left\langle
 \mathcal O_{A,G_A^{-1}}(\lambda_A)\,
 \mathcal O_{B,G_B}(\lambda_B^{-1})\,
 \mathcal O_C(0)
 \right\rangle\,.
 \label{eq:HHL-bracket-correlator}
\end{equation}

\paragraph{HHL overlap from Q-functions.}
All of the ingredients required to compute \eqref{eq:HHL-definition} in terms of the glued Q-functions $\mathbb{Q}_{+,A}$ and $\mathbb{Q}_{+,B}$ were computed in the previous section. The projection of state $B$ is~\eqref{eq:sov-projected-Qplus}, while the weighted infinity projection of state $A$ is~\eqref{eq:HHL-bra-transform}. Substituting these transforms into
the length-$\ell$ scalar product detailed in Section~\ref{sec:overlaps-resolution} then gives
\begin{equation}
 \boxed{\;
 \frac{\LRpair{\psi_A^{\rm C}}{\psi_B^{\rm C}}_\ell}
      {\psi_A^{\rm C}(\boldsymbol0)\psi_B^{\rm C}(\boldsymbol0)}
 =
 \frac{\mathcal N_\ell^{\rm SoV}\ell! \prod_{\alpha=\ell+1}^{J_A}\kappa_{A,\alpha}}
 {\prod_{\alpha=1}^{\ell}
  \mathbb Q_{+,A}(\svx_{\alpha,0},\dot\svx_{\alpha,0})
  \mathbb Q_{+,B}(\svx_{\alpha,0},\dot\svx_{\alpha,0})}\times\det\limits_{1 \leq n,m \leq \ell}\mathcal M_\ell^{A,B}}
 \label{eq:HHL-formula}
\end{equation}
where the moment matrix is
\begin{equation}
 (\mathcal M_\ell^{A,B})_{nm}=
 \int\dd^2\svx\,
 \svx^{n-1}\dot\svx^{m-1}
 \mu_\ell(\svx,\dot\svx)
 \mathbb Q_{+,A}(\svx,\dot\svx)
 \mathbb Q_{+,B}(\svx,\dot\svx).
 \label{eq:HHL-moments}
\end{equation}
Only the $\ell$ glued sites enter the measure,
\begin{equation}
 \mu_\ell(u,\dot u)=
 \prod_{\alpha=1}^{\ell}
 \mu_s(u-\theta_\alpha,\dot u-\dot\theta_\alpha).
 \label{eq:endpoints-short-measure}
\end{equation}

Note that the origin factors $\psi_A^{\rm C}(\boldsymbol0)$ and $\psi_B^{\rm C}(\boldsymbol0)$ in \eqref{eq:HHL-formula} remove the overall normalisations of the wave functions of the two states, so that the state dependence of the r.h.s. enters only through the two Q-functions $\mathbb{Q}_{+,A}$ and $\mathbb{Q}_{+,B}$.

To spell out the remaining factor $\kappa_{A,\alpha}$ explicitly we need to comment on the inhomogeneities of the two states. For $\ell=J_A=J_B$ the overlap above reduces to the planar bracket of~\eqref{eq:overlaps-bilinear-determinant} involving all spin chain sites. For generic $\ell$ we label the glued lines $\alpha=1,\ldots,\ell$ and match their
inhomogeneities along the seam $\theta_\alpha^A=\theta_\alpha^B=\theta_\alpha$, which are all distinct. The spectator inhomogeneities are independent and we also consider all unshared values to be distinct from one another and from those on the glued sites. This distinction is crucial to separate all pole towers entering the SoV formulation below.
With this in mind, we can now write $\kappa_{A,\alpha}$, which is the ratio of the Q-function $\mathbb{Q}_{+,A}$ at the upper and lower vacuum points
\begin{equation}
 \kappa_{A,\alpha}=
 \frac{\mathbb Q_{+,A}(\theta^A_\alpha+\ii s,\dot\theta^A_\alpha+\ii s)}
      {\mathbb Q_{+,A}(\theta^A_\alpha-\ii s,\dot\theta^A_\alpha-\ii s)}.
 \label{eq:HHL-kappa}
\end{equation}

\paragraph{Twist structure.}
The result \eqref{eq:HHL-formula} assumes that both heavy states have distinct values of twists $\lambda_A$ and $\lambda_B$. Removing the twist cuts leaves a graph with three local insertions, which is the three-point function that we ultimately want to compute. However, this untwisting limit is subtle and it will only be considered in Section~\ref{sec:untwisted} after we establish the twisted framework below.

The relations \eqref{eq:HHL-bracket-correlator} and \eqref{eq:HHL-formula} already express all information about the three-point function \eqref{eq:HHL-target-correlator} in terms of Q-functions. However, at the moment the integrals in \eqref{eq:HHL-moments} are rather unwieldy, as they are double integrals of the glued Q-functions $\mathbb{Q}_{+,A}$ and $\mathbb{Q}_{+,B}$. In the next subsection we will simplify these expressions substantially by rewriting them in terms of one-dimensional integrals of the chiral (i.e. undotted and dotted) Q-functions, or equivalently as sums over poles.

\subsection{The q-bilinear representation}
\label{sec:HHL-poles}

\begin{figure}[t]
\centering
\includegraphics[width=0.8\linewidth]{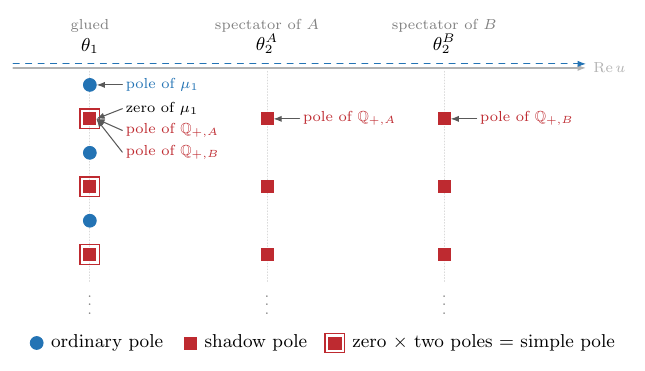}
\caption{Lower-half-plane poles of the HHL integrand~\eqref{eq:HHL-moments}
for $J_A=J_B=2$, $\ell=1$. Positions are drawn at $s=\tfrac14$; the dashed line is the integration contour. Arrows indicate which factor produces each tower. At the glued site $\mu_1$ gives the ordinary tower; on the shadow level its zero meets the poles of both Q-functions and leaves a simple pole (framed squares). Each spectator carries the shadow tower of its own state. All four towers are picked up.}
\label{fig:HHL-ell1-poles}
\end{figure}

To evaluate the moments~\eqref{eq:HHL-moments} we need to compute the
integral
$\int\dd^2\svx\,\svx^{n-1}\dot\svx^{m-1}\mu_\ell\,\mathbb Q_{+,A}\mathbb Q_{+,B}$.
Recalling from~\eqref{eq:sov-sum-integral-notation} that
$\svx=\nu+\frac{\ii k}{2}$, $\dot\svx=\nu-\frac{\ii k}{2}$ and
$\int\dd^2\svx=\sum_{k\in\mathbb Z}\int_{\mathbb R}\dd\nu$ (with $m$ renamed $k$, since $m$ now labels the moments), we compute it,
as in Section~\ref{sec:overlaps-resolution}, by closing the $\nu$-contour
downwards at fixed $k$. Before this we need to understand which poles
are picked up.

\paragraph{Pole structure.} At fixed $k$, the integrand has a pole in $\nu$ only where $\svx$ and $\dot\svx=\svx-\ii k$ sit on the same tower, with tower indices $n\ge\max(0,-k)$ and $\dot n=n+k$, exactly as for $J=1$. It is therefore enough to list the towers in $\svx$. There are four families of them in the region where the contour is closed, as depicted in Figure~\ref{fig:HHL-ell1-poles}. The Q-functions of both states have poles at the shadow tower
\begin{equation}
  \widetilde\svx_{\alpha,n}^A=\theta_\alpha^A-\ii(1-s+n), \qquad \widetilde\svx_{\alpha,n}^B=\theta_\alpha^B-\ii(1-s+n),
\end{equation}
while the measure has poles on the ordinary tower
\begin{equation}
 \svx_{\alpha,n}=\theta_\alpha-\ii(s+n), \qquad \alpha = 1,\dots,\ell,
\end{equation}
where both states have the same inhomogeneities. The spectators give two further families: the shadow towers of $\mathbb Q_{+,A}$ at the spectators of $A$, and those of $\mathbb Q_{+,B}$ at the spectators of $B$. For the glued sites $\alpha \leq \ell$, there are also two families. One consists of the ordinary poles of the measure $\mu_\ell$. The other is the combination of the single poles of each Q-function and the zeros of the measure which results again in a simple pole. These pole families are summarised in Table~\ref{tab:poleFamiliesAB} below.

\begin{table}[H]
\begin{center}
\begingroup
\renewcommand{\arraystretch}{1.25}
\begin{tabular}{@{}lll@{}}
 \toprule
 Family & Pole positions & Origin of the pole \\
 \midrule
 Glued, ordinary & $\theta_\alpha-\ii(s+n)$ &
 Poles of $\mu_\ell$. \\
 Glued, shadow & $\theta_\alpha-\ii(1-s+n)$ &
 Zero of $\mu_\ell$ $\times$ poles of both $\mathbb Q_+$. \\
 Spectator of $A$ & $\theta^A_\alpha-\ii(1-s+n)$ &
 Poles of $\mathbb Q_{+,A}$ only. \\
 Spectator of $B$ & $\theta^B_\alpha-\ii(1-s+n)$ &
 Poles of $\mathbb Q_{+,B}$ only. \\
 \bottomrule
\end{tabular}
\endgroup
\end{center}
\vspace{-1em}
\caption{Pole families of the HHL moment integrand~\eqref{eq:HHL-moments}.}
\label{tab:poleFamiliesAB}
\end{table}
\vspace{-0.5em}

\paragraph{Partial shadow transform.}
To put the spectator poles and the glued ordinary poles in a common
measure we will use a partial shadow transform. We index the sites by $\alpha=1,\ldots,J_A+J_B-\ell$: the glued
sites $\alpha\le\ell$ first and counted once, then the spectators of $A$
and of $B$, with $\theta_\alpha$ the corresponding inhomogeneities. We then
shadow transform \textit{only} the spectator sites, $\hat s_\alpha=s$ for
$\alpha\le\ell$ and $\hat s_\alpha=1-s$ otherwise, and define the
mixed-chain measure
\begin{equation}
 \hat\mu^\downarrow(u)
 \equiv\prod_{\alpha=1}^{J_A+J_B-\ell}
 \frac{\Gamma(\hat s_\alpha-\ii(u-\theta_\alpha))}
      {\Gamma(1-\hat s_\alpha-\ii(u-\theta_\alpha))},\quad
 \hat\mu^\uparrow(u)
 \equiv\prod_{\alpha=1}^{J_A+J_B-\ell}
 \frac{\Gamma(\hat s_\alpha+\ii(u-\theta_\alpha))}
      {\Gamma(1-\hat s_\alpha+\ii(u-\theta_\alpha))}.
 \label{eq:HHL-partial-shadow-measure}
\end{equation}
The Q-functions of each mixed chain are related to the original ones by
\begin{equation}
 \begin{aligned}
 \hat q_i^{\downarrow}(u)
 &=e^{\frac{\ii\pi}{2}(J-\ell)(2s-1)}
 \prod_{\alpha=\ell+1}^{J}
 \frac{\Gamma(s-\ii(u-\theta_\alpha))}
      {\Gamma(1-s-\ii(u-\theta_\alpha))}\,
 q_i^{\downarrow}(u),\\
 \hat q_i^{\uparrow}(u)
 &=e^{-\frac{\ii\pi}{2}(J-\ell)(2s-1)}
 \prod_{\alpha=\ell+1}^{J}
 \frac{\Gamma(s+\ii(u-\theta_\alpha))}
      {\Gamma(1-s+\ii(u-\theta_\alpha))}\,
 q_i^{\uparrow}(u),
 \end{aligned}
 \label{eq:endpoints-partial-shadow-q}
\end{equation}
where $J$ stands for $J_A$ or $J_B$ and, similarly, $\theta_\alpha$
stands for $\theta_\alpha^A$ or $\theta_\alpha^B$.

\paragraph{Simplifying the integrand.}
In the mixed-chain variables the moment integrand takes the same form as
for the planar overlap. Writing $\mathbb Q_{+,A}$ as~\eqref{eq:glued-Qplus}
and $\mathbb Q_{+,B}$ as~\eqref{eq:glued-Qplus-arrow-switch}, we find, on the lattice $u-\dot u\in\ii\mathbb Z$,
\begin{equation}
 \mu_\ell(u,\dot u)\,\mathbb Q_{+,A}(u,\dot u)\,\mathbb Q_{+,B}(u,\dot u)
 =\Gamma_A^{ik}\Gamma_B^{jl}\,
 \hat\mu^\downarrow(u)\,\hat q_i^{A\downarrow}(u)\,\hat q_j^{B\uparrow}(u)\,
 \hat\mu^\uparrow(\dot u)\,\dot{\hat q}_k^{A\uparrow}(\dot u)\,\dot{\hat q}_l^{B\downarrow}(\dot u)\,.
 \label{eq:HHL-hatted-integrand}
\end{equation}
Indeed, the spectator factors of $A$ in~\eqref{eq:HHL-partial-shadow-measure}
and~\eqref{eq:endpoints-partial-shadow-q} cancel identically. Those of $B$
leave $\sin\pi(s-\ii(u-\theta_\alpha))/\sin\pi(s+\ii(u-\theta_\alpha))$ in
the undotted sector and the inverse of its dotted counterpart, and the two
cancel on the lattice $u-\dot u\in\ii\mathbb Z$, as
in~\eqref{eq:overlaps-one-pair-measure}. The phases
in~\eqref{eq:endpoints-partial-shadow-q} cancel between the two
chiralities.

In the mixed chain the four families of Table~\ref{tab:poleFamiliesAB}
sit on two towers at each glued site and on a single tower at each spectator. Since $\hat s_\alpha=1-s$ at the spectators,
the down tower $\theta_\alpha-\ii(\hat s_\alpha+n)$ of $\hat\mu^\downarrow$
contains the glued ordinary poles for $\alpha\le\ell$ and the spectator
poles of $A$ and $B$ for $\alpha>\ell$. The $\hat\mu^\uparrow$ tower
$\theta_\alpha-\ii(1-\hat s_\alpha+n)$ contains the glued shadow poles. A spectator has no second tower, since its only poles are
the shadow tower of its own Q-function, which is now the down tower.

Let $p_\alpha=\theta_\alpha-\ii\hat s_\alpha$ be the base of the down
tower at site $\alpha$. As in~\eqref{eq:J1-measure-conversion-coefficients},
the conversion coefficients are
\begin{equation}
 \hat k_\alpha \equiv \hat k_\alpha(\hat{\boldsymbol s})
 =\frac{\hat\mu^\uparrow(p_\alpha)}
 {\operatorname*{Res}_{u=p_\alpha}\hat\mu^\downarrow(u)},
 \qquad
 \hat k_\alpha^{\rm sh}=
 \begin{cases}
  \hat k_\alpha(1-\hat{\boldsymbol s}),&\alpha\le\ell,\\
  0,&\alpha>\ell,
 \end{cases}
 \label{eq:khatAndKShHat}
\end{equation}
where $\hat{\boldsymbol s}=(\hat s_1,\ldots,\hat s_{J_A+J_B-\ell})$, so
that $1-\hat{\boldsymbol s}$ flips all spins. For a spectator of $A$
(or $B$) we also write $\hat k_{\alpha,A}$ ($\hat k_{\alpha,B}$), with
$\alpha$ the site index of that state.
Appendix~\ref{app:HHL-residues} gives the individual residue products
for the four families.

\paragraph{Computing the moments.}
To combine the four families in one sum, we use the contour
brackets~\eqref{eq:lower-tower-contour} and define the mixed-chain
version of~\eqref{eq:paired-tower-bracket},
\begin{equation}
 \towerbracket{F}{G}_\alpha^{\hat\mu}
 \equiv \hat k_\alpha
   \bigl(\hat\mu^\downarrow(u)F(u)\bigr)_\alpha
   \bigl(\hat\mu^\downarrow(\dot u)G(\dot u)\bigr)_\alpha
 +\hat k_\alpha^{\rm sh}
   \bigl(\hat\mu^\uparrow(u)F(u)\bigr)_\alpha
   \bigl(\hat\mu^\uparrow(\dot u)G(\dot u)\bigr)_\alpha .
 \label{eq:HHL-site-bracket}
\end{equation}
The first term collects the down towers at every site, and the second
the glued shadow towers. With~\eqref{eq:HHL-hatted-integrand}, the
moments~\eqref{eq:HHL-moments} become
\begin{equation}
 (\mathcal M_\ell^{A,B})_{nm}
 =-2\pi\ii\Gamma_A^{ik}\Gamma_B^{jl}\!
 \sum_{\alpha=1}^{J_A+J_B-\ell}
 \towerbracket{u^{n-1}\hat q_i^{A\downarrow}\hat q_j^{B\uparrow}}
              {\dot u^{m-1}\dot{\hat q}_k^{A\uparrow}\dot{\hat q}_l^{B\downarrow}}
              _\alpha^{\hat\mu},
 \label{eq:HHL-moment-towers}
\end{equation}
with the contour brackets $\zeta$-regularised as described
below~\eqref{eq:lower-tower-contour}.

As a cross-check, notice that for $\ell=J_A=J_B=J$, all sites are glued and the measure coefficients
reduce to $k_\alpha(s)$ and $k_\alpha(1-s)$. Rewriting the integrands
with canonical shadow Q-functions via~\eqref{eq:shadow-q-map} gives
the coefficients $k_\alpha^{\mathrm{c}}$ in~\eqref{eq:overlaps-moment-residues}.
The spectator product of $\kappa_{A,\alpha}$ is empty, and
\eqref{eq:HHL-formula} reduces to the planar overlap
\eqref{eq:overlaps-bilinear-determinant}.

Together, \eqref{eq:HHL-formula} and~\eqref{eq:HHL-moment-towers}
express the twisted HHL three-point function entirely through the
Q-functions of the two heavy states. The two-dimensional SoV integrals
have collapsed into an $\ell\times\ell$ determinant whose entries are
finite sums, over the $J_A+J_B-\ell$ sites, of products of
one-dimensional contour brackets, one for each chirality. In the next subsection we
take one of the two states to be a reference state, which gives the
HLL overlap.

\subsection{HLL overlap}
\label{sec:HLL}
\begin{figure}[t]
 \centering
 \includegraphics[width=0.65\linewidth]{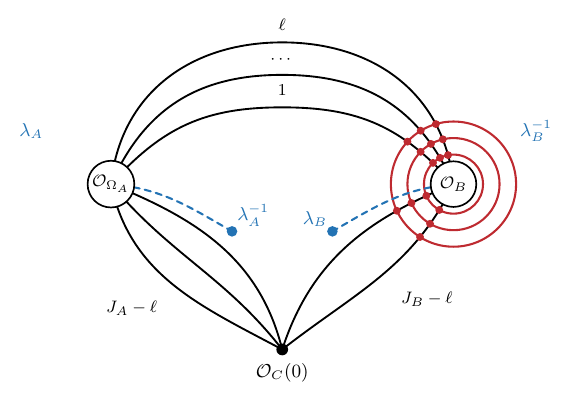}
 \caption{The HLL correlator~\eqref{eq:endpoints-three-point-correlator}.
 The free $A$ insertion and the
 interacting $B$ insertion share $\ell$ propagators; their remaining
 legs meet at $\mathcal O_C(0)$. Only $B$ is dressed by wheels (red).
 Dashed blue lines denote twist cuts.}
 \label{fig:HLL-three-point}
\end{figure}

The coupling enters the HHL overlap~\eqref{eq:HHL-formula} only through the Q-functions of states $A$ and $B$, so the two states can be taken at independent values $\xi_A$ and $\xi_B$ of the coupling. A particularly simple and important case is depicted in Figure~\ref{fig:HLL-three-point}: state $A$ is the reference state $\Omega_A$, which has no wheels and is evaluated at $\xi_A=0$, while state $B$ is at arbitrary coupling. We refer to
\begin{equation}
 \LRpair{\Omega_{A}^{\rm C}}{\psi_B^{\rm C}}_\ell\equiv \left.\left\langle
 \mathcal O_{A,G_A^{-1}}(\lambda_A)\,
 \mathcal O_{B,G_B}(\lambda_B^{-1})\,
 \mathcal O_C(0)
 \right\rangle\right|_{\xi_A=0,\;\xi_B=\xi},
 \label{eq:endpoints-three-point-correlator}
\end{equation}
as the HLL overlap. Because the reference state is so simple, the wave-function form~\eqref{eq:HHL-definition} of this overlap reduces to an evaluation,
\begin{equation}
 \LRpair{\Omega_A^{\rm C}}{\psi_B^{\rm C}}_\ell
 =[-\lambda_A]^{-2s(J_A-\ell)}\mathfrak n_{\lambda_A}^{(J_A)}
 \psi_B^{\rm C}(\lambda_A^\ell,0^{J_B-\ell}),
 \label{eq:HLL-overlap-wavefunction}
\end{equation}
where the first two factors are the value and normalisation of the reference state~\eqref{eq:reference-state-frames}, and the last is the wave function of state $B$ evaluated at the twist $\lambda_A$ of the reference state.

 Apart from these overall factors, the HLL overlap is therefore the correlator of $B$ with the operator
\begin{equation}
 \operatorname{tr}\!\left[
 \mathcal T_{G_B^{-1}}\,
 (\phi_1^\dagger(\lambda_A))^\ell
 (\phi_1^\dagger(0))^{J_B-\ell}
 \right],
 \label{eq:HLL-probe-operator}
\end{equation}
namely
\begin{equation}
 \psi_B^{\rm C}(\lambda_A^\ell,0^{J_B-\ell})
 =\left\langle
 \mathcal O_{B,G_B}(\lambda_B^{-1})\,
 \operatorname{tr}\!\left[
 \mathcal T_{G_B^{-1}}\,
 (\phi_1^\dagger(\lambda_A))^\ell
 (\phi_1^\dagger(0))^{J_B-\ell}
 \right]
 \right\rangle.
 \label{eq:endpoints-probe-correlator}
\end{equation}
Such operators appear in the OPE decomposition of the four-point function considered in~\cite{Kazakov:2018qbr}, from which the overlap can be extracted at $J_B=2$ for arbitrary coupling~\cite{Kazakov:2018qbr}, and at $J_B=3$ to first perturbative order, see Appendix~\ref{app:4pt}. Besides being interesting correlators in their own right, the HLL overlaps~\eqref{eq:endpoints-probe-correlator} therefore provide a natural testing ground for our SoV framework, for both twisted and untwisted structure constants.

The SoV formulation of the HLL overlap combines the moment determinant~\eqref{eq:HHL-formula} with the Q-bilinear representation of its moments~\eqref{eq:HHL-moment-towers}. Both simplify for the reference state. Substituting $\mathbb Q_{+,A}(\svx,\dot\svx)=[\lambda_A]^{\{\ii\svx,\ii\dot\svx\}}$ into the moment determinant gives
\begin{equation}
\boxed{
 \frac{\LRpair{\Omega_A^{\rm C}}{\psi_B^{\rm C}}_\ell}
      {\psi_B^{\rm C}(\boldsymbol0)}
 =\frac{\mathcal N_\ell^{\rm SoV}\ell!\,
 [-\lambda_A]^{-2s(J_A-\ell)}\mathfrak n_{\lambda_A}^{(J_A)}}
 {\prod_{\alpha=1}^{\ell}
 [\lambda_A]^{\{\ii\svx_{\alpha,0},\ii\dot\svx_{\alpha,0}\}}
 \mathbb Q_{+,B}(\svx_{\alpha,0},\dot\svx_{\alpha,0})}\times \det\limits_{1\le n,m\le\ell}\mathcal M_\ell^{\Omega_A,B}.}
 \label{eq:HLL-overlap-moment-determinant}
\end{equation}

For the reference state $\Gamma_A=\operatorname{diag}(1,0)$ and
$q_1^{A\downarrow}(u)=q_1^{A\uparrow}(u)=\lambda_A^{\ii u}$, so only the
first Q-function of $A$ enters, and~\eqref{eq:endpoints-partial-shadow-q}
gives its mixed-chain version
\begin{equation}
 \begin{aligned}
 \hat q_1^{A\downarrow}(u)
 &=e^{\frac{\ii\pi}{2}(J_A-\ell)(2s-1)}
 \prod_{\beta=\ell+1}^{J_A}
 \frac{\Gamma(s-\ii(u-\theta_\beta^A))}
      {\Gamma(1-s-\ii(u-\theta_\beta^A))}\,
 \lambda_A^{\ii u},\\
 \hat q_1^{A\uparrow}(u)
 &=e^{-\frac{\ii\pi}{2}(J_A-\ell)(2s-1)}
 \prod_{\beta=\ell+1}^{J_A}
 \frac{\Gamma(s+\ii(u-\theta_\beta^A))}
      {\Gamma(1-s+\ii(u-\theta_\beta^A))}\,
 \lambda_A^{\ii u},
 \end{aligned}
 \label{eq:HLL-reference-partial-shadow-q}
\end{equation}
with the dotted functions obtained by replacing $u$, $\theta_\beta^A$
and $\lambda_A$ with $\dot u$, $\dot\theta_\beta^A$ and $\dot\lambda_A$.
Since $\lambda_A^{\ii u}$ is entire, the Q-function of $A$ has no poles.
Of the four families in Table~\ref{tab:poleFamiliesAB}, only the glued
ordinary towers and the spectator towers of $B$ therefore survive. In
the mixed chain, this is seen from the fact that at a spectator site of
$A$ the down measure $\hat\mu^\downarrow$ cancels the $\Gamma$-function
factor of $\hat q_1^{A\downarrow}$, so the undotted integrand is regular
there and its contour bracket vanishes. The moments~\eqref{eq:HHL-moment-towers}
therefore become
\begin{equation}
 (\mathcal M_\ell^{\Omega_A,B})_{nm}
 =-2\pi\ii\,\Gamma_B^{ij}\!
 \sum_{\alpha=1}^{J_A+J_B-\ell}
 \towerbracket{u^{n-1}\hat q_1^{A\downarrow}\hat q_i^{B\uparrow}}
              {\dot u^{m-1}\dot{\hat q}_1^{A\uparrow}\dot{\hat q}_j^{B\downarrow}}
              _\alpha^{\hat\mu}.
 \label{eq:HLL-overlap-moment-bilinears}
\end{equation}

The remaining ingredient to turn both HLL~\eqref{eq:HLL-overlap-moment-determinant} and HHL~\eqref{eq:HHL-formula} correlators into proper normalised local structure constants is their two-point function normalisation. Evaluating this norm in terms of Q-bilinears is what we turn to next.

\section{Norms: the equal-twist logarithm}
\label{sec:norms}

In this section we recall the norm of the CFT wave function and evaluate it via the SoV formalism. The starting point is two CFT wave functions $\psi_A$ and $\psi_B$, with equal conformal weights $h_A = h_B$ and $\dot{h}_A = \dot{h}_B$ in the limit where their twists approach each other $\lambda_B\rightarrow \lambda_A$ and $\dot{\lambda}_B\rightarrow \dot{\lambda}_A$. As discussed in Section~\ref{sec:coordinate-log-norm}, their inner product diverges logarithmically~\eqref{eq:coordinate-logarithmic-overlap},
\begin{equation}
 \LRpair{\psi_A}{\psi_B}
 =2\,\mathcal C_{\log}\log(1/\varepsilon)+O(1),
 \qquad
 \varepsilon=\left|\frac{\lambda_A-\lambda_B}{1-\lambda_A\lambda_B}\right|,
 \label{eq:norms-twist-log-overlap}
\end{equation}
with the twist difference acting as a point-splitting regularisation and the coefficient of the divergence giving the CFT norm $\mathcal N_2$
through~\eqref{eq:coordinate-general-norm-relation},
\begin{equation}
 \mathcal C_{\log}
 =-\frac{1}{J}\mathcal D_\xi\,\mathcal N_2,
 \qquad
 \mathcal D_\xi\equiv \xi^2\frac{\partial\Delta}{\partial\xi^2}.
 \label{eq:norms-general-norm-relation-recalled}
\end{equation}

The goal of this section is to start from the SoV representation for the CFT overlap given in~\eqref{eq:overlaps-moment-determinant} and recalled below,
\begin{equation}
 \LRpair{\psi_A}{\psi_B}
 =\mathcal N_J^{\mathrm{SoV}}\mathcal A_A\mathcal A_B
 J!\det\limits_{1 \leq n,m \leq J}\mathcal M^{A,B}_{n,m}\,,
 \label{eq:norms-overlap-recalled}
\end{equation}
and show how logarithmic divergences arise and compute the CFT norm $\mathcal{N}_2$. The outcome is the following SoV representation for the two-point function normalisation

\begin{equation}
 \boxed{
 \mathcal N_2
 =-\frac{4\pi c\,J\,J!\,\mathcal N_J^{\mathrm{SoV}}
          \mathcal A^2(-2\pi\ii)^{J-1}}{\mathcal D_\xi}
 \det\limits_{1\le n,m\le J-1}\!\left[
 \sum_{\alpha=1}^{J}\Gamma^{ik}\Gamma^{jl}\,
 \towerbracket{u^{n-1}q_i^\downarrow q_j^\uparrow}
              {\dot u^{m-1}\dot q_k^\uparrow\dot q_l^\downarrow}_\alpha
 \right].}
 \label{eq:norms-two-point-normalisation}
\end{equation}
Here $\Gamma=\operatorname{diag}(1,c)$, with $c=\Gamma^{22}$.
All Q-functions belong to the same on-shell state at a fixed
generic common twist, and their state labels are suppressed. The determinant
uses the site bracket~\eqref{eq:paired-tower-bracket} for the upper-left
$(J-1)\times(J-1)$ block of~\eqref{eq:overlaps-moment-residues}, with a
factor $-2\pi\ii$ extracted from each row. The Q-functions, conversion
coefficients and site sum refer to the full $J$-site chain.
The brackets use the regularisation of
Section~\ref{sec:J1-overlap-q-bilinears}. For $J=1$, the minor is
empty and its determinant is one.\par

The tower sums in this minor can be carried out completely, and each entry collapses to a finite sum of residues of the connection matrix $\Omega$ of Section~\ref{sec:setup-omega} and its dotted partner $\dot\Omega$:
\begin{equation}
 \boxed{\displaystyle
 \mathcal M_{nm}
 =-2\pi\ii(\lambda-\lambda^{-1})(\dot\lambda-\dot\lambda^{-1})
 \sum_{p}\operatorname*{Res}_{u=p}\,
 \frac{\Omega_{22}^2}{\det\Omega}\,
 \partial_{a_{n-1}}\!\left(\frac{\Omega_{12}}{\Omega_{22}}\right)
 \partial_{\dot a_{m-1}}\!\left(\frac{\dot\Omega_{12}}{\dot\Omega_{22}}\right)}
 \label{eq:norms-gaudin-form}
\end{equation}
for $1\le n,m\le J-1$. Here $p$ runs over the $2J$ ordinary and shadow points $\theta_\alpha-\ii s$ and $\theta_\alpha-\ii(1-s)$ in one period strip, and the charges are varied off shell at fixed twist and site data.\par

This is the analogue of the Gaudin formula~\cite{Gaudin:1981,Korepin:1982gg}, which gives the norm as the Jacobian of the Bethe equations. Here the gluing condition~\eqref{eq:setup-omega-gluing}, $\dot\Omega=\Gamma^{-1}\Omega^{\mathsf T}\Gamma$, plays the role of the Bethe equations. Its diagonal gives $\dot\Omega_{22}=\Omega_{22}$, and its off-diagonal entries state that
\begin{equation}
 F\equiv\frac{\Omega_{12}}{\Omega_{22}}-c\,\frac{\dot\Omega_{21}}{\dot\Omega_{22}}=0,
 \qquad
 \dot F\equiv\frac{\dot\Omega_{12}}{\dot\Omega_{22}}-c\,\frac{\Omega_{21}}{\Omega_{22}}=0
 \label{eq:norms-gaudin-quantisation-functions}
\end{equation}
for every $u$ on shell. Since $\Omega$ depends only on the undotted charges and $\dot\Omega$ only on the dotted ones, the derivatives in~\eqref{eq:norms-gaudin-form} are $\partial_{a_{n-1}}F$ and $\partial_{\dot a_{m-1}}\dot F$. The formula therefore pairs the responses of the two quantisation conditions to the free charges.\par

Even though the result is neat and simple, some steps in the derivation are quite involved. We found, however, that the derivation can be made straightforward after we rewrite the expressions in terms of the connection matrix $\Omega$ as defined in Section~\ref{sec:setup-omega}. The derivation has four steps, corresponding to the subsections below:
\begin{enumerate}
 \item[Sec~\ref{sec:norms-tails}.] We write the tower sums in connection form and show that logarithmic divergences arise only in the last row and column of $\mathcal M$, with the double logarithm coming from the corner entry $\mathcal M_{JJ}$.
 \item[Sec~\ref{sec:norms-finite}.] Using the Baxter equation, we reduce the finite parts of the tower sums to charge derivatives of first-pole data, and assemble the moments into a single residue formula.
 \item[Sec~\ref{sec:norms-onshell}.] The coordinate-space result has a single logarithm~\eqref{eq:norms-twist-log-overlap}, so double logarithms must vanish; we show this and, moreover, the single logarithms outside the corner entry all cancel as well. The remaining single logarithm at $\mathcal M_{JJ}$ multiplies the $(J-1)\times(J-1)$ minor, which results in~\eqref{eq:norms-two-point-normalisation}.
 \item[Sec~\ref{sec:norms-collapse}.] The remaining minor has no logarithms, so its tower sums reduce to first-pole data, giving the Gaudin-like form~\eqref{eq:norms-gaudin-form}.
\end{enumerate}

Each moment $\mathcal M_{nm}$ is built from products of undotted and
dotted contour brackets~\eqref{eq:lower-tower-contour}, summed over sites
and over the ordinary and shadow towers as in~\eqref{eq:paired-tower-bracket}.
To analyse the logarithmic divergences, we focus on the undotted bracket
on the ordinary tower.

\subsection{Logarithms of the tower sums}
\label{sec:norms-tails}

\paragraph{Poles from the connection matrix.} The first step is to write the moments $\mathcal{M}_{nm}$ in connection form, which makes explicit where the poles of each contour bracket come from. Consider first the ordinary-tower bracket $\bigl(\mu^\downarrow u^{n-1}q_i^{A\downarrow}q_j^{B\uparrow}\bigr)_\alpha$, the first term of~\eqref{eq:paired-tower-bracket}. By the shadow map~\eqref{eq:shadow-q-map}, the product
$\mu^\downarrow q_i^{A\downarrow}$ equals
$e^{-\frac{\ii\pi}{2}(2Js-J)}\widetilde q_i^{A\downarrow}$, so the pole comes from the down-arrow shadow function of state $A$,
while $q_j^{B\uparrow}$ is regular. We express the shadow function
in the up-arrow basis using~\eqref{eq:baxter-connection-matrix},
\begin{equation}
 \widetilde q^{\,\downarrow}(u)=\widetilde\Omega^{-1}(u)\,
 \widetilde q^{\,\uparrow}(u).
\end{equation}
Both up-arrow functions are regular on the lower tower, so the poles now come entirely from the $\ii$-periodic inverse connection
matrix $\widetilde\Omega^{-1}$. It has a simple pole at every ordinary point $\svx_{\alpha,l}=\theta_\alpha-\ii(s+l)$, with the same residue for all $l$ by periodicity. This residue is the matrix $\widetilde{\mathsf R}_\alpha$ of~\eqref{eq:setup-omega-residue-matrices},
\begin{equation}
 \operatorname*{Res}_{u=\svx_{\alpha,l}}\widetilde\Omega^{-1}(u)
 =-\ii\,e^{+\frac{\ii\pi}{2}(2Js-J)}\,\widetilde{\mathsf R}_\alpha ,
\end{equation}
where the phase in the normalisation of $\widetilde{\mathsf R}_\alpha$ was chosen to cancel the one from the shadow map. Since the contour bracket~\eqref{eq:lower-tower-contour} is the sum of the residues at site $\alpha$, we obtain
\begin{equation}
 \bigl(\mu^\downarrow u^{n-1}q_i^{A\downarrow}q_j^{B\uparrow}\bigr)_\alpha
 =-\ii\sum_{l\geq0}\sum_{k=1}^{2}(\widetilde{\mathsf R}_\alpha)_{ik}\,
 \bigl[u^{n-1}\widetilde q_k^{A\uparrow}(u)q_j^{B\uparrow}(u)\bigr]_{u=\svx_{\alpha,l}}.
 \label{eq:norms-tower-residues}
\end{equation}
All the pole information is now carried by the constant matrix $\widetilde{\mathsf R}_\alpha$, while the remaining factor is a product of up-arrow functions evaluated along the tower.

The shadow-tower bracket, the second term of~\eqref{eq:paired-tower-bracket}, works in the same way with the two towers exchanged. There $\mu^\uparrow$ is regular and the pole comes from $q_i^{A\downarrow}=\bigl(\Omega^{-1}q^{A\uparrow}\bigr)_i$, whose residue at the shadow points defines $\mathsf R_\alpha$ in~\eqref{eq:setup-omega-residue-matrices}. Writing $\mu^\uparrow q_j^{B\uparrow}=e^{+\frac{\ii\pi}{2}(2Js-J)}\widetilde q_j^{B\uparrow}$ again cancels the phase, and
\begin{equation}
 \bigl(\mu^\uparrow u^{n-1}q_i^{A\downarrow}q_j^{B\uparrow}\bigr)_\alpha
 =-\ii\sum_{l\geq0}\sum_{k=1}^{2}(\mathsf R_\alpha)_{ik}\,
 \bigl[u^{n-1}q_k^{A\uparrow}(u)\widetilde q_j^{B\uparrow}(u)\bigr]_{u=\widetilde\svx_{\alpha,l}}.
 \label{eq:norms-shadow-tower-residues}
\end{equation}

\paragraph{Extracting the log term.} Divergences come from the large-$l$ tail of the tower sum in~\eqref{eq:norms-tower-residues}. When $h_A=h_B$, the terms with $k\ne j$ pair a solution growing like $u^{h}$ with one decaying like $u^{-h}$, so the conformal weights drop out. By~\eqref{eq:baxter-asymptotics} and~\eqref{eq:shadow-baxter-asymptotics}, these terms behave as
\begin{equation}
 \bigl[u^{n-1}\widetilde q_k^{A\uparrow}(u)\,q_j^{B\uparrow}(u)\bigr]_{u=\svx_{\alpha,l}}
 \sim\Bigl(\frac{\lambda_A}{\lambda_B}\Bigr)^{\pm l}\,l^{\,n-1-J},
 \qquad l\to\infty,\quad k\ne j.
\end{equation}
For $n<J$ the sum converges even at equal twists. For $n=J$ it is a harmonic sum, regularised by the twist ratio, which we define with a constant chosen to match the cutoff convention below:
\begin{equation}
 \mathbb L\equiv\sum_{l=1}^{\infty}\frac{(\lambda_A/\lambda_B)^{l}}{l}+\log(-\ii)
 =-\log\left(1-\frac{\lambda_A}{\lambda_B}\right)+\log(-\ii).
 \label{eq:norms-harmonic-twist-sum}
\end{equation}
The terms with $k=j$ instead carry $(\lambda_A\lambda_B)^{\pm l}$, which still oscillates at equal twists, and are finite with the $\zeta$-prescription of Section~\ref{sec:J1-overlap-q-bilinears}. Thus, the leading behaviour of the ordinary tower of residues~\eqref{eq:norms-tower-residues} can be compactly written as
\begin{equation}
 \bigl(\mu^\downarrow u^{n-1}q_i^\downarrow q_j^\uparrow\bigr)_\alpha
 =\delta_{nJ}(\widetilde{\mathsf R}_\alpha\sigma_1)_{ij}\,\mathbb L+O(1),
 \label{eq:norms-log-coefficient}
\end{equation}
where we introduced
\begin{equation}
\sigma_1=\begin{pmatrix}0&1\\1&0\end{pmatrix}=\sigma_3\epsilon,
 \label{eq:norms-pairing-tensors}
\end{equation}
to encode the $k \neq j$ condition on the divergences of~\eqref{eq:norms-tower-residues} discussed above. Here $\epsilon$ is the antisymmetric matrix of~\eqref{eq:wronskian-anti-sym}.
 On the shadow tower, the bracket uses $\mu^\uparrow$ and the log coefficient uses $\mathsf R_\alpha$. In the dotted brackets the down-arrow function belongs to state $B$, so
the logarithm depends on the opposite ratio $\hspace{-0.25em}\dot{\hspace{0.25em}\mathbb L}\equiv-\log(1-\dot\lambda_B/\dot\lambda_A)+\log(-\ii)$, with
the residue matrices dotted in the same way.

\subsection{Finite parts and the moment matrix}
\label{sec:norms-finite}

The logarithm only captures the leading large-$l$ behaviour of the tower sums~\eqref{eq:norms-tower-residues}; we now compute the sums in full. Only the harmonic sum at $n=J$ needs the twist split. All other contributions have a smooth limit as $\lambda_B\to\lambda_A$, so we compute the finite parts at a common twist $\lambda$ and keep the twist split only in $\mathbb L$. The idea is to keep the state $A$ fixed and move the state $B$ slightly off shell, by changing one of its charges~\eqref{eq:transfer-matrix-expansion}. The difference between the two Baxter equations is then a total finite difference, so the tower sum telescopes to two endpoint terms, and differentiating at coinciding states turns the infinite sum into a derivative of first-pole data. We now work this out in detail.

\paragraph{A mixed connection matrix.} Let $\widetilde q^{A\downarrow}$ solve the shadow Baxter equation with transfer-matrix eigenvalue $t_A$, and $q^{B\uparrow}$ the Baxter equation with $t_B$, with the same twist $\lambda$ and site data. As in~\eqref{eq:norms-tower-residues}, we express the down-arrow solution of $A$ in an up-arrow basis, now that of $B$, since $B$ is the state we vary. The shadow version of~\eqref{eq:omega-cramer} gives the mixed inverse connection matrix
\begin{equation}
 (\widetilde\Omega^{-1}_{(AB)})_{ij}(u)\equiv
 \sum_{k=1}^{2}\epsilon_{jk}
 \frac{\mathcal W[\widetilde q_i^{A\downarrow},\widetilde q_k^{B\uparrow}](u)}
      {\mathcal W[\widetilde q_1^\uparrow,\widetilde q_2^\uparrow](u)}.
 \label{eq:norms-mixed-connection-wronskian}
\end{equation}
The down-arrow basis belongs to $A$ and the up-arrow basis to $B$. The denominator carries no state label: by~\eqref{eq:setup-wronskian-normalisation} and its shadow version, its canonical normalisation is fixed by the twist and site data alone. It is independent of all charges, including $h$, so only the numerator varies with $B$. Since the shifted Wronskian~\eqref{eq:omega-casoratian-definition} only involves the values at $u$ and $u+\ii$, \eqref{eq:norms-mixed-connection-wronskian} is just Cramer's rule for expanding these two values of $\widetilde q^{A\downarrow}$ in the two independent solutions $\widetilde q_k^{B\uparrow}$,
\begin{equation}
 \widetilde q^{A\downarrow}(u+r\ii)
 =\widetilde\Omega^{-1}_{(AB)}(u)\widetilde q^{B\uparrow}(u+r\ii),
 \qquad r=0,1.
 \label{eq:norms-mixed-connection}
\end{equation}
At $A=B$ both sides solve the same second-order difference equation, so the expansion propagates to every shift $u+r\ii$, $r\in\mathbb Z$, and $\widetilde\Omega^{-1}_{(AB)}$ becomes the $\ii$-periodic $\widetilde\Omega^{-1}$. For $A\ne B$ the two sides obey different recursions, so the expansion holds only at $r=0,1$, and $\widetilde\Omega^{-1}_{(AB)}$ need not be periodic.

\paragraph{Telescoping.} Combining the shadow Baxter equation for $\widetilde q_i^{A\downarrow}$ with the Baxter equation for $q_j^{B\uparrow}$, the difference of the two equations becomes a total finite difference,
\begin{equation}
 (t_A-t_B)\,\widetilde q_i^{A\downarrow}q_j^{B\uparrow}
 =D_{ij}(u-\ii)-D_{ij}(u),
 \label{eq:norms-baxter-sum-by-parts}
\end{equation}
of the Wronskian-type bilinear
\begin{equation}
 D_{ij}(u)\equiv P_+(u)\,\widetilde q_i^{A\downarrow}(u)\,q_j^{B\uparrow}(u+\ii)
 -P_-(u+\ii)\,\widetilde q_i^{A\downarrow}(u+\ii)\,q_j^{B\uparrow}(u).
 \label{eq:norms-D-bilinear}
\end{equation}
By the change of basis~\eqref{eq:norms-mixed-connection} and the canonical
pairing~\eqref{eq:setup-wronskian-normalisation}, $D$ is simply the mixed connection matrix,
\begin{equation}
 D(u)=(\lambda-\lambda^{-1})\,\widetilde\Omega^{-1}_{(AB)}(u)\,\epsilon .
\end{equation}
The identity~\eqref{eq:norms-baxter-sum-by-parts} holds for any two states with the same twist and site data; their conformal weights need not coincide. Summing the residues over the first $N+1$ points of the tower therefore telescopes, leaving only the two endpoints:
\begin{equation}
 \sum_{l=0}^{N}\operatorname*{Res}_{u=\svx_{\alpha,l}}
 \bigl[(t_A-t_B)\widetilde q_i^{A\downarrow}q_j^{B\uparrow}\bigr]=-\operatorname*{Res}_{u=\svx_{\alpha,0}}D_{ij}(u)
 +\operatorname*{Res}_{u=\svx_{\alpha,N}}D_{ij}(u-\ii).
 \label{eq:norms-baxter-tower-sum}
\end{equation}
Let $\widetilde{\mathsf R}_{\alpha,(AB)}$ be the residue of $\widetilde\Omega^{-1}_{(AB)}$ at $\svx_{\alpha,0}$, normalised as in~\eqref{eq:setup-omega-residue-matrices}. The first-pole term is then
\begin{equation}
 e^{-\frac{\ii\pi}{2}(2Js-J)}
 \operatorname*{Res}_{u=\svx_{\alpha,0}}D_{ij}
 =-\ii(\lambda-\lambda^{-1})(\widetilde{\mathsf R}_{\alpha,(AB)}\epsilon)_{ij}.
 \label{eq:norms-first-pole-boundary}
\end{equation}

\paragraph{Charge derivatives.} We choose the deformation
\begin{equation}
 t_A(u)=t(u),\qquad t_B(u)=t(u)+\eta u^{n-1},\qquad n=1,\ldots,J,
 \label{eq:norms-one-sided-deformation}
\end{equation}
which shifts only the charge $a_{n-1}$. By~\eqref{eq:leading-transfer-eigenvalues}, the weight $h$ changes only for $n=J$. We keep the canonical normalisation fixed and impose no quantisation condition on $B$ away from $\eta=0$. The one-sided derivative of the residue matrix is then defined by
\begin{equation}
 \widetilde{\mathsf R}_{\alpha,(AB)}
 =\widetilde{\mathsf R}_\alpha
 +\eta\,\partial^B_{a_{n-1}}\widetilde{\mathsf R}_\alpha
 +O(\eta^2).
 \label{eq:norms-one-sided-residue-variation}
\end{equation}
We now expand~\eqref{eq:norms-baxter-tower-sum} to first order in $\eta$. Its left-hand side is $-\eta$ times the residue sum we want. On the right, the two endpoint residues cancel at $\eta=0$, where $\widetilde\Omega^{-1}$ is $\ii$-periodic, and for $n<J$ the far endpoint does not vary as $N\to\infty$, since the deformation only changes subleading asymptotics. Hence
\begin{equation}
 \sum_{l\geq0}\operatorname*{Res}_{u=\svx_{\alpha,l}}
 \bigl[u^{n-1}\widetilde q_i^{\downarrow}q_j^{\uparrow}\bigr]
 =\partial_\eta\operatorname*{Res}_{u=\svx_{\alpha,0}}D_{ij}\Big|_{\eta=0},
 \qquad n<J.
\end{equation}
With the shadow-map phase, \eqref{eq:norms-first-pole-boundary} and~\eqref{eq:norms-one-sided-residue-variation} then give
\begin{equation}
 \bigl(\mu^\downarrow u^{n-1}q_i^\downarrow q_j^\uparrow\bigr)_\alpha
 =-\ii(\lambda-\lambda^{-1})\,
 (\partial^B_{a_{n-1}}\widetilde{\mathsf R}_\alpha\,\epsilon)_{ij},
 \qquad n<J.
 \label{eq:norms-finite-moment-connection}
\end{equation}
The infinite residue sum has thus reduced to a charge derivative of first-pole data. In the dotted sector $B$ again labels the off-shell copy, so the dotted formulas follow by dotting all chain data while keeping the label $B$.

\paragraph{The highest moment.} At $n=J$ the weight varies, and the derivative of $u^{\pm h_B}$ produces $\log u$, so the far endpoint now contributes. Its coefficient is the one already found in~\eqref{eq:norms-log-coefficient}, while the first-pole term is unchanged. Since $\log\svx_{\alpha,N}=\log N+\log(-\ii)+O(1)$, the far endpoint splits into two pieces. The $\log N$ is the harmonic divergence that the twist split replaces by $\mathbb L$. The remaining $\log(-\ii)$ is the constant we included in $\mathbb L$ and $\hspace{-0.25em}\dot{\hspace{0.25em}\mathbb L}$. Appendix~\ref{app:omega-highest-moment} records a direct check of this finite part,~\eqref{eq:norms-highest-moment-endpoint}.

\paragraph{The moment matrix in connection form.}
We now have both pieces of every bracket: the logarithm
\eqref{eq:norms-log-coefficient} and the finite part
\eqref{eq:norms-finite-moment-connection}, each an entry of a residue matrix.
It remains to assemble them into the moments. We substitute~\eqref{eq:norms-log-coefficient}
and~\eqref{eq:norms-finite-moment-connection} into the site bracket
\eqref{eq:paired-tower-bracket} at coinciding states, with diagonal
$\Gamma$. The general conversion~\eqref{eq:setup-residue-value-conversion}
turns the products of residue matrices, including their measure factors,
into residues of a single matrix function:
\begin{equation}
 \boxed{
 \mathcal M_{nm}
 \simeq-2\pi\ii\sum_{p}\operatorname*{Res}_{u=p}\,
 \det\Omega(u)\,
 \operatorname{tr}\!\bigl[\Gamma\,\mathsf B_{n}(u)\,\Gamma\,\dot{\mathsf B}_{m}(u)\bigr],}
 \label{eq:norms-moment-connection-form}
\end{equation}
\begin{equation}
 \begin{aligned}
 \mathsf B_{n}&\equiv\ii\,\delta_{nJ}\,\mathbb L\;\Omega^{-1}\sigma_1
 +(\lambda-\lambda^{-1})\,(\partial^B_{a_{n-1}}\Omega^{-1})\,\epsilon,
 \\
 \dot{\mathsf B}_{m}&\equiv\ii\,\delta_{mJ}\,\hspace{-0.25em}\dot{\hspace{0.25em}\mathbb L}\;\dot\Omega^{-1}\sigma_1
 +(\dot\lambda-\dot\lambda^{-1})\,(\partial^B_{\dot a_{m-1}}\dot\Omega^{-1})\,\epsilon.
 \end{aligned}
 \label{eq:norms-B-matrices}
\end{equation}
The sum runs over the $2J$ poles at
$p=\svx_{\alpha,0},\widetilde\svx_{\alpha,0}$.
At an ordinary point the pole comes from $\det\Omega$. At a shadow point
$\det\Omega$ vanishes and both inverse matrices have poles. Both families are
residues of the same function, by the shadow
map~\eqref{eq:setup-omega-shadow} (Appendix~\ref{app:omega-mixed-log}).
Here $\simeq$ means equality as $\lambda_B\to\lambda_A$ and
$\dot\lambda_B\to\dot\lambda_A$. For $n,m<J$ the entries agree in this
limit. In the last row and column all logarithmic terms agree, while
bounded terms are dropped; these do not enter the divergent part of the
determinant once the other logarithms have cancelled.

\subsection{Cancellation of the logarithms}
\label{sec:norms-onshell}

By~\eqref{eq:norms-B-matrices}, $\mathbb L$ appears only in the last row of
$\mathcal M$ and $\hspace{-0.25em}\dot{\hspace{0.25em}\mathbb L}$ only in its last column. Since a
determinant is linear in each row and column, $\det\mathcal M$ contains at
most one logarithm of each kind. A term $\mathbb L\hspace{-0.25em}\dot{\hspace{0.25em}\mathbb L}$ would
contradict the single logarithm of the coordinate-space
argument~\eqref{eq:coordinate-logarithmic-overlap}. In this subsection we show
that on shell the logarithmic part of the moment matrix is
\begin{equation}
 \begin{aligned}
 \mathcal M_{Jm}&={\mathbb L}\times 0+\text{finite},\qquad
 \mathcal M_{nJ}=\hspace{-0.25em}\dot{\hspace{0.25em}\mathbb L}\times 0+\text{finite}\qquad(n,m<J),\\
 \mathcal M_{JJ}&={\mathbb L}\;\hspace{-0.25em}\dot{\hspace{0.25em}\mathbb L}\times 0
 +4\pi c(\mathbb L+\hspace{-0.25em}\dot{\hspace{0.25em}\mathbb L})+\text{finite}.
 \end{aligned}
 \label{eq:norms-onshell-glued-relations}
\end{equation}
Thus the double logarithm and all single logarithms outside
$\mathcal M_{JJ}$ cancel. The key algebraic input is gluing~\eqref{eq:setup-omega-gluing}.

\paragraph{A residue lemma.}
Both cancellations below rest on the following property.
Let $f$ be an $\ii$-periodic meromorphic function with finite limits
$f(\pm\infty)$ as $\Re u\to\pm\infty$. Integrating $f$ around the boundary
of the rectangle $|\Re u|\le R$, $0\le\Im u\le1$, and taking $R\to\infty$,
gives
\begin{equation}
 \sum_{p}\operatorname*{Res}_{u=p}f(u)
 =\frac{f(+\infty)-f(-\infty)}{2\pi}.
 \label{eq:setup-omega-residue-lemma}
\end{equation}
The sum runs over the poles in one period strip. The horizontal sides
cancel by periodicity, and the vertical sides give
$\ii[f(+\infty)-f(-\infty)]$. We apply the lemma to functions
built from $\Omega$ and $\dot\Omega$. By~\eqref{eq:setup-omega-pole-table}, their only poles are the $2J$ ordinary
and shadow points, and their limits follow
from~\eqref{eq:setup-omega-limits}.

\paragraph{The double logarithm.}
By~\eqref{eq:norms-moment-connection-form}, the coefficient of the double logarithm
$\mathbb L\hspace{-0.25em}\dot{\hspace{0.25em}\mathbb L}$ in $\mathcal M_{JJ}$ is $2\pi\ii$ times the
residue sum of
\begin{equation}
 \det\Omega\,
 \operatorname{tr}\!\bigl[\Gamma\Omega^{-1}\sigma_1\Gamma\dot\Omega^{-1}\sigma_1\bigr]
 =2c\,\frac{\Omega_{11}\Omega_{22}+\Omega_{12}\Omega_{21}}{\det\Omega}
 =2c+4c\,\frac{\Omega_{12}\Omega_{21}}{\det\Omega}.
 \label{eq:norms-glued-two-by-two}
\end{equation}
Here we used gluing~\eqref{eq:setup-omega-gluing} and the inverse
matrix~\eqref{eq:omega-two-by-two-inverse}, leaving a function of $\Omega$
alone. The constant has no residues and thus does not contribute to \eqref{eq:norms-moment-connection-form}. The second term is $\ii$-periodic, and its
only poles in the periodicity strip are the $2J$ first poles. Furthermore, by~\eqref{eq:setup-omega-limits} this combination
vanishes at both ends: $\Omega_{12}\to0$ at $+\infty$ and
$\Omega_{21}\to0$ at $-\infty$. The residue
lemma~\eqref{eq:setup-omega-residue-lemma} then sets the residue sum to zero:
\begin{equation}
 \left.\partial_{\mathbb L}\partial_{\hspace{-0.25em}\dot{\hspace{0.25em}\mathbb L}}\mathcal M_{JJ}\right.=0.
 \label{eq:norms-double-log-zero}
\end{equation}
Neither tower vanishes on its own: the cancellation needs the sum over
all $2J$ poles in the strip, ordinary and shadow together.

\paragraph{The single logarithms.}
The coefficient of $\mathbb L$ in $\mathcal M_{Jm}$ pairs the
$\mathbb L$-term of $\mathsf B_J$ with the finite part of $\dot{\mathsf B}_{m}$
in~\eqref{eq:norms-B-matrices}. Write $\partial\equiv\partial_{\dot a_{m-1}}$, with the twists and site
data fixed. We evaluate this coefficient in three steps.

\paragraph{Completing the variation.}
The finite part of $\dot{\mathsf B}_m$ contains the one-sided variation
$\partial^B$, which acts on the up-arrow factor of the residue matrix. Its
complement $\partial^A\equiv\partial-\partial^B$ acts on the down-arrow
factor and drops out of the contraction on shell: the first-pole
residues factorise, and the $\partial^A$ term is proportional to
$\operatorname*{Res}\Qp(u,u+\ii)$, which vanishes by the analyticity of
$\mathbb Q_+$~\eqref{eq:setup-Qplus-first-pole-regularity}
(Appendix~\ref{app:omega-single-log}). We may therefore replace
$\partial^B$ by $\partial$ to write
\begin{equation}
 \begin{aligned}
 \partial_{\mathbb L}\mathcal M_{Jm}
 ={}&2\pi(\dot\lambda-\dot\lambda^{-1})\sum_p\operatorname*{Res}_{u=p}\det\Omega\,
 \operatorname{tr}\!\bigl[\epsilon\,\Gamma\Omega^{-1}\sigma_1\Gamma
 (\partial\dot\Omega^{-1})\bigr].
 \end{aligned}
 \label{eq:norms-single-log-connection}
\end{equation}

\paragraph{Reducing the trace.}
Gluing~\eqref{eq:setup-omega-gluing}, the inverse
formula~\eqref{eq:omega-two-by-two-inverse} and
$\det\Omega=\det\dot\Omega$ from~\eqref{eq:setup-omega-determinant}
collapse the trace in~\eqref{eq:norms-single-log-connection} to a
$\sigma_3$-graded trace of $\dot\Omega$ alone
(Appendix~\ref{app:omega-single-log}),
\begin{equation}
 \partial_{\mathbb L}\mathcal M_{Jm}
 =-2\pi c(\dot\lambda-\dot\lambda^{-1})
 \sum_p\operatorname*{Res}_{u=p}
 \operatorname{tr}\!\bigl[\sigma_3\dot\Omega^{-1}\partial\dot\Omega\bigr].
 \label{eq:norms-graded-trace}
\end{equation}

\paragraph{Evaluating the residue sum.}
It only remains to compute the residue sum in~\eqref{eq:norms-graded-trace}. Since it is built from connection matrices, it is $\ii$-periodic and by~\eqref{eq:setup-omega-pole-table},
its only possible poles are the $2J$ ordinary and shadow points
$u=\theta_\alpha-\ii s$ and $u=\theta_\alpha-\ii(1-s)$ modulo $\ii$,
$\alpha=1,\ldots,J$. Charge variations keep these positions fixed and
only the diagonal entries of $\dot\Omega^{-1}\partial\dot\Omega$
contribute to the $\sigma_3$-graded trace. Using~\eqref{eq:setup-omega-limits},
these reduce to logarithmic derivatives of the diagonal phases, giving
zero at $+\infty$ and $-4\pi\ii\,\partial\dot h$ at $-\infty$. The residue
lemma~\eqref{eq:setup-omega-residue-lemma} therefore evaluates the sum
in~\eqref{eq:norms-graded-trace} to $2\ii\,\partial\dot h$, giving
\begin{equation}
 \partial_{\mathbb L}\mathcal M_{Jm}
 =-4\pi\ii\,c\,(\dot\lambda-\dot\lambda^{-1})\,
 \frac{\partial\dot h}{\partial\dot a_{m-1}}
 =4\pi c\,\delta_{mJ}.
 \label{eq:norms-single-log}
\end{equation}
In the last step we differentiated the leading transfer-matrix
coefficient~\eqref{eq:leading-transfer-eigenvalues} at fixed twists and
site data. It ties $\dot h$ to $\dot a_{J-1}$ alone, with
$\partial\dot h/\partial\dot a_{J-1}=\ii/(\dot\lambda-\dot\lambda^{-1})$.
The lower charges leave the asymptotics, and hence the phases, unchanged.
Thus the single logarithm measures the variation of the relative
asymptotic phase of the two Baxter solutions. Exchanging dotted and
undotted quantities gives the same result for $\dot{\mathbb L}$ in the
last column.

\paragraph{The single logarithm in the norm.}
The logarithmic part of the moment matrix therefore sits entirely in its
corner entry. Expanding the
determinant along this entry gives
\begin{equation}
 \det\mathcal M=4\pi c(\mathbb L+\hspace{-0.25em}\dot{\hspace{0.25em}\mathbb L}) \det_{1\le n,m \le J-1}\mathcal M_{nm}\,+\text{finite}.
 \label{eq:norms-single-log-determinant}
\end{equation}
The dotted and undotted logarithms contribute equally, so the coefficient
of the common twist logarithm is twice the coordinate-space coefficient
$\mathcal C_{\log}$, as in~\eqref{eq:norms-twist-log-overlap}. Restoring the state normalisation
from~\eqref{eq:norms-overlap-recalled}, we obtain
\begin{equation}
 \mathcal C_{\log}=4\pi c\,\mathcal N_J^{\mathrm{SoV}}\mathcal A^2 J!\det_{1\le n,m \le J-1}\mathcal M_{nm}.
 \label{eq:norms-coordinate-log-coefficient}
\end{equation}
Finally, combining this coefficient with~\eqref{eq:norms-general-norm-relation-recalled}
gives~\eqref{eq:norms-two-point-normalisation}.

\subsection{Collapsing of the norm}
\label{sec:norms-collapse}
The finite minor in~\eqref{eq:norms-coordinate-log-coefficient} is
determined entirely by connection data at the first poles. For $b,d=1,\ldots,J-1$,
the charge variation of the telescoping
identity~\eqref{eq:norms-baxter-tower-sum} has no boundary contribution
at infinity, so each convergent tower sum reduces to its first-pole
term. The ordinary and shadow contributions can then be collected in a
single residue formula. Define
\begin{equation}
 \mathsf E_b(u)\equiv\Omega^{-1}(u)\partial_{a_{b-1}}\Omega(u),
 \qquad
 \rho(u)\equiv\frac{\Omega_{22}(u)^2}{\det\Omega(u)},
 \label{eq:norms-collapsed-data}
\end{equation}
with $\dot{\mathsf E}_d=\dot\Omega^{-1}\partial_{\dot a_{d-1}}\dot\Omega$
in the dotted sector. The derivatives act off shell at fixed twist and
site data, with the canonical normalisation
held fixed, and are evaluated at the physical state. As derived in
Appendix~\ref{sec:norms-boundary-variation}, the finite norm matrix can be written as
\begin{equation}
 \boxed{\displaystyle
 \mathcal M_{bd}
 =-2\pi\ii(\lambda-\lambda^{-1})(\dot\lambda-\dot\lambda^{-1})
 \sum_p\operatorname*{Res}_{u=p}
 \frac{(\mathsf E_b(u))_{12}(\dot{\mathsf E}_d(u))_{12}}{\rho(u)}.}
 \label{eq:norms-connection-matrix}
\end{equation}
For generic site data the sum contains the $2J$ ordinary and shadow
first poles in one period strip. Since
$(\mathsf E_b)_{12}=(\Omega_{22}^2/\det\Omega)\,\partial_{a_{b-1}}(\Omega_{12}/\Omega_{22})$,
and on shell $\dot\Omega_{22}=\Omega_{22}$ by gluing~\eqref{eq:setup-omega-gluing}
and $\det\dot\Omega=\det\Omega$ by~\eqref{eq:setup-omega-determinant},
this is the Gaudin-like form~\eqref{eq:norms-gaudin-form}: the derivatives
are those of the quantisation
conditions~\eqref{eq:norms-gaudin-quantisation-functions}. Thus the infinite
towers have collapsed to a finite pairing of the undotted and dotted
charge responses of the gluing condition.

This form reduces the untwisting problem to the limiting behaviour of
$\Omega$ and its charge derivatives. In the untwisting limit of Section~\ref{sec:untwisting-norm}
we separate the charge variations that preserve the zero-twist
conformal weights from the singular Casimir direction.

\subsection{Explicit \texorpdfstring{$J=1$}{J=1} example}
\label{sec:norms-J1-example}

As a check of our formula \eqref{eq:norms-coordinate-log-coefficient} we can perform an explicit computation for the twisted $J=1$ chain. Here every quantity is explicit, and the result can be compared with the coordinate-space computation in~\eqref{eq:coordinate-J1-logarithmic-overlap}.

\paragraph{The logarithm.}
For $J=1$ the moment matrix has a single entry, $\mathcal M_{11}=\mathcal M_{JJ}$, and the minor in~\eqref{eq:norms-single-log-determinant} is empty. By~\eqref{eq:norms-onshell-glued-relations}, the overlap~\eqref{eq:norms-overlap-recalled} is therefore
\begin{equation}
 \LRpair{\psi_A}{\psi_B}
 =4\pi c\,\mathcal N_1^{\mathrm{SoV}}\mathcal A^2\,
 \bigl(\mathbb L+\hspace{-0.25em}\dot{\hspace{0.25em}\mathbb L}\bigr)+O(1).
\end{equation}
Both $\mathbb L$ and $\hspace{-0.25em}\dot{\hspace{0.25em}\mathbb L}$ diverge as $\log(1/\varepsilon)$. Comparing with~\eqref{eq:norms-twist-log-overlap} gives
\begin{equation}
 \mathcal C_{\log}=4\pi c\,\mathcal N_1^{\mathrm{SoV}}\mathcal A^2 ,
 \label{eq:norms-J1-Clog-SoV}
\end{equation}
which is~\eqref{eq:norms-coordinate-log-coefficient} at $J=1$.

\paragraph{The three constants.}
Each factor in~\eqref{eq:norms-J1-Clog-SoV} is known in closed form:
\begin{itemize}
 \item The SoV normalisation $\mathcal N_1^{\mathrm{SoV}}=\Gamma(1-2s)/(2\pi\Gamma(2s))$ is given in~\eqref{eq:J1-sov-resolution-normalisation}.
 \item The gluing constant $c=\Gamma^{22}$ is given in~\eqref{eq:J1-Qplus-closed}:
 \begin{equation}
  c=\frac{1}{[\lambda-\lambda^{-1}]^{\{2h,2\dot h\}}}\,
  \frac{\Gamma(1-s+h)\,\Gamma(s+\dot h)}{\Gamma(1-s-h)\,\Gamma(s-\dot h)}.
 \end{equation}
 \item The state normalisation $\mathcal A=\psi^{\rm C}(0)/\mathbb Q_+(-\ii s,-\ii s)$ is given in~\eqref{eq:J1-mellin-normalisation}.
\end{itemize}
To evaluate $\mathcal A$, we need $\mathbb Q_+$ at the first ordinary point. At $u=-\ii s$ the down-arrow functions~\eqref{eq:J1-q-closed} become elementary, since ${}_2F_1(a,b;a;x)=(1-x)^{-b}$. The up-arrow functions do not, but the glued combination does, and one finds
\begin{equation}
 \mathcal A^2
 =[\lambda-\lambda^{-1}]^{\{2h,2\dot h\}}\,
 \frac{\Gamma(2s)^2}{\Gamma(1-2s)^2}\,
 \frac{\Gamma(1-s-h)\,\Gamma(1-s-\dot h)}{\Gamma(s+h)\,\Gamma(s+\dot h)}.
 \label{eq:norms-J1-A}
\end{equation}

\paragraph{The coefficient of the logarithm.}
In the product~\eqref{eq:norms-J1-Clog-SoV}, the twist dependence of $c$ cancels against that of $\mathcal A^2$, leaving
\begin{equation}
 \mathcal C_{\log}
 =\frac{2\,\Gamma(2s)}{\Gamma(1-2s)}\,
 \frac{\Gamma(1-s+h)\,\Gamma(1-s-\dot h)}{\Gamma(s+h)\,\Gamma(s-\dot h)}.
 \label{eq:norms-J1-Clog}
\end{equation}
This is independent of $\lambda$, as it should be. For the scalar fishnet state, $s=\tfrac14$ and $h=\dot h$, the reflection formula gives
\begin{equation}
 \mathcal C_{\log}
 =\frac{4\pi^2}{\cos(2\pi h)\,\Gamma(\tfrac14-h)^2\,\Gamma(\tfrac14+h)^2},
\end{equation}
which agrees with the coordinate-space logarithm~\eqref{eq:coordinate-J1-logarithmic-overlap}.

\paragraph{The norm.}
Finally, at $h=\dot h$ the exact coupling~\eqref{eq:J1-exact-coupling} gives $\mathcal D_\xi=2\xi^2/\partial_h\xi^2$, and $\xi^2$ is proportional to the Gamma ratio in~\eqref{eq:norms-J1-Clog}. The relation~\eqref{eq:norms-general-norm-relation-recalled} at $J=1$ then gives
\begin{equation}
 \mathcal N_2
 =-\frac{\mathcal C_{\log}}{\mathcal D_\xi}
 =-\partial_h
 \frac{\Gamma(\tfrac34-h)\,\Gamma(\tfrac34+h)}{\Gamma(\tfrac14-h)\,\Gamma(\tfrac14+h)},
\end{equation}
which reproduces~\eqref{eq:coordinate-J1-two-point-normalisation}. At the free point $h=\tfrac14$ it equals $1$, the norm of the free field.

\section{Removing the twist: norms and OPE coefficients}
\label{sec:untwisted}

A normalised three-point function of scalar primaries is built from a
three-point amplitude and the two-point normalisations of its three
operators,
\begin{equation}
 \begin{aligned}
 \bigl\langle\mathcal O_A(x_A)\mathcal O_B(x_B)\mathcal O_C(x_C)\bigr\rangle
 &=\frac{\mathcal C_{ABC}\,\sqrt{\mathcal N^{\theta=0}_{2,A}\mathcal N^{\theta=0}_{2,B}\mathcal N^{\theta=0}_{2,C}}}
 {|x_{AB}|^{\Delta_A+\Delta_B-\Delta_C}
  |x_{AC}|^{\Delta_A+\Delta_C-\Delta_B}
  |x_{BC}|^{\Delta_B+\Delta_C-\Delta_A}},\\[3pt]
 \bigl\langle\mathcal O_X(x)\bar{\mathcal O}_X(0)\bigr\rangle
 &=\frac{\mathcal N^{\theta=0}_{2,X}}{|x|^{2\Delta_X}}.
 \end{aligned}
 \label{eq:untwisting-normalised-3pt}
\end{equation}

\begin{figure}[htbp]
 \centering
 \includegraphics[width=.98\linewidth]{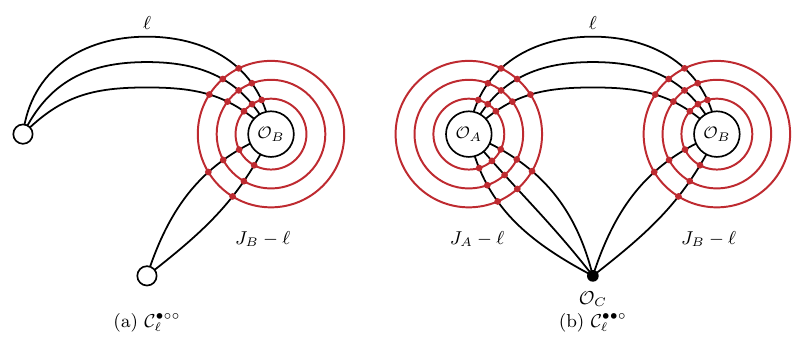}
 \caption{The two three-point coefficients obtained after removing the
 twist. (a) $\mathcal C_\ell^{\bullet\circ\circ}$ couples one interacting
 operator to two groups of $\ell$ and $J_B-\ell$ scalar probes in the
 same ordered trace. (b)
 $\mathcal C_\ell^{\bullet\bullet\circ}$ couples two interacting operators
 sharing $\ell$ lines to the wheel-free mixed trace $\mathcal O_C$.
 Black lines are scalar propagators and red circles are fishnet wheels.
 The external two-point normalisations are understood.}
 \label{fig:untwisting-three-point-overview}
\end{figure}

We consider the two cases of
Figure~\ref{fig:untwisting-three-point-overview}. A bullet $\bullet$ marks an
operator dressed by fishnet wheels, while a circle $\circ$ refers to an operator at zero coupling. In terms of the normalised operators
$\widehat{\mathcal O}_X=\mathcal O_X/\sqrt{\mathcal N^{\theta=0}_{2,X}}$ placed at $\infty$, $1$ and $0$, the $3$-point coefficients are
\begin{equation}
 \begin{aligned}
 \mathcal C_\ell^{\bullet\circ\circ}&\equiv
 \left.\bigl\langle\widehat{\mathcal O}_{\Omega_A}(\infty)
 \widehat{\mathcal O}_B(1)\widehat{\mathcal O}_C(0)\bigr\rangle
 \right|_{\xi_A=0,\;\xi_B=\xi},\\[3pt]
 \mathcal C_\ell^{\bullet\bullet\circ}&\equiv
 \left.\bigl\langle\widehat{\mathcal O}_A(\infty)
 \widehat{\mathcal O}_B(1)\widehat{\mathcal O}_C(0)\bigr\rangle
 \right|_{\xi_A=\xi_B=\xi},
 \end{aligned}
 \label{eq:untwisting-HHL-OPE-definition}
\end{equation}
where the bridge length $\ell$ must satisfy $1\le\ell<\min(J_A,J_B)$ and the length of the third operator $\mathcal{O}_C$ must satisfy $J_C=J_A+J_B-2\ell$. The two operators at zero coupling are
\begin{equation}
 \mathcal O_{\Omega_A}=\operatorname{tr}[(\phi_1^\dagger)^{J_A}],
 \qquad
 \mathcal O_C=\operatorname{tr}
 [\phi_1^{J_A-\ell}\phi_1^{\dagger\,J_B-\ell}].
\end{equation}
Both operators are free, with planar two-point normalisations
$\mathcal N^{\theta=0}_{2,\Omega_A}=J_A$ and $\mathcal N^{\theta=0}_{2,C}=\mathcal N_{2,C}=1$;
the mixed trace has no cyclic symmetry. At generic $\theta_\alpha$ the
planar brackets and the norms
of~\eqref{eq:coordinate-general-norm-relation} refer to one fixed
ordering of the legs, and $\Omega_{J_A}$ has norm $1$. As
$\theta_\alpha\to0$ the three-point function receives all $J_AJ_B$
cyclic alignments of the two traces, and the norms
become~\eqref{eq:norm-theta-zero}.

In terms of the planar brackets~\eqref{eq:endpoints-three-point-correlator}
and~\eqref{eq:HHL-target-correlator} of the twisted CFT wave functions with the operators at
$x_A=\lambda$, $x_B=\lambda^{-1}$, $x_C=0$ and
$\lambda=e^{\ii\phi}$, the structure constants can be computed as\footnote{Setting the inhomogeneities $\theta_\alpha$ to zero slightly changes the combinatorics, since the operators become cyclic and the pole contributions of different sites coincide, giving extra symmetry factors. At $\theta_\alpha=0$ each of the $J_AJ_B$ cyclic alignments of the two traces contributes equally to the three-point function, and the two-point normalisations are $\mathcal N_2^{\theta=0}$ of~\eqref{eq:norm-theta-zero}. Relative to the generic-$\theta_\alpha$ norms the net factor is $\sqrt{J_AJ_B}$, the same in both lines, so the HLL line is the HHL line evaluated at $\psi_A=\Omega_{J_A}$. At $\xi=0$, where $\psi_{A,B}\propto\Omega_{J_{A,B}}$, both lines reproduce the free planar Wick count $\sqrt{J_AJ_B}$ for every $J$; at $J=2$ this was also checked against the closed-form results below.}
\begin{equation}
 \boxed{\begin{aligned}
 \mathcal C_\ell^{\bullet\circ\circ}
 &=\lim_{\theta_\alpha^A,\theta_\alpha^B\to0}\lim_{\phi\to0^+}
 (2\sin\phi)^{-J_C/2}J_AJ_B\,
 \frac{\LRpair{\Omega_{J_A}^{\rm C}}{\psi_B^{\rm C}}_\ell}
 {\sqrt{\mathcal N^{\theta=0}_{2,\Omega_A}\mathcal N^{\theta=0}_{2,B}\mathcal N_{2,C}}},\\[5pt]
 \mathcal C_\ell^{\bullet\bullet\circ}
 &=\lim_{\theta_\alpha^A,\theta_\alpha^B\to0}\lim_{\phi\to0^+}
 (2\sin\phi)^{-J_C/2}J_AJ_B\,
 \frac{\LRpair{\psi_A^{\rm C}}{\psi_B^{\rm C}}_\ell}
 {\sqrt{\mathcal N^{\theta=0}_{2,A}\mathcal N^{\theta=0}_{2,B}\mathcal N_{2,C}}}.
 \end{aligned}}
 \label{eq:untwisting-normalised-limits}
\end{equation}
Here $(2\sin\phi)^{J_C/2}$ is the position factor
of~\eqref{eq:untwisting-normalised-3pt} at these points, divided by
the two-point factors of $A$ and $B$. By~\eqref{eq:norm-theta-zero}
the prefactor $J_AJ_B/\sqrt{\mathcal N^{\theta=0}_{2,A}\mathcal N^{\theta=0}_{2,B}}$
equals $\sqrt{J_AJ_B}/\sqrt{\mathcal N_{2,A}\mathcal N_{2,B}}$ in terms of
the generic-$\theta_\alpha$ norms, with $\mathcal N_{2,\Omega_A}=1$; the
formulas below use this form.

In the previous sections we described how to compute each of these ingredients in
the twisted theory. What is left is to remove the twist. We stress that this is not just a matter of setting $\phi=0$. Both the numerators and denominators in \eqref{eq:untwisting-normalised-limits} diverge as powers of $\phi$, and the powers cancel only in the
ratio. The goal of this section is
to extract the finite coefficients of these divergences.

The section is organised as follows. In
Section~\ref{sec:untwisting-whittaker} we relate the twisted Q-functions to the zero-twist Q-functions. We then extract the leading divergence of the norm
(Section~\ref{sec:untwisting-norm}) and of the HHL moment determinant
(Section~\ref{sec:untwisting-moment-reduction}). Their ratio gives the OPE coefficients in
Section~\ref{sec:untwisting-final-result}, which we make fully explicit
for two sites in Section~\ref{sec:untwisting-J2-example}. Descendants
and numerical checks are deferred to
Appendices.

\subsection{Q-functions at zero twist}
\label{sec:untwisting-whittaker}

First, we fix the relative normalisation of the twisted and zero-twist
Q-functions and study how they translate into each other in the zero-twist limit. Throughout this section, boldface denotes zero-twist
quantities, such as $\boldsymbol q$, $\boldsymbol\Omega$ and
$\mathbf Q_+$.

\paragraph{Untwisting the transfer matrix.}
Recall that the twist enters the Baxter
equation~\eqref{eq:baxter-equation} through the transfer matrix eigenvalue
$t(u)$, with
\begin{equation}
 t(u)=\sum_{\beta=0}^{J}u^\beta a_\beta,
\end{equation}
see~\eqref{eq:transfer-matrix-expansion}. In particular, we have
\begin{equation}
  a_J = 2\cos\phi,\qquad a_{J-1} = -2\cos\phi\,\Theta+2h\sin\phi,
\end{equation}
with $\Theta=\sum_\alpha\theta_\alpha$, see~\eqref{eq:leading-transfer-eigenvalues}.
As $\phi\to0$, the coefficients approach their zero-twist values $\boldsymbol{a}_\beta$, and the two above become state-independent,
\begin{equation}
\boldsymbol{a}_J=2, \quad \boldsymbol{a}_{J-1} = -2\Theta\,.
\end{equation}
Instead, the dependence on the conformal weight now first appears in $\boldsymbol{a}_{J-2}$. One integral of motion is thus absorbed by the enlarged global symmetry, which is now restored to the full $\mathfrak{so}(3,1)$, or $\mathfrak{sl}(2)\oplus\mathfrak{sl}(2)$ after complexification. In addition, the exponential factors in the Q-functions~\eqref{eq:baxter-asymptotics} disappear,
so they no longer distinguish the two solutions at infinity. In this subsection we study this dramatic change in structure in detail.

\paragraph{Primaries and descendants.}
At finite twist $\phi$, the global $\mathfrak{so}(3,1)$ symmetry is broken and there is no notion of highest-weight states or descendants. As $\phi\rightarrow 0$ and the symmetry is restored, states organise themselves into multiplets of the full conformal group. A finite-twist state then approaches either a conformal primary or a descendant as $\phi\rightarrow 0$.

A descendant of a primary can be labelled by how many times $(n_{\mathrm d},\dot n_{\mathrm d})$ we apply the global $\mathfrak{sl}(2)$ raising operators in the undotted and dotted sectors. For a conformal primary of weight $(\boldsymbol{h},\dot{\boldsymbol h})$, a state untwisting to a member of its multiplet behaves as
\begin{equation}
  h(\phi)\to\boldsymbol h+n_{\mathrm d},\quad \dot h(\phi)\to\dot{\boldsymbol h}+\dot n_{\mathrm d}\,.
\end{equation}
In the main text we consider only states which untwist to conformal primaries, i.e. $n_{\mathrm d}=\dot n_{\mathrm d}=0$, so that $h\to\boldsymbol h$ and
$\dot h\to\dot{\boldsymbol h}$. For these states we also write $h\equiv\boldsymbol h$ and $\dot h\equiv\dot{\boldsymbol h}$ for the zero-twist weights, and keep $h(\phi)$ for the twisted weight where the distinction matters. The general case, which includes descendants, is considered in Appendix~\ref{app:untwisting-descendants}.

\paragraph{Conformal Casimir.}
At zero twist, the conformal weight $\boldsymbol h$ is encoded in the quadratic Casimir
$\boldsymbol h(\boldsymbol h-1)$. This appears in the first non-trivial charge $\boldsymbol{a}_{J-2}$ of the transfer matrix:
\begin{equation}
 \boldsymbol{a}_{J-2}=-\boldsymbol{h}(\boldsymbol{h}-1)+Js(s-1)
 +2\sum_{\alpha<\beta}\theta_\alpha\theta_\beta\,,
 \label{eq:untwisting-casimir-charge}
\end{equation}
where we have assumed equal spins $s_\alpha=\dot s_\alpha=s$ on all sites.

\paragraph{Q-function asymptotics.}
In the twisted case, Q-functions were labelled by their asymptotics, and we do the same here. Inserting $q(u)\sim u^\nu$ into the Baxter equation at $\phi=0$ gives
\begin{equation}
  (\nu+Js-\boldsymbol{h})(\nu+Js+\boldsymbol{h}-1)=0\,.
\end{equation}
The two roots give the two possible asymptotics, and we label the corresponding solutions as
\begin{equation}
 \boldsymbol{q}_1(u)\simeq u^{\boldsymbol{h}-Js},\qquad
 \boldsymbol{q}_2(u)\simeq u^{1-\boldsymbol{h}-Js}.
 \label{eq:untwisting-zero-twist-asymptotics}
\end{equation}
As in the twisted case, both solutions come in UHPA and LHPA versions, $\boldsymbol{q}^\downarrow_a(u)$ and $\boldsymbol{q}^\uparrow_a(u)$.

It is useful to compare with the twisted asymptotics~\eqref{eq:baxter-asymptotics}, where the power of $u$ is $h-Js$ for $q_1$ and $-h-Js$ for $q_2$. The power of $q_1$ flows smoothly to that of $\boldsymbol{q}_1$ as $\phi\to0$, but the power of $q_2$ jumps \textit{up} by one. This phenomenon is well understood for highest-weight representations~\cite{Kazakov:2015efa}, where one can work almost exclusively with polynomial Q-functions. In our set-up the situation is more intricate, as we will see.

\paragraph{Wronskian.}
We will also use the Wronskian at zero twist. The finite-difference equation~\eqref{eq:setup-wronskian-recurrence}, which follows from the Baxter equation, fixes its $u$-dependence, while the large-$u$ asymptotics of the Q-functions give
$\mathcal W[\boldsymbol{q}_1,\boldsymbol{q}_2](u)\simeq\ii(1-2\boldsymbol{h})u^{-2Js}$.
Hence we obtain
\begin{equation}
 \begin{aligned}
 P_+(u)\mu^\uparrow(u)
 \mathcal W[\boldsymbol{q}_1^{\uparrow},\boldsymbol{q}_2^{\uparrow}](u)
 &=\ii^{1+(2s-1)J}(1-2\boldsymbol{h}),\\
 P_+(u)\mu^\downarrow(u)
 \mathcal W[\boldsymbol{q}_1^{\downarrow},\boldsymbol{q}_2^{\downarrow}](u)
 &=\ii^{1-(2s-1)J}(1-2\boldsymbol{h}).
 \end{aligned}
 \label{eq:untwisting-zero-twist-wronskians}
\end{equation}
As usual, the dotted identities follow by replacing $\boldsymbol{h}$ by $\dot{\boldsymbol{h}}$.

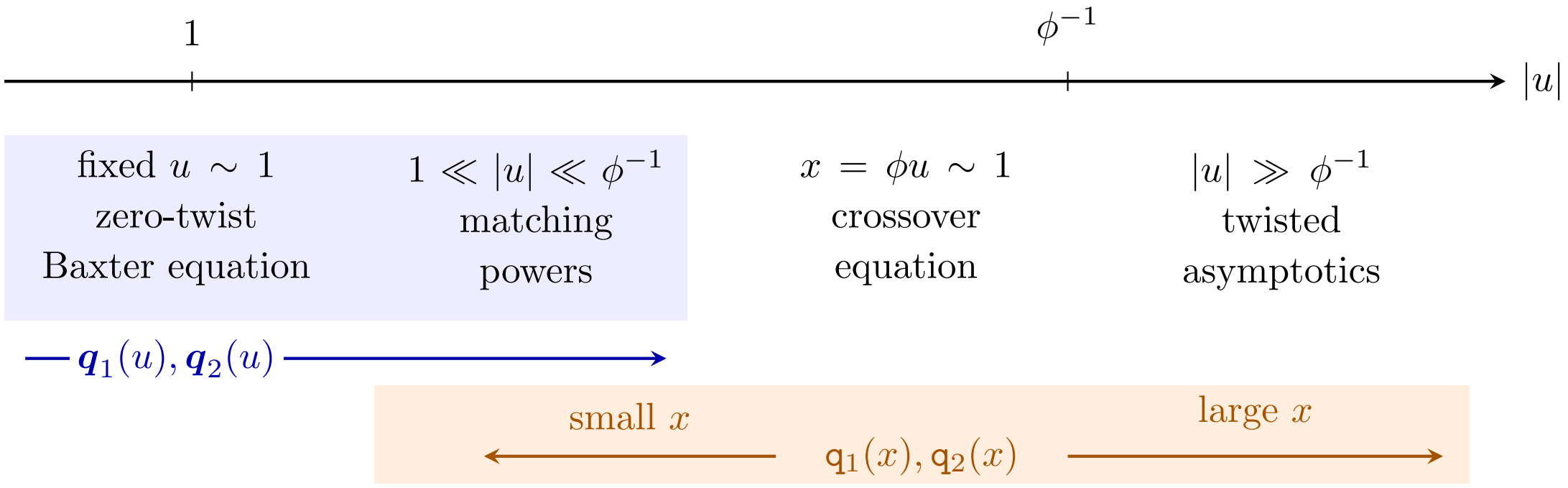
\begin{figure}[tb]
 \centering
 \begin{tikzpicture}[x=0.96cm,y=1cm,>=stealth,font=\small]
  \fill[blue!7] (0,-0.55) rectangle (6.55,1.15);
  \fill[shift={(-0.05,+.20)}, orange!13] (3.6,-2.25) rectangle (14.1,-1.35);
  \draw[->,thick] (0,1.65)--(14.4,1.65) node[right] {$|u|$};
  \draw (1.8,1.56)--(1.8,1.74) node[above=3pt] {$1$};
  \draw (10.2,1.56)--(10.2,1.74) node[above=3pt] {$\phi^{-1}$};
  \node[anchor=north,align=center,text width=2.8cm,inner sep=0pt]
       at (1.65,1.02) {fixed $u\sim1$\\zero-twist\\Baxter equation};
  \node[anchor=north,align=center,text width=2.8cm,inner sep=0pt]
       at (5.1,1.02) {$1\ll|u|\ll\phi^{-1}$\\matching\\powers};
  \node[anchor=north,align=center,text width=2.8cm,inner sep=0pt]
       at (8.65,1.02) {$x=\phi u\sim1$\\crossover\\equation};
  \node[anchor=north,align=center,text width=2.8cm,inner sep=0pt]
       at (12.25,1.02) {$|u|\gg\phi^{-1}$\\twisted\\asymptotics};
  \draw[shift={(-0.05,0)}, blue!65!black, thick, ->] (0.25,-0.9)--(6.4,-0.9);
  \node[blue!65!black,fill=white,inner sep=2pt]
       at (1.65,-0.9) {$\boldsymbol{q}_1(u),\boldsymbol{q}_2(u)$};
  \node[orange!65!black] at (8.8,-1.8) {$\mathtt{q}_1(x),\mathtt{q}_2(x)$};
  \draw[orange!65!black,thick,->] (7.4,-1.8)--(4.6,-1.8)
       node[midway,above=3pt] {small $x$};
  \draw[orange!65!black,thick,->] (10.2,-1.8)--(13.8,-1.8)
       node[midway,above=3pt] {large $x$};
 \end{tikzpicture}
 \caption{Matching at small twist. The zero-twist Baxter functions
 describe fixed $u$, while $\mathtt{q}_1,\mathtt{q}_2$ resolve $u\sim\phi^{-1}$ and
 connect to the twisted asymptotics~\eqref{eq:baxter-asymptotics}.
 Their small-$x$ powers in~\eqref{eq:untwisting-crossover-powers}
 match the large-$u$ zero-twist powers~\eqref{eq:untwisting-zero-twist-asymptotics}
 in the common range.
 Descendant levels change the crossover functions, but the matching
 powers are set by the parent weight. The scale is schematic.}
 \label{fig:untwisting-matching-regions}
\end{figure}

\paragraph{Untwisting the Q-functions.} We now discuss how to take the twisted Q-functions $q_1$ and $q_2$ and untwist them. At fixed $u$, we can try to take the limit $\phi\rightarrow 0$ in the Baxter equation. However, the
twisted asymptotics~\eqref{eq:baxter-asymptotics} apply only for $|u|\gg\phi^{-1}$. To probe the region
\begin{equation}
 1\ll|u|\ll\phi^{-1},
 \label{eq:untwisting-crossover-region}
\end{equation}
we introduce $x\equiv\phi u$ and write
$q_i(u)\simeq \mathtt{q}_i(\phi u)$, keeping the
unit-leading twisted normalisation~\eqref{eq:baxter-asymptotics}.
Figure~\ref{fig:untwisting-matching-regions} shows the overlap
$1\ll|u|\ll\phi^{-1}$: small $x$ in the crossover solution must
match large $u$ in the zero-twist solution, rather than its value
near $u=0$. This is the same matching of spectral parameter ranges used
in~\cite{Chernikov:2026lcv}.

The details of this matching are given in Appendix~\ref{app:untwisting-crossover-matching}. The result is that the Q-functions untwist as
\begin{equation}
 q_1\to \boldsymbol{q}_1,\qquad \frac{q_2}{2\phi}\to\frac{\boldsymbol{q}_2}{1-2h},
 \qquad
 \dot q_1\to\dot{\boldsymbol{q}}_1,\qquad
 \frac{\dot q_2}{2\phi}\to-\frac{\dot{\boldsymbol{q}}_2}{1-2\dot h}.
 \label{eq:untwisting-basis-limit}
\end{equation}
These limits hold for both $q^\downarrow$ and $q^\uparrow$. We will see below, however, that when summing a divergent q-bilinear the crossover region
$u\sim\phi^{-1}$ must be taken into account, and the replacement~\eqref{eq:untwisting-basis-limit} alone is not enough.

To close this subsection we discuss the untwisting of two more objects.

\paragraph{The gluing coefficient.}
\label{sec:untwisting-gluing}
First we recall the glued bilinear Q-function~\eqref{eq:glued-Qplus},
\begin{equation}
  \mathbb{Q}_+ = q_1^\downarrow \dot q_1^\uparrow +c\,q_2^\downarrow \dot q_2^\uparrow ,
\end{equation}
where the undotted Q-functions are evaluated at $u$ and the dotted ones at $\dot{u}$.
Its untwisting is regular, $\mathbb{Q}_+\rightarrow\mathbf{Q}_+$, with~\cite{Ekhammar:2025fvn}
\begin{equation}
 \mathbf{Q}_+
 =\boldsymbol{q}_1^{\downarrow}\dot{\boldsymbol{q}}_1^{\uparrow}
  +\boldsymbol{c}\,\boldsymbol{q}_2^{\downarrow}\dot{\boldsymbol{q}}_2^{\uparrow},
\end{equation}
where $\boldsymbol{c}$ is the $\phi=0$ gluing coefficient. Comparing with the untwisting of $q_2$ in~\eqref{eq:untwisting-basis-limit}, we see that the gluing coefficient $c$ must untwist as
\begin{equation}
 (2\phi)^2c\to-(2h-1)(2\dot h-1)\boldsymbol{c}.
 \label{eq:untwisting-gluing-limit}
\end{equation}
The divergence of $c$ thus cancels against the vanishing of $q_2$ as $\phi\rightarrow 0$.

\paragraph{The connection matrices.} Finally, we discuss the untwisting of the connection matrices $\Omega$ relating the UHPA and LHPA Q-functions. Since the rescaling~\eqref{eq:untwisting-basis-limit} is the same for
both $q^\downarrow$ and $q^\uparrow$, we simply need to insert it into $q^\uparrow=\Omega q^\downarrow$. At fixed $u$ this gives
\begin{equation}
 \begin{pmatrix}
 \Omega_{11}&\dfrac{2\phi}{1-2h}\Omega_{12}\\[4pt]
 \dfrac{1-2h}{2\phi}\Omega_{21}&\Omega_{22}
 \end{pmatrix}
 \longrightarrow\boldsymbol{\Omega}\,,
 \label{eq:untwisting-omega-limit}
\end{equation}
where $\boldsymbol{q}^\uparrow=\boldsymbol{\Omega}\boldsymbol{q}^\downarrow$.
The inverse matrix undergoes the same conjugation, as does the
shadow matrix by~\eqref{eq:setup-omega-shadow}.
The dotted limits follow with $(h,\phi)\mapsto(\dot h,-\phi)$.

\subsection{Untwisting the norm}
\label{sec:untwisting-norm}

Now we remove the twist from the norm. Recall from
Section~\ref{sec:norms} that the two-point normalisation $\mathcal N_2$ is
proportional to a minor of the moment matrix $\mathcal N_2\propto\det_{1\le b,c\le J-1}\mathcal M_{bc}$.
Each entry $\mathcal M_{bc}$ pairs two variations of the Baxter data: the
undotted one with respect to the charge $a_{b-1}$, and the dotted one with
respect to $\dot a_{c-1}$. As discussed above in Section~\ref{sec:norms} and derived in
 Appendix~\ref{sec:norms-boundary-variation}, this pairing
can be written entirely in terms of the connection matrix
$q^\uparrow=\Omega\,q^\downarrow$ (cf.~\eqref{eq:norms-connection-matrix}):
\begin{equation}
 \mathcal M_{bc}
 =-2\pi\ii(\lambda-\lambda^{-1})(\dot\lambda-\dot\lambda^{-1})
 \sum_p\operatorname*{Res}_{u=p}
 \frac{(\mathsf E_b)_{12}(\dot{\mathsf E}_c)_{12}}{\rho}\,.
 \label{eq:untwisting-norm-connection-matrix}
\end{equation}
Here we used the shorthand notation
\begin{equation}
 \mathsf E_b\equiv\Omega^{-1}\partial_{a_{b-1}}\Omega,
 \qquad
 \rho\equiv\frac{\Omega_{22}^2}{\det\Omega},
 \label{eq:untwisting-norm-shorthands}
\end{equation}
and the sum runs over the ordinary and shadow poles in one period strip,
\begin{equation}
 p\in\bigl\{\theta_\alpha+\ii(1-s),\ \theta_\alpha+\ii s\bigr\},
 \qquad \alpha=1,\ldots,J.
 \label{eq:untwisting-norm-poles}
\end{equation}
This form is convenient because we already know the zero-twist limit of
$\Omega$ from~\eqref{eq:untwisting-omega-limit}. We now take the
untwisting limit $\phi\to0$.

\paragraph{Block structure.}
The moment matrix splits naturally into two parts, according to which
charge is varied.
\begin{itemize}
 \item The charges $a_0,\dots,a_{J-3}$ do not change the conformal
 weight. Their variations stay regular at zero twist and give a regular
 $(J-2)\times(J-2)$ block.
 \item The last charge $a_{J-2}$ contains the Casimir at zero
 twist~\eqref{eq:untwisting-casimir-charge}, so varying it changes the
 weight $\boldsymbol h$. Just as $a_{J-1}$ did in the twisted norm, it
 produces a singular last row and column.
\end{itemize}
We treat the regular block first. We then show that the singular part
reduces to a single corner entry, leaving a $(J-2)\times(J-2)$ minor.
For $J=1$ there is no non-trivial untwisted spectrum, so we take $J\ge2$
throughout.

As an example, take $J=4$. The matrix is then $3\times3$. The first two
rows and columns vary the regular charges $a_0$ and $a_1$, and the last
varies the Casimir charge $a_2$:
\begin{equation}
 \mathcal M=
 \left(\begin{array}{cc|c}
  \mathcal M_{11} & \mathcal M_{12} & \mathcal M_{13}\\
  \mathcal M_{21} & \mathcal M_{22} & \mathcal M_{23}\\ \hline
  \mathcal M_{31} & \mathcal M_{32} & \mathcal M_{33}
 \end{array}\right).
 \label{eq:untwisting-norm-J4-example}
\end{equation}
As $\phi\to0$ we will find three kinds of behaviour:
\begin{itemize}
 \item the regular block has a finite limit,
 $\mathcal M_{bc}\to\boldsymbol{\mathcal M}_{bc}$ for $b,c\le2$;
 \item the corner diverges, $\mathcal M_{33}\sim(2\phi)^{2-2\Delta}$;
 \item in the mixed entries $\mathcal M_{3c}$ and $\mathcal M_{c3}$,
 the would-be leading singular parts cancel.
\end{itemize}
The determinant therefore factorises into the corner times the regular
block,
\begin{equation}
 \det\mathcal M\simeq\mathcal M_{33}\,
 \det_{1\leq b,c\le2}\boldsymbol{\mathcal M}_{bc}.
\end{equation}
Now we examine the blocks in detail.

\paragraph{The regular block.}

First consider the $(J-2)\times(J-2)$ upper-left block of the moment matrix $\mathcal M$ in~\eqref{eq:untwisting-norm-connection-matrix}, i.e.\ $1\leq b,c\le J-2$. Its entries are variations with respect to the higher charges, which do not affect the $a_{J-2}$ component of $t(u)$ and thus preserve the parent weights $\boldsymbol{h}$. We can therefore use the zero-twist limit of the Q-functions~\eqref{eq:untwisting-basis-limit}, which also holds off shell, i.e.\ under the charge variation. In particular, the factors $\frac{1-2h}{2\phi}$ in~\eqref{eq:untwisting-omega-limit} are constant under the variation. Differentiating therefore simply rescales $\mathsf E_b=\Omega^{-1}\partial_{a_{b-1}}\Omega$
by the same factors, while $\rho=\Omega_{22}^2/\det\Omega$
is unchanged. Hence for this block we simply replace
\begin{equation}
 \rho\to\boldsymbol{\rho},\qquad
 (\mathsf E_b)_{12}\to\frac{1-2h}{2\phi}(\boldsymbol{\mathsf E}_b)_{12},\qquad
 (\dot{\mathsf E}_c)_{12}\to\frac{1-2\dot h}{-2\phi}(\dot{\boldsymbol{\mathsf E}}_c)_{12}.
 \label{eq:untwisting-norm-regular-connection-limit}
\end{equation}
Since $(\lambda-\lambda^{-1})(\dot\lambda-\dot\lambda^{-1})
=4\sin^2\phi\sim(2\phi)^2$, substitution into
\eqref{eq:untwisting-norm-connection-matrix} gives a finite $\phi$-independent limit
$\mathcal M_{bc}\to\boldsymbol{\mathcal M}_{bc}$, where
\begin{equation}
 \boldsymbol{\mathcal M}_{bc}\equiv
 2\pi\ii(2h-1)(2\dot h-1)
 \sum_p\operatorname*{Res}_{u=p}
 \frac{(\boldsymbol{\mathsf E}_b(u))_{12}
       (\dot{\boldsymbol{\mathsf E}}_c(u))_{12}}
      {\boldsymbol{\rho}(u)},
 \qquad 1\leq b,c\le J-2.
 \label{eq:untwisting-norm-boundary-matrix}
\end{equation}

It remains to compute the last row and column of the moment matrix $\mathcal{M}$, generated by varying
$a_{J-2}$. This changes the zero-twist weight $h$, so we must return
to the crossover region~\eqref{eq:untwisting-crossover-region} of the Q-functions, to determine its variation.

\paragraph{Varying the Casimir charge.}
The last row varies $a_{J-2}$ at fixed $h(\phi)$ and $\phi$, via
\begin{equation}
 \delta t(u)=\epsilon\,u^{J-2}.
 \label{eq:untwisting-norm-casimir-variation}
\end{equation}
By~\eqref{eq:untwisting-casimir-charge}, this shifts the zero-twist
Casimir weight to
\begin{equation}
 \boldsymbol{h}_\epsilon=h-\frac{\epsilon}{2h-1}+O(\epsilon^2),
 \label{eq:untwisting-norm-shifted-weight}
\end{equation}
while the twisted weight stays unchanged.

We now need to understand how this affects the two solutions. We leave the details of how this changes the Q-functions in the crossover
region~\eqref{eq:untwisting-crossover-region}
to Appendix~\ref{app:untwisting-crossover-matching} and simply state
the main result: the variation mixes a $\boldsymbol q_2$ component into
$q_1$,
\begin{equation}
 \begin{aligned}
 \partial_\epsilon q_1^\uparrow
 &\simeq-\frac{\Gamma(2h-1)}{(2h-1)(+2\phi)^{2h-1}}\,
 \boldsymbol{q}_2^{\uparrow},\\[4pt]
 \partial_{\dot\epsilon}\dot q_1^\uparrow
 &\simeq-\frac{\Gamma(2\dot h-1)}{(2\dot h-1)(-2\phi)^{2\dot h-1}}\,
 \dot{\boldsymbol{q}}_2^{\uparrow}.
 \end{aligned}
 \label{eq:untwisting-norm-weight-variation}
\end{equation}

The same holds for the down-arrow Q-functions and for the shadow
Q-functions:
by~\eqref{eq:shadow-q-map} they differ from $q$ only by the factor
$\mu(u)$, which does not depend on the charges, so their mixing
coefficients are the same.

\paragraph{The singular connection variation.}
Using~\eqref{eq:untwisting-basis-limit} to write $\boldsymbol q_2$ in
terms of the twisted $q_2$ changes \eqref{eq:untwisting-norm-weight-variation} to
\begin{equation}
 \partial_\epsilon q_1\simeq\frac{\Gamma(2h-1)}{(2\phi)^{2h}}\,q_2,
 \label{eq:untwisting-norm-q1-twisted-variation}
\end{equation}
for both $q^\uparrow$ and $q^\downarrow$, while the variation of $q_2$
is negligible. The variation therefore adds a large multiple, of order
$\phi^{-2h}$, of $q_2$ to $q_1$, and it does so in the same way for both
arrows. Writing $q=(q_1,q_2)^{\mathsf T}$, the varied Q-functions are
thus, to first order in $\epsilon$,
\begin{equation}
 q^{\uparrow,\downarrow}_\epsilon\simeq(1+\epsilon K)\,q^{\uparrow,\downarrow},
 \qquad
 K=\frac{\Gamma(2h-1)}{(2\phi)^{2h}}
 \begin{pmatrix}0&1\\0&0\end{pmatrix}.
 \label{eq:untwisting-norm-conjugation}
\end{equation}

Inserting this into $q^\uparrow=\Omega q^\downarrow$ gives
\begin{equation}
 q^\uparrow_\epsilon\simeq(1+\epsilon K)\,\Omega\,q^\downarrow
 =(1+\epsilon K)\,\Omega\,(1+\epsilon K)^{-1}\,q^\downarrow_\epsilon,
\end{equation}
so the varied connection matrix is the conjugate
$\Omega_\epsilon\simeq(1+\epsilon K)\,\Omega\,(1-\epsilon K)$.
Differentiating at $\epsilon=0$ gives
\begin{equation}
 \partial_\epsilon\Omega\simeq
 \frac{\Gamma(2h-1)}{(2\phi)^{2h}}
 \begin{pmatrix}
  \Omega_{21}&\Omega_{22}-\Omega_{11}\\
  0&-\Omega_{21}
 \end{pmatrix}.
 \label{eq:untwisting-norm-omega-mixing}
\end{equation}

Multiplying by $\Omega^{-1}$ on the left gives
$\mathsf E_{J-1}=\Omega^{-1}\partial_\epsilon\Omega$.
By~\eqref{eq:omega-two-by-two-inverse}, its $12$ entry is
\begin{equation}
 (\mathsf E_{J-1})_{12}
 \simeq\frac{\Gamma(2h-1)}{(2\phi)^{2h}}
 \left(\frac{\Omega_{22}^2}{\det\Omega}-1\right)
 =\frac{\Gamma(2h-1)}{(2\phi)^{2h}}(\rho-1).
 \label{eq:untwisting-norm-omega-variation}
\end{equation}
For the dotted expression, recall that $\dot\phi=-\phi$ and
$\dot\rho=\rho$.

The situation is similar to the twisted norm: the last row and column
are large again. We would therefore expect only the last diagonal
element to contribute, leaving a $(J-2)\times(J-2)$ minor. We now show
that this is indeed the case, by
inserting~\eqref{eq:untwisting-norm-omega-variation}
into~\eqref{eq:untwisting-norm-connection-matrix}.

\paragraph{Cancellations and the block structure.}
The mechanism for the cancellation is again very similar to that for
the twisted norm.

Consider a mixed entry $\mathcal M_{J-1,c}$ with $c\le J-2$. Inserting~\eqref{eq:untwisting-norm-omega-variation} produces the factor $(\boldsymbol\rho-1)/\boldsymbol\rho=1-1/\boldsymbol\rho$. Since $\boldsymbol{\rho}$ has a simple pole at each point $p$ of~\eqref{eq:untwisting-norm-poles}, the term $1/\boldsymbol\rho$ vanishes there and does not contribute.
 We are left with
the sum of all residues of $(\dot{\boldsymbol{\mathsf E}}_c)_{12}$ in
one period strip:
\begin{equation}
 \mathcal M_{J-1,c}
 \propto\sum_p\operatorname{Res}_{u=p}
 \frac{(\boldsymbol{\rho}-1)(\dot{\boldsymbol{\mathsf E}}_c)_{12}}
      {(2\phi)^{2h-1}\boldsymbol{\rho}}
 =\frac{1}{(2\phi)^{2h-1}}\sum_p\operatorname{Res}_{u=p}
 (\dot{\boldsymbol{\mathsf E}}_c)_{12}.
 \label{eq:untwisting-norm-mixed-pairing}
\end{equation}

We now show that this sum of residues vanishes, as before. What
matters is the behaviour of this $\ii$-periodic function at infinity.
From the asymptotics~\eqref{eq:untwisting-zero-twist-asymptotics} we
find
\begin{equation}
 \boldsymbol{\Omega}(+\infty)=\mathbf1,\qquad
 \boldsymbol{\Omega}(-\infty)=\operatorname{diag}\!\left(
 e^{-2\pi\ii(h-Js)},e^{2\pi\ii(h+Js)}\right).
 \label{eq:untwisting-omega-asymptotics}
\end{equation}
This is even simpler than in the twisted case: without the twist there
is no constant mixing at either end of the strip.

These limits do not depend on the non-trivial charges. Hence
$\boldsymbol{\mathsf E}_b$ and $\dot{\boldsymbol{\mathsf E}}_c$ vanish
at both ends, and the singular part of $\mathcal M_{J-1,c}$ in $\phi$
vanishes. The same argument holds for $\mathcal M_{c,J-1}$. The only
remaining divergent entry is the corner $\mathcal M_{J-1,J-1}$.

\paragraph{The corner entry.}
For the corner, the same substitution
into~\eqref{eq:untwisting-norm-connection-matrix} gives
\begin{equation}
 \lim_{\phi\to0}(2\phi)^{2\Delta-2}\mathcal M_{J-1,J-1}
 =2\pi\ii e^{\pi\ii(1-2\dot h)}
 \Gamma(2h-1)\Gamma(2\dot h-1)
 \sum_p\operatorname{Res}_{u=p}\boldsymbol{\rho}(u).
 \label{eq:untwisting-norm-corner-limit}
\end{equation}
The same contour argument, now using the limits
\eqref{eq:untwisting-omega-asymptotics}, yields
\begin{equation}
 \sum_p\operatorname{Res}_{u=p}\boldsymbol{\rho}(u)
 =\frac{\boldsymbol{\rho}(+\infty)-\boldsymbol{\rho}(-\infty)}{2\pi}
 =\frac{1-e^{4\pi\ii h}}{2\pi}.
 \label{eq:untwisting-norm-corner-residues}
\end{equation}

Using $\dot h-h=S\in\mathbb Z$, the surviving corner multiplies the
regular minor:
\begin{equation}
 \boxed{\displaystyle
 \det_{1\le b,c\le J-1}\mathcal M_{bc}
 \simeq\frac{2(-1)^{S+1}\sin(\pi\Delta)
 \Gamma(2h-1)\Gamma(2\dot h-1)}
 {(2\phi)^{2\Delta-2}}
 \det_{1\le b,c\le J-2}\boldsymbol{\mathcal M}_{bc}.}
 \label{eq:untwisting-norm-minor}
\end{equation}
The determinant now runs only over the $J-2$ charges that remain
independent at fixed conformal weights. For $J=2$ it is empty and
equals one.

\paragraph{The norm coefficient.}
For general $s$, the factor $\mathcal A^2$
in~\eqref{eq:sov-state-normalisation} supplies $(2\phi)^{4Js}$ and the
inverse product of squared $\mathbf Q_+$-values. We combine this with~\eqref{eq:untwisting-gluing-limit}
and~\eqref{eq:untwisting-norm-minor}
in~\eqref{eq:norms-two-point-normalisation}, with
$\mathcal N_J^{\mathrm{SoV}}$ defined
by~\eqref{eq:overlaps-resolution-normalisation}. This gives the leading small-twist norm
\begin{equation}
 \boxed{\displaystyle
 \boldsymbol{\mathcal{N}}_2
 \simeq8\pi J\,J!\mathcal N_J^{\mathrm{SoV}}
 (-1)^{S+1}\frac{\sin(\pi\Delta)\Gamma(2h)\Gamma(2\dot h)}
 {(2\phi)^{2\Delta-4Js}}
 \frac{\boldsymbol{c}}{\boldsymbol{\mathcal D}_\xi}
 \frac{\det\limits_{1\le b,c\le J-2}\boldsymbol{\mathcal M}_{bc}}
 {\prod_{\alpha=1}^J
  \mathbf{Q}_+(\svx_{\alpha,0},
                \dot\svx_{\alpha,0})^2}.}
 \label{eq:untwisting-norm-limit}
\end{equation}

Here $\boldsymbol{\mathcal D}_\xi=\xi^2\partial\Delta/\partial\xi^2$ is the spectral Jacobian~\eqref{eq:coordinate-general-norm-relation} at zero twist, and $\boldsymbol{\mathcal M}_{bc}$ is the regular matrix~\eqref{eq:untwisting-norm-boundary-matrix}, built from $\boldsymbol{\mathsf E}_b=\boldsymbol{\Omega}^{-1}\partial_{a_{b-1}}\boldsymbol{\Omega}$ at fixed weights and $\boldsymbol{\rho}=\boldsymbol{\Omega}_{22}^2/\det\boldsymbol{\Omega}$. The spectral Jacobian and the
Q-values in the denominator are assumed finite and nonzero. An
equivalent formula in terms of ordinary q-values is given in
Appendix~\ref{app:untwisting-evaluation}.

\subsection{The numerator: reducing the HHL moment determinant}
\label{sec:untwisting-moment-reduction}

We now turn to the numerator of~\eqref{eq:HHL-formula} and repeat the
analysis we did for the norm. As before, the last row and column of
the moment matrix carry the divergent terms in the zero-twist limit,
and separating them reduces the size of the determinant by one. We
keep the site weight $s$ generic and set $s=1/4$ only at the very end.
It is convenient to introduce the shorthands
\begin{equation}
 \mathcal M\equiv\mathcal M_\ell^{A,B},\qquad
 J_C\equiv J_A+J_B-2\ell,\qquad
 \eta\equiv h_A+h_B-sJ_C.
 \label{eq:untwisting-numerator-exponent}
\end{equation}
The dotted exponent is $\dot\eta\equiv\dot h_A+\dot h_B-sJ_C$, so that
$\eta+\dot\eta=\Delta_A+\Delta_B-2sJ_C$.
To start, we take $h,\dot h>1/2$ for both states and
$0<\eta,\dot\eta<1$. In this range only the last row and column can
diverge. Other ranges are reached by analytic continuation.

We already know which power to expect. The norm
result~\eqref{eq:untwisting-norm-limit} gives
$\mathcal N_2\propto(2\phi)^{4Js-2\Delta}$, and combining this with the
prefactors in~\eqref{eq:HHL-formula} suggests
$\det\mathcal M\propto(2\phi)^{-\eta-\dot\eta}$.
What we need is its leading coefficient.

The strategy is the same as for~\eqref{eq:untwisting-norm-minor}. We
show that the mixed pairings cancel, so that the surviving corner
multiplies a regular determinant of size $\ell-1$. The new feature is
that the two states now have different transfer polynomials, so we
work directly with the moments~\eqref{eq:HHL-moments}. In the original
sum--integral the measure contains only the $\ell$ glued sites, while the Q-functions
see the full chains, including all the spectator towers of
Section~\ref{sec:HHL-poles}. The way we separate the contribution from
infinity is similar to Appendix N of~\cite{Bargheer:2026kon}.

\paragraph{Locating the divergent part.}
As in the norm calculation, we first write all down-arrow functions in
the up-arrow basis using~\eqref{eq:baxter-connection-matrix}. For an
ordinary tower this gives
\begin{equation}
 \begin{aligned}
 &\operatorname{Res}_{u=\svx_{\alpha,n}}
 [\mu_\ell^\downarrow q_i^{A\downarrow}q_j^{B\uparrow}]=\sum_{k=1}^2
 \operatorname{Res}_{u=\svx_{\alpha,n}}
 [\mu_\ell^\downarrow(\Omega_A^{-1})_{ik}]
 q_k^{A\uparrow}(\svx_{\alpha,n})
 q_j^{B\uparrow}(\svx_{\alpha,n}).
 \end{aligned}
 \label{eq:untwisting-example-tower}
\end{equation}
The most dangerous terms come from the first index of both
Q-functions,
$q_1^{A\uparrow}(u)q_1^{B\uparrow}(u)
\simeq e^{-2\phi u}u^{h_A+h_B-s(J_A+J_B)}$.
On a lower tower, at fixed inhomogeneities,
$e^{-2\phi u}=e^{2\ii\phi n}(1+O(\phi))$. The
measure~\eqref{eq:shadow-chain-mu-factors}, restricted to the $\ell$
glued sites, has residues proportional to $n^{\ell(2s-1)}$. Including
$u^{b-1}$, the tail of this term therefore goes like
$n^{\eta+b-\ell-1}e^{2\ii\phi n}$. The dotted first channel looks the
same, with $c,\dot\eta$ and the opposite exponential, $e^{+2\phi\dot u}$.
For $0<\eta,\dot\eta<1$ the block $b,c<\ell$ is thus convergent, and we
can take the limit directly. The first solutions come with unit
relative coefficients, while the second solutions carry
$2\phi/(1-2h)$ and $-2\phi/(1-2\dot h)$
by~\eqref{eq:untwisting-basis-limit}. Their product is exactly
cancelled by the scaling of $c$ in~\eqref{eq:untwisting-gluing-limit},
so $cq_2\dot q_2\to\boldsymbol c\,\boldsymbol q_2\dot{\boldsymbol q}_2$.
In this block we can therefore simply replace $q,\Gamma$ by their
zero-twist counterparts $\boldsymbol q,\boldsymbol\Gamma$, with no
leftover prefactor.
The last row and column are a different story, since they can still
carry divergent first-channel tails. As in the norm
reduction~\eqref{eq:untwisting-norm-minor}, we keep only their leading
divergent parts. Writing the tails in the spectral coordinate
$u\sim-\ii n$, we need, as $\phi\to0^+$,
\begin{equation}
 \sum_{n\ge1}(-\ii n)^{a-1}e^{\pm2\ii\phi n}
 =\frac{\ii\Gamma(a)}{(2\phi)^a}
 \begin{cases}
  1,&(+),\\
  e^{-\pi\ii a},&(-),
 \end{cases}
 +\mathcal O(1),
 \qquad \Re a>0.
 \label{eq:untwisting-Abel-tail}
\end{equation}
We will again find that only the $(\ell,\ell)$ entry supplies the
leading singular factor, which multiplies the regular
$(\ell-1)\times(\ell-1)$ minor.

\paragraph{Cancelling the mixed entries.}
Next we show that the leading divergent terms in
$\mathcal M_{b\ell}$ and $\mathcal M_{\ell c}$, with $b,c<\ell$,
cancel once all pole towers are summed, just as
in~\eqref{eq:untwisting-norm-mixed-pairing}. All four tower families of
Section~\ref{sec:HHL-poles} enter: the ordinary and shared shadow
towers on the glued sites, and the shadow towers of the $A$ and $B$
spectators. Since the mechanism is the same as for the norm, we keep
the discussion brief.
Recall from~\eqref{eq:HHL-four-residue-products} that each tower contributes a
product of two sums. Take an ordinary glued tower as an example. Its
contribution to $\mathcal M_{b\ell}$ is
\[
 \begin{aligned}
 &-2\pi\ii k_\alpha(s)\sum_{i,j}\Gamma_A^{ii}\Gamma_B^{jj}\left(\sum_{n\ge0}\operatorname{Res}_{u=\svx_{\alpha,n}}
 [\mu_\ell^\downarrow(u)u^{b-1}
 q_i^{A\downarrow}(u)q_j^{B\uparrow}(u)]\right)\\
 &\quad\times
 \left(\sum_{\dot n\ge0}
 \operatorname{Res}_{\dot u=\dot\svx_{\alpha,\dot n}}
 [\mu_\ell^\downarrow(\dot u)\dot u^{\ell-1}
 \dot q_i^{A\uparrow}(\dot u)\dot q_j^{B\downarrow}(\dot u)]\right).
 \end{aligned}
\]
where $k_\alpha(s)$ is~\eqref{eq:J1-measure-conversion-coefficients}
for the $\ell$-site measure. For $b<\ell$ the first sum is harmless, so
we leave it as it is. In the second sum, we convert
$\dot q_j^{B\downarrow}$ to the up-arrow basis as
in~\eqref{eq:untwisting-example-tower} and replace the Q-functions and
measure by their large-$\dot n$ asymptotics. The $\dot n$ sum is then
given by~\eqref{eq:untwisting-Abel-tail}. Its leading term picks $i=1$
and, up to a common scalar factor proportional to
$(2\phi)^{-\dot\eta}$, leaves the vector
$(\dot\Omega_B^{-1})_{j1}$, which does not depend on the undotted tower index $n$. The gluing
relation~\eqref{eq:setup-omega-gluing} turns
$\Gamma_B^{jj}(\dot\Omega_B^{-1})_{j1}$ into $(\Omega_B^{-1})_{1j}$, so
the $j$ sum simply flips the arrow and gives $q_1^{B\downarrow}$.

The other three tower families work the same way, as one can see
from~\eqref{eq:HHL-moment-towers}. After the zero-twist limit, the
coefficient of $(2\phi)^{-\dot\eta}$ is, up to an overall factor,
\[
 \sum_{\substack{p\text{ in the}\\\text{lower half-plane}}}
 \operatorname{Res}_{u=p}
 \left[\mu_\ell^\downarrow(u)u^{b-1}
 \boldsymbol q_1^{A\downarrow}(u)
 \boldsymbol q_1^{B\downarrow}(u)\right].
\]
We can write this sum over residues as a contour integral. The
integrand has no poles in the upper half-plane and decays as
$u^{\eta+b-\ell-1}$, i.e.\ at least as fast as $u^{\eta-2}$. Since
$\Re\eta<1$, we can close the contour upwards, and the residue sum
vanishes. The divergent part of the last row $\mathcal M_{\ell c}$
cancels in the same way.

\paragraph{Simplifying the corner entry.}
For the corner, the dotted tail and the gluing work exactly as above.
Up to the common dotted Abel factor, we are left with
\[
 \sum_{\substack{p\text{ in the}\\\text{lower half-plane}}}
 \operatorname{Res}_{u=p}
 \left[\mu_\ell^\downarrow(u)u^{\ell-1}
 q_1^{A\downarrow}(u)q_1^{B\downarrow}(u)\right].
\]
The difference is that this sum no longer converges at zero twist. Its
leading power is now $u^{\eta-1}$, rather than $u^{\eta-2}$ or faster,
so we cannot yet replace the Q-functions by $\boldsymbol q$ or close the
contour upwards. Instead, we extract this second divergent tail as
well, using~\eqref{eq:untwisting-example-tower}
and~\eqref{eq:untwisting-Abel-tail}.
Converting both down-arrow functions to the up-arrow basis, the leading
first-first channel carries $(\Omega_A^{-1})_{11}(\Omega_B^{-1})_{11}$.
This coefficient is periodic, so we can take its zero-twist limit
directly. In the same way, we can flip the arrow of the
$\mu_\ell^\downarrow$ factor using
\begin{equation}
 \frac{\mu_\ell^\downarrow(u)}{\mu_\ell^\uparrow(u)}
 =\prod_{\alpha=1}^{\ell}
 \frac{\sin\pi\bigl(s+\ii(u-\theta_\alpha)\bigr)}
      {\sin\pi\bigl(s-\ii(u-\theta_\alpha)\bigr)}.
 \label{eq:untwisting-short-measure-ratio}
\end{equation}
Combining the Q-function asymptotics~\eqref{eq:baxter-asymptotics}
with the measure~\eqref{eq:shadow-chain-mu-factors}, we find on the
lower towers
\[
 \mu_\ell^\uparrow(u)u^{\ell-1}
 q_1^{A\uparrow}(u)q_1^{B\uparrow}(u)
 \simeq e^{\frac{\pi\ii}{2}\ell(2s-1)}
 u^{\eta-1}e^{-2\phi u}.
\]
The dotted factor has the same measure phase, with $\eta\to\dot\eta$
and the opposite exponential. Everything else is collected in a
periodic factor, analogous to the norm weight $\boldsymbol\rho$,
\begin{equation}
 \boldsymbol{\rho}_{AB}(u)\equiv
 \frac{\mu_\ell^\downarrow(u)}{\mu_\ell^\uparrow(u)}
 (\boldsymbol\Omega_A^{-1})_{11}(u)(\boldsymbol\Omega_B^{-1})_{11}(u).
 \label{eq:untwisting-numerator-period}
\end{equation}
Applying~\eqref{eq:untwisting-Abel-tail} to both sums, all four tower
families combine into residues of $\boldsymbol\rho_{AB}$ in one period
strip,
\begin{equation}
 \mathcal M_{\ell\ell}\simeq
 \frac{2\pi\ii e^{\pi\ii[\ell(2s-1)-\dot\eta]}\Gamma(\eta)\Gamma(\dot\eta)}
 {(2\phi)^{\eta+\dot\eta}}
 \sum_p\operatorname{Res}_{u=p}\boldsymbol{\rho}_{AB}(u).
 \label{eq:untwisting-ordinary-joint-tail}
\end{equation}
Here $p$ runs over the first poles in one period strip. The prefactor
collects the two measure phases $e^{\pi\ii\ell(2s-1)}$, the factor
$-e^{-\pi\ii\dot\eta}$ from~\eqref{eq:untwisting-Abel-tail}, and the
clockwise contour factor $-2\pi\ii$.
As for the norm, the sum of residues over a period strip is fixed by
the asymptotics at $\pm\infty$.
Using~\eqref{eq:untwisting-omega-asymptotics}
and~\eqref{eq:untwisting-short-measure-ratio}, we get
\[
 \sum_p\operatorname{Res}_{u=p}\boldsymbol{\rho}_{AB}(u)
 =\frac{\boldsymbol{\rho}_{AB}(+\infty)-\boldsymbol{\rho}_{AB}(-\infty)}{2\pi}
 =e^{-\pi\ii\ell(2s-1)}\frac{1-e^{2\pi\ii\eta}}{2\pi}.
\]
So far we have not used any relation between $\eta$ and $\dot\eta$.
Substituting into~\eqref{eq:untwisting-ordinary-joint-tail} gives
\begin{equation}
 \mathcal M_{\ell\ell}\simeq
 \frac{2e^{\pi\ii(\eta-\dot\eta)}
 \Gamma(\eta)\Gamma(\dot\eta)\sin(\pi\eta)}
 {(2\phi)^{\eta+\dot\eta}}.
 \label{eq:untwisting-numerator-diagonal}
\end{equation}

We have thus arrived at the same reduction as for the norm: the
rescaled mixed entries vanish, and the corner multiplies the regular
block. With~\eqref{eq:untwisting-numerator-diagonal}, we obtain
\begin{equation}
 \boxed{\displaystyle
 \det_{1\le b,c\le\ell}\mathcal M
 \simeq
 \frac{2e^{\pi\ii(\eta-\dot\eta)}
 \Gamma(\eta)\Gamma(\dot\eta)\sin(\pi\eta)}
 {(2\phi)^{\eta+\dot\eta}}
 \det_{1\le b,c\le\ell-1}\boldsymbol{\mathcal M}_{bc}.}
 \label{eq:untwisting-numerator-minor}
\end{equation}
\begingroup
\setlength{\emergencystretch}{1em}
The reduced moments are simply those of~\eqref{eq:HHL-moment-towers},
with zero-twist Q-functions,
$\boldsymbol\Gamma_{A/B}=\operatorname{diag}(1,\boldsymbol c_{A/B})$,
and insertion degrees $0,\ldots,\ell-2$. For $\ell=1$ the determinant
is empty and equals one.
Both the coefficient and the reduced moments are meromorphic in the
weights, so outside $0<\eta,\dot\eta<1$ we define the result by
analytic continuation. For $\boldsymbol{\mathcal M}$ this amounts to
zeta regularisation of each q-bilinear sum before contracting the
Baxter indices and adding the towers, as implemented in
\texttt{BSolver2D} in Appendix~\ref{app:untwisting-evaluation}.
The determinant reduction at nonzero spin is derived in
Appendix~\ref{app:untwisting-spinning-reduction}.
Finally, for physical states $\dot\eta-\eta=S_A+S_B\in\mathbb Z$, so
$e^{\pi\ii(\eta-\dot\eta)}\sin(\pi\eta)=\sin(\pi\dot\eta)$.
\par\endgroup

\subsection{The untwisted HHL OPE coefficient}
\label{sec:untwisting-numerators}
\label{sec:untwisting-endpoint}
\label{sec:untwisting-final-result}

We now have both ingredients: the reduced numerator and the untwisted
norms. In this subsection we set $s=1/4$ and put them together to get
the OPE coefficient. We first give the general formula; in the next
subsection we make it completely explicit for $J=2$.

\Needspace{7\baselineskip}
\paragraph{The general formula.}
We take $1\le\ell<\min(J_A,J_B)$, so that both blocks of the light
operator are nonempty, and use $J_C,\eta,\dot\eta$
from~\eqref{eq:untwisting-numerator-exponent} at $s=1/4$.
We start with the numerator. Inserting the reduced
determinant~\eqref{eq:untwisting-numerator-minor}
into~\eqref{eq:HHL-formula} gives, at finite inhomogeneities,
\begin{equation}
 \lim_{\phi\to0}(2\phi)^{\eta+\dot\eta}
 \frac{\LRpair{\psi_A^{\rm C}}{\psi_B^{\rm C}}_\ell}
 {\psi_A^{\rm C}(\boldsymbol0)\psi_B^{\rm C}(\boldsymbol0)}
 =\frac{\Gamma(\eta)\det\boldsymbol{\mathcal M}}
 {(2\pi)^{\ell-1}\Gamma(1-\dot\eta)}\,
 \frac{\displaystyle\prod\nolimits_{\alpha=\ell+1}^{J_A}
  \kappa_{A,\alpha}}
 {\displaystyle\prod\nolimits_{\alpha=1}^{\ell}
  \mathbf{Q}_{+,A}(\svx_{\alpha,0},\dot\svx_{\alpha,0})
  \mathbf{Q}_{+,B}(\svx_{\alpha,0},\dot\svx_{\alpha,0})},
 \label{eq:untwisting-HHL-unnormalised}
\end{equation}
with all quantities on the right collected at the end of this
subsection. Next we divide by the heavy norms. Their zero-twist coefficients
$\boldsymbol{\mathcal N}_{2,A/B}\equiv
\lim_{\phi\to0}(2\phi)^{2\Delta_{A/B}-J_{A/B}}\mathcal N_{2,A/B}$
were found in~\eqref{eq:untwisting-norm-limit}.
Since $\eta+\dot\eta=\Delta_A+\Delta_B-J_C/2$, the powers of the twist
cancel against the prefactor in~\eqref{eq:untwisting-normalised-limits}:
\begin{equation}
 \begin{aligned}
 &\lim_{\phi\to0}(2\phi)^{-J_C/2}
 \frac{\LRpair{\psi_A^{\rm C}}{\psi_B^{\rm C}}_\ell}
 {\sqrt{\mathcal N_{2,A}\mathcal N_{2,B}\mathcal N_{2,C}}}=\frac{1}
 {\sqrt{\boldsymbol{\mathcal N}_{2,A}\boldsymbol{\mathcal N}_{2,B}\mathcal N_{2,C}}}
 \lim_{\phi\to0}(2\phi)^{\eta+\dot\eta}
 \frac{\LRpair{\psi_A^{\rm C}}{\psi_B^{\rm C}}_\ell}
 {\psi_A^{\rm C}(\boldsymbol0)\psi_B^{\rm C}(\boldsymbol0)}.
 \end{aligned}
 \label{eq:untwisting-HHL-OPE-conversion}
\end{equation}
Finally, we take the homogeneous limit. For cyclic on-shell states at
a common coupling, including the cyclic factor $\sqrt{J_AJ_B}$
of~\eqref{eq:untwisting-normalised-limits}, this gives the OPE coefficient
\begin{equation}
 \boxed{\begin{aligned}
 \mathcal C_\ell^{\bullet\bullet\circ}
 =\lim_{\theta_\alpha^A,\theta_\alpha^B\to0}
 &\frac{\sqrt{J_AJ_B}\,\Gamma(\eta)\det\boldsymbol{\mathcal M}}
 {(2\pi)^{\ell-1}\Gamma(1-\dot\eta)
  \sqrt{\boldsymbol{\mathcal N}_{2,A}\boldsymbol{\mathcal N}_{2,B}\mathcal N_{2,C}}}\\
 &\times\frac{\displaystyle\prod\nolimits_{\alpha=\ell+1}^{J_A}
  \kappa_{A,\alpha}}
 {\displaystyle\prod\nolimits_{\alpha=1}^{\ell}
  \mathbf{Q}_{+,A}(\svx_{\alpha,0},\dot\svx_{\alpha,0})
  \mathbf{Q}_{+,B}(\svx_{\alpha,0},\dot\svx_{\alpha,0})}.
 \end{aligned}}
 \label{eq:untwisting-HHL-final-result}
\end{equation}
In practice, we regulate with distinct inhomogeneities, keeping
$\theta_\alpha^A=\theta_\alpha^B$ on the glued sites $\alpha\le\ell$,
and follow the on-shell branches back to homogeneity.
A convenient range to start from is $1/2<h_{A/B}<1$,
$S_{A/B}=\dot h_{A/B}-h_{A/B}\in\mathbb Z_{\ge0}$,
$0<\eta<1$ and, for $\ell>1$,
$\eta+\dot\eta<2\min(\Delta_A,\Delta_B)$,
where $\Delta_{A/B}=h_{A/B}+\dot h_{A/B}$.
For $S_A+S_B>0$ the dotted exponent $\dot\eta=\eta+S_A+S_B$ then lies
outside the range $0<\dot\eta<1$ used in the derivation; it is reached,
like all other weights, by the analytic continuation described
after~\eqref{eq:untwisting-numerator-minor}.

\paragraph{Quantities entering the result.}
Let us now go through the ingredients
of~\eqref{eq:untwisting-HHL-final-result} one by one. The heavy norms are built from the first-pole
matrix~\eqref{eq:untwisting-norm-boundary-matrix}. The light norm
$\mathcal N_{2,C}$ is simply the free planar two-point coefficient of
$\mathcal O_C$. We assume that the norms and the Q-values in the
denominator are nonzero. The square roots follow the phases of the
external states; we never take absolute values of the norms.

For the Q-values we use the points of
Section~\ref{sec:endpoints-collapse}, now at $s=1/4$:
$\svx_{\alpha,0}=\theta_\alpha-\ii/4$ and
$\svx^\uparrow_{\alpha,0}=\theta_\alpha+\ii/4$ are the lower and upper
vacuum points. Each spectator site of $A$ contributes the
upper-vacuum ratio~\eqref{eq:HHL-bra-transform},
$\kappa_{A,\alpha}=\mathbf{Q}_{+,A}(\svx^\uparrow_{\alpha,0},\dot\svx^\uparrow_{\alpha,0})
/\mathbf{Q}_{+,A}(\svx_{\alpha,0},\dot\svx_{\alpha,0})$,
evaluated at its inhomogeneity $\theta^A_\alpha$. All of these are
values of the zero-twist Q-function
\begin{equation}
 \mathbf{Q}_{+,A}(u,\dot u)
 =\boldsymbol{q}_1^{A\downarrow}(u)\dot{\boldsymbol{q}}_1^{A\uparrow}(\dot u)
  +\boldsymbol{c}_A\boldsymbol{q}_2^{A\downarrow}(u)\dot{\boldsymbol{q}}_2^{A\uparrow}(\dot u),
 \label{eq:untwisting-HHL-final-Q}
\end{equation}
and analogously for $\mathbf Q_{+,B}$. Here the Q-functions are the
unit-leading zero-twist solutions, and $\boldsymbol c_{A/B}$ are their
on-shell gluing coefficients.

For $b,c=1,\ldots,\ell-1$, the reduced moments are the zero-twist
version of~\eqref{eq:HHL-moment-towers},
\begin{equation}
 \boldsymbol{\mathcal M}_{bc}
 =-2\pi\ii\,\boldsymbol\Gamma_A^{ik}\boldsymbol\Gamma_B^{jl}\!
 \sum_{\alpha=1}^{J_A+J_B-\ell}
 \towerbracket{u^{b-1}\hat{\boldsymbol q}_i^{A\downarrow}\hat{\boldsymbol q}_j^{B\uparrow}}
              {\dot u^{c-1}\dot{\hat{\boldsymbol q}}_k^{A\uparrow}\dot{\hat{\boldsymbol q}}_l^{B\downarrow}}
              _\alpha^{\hat\mu}.
 \label{eq:untwisting-HHL-final-moments}
\end{equation}
Here $\boldsymbol\Gamma_{A/B}=\operatorname{diag}(1,\boldsymbol c_{A/B})$,
and the hats on the zero-twist Q-functions denote the partial shadow
transform~\eqref{eq:endpoints-partial-shadow-q} at the spectator sites.
The bracket~\eqref{eq:HHL-site-bracket} and its conversion
coefficients~\eqref{eq:khatAndKShHat} use the common measure
\eqref{eq:HHL-partial-shadow-measure} at $s=\tfrac14$.
Appendix~\ref{app:HHL-residues} gives the four residue products that
enter the site sum. Each chiral sum is zeta-regularised separately
before forming products and contracting the Baxter indices.
For $\ell=1$ the matrix is empty and its determinant is one.

\paragraph{HLL as a reference-state specialisation.}
The HLL coefficient is a special case of this result. We take
$A=\Omega_{J_A}$ before normalising the external states.
Then $h_A=\dot h_A=J_A/4$ and $\mathbf Q_{+,A}=1$, so the spectator
ratios of the bra disappear and $\eta=h_B+(2\ell-J_B)/4$.
The reduced moments become~\eqref{eq:HLL-overlap-moment-bilinears}.
By the first line of~\eqref{eq:untwisting-normalised-limits}, we then
obtain $\mathcal C_\ell^{\bullet\circ\circ}$ by dividing by the square root
$\sqrt{\boldsymbol{\mathcal N}_{2,B}}$ of the heavy norm, with the same
cyclic factor $\sqrt{J_AJ_B}$ as for HHL.

\paragraph{Evaluation.}
To evaluate the numerator and the heavy norm coefficients, one
also needs to supply $\boldsymbol{\mathcal D}_{\xi}=\xi^2\partial_{\xi^2}\Delta$
and the free light norm; examples are given in
Appendix~\ref{app:untwisting-evaluation}.
Finally, global descendants do not give new OPE data. Their matrix
elements follow from the highest-weight coefficient
by~\eqref{eq:untwisting-descendant-normalised-factor}, which we derive
and match to the untwisting limit in
Appendix~\ref{app:untwisting-descendants}.

\subsection{The two-site example: norms, HHL and HLL}
\label{sec:untwisting-J2-example}

To close this section we give explicit examples of the main untwisting results in action. We begin by showing how to compute the norm for the $J=2$ chain at $\phi\rightarrow 0$.

\Needspace{7\baselineskip}
\paragraph{Explicit norm for $J=2$.}
All Q-functions for this chain were worked out explicitly
in~\cite{Ekhammar:2025fvn}, and we refer the reader to that paper
for the details. Here we only repeat what we need.
For the homogeneous $J=2$ chain at $s=1/4$, the regular determinant in~\eqref{eq:untwisting-norm-limit} is empty and all remaining factors are completely explicit, including at nonzero spin $S$.
For $\theta_1=\theta_2=0$ all vacuum points coincide,
$\svx_{\alpha,0}=\dot\svx_{\alpha,0}=-\ii/4$.
Setting $J=2$ and $4Js=2$ in~\eqref{eq:untwisting-norm-limit}, what is
left to compute is
\begin{equation}
 \mathcal N_2
 \simeq32\pi\,\mathcal N_{J=2}^{\mathrm{SoV}}
 (-1)^{S+1}\frac{\sin(\pi\Delta)\Gamma(2h)\Gamma(2\dot h)}
 {(2\phi)^{2\Delta-2}}\,
 \frac{\boldsymbol{c}}{\boldsymbol{\mathcal D}_\xi}\,
 \frac{1}{\mathbf{Q}_+(-\tfrac{\ii}{4},-\tfrac{\ii}{4})^4}.
 \label{eq:untwisting-J2-norm-left}
\end{equation}
Here $\boldsymbol{\mathcal D}_\xi$ is the spectral Jacobian defined
in~\eqref{eq:coordinate-general-norm-relation}.
 For $J=2$ the spectrum is
known exactly~\cite{Kazakov:2018qbr},
\begin{equation}
 \xi^4=\tfrac14\,(1+S-\Delta)(\Delta+S-1).
 \label{eq:untwisting-J2-spectrum}
\end{equation}
Differentiating at fixed spin $S$ gives
\begin{equation}
 \boldsymbol{\mathcal D}_\xi
 =\xi^2\frac{\partial\Delta}{\partial\xi^2}
 =\frac{(\Delta-1)^2-S^2}{\Delta-1}
 =\frac{(2h-1)(2\dot h-1)}{\Delta-1}.
 \label{eq:untwisting-J2-jacobian}
\end{equation}
It remains to compute the gluing coefficient $\boldsymbol{c}$ and
$\mathbf{Q}_+(-\tfrac{\ii}{4},-\tfrac{\ii}{4})$. The gluing coefficient was
found in~\cite{Ekhammar:2025fvn},
\begin{equation}
 \boldsymbol{c}=2^{4-4h-4\dot h}\,
 \frac{\sin(\pi h)}{\sin(\pi\dot h)}\,
 \frac{\Gamma(h)\Gamma(\dot h)}{\Gamma(1-h)\Gamma(1-\dot h)}.
 \label{eq:untwisting-J2-gluing-exact}
\end{equation}
We now compute $\mathbf Q_+(-\tfrac{\ii}{4},-\tfrac{\ii}{4})$. The key
is what happens to the connection matrix at a vacuum point. There, each entry of $\boldsymbol\Omega^{-1}$ factorises into a piece
that depends only on its row and a piece that depends only on its
column, see~\eqref{eq:norms-connection-endpoint-factors}:
\begin{equation}
 (\boldsymbol\Omega^{-1})_{ij}(-\tfrac{\ii}{4})\;\propto\;
 \boldsymbol q_i^\downarrow(-\tfrac{\ii}{4})\,
 \operatorname*{Res}_{u=-\ii/4+\ii}\boldsymbol q_j^\uparrow(u).
 \label{eq:untwisting-J2-rank-one}
\end{equation}
In other words, $\boldsymbol\Omega^{-1}$ is the product of a column and
a row vector, so it has rank one there. Gluing relates the
$\mathbf Q_+$-value to these same building blocks. The rank-one structure
then lets us fix it from a single entry,
\begin{equation}
 \mathbf{Q}_+(-\tfrac{\ii}{4},-\tfrac{\ii}{4})
 =\frac{\boldsymbol{q}_1^{\downarrow}(-\tfrac{\ii}{4})\,
        \dot{\boldsymbol{q}}_1^{\downarrow}(-\tfrac{\ii}{4})}
       {(\boldsymbol{\Omega}^{-1})_{11}(-\tfrac{\ii}{4})}.
 \label{eq:untwisting-J2-Q-from-q}
\end{equation}
Evaluating the right-hand side with the exact Q-functions
of~\cite{Ekhammar:2025fvn} gives
\begin{equation}
 \mathbf{Q}_+(-\tfrac{\ii}{4},-\tfrac{\ii}{4})
 =2^{3-2\Delta}e^{\frac{\pi\ii}{2}S}
  \frac{\Gamma(h)}{\Gamma(1-\dot h)}.
 \label{eq:untwisting-J2-Q-values}
\end{equation}
Substituting~\eqref{eq:untwisting-J2-jacobian},~\eqref{eq:untwisting-J2-gluing-exact} and~\eqref{eq:untwisting-J2-Q-values}
into~\eqref{eq:untwisting-J2-norm-left} and simplifying with the Gamma reflection and duplication formulas gives
the symmetric norm
\begin{equation}
 \boxed{\displaystyle
 \mathcal N_2
 \simeq\frac{2^{6\Delta-10}(1-\Delta)\sin(\pi\Delta)}
 {\pi^2(2\phi)^{2\Delta-2}}
 \Gamma\!\big(h-\tfrac12\big)
 \Gamma\!\big(\dot h-\tfrac12\big)
 \Gamma\big(1-h\big)\Gamma\big(1-\dot h\big).}
 \label{eq:untwisting-J2-norm}
\end{equation}
This is the single-ordering norm in the limit $\theta_\alpha\to0$; the
physical norm is $\mathcal N_2^{\theta=0}=2\mathcal N_2$.

With the norm explicit, we now compute the HHL and HLL three-point coefficients.

\Needspace{7\baselineskip}
\paragraph{HHL coefficients.}

We take two $J=2$ states, $J_A=J_B=2$, with bridge length $\ell=1$. As in the norm, for this set-up the reduced determinant in~\eqref{eq:untwisting-HHL-unnormalised} is empty
and the result is simply
\begin{equation}
 \lim_{\phi\to0}(2\phi)^{\eta+\dot\eta}
 \frac{\LRpair{\psi_A^{\rm C}}{\psi_B^{\rm C}}_1}
 {\psi_A^{\rm C}(\boldsymbol0)\psi_B^{\rm C}(\boldsymbol0)}
 =\frac{\Gamma(\eta)}{\Gamma(1-\dot\eta)}
 \frac{\kappa_{A,2}}
 {\mathbf{Q}_{+,A}(\theta_1-\tfrac{\ii}{4},\theta_1-\tfrac{\ii}{4})
  \mathbf{Q}_{+,B}(\theta_1-\tfrac{\ii}{4},\theta_1-\tfrac{\ii}{4})},
 \label{eq:untwisting-J2-HHL}
\end{equation}
where the spectator site of $A$ contributes the ratio
\begin{equation}
 \kappa_{A,2}
 =\frac{\mathbf{Q}_{+,A}(\theta^A_2+\tfrac{\ii}{4},\theta^A_2+\tfrac{\ii}{4})}
        {\mathbf{Q}_{+,A}(\theta^A_2-\tfrac{\ii}{4},\theta^A_2-\tfrac{\ii}{4})},
 \label{eq:untwisting-J2-kappa}
\end{equation}
and the exponents $\eta,\dot\eta$ are given by
\begin{equation}
 \eta=h_A+h_B-\tfrac12,
 \qquad
 \dot\eta=\dot h_A+\dot h_B-\tfrac12.
 \label{eq:untwisting-J2-eta}
\end{equation}
Here $\theta_1$ is the common inhomogeneity of the glued site.

Everything on the right is now explicit. In the homogeneous limit
$\theta_1=\theta^A_2=0$, parity of the exact Q-functions
of~\cite{Ekhammar:2025fvn} relates the values at the two vacuum points
\begin{equation}
 \mathbf{Q}_+(\tfrac{\ii}{4},\tfrac{\ii}{4})
 =(-1)^S\,\mathbf{Q}_+(-\tfrac{\ii}{4},-\tfrac{\ii}{4}),
 \label{eq:untwisting-J2-parity}
\end{equation}
so $\kappa_{A,2}=(-1)^{S_A}$. Using the Q-function value~\eqref{eq:untwisting-J2-Q-values}, the result becomes
\begin{equation}
 \begin{aligned}
 &\lim_{\theta\to0}\,
 \lim_{\phi\to0}(2\phi)^{\Delta_A+\Delta_B-1}
 \frac{\LRpair{\psi_A^{\rm C}}{\psi_B^{\rm C}}_1}
 {\psi_A^{\rm C}(\boldsymbol0)\psi_B^{\rm C}(\boldsymbol0)}\\
 &\qquad=2^{2(\Delta_A+\Delta_B)-6}
 e^{\frac{\pi\ii}{2}(S_A-S_B)}
 \frac{\Gamma(h_A+h_B-\tfrac12)
 \Gamma(1-\dot h_A)\Gamma(1-\dot h_B)}
 {\Gamma(\tfrac32-\dot h_A-\dot h_B)\Gamma(h_A)\Gamma(h_B)}.
 \end{aligned}
 \label{eq:untwisting-J2-HHL-homogeneous}
\end{equation}

To turn this into a structure constant we divide by the norms. The
third operator is $\mathcal O_C=\operatorname{tr}[\phi_1\phi_1^\dagger]$,
with $\mathcal N_{2,C}=1$. Multiplying by the cyclic factor
$\sqrt{J_AJ_B}=2$ of~\eqref{eq:untwisting-normalised-limits} and dividing
by the two norm coefficients~\eqref{eq:untwisting-J2-norm} then gives
\begin{equation}
 \boxed{\begin{aligned}
 \mathcal C_1^{\bullet\bullet\circ}
 &=2e^{-\frac{\pi\ii}{2}(S_A-S_B)}
 \frac{\Gamma(h_A+h_B-\tfrac12)}
 {\Gamma(\tfrac32-\dot h_A-\dot h_B)}\\
 &\quad\times
 \left[\frac{\Gamma(2-2\dot h_A)\Gamma(2-2\dot h_B)}
 {(\Delta_A-1)(\Delta_B-1)
 \Gamma(2h_A-1)\Gamma(2h_B-1)}\right]^{1/2}.
 \end{aligned}}
 \label{eq:untwisting-J2-HHL-OPE}
\end{equation}

The four-dimensional analogue of this result, with two unprotected length-two
zero-magnon operators and one protected operator, was obtained
in~\cite{Gromov:2018hut}.

Two remarks are in order. First,~\eqref{eq:untwisting-J2-HHL-OPE}
holds for any two on-shell states, but it is a CFT OPE coefficient only
when both states are cyclic and at the same coupling. For two-site
primaries, cyclicity requires even spin.
Second, the square root is positive for real $1/2<h_A,h_B<1$; elsewhere
we take the branch obtained by analytic continuation in the weights.
Since the norms are not positive in general, this amounts to a choice
of phase for the normalised external operators. In particular, the phase
$e^{-\frac{\pi\ii}{2}(S_A-S_B)}$ in~\eqref{eq:untwisting-J2-HHL-OPE}
depends on this choice only through a factor $(-1)^{S_A-S_B}$, which
equals one for the even-spin cyclic states.
Appendix~\ref{sec:J2-untwisted} gives an independent coordinate-space
derivation of the HHL coefficient, see~\eqref{eq:J2-untwisted-direct-C}.

\paragraph{The HLL limit.}
The HLL coefficient is defined by the first line
of~\eqref{eq:untwisting-normalised-limits}. Its numerator follows from
the HHL one by taking $A$ to be the reference state $\Omega_2$, as
described in Section~\ref{sec:untwisting-final-result}. Then
$h_A=\dot h_A=\tfrac12$ and $\mathbf Q_{+,A}=1$, so
$\kappa_{A,2}=1$, $\eta=h_B$, $\dot\eta=\dot h_B$,
and~\eqref{eq:untwisting-J2-HHL} becomes
\begin{equation}
 \lim_{\phi\to0}(2\phi)^{\Delta_B}
 \frac{\LRpair{\Omega_2^{\rm C}}{\psi_B^{\rm C}}_1}
 {\Omega_2^{\rm C}(\boldsymbol0)\psi_B^{\rm C}(\boldsymbol0)}
 =\frac{\Gamma(h_B)}
 {\Gamma(1-\dot h_B)\,\mathbf{Q}_{+,B}(\theta_1-\tfrac{\ii}{4},\theta_1-\tfrac{\ii}{4})}.
 \label{eq:untwisting-J2-HLL}
\end{equation}
In the homogeneous limit the Q-value~\eqref{eq:untwisting-J2-Q-values}
turns the right-hand side into $2^{2\Delta_B-3}e^{-\frac{\pi\ii}{2}S_B}$.
Dividing by the square root of the norm~\eqref{eq:untwisting-J2-norm}
and multiplying by the cyclic factor $\sqrt{J_AJ_B}=2$
gives
\begin{equation}
 \boxed{\displaystyle
 \mathcal C_1^{\bullet\circ\circ}
 =e^{-\frac{\pi\ii}{2}S_B}
 \left[\frac{4\Gamma(2-2h_B)\Gamma(\dot h_B)^2}
 {(\Delta_B-1)\Gamma(2\dot h_B-1)\Gamma(1-h_B)^2}\right]^{1/2}.}
 \label{eq:untwisting-J2-HLL-OPE}
\end{equation}
At the free point $h_B=\dot h_B=\tfrac12$ this gives
$\mathcal C_1^{\bullet\circ\circ}=2$, the planar Wick contraction of the
three unit-normalised traces.
For the even spins allowed by cyclicity, half the algebraic square of
\eqref{eq:untwisting-J2-HLL-OPE} reproduces the point-split block
coefficient obtained in~\cite[Eq.~(19)]{Kazakov:2018qbr}.

Having described how to compute three-point structure constants in the simplest $J=2$ set-up, we now use our developed technology in the next section to study
the CFT data of the theory in depth.

\section{CFT data}\label{sec:CFTData}

In this section, we put our SoV framework to work by numerically evaluating the structure constants of operators with various values of $J$. For general $J$, the computational complexity grows only as $J^3$, largely due to the increasing size of the determinants, so that, in principle, any finite $J$ should be within reach. The advantage of SoV becomes apparent when compared with the exponential growth in complexity of multidimensional integrals in coordinate space, where we were only able to reach $J=3$ with rather limited precision. Since $J=3$ is the first non-trivial case that is not amenable to an analytical solution, we use it as an illustrative example for which a direct comparison with the coordinate-space representation is still feasible. But we first analyse the $J=2$ case, which at zero twist we computed analytically.

\subsection{Length two}
\label{sec:plots-J2}

The only non-protected operators with $J=2$ are of schematic form $\mathcal O_S\sim\operatorname{tr}[\phi_1\bar\partial^{S}\phi_1]$ or with $\partial \leftrightarrow \bar{\partial}$. Their exact spectrum and a number of structure constants are already available in the literature \cite{Grabner:2017pgm,Gromov:2018hut,Kazakov:2018qbr}. Let us briefly recall the main results.

\begin{figure}[t]
  \centering
  \includegraphics[width=0.6\textwidth]{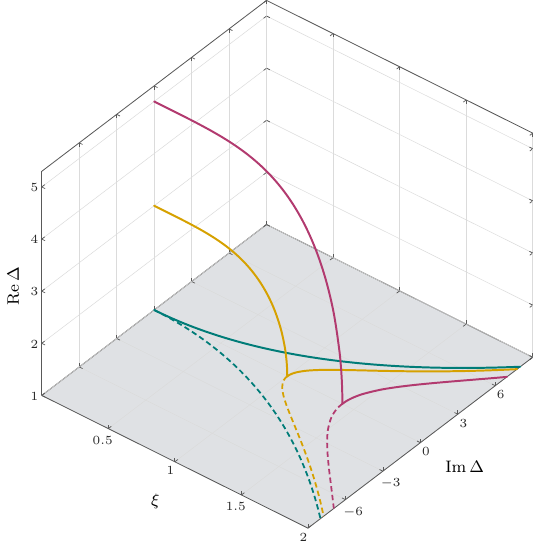}
  \caption{Real and imaginary parts of the scaling dimension $\Delta$ of the lowest $J=2$ operators with spin $S=0$ (teal), $S=2$ (gold) and $S=4$ (raspberry), as functions of the coupling $\xi$. The grey plane $\Re \Delta=1$ marks where each operator meets its shadow, at $\xi=\sqrt{S/2}$; beyond this point the two dimensions form a complex-conjugate pair. Solid lines follow the physical operator and dashed lines its shadow.}
  \label{plot:J2-Dimensions}
\end{figure}

\begin{figure}[t]
  \centering
  \includegraphics[width=0.6\textwidth]{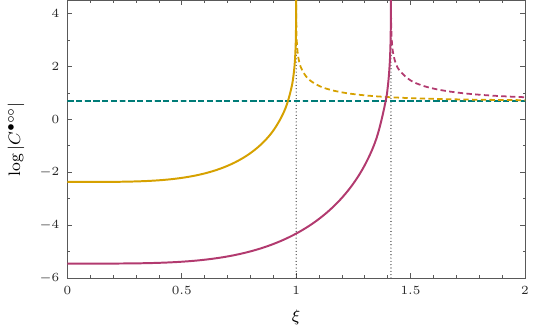}
  \caption{Logarithm of the absolute value of the structure constant $C_{2,S}^{\bullet\circ\circ}$ as a function of the coupling for the same $J=2$ operators of Figure~\ref{plot:J2-Dimensions}, with the same colour code. Solid lines indicate a real structure constant and dashed lines a complex one. The dotted vertical lines mark the collisions with the shadow at $\xi=1$ for $S=2$ and $\xi=\sqrt2$ for $S=4$, where the structure constant diverges before becoming complex. For $S=0$ the absolute value is a coupling-independent constant.}
  \label{plot:J2-LogAbsC}
\end{figure}

\paragraph{The $J=2$ spectrum.}
The exact spectrum is fixed by \cite{Kazakov:2018qbr}
\begin{equation}
  \xi^4=\frac14\,(1+S-\Delta)(\Delta+S-1)\,,
  \label{eq:J2-spectrum}
\end{equation}
where cyclicity requires $S\in 2\mathbb{Z}$ for a physical single-trace
primary. The map $\Delta\to 2-\Delta$ exchanges an operator with its shadow. As we increase the coupling, the two branches meet at $\Delta=1$ and continue as a complex-conjugate pair, see Figure~\ref{plot:J2-Dimensions}.
\paragraph{HLL structure constants.}
For even spin, half the square of the normalised HLL three-point coefficient
\eqref{eq:untwisting-J2-HLL-OPE} gives the point-split block coefficient
\begin{equation}
  C_{2,S}^{\bullet\circ\circ}
  =\tfrac12\bigl(\mathcal C_1^{\bullet\circ\circ}\bigr)^2
  =\frac{1}{\Delta-1}\left[
  \frac{\Gamma(2-2h)\Gamma(\dot h)^2}
       {\Gamma(1-h)^2\Gamma(2\dot h-1)}
  +\frac{\Gamma(2-2\dot h)\Gamma(h)^2}
        {\Gamma(1-\dot h)^2\Gamma(2h-1)}
  \right].
  \label{eq:J2-Cpt}
\end{equation}
The two terms are exchanged by $h\leftrightarrow\dot h$ and are equal by
Euler reflection, since $\dot h-h$ is an integer.
The block coefficient~\eqref{eq:J2-Cpt} diverges at the shadow collision
$\Delta=1$ for nonzero spin and remains finite for $S=0$, as shown in
Figure~\ref{plot:J2-LogAbsC}.

\begin{figure}[t]
  \centering
  \includegraphics[width=0.6\textwidth]{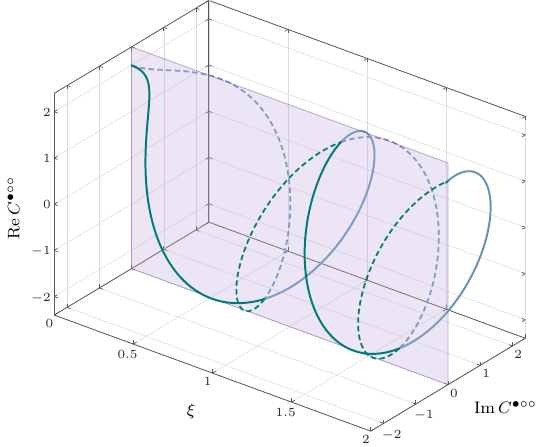}
  \caption{Real and imaginary parts of the structure constant $C_{2,0}^{\bullet\circ\circ}$ of the $J=2$ scalar operator as functions of the coupling. The solid line (physical operator) and the dashed line (shadow) are complex conjugates and both lie on the cylinder $|C_{2,0}^{\bullet\circ\circ}|=2$. The shaded plane at zero imaginary value is drawn to guide the eye.}
  \label{plot:J2-S0-ComplexC}
\end{figure}

\begin{figure}[t!]
\centering
\begin{subfigure}[b]{0.45\linewidth}
\centering
\includegraphics[width=\linewidth]{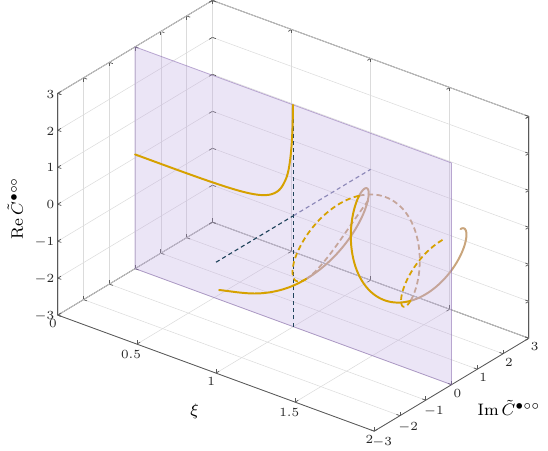}
\caption{$S=2$}
\label{fig:J2-S2}
\end{subfigure}\hfill
\begin{subfigure}[b]{0.45\linewidth}
\centering
\includegraphics[width=\linewidth]{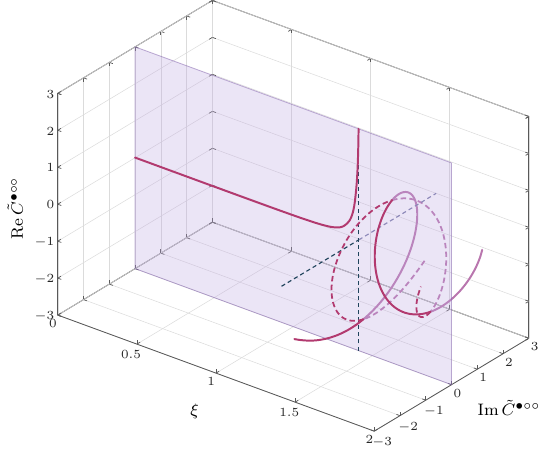}
\caption{$S=4$}
\label{fig:J2-S4}
\end{subfigure}
\caption{Real and imaginary parts of the normalised structure constant $\tilde C_{2,S}^{\bullet\circ\circ}$ of the $J=2$ operators with (a)~$S=2$ and (b)~$S=4$, as functions of the coupling. Before the collision with the shadow, $\tilde C_{2,S}^{\bullet\circ\circ}$ is real and grows logarithmically as the structure constant diverges at $\xi=\sqrt{S/2}$. Beyond the collision, the physical operator (solid line) and its shadow (dashed line) have complex-conjugate structure constants, which return from infinity along $\mp\ii$. The shaded plane $\Im \tilde C=0$ is drawn to guide the eye.}
\label{plot:J2-S24-ComplexC}
\end{figure}

For the spinning operators the structure constants are real and grow with the coupling until the operator meets its shadow, where they diverge. Beyond the collision, they become complex, and the physical and shadow states acquire complex-conjugate structure constants.
 To plot them, we use the normalised structure constant
\begin{equation}
  \tilde{C} = \log(1+|C|)\,\frac{C}{|C|},
  \label{eq:Ctilde}
\end{equation}
which preserves the phase and compresses large values of $|C|$ logarithmically. After the structure constants of the spinning states diverge, the normalised structure constants of the physical and shadow states return from infinity respectively along $\mp \ii$, see Figure~\ref{plot:J2-S24-ComplexC}.

\paragraph{HHL structure constants.}
For two heavy operators of length $J=2$ and one light operator, we use
the closed form~\eqref{eq:untwisting-J2-HHL-OPE} derived in
Section~\ref{sec:untwisting-J2-example}. For even spins, Euler reflection
gives the following form up to the phase convention for the square roots,
\begin{equation}
 C_{J=2}^{\bullet\bullet\circ} =2\,\frac{\Gamma(h_A+h_B-\tfrac12)}
 {\Gamma(\tfrac32-\dot h_A-\dot h_B)}\times
 \left[
 \frac{\Gamma(2-2h_A)\Gamma(2-2h_B)}
 {(\Delta_A-1)(\Delta_B-1)
  \Gamma(2\dot h_A-1)\Gamma(2\dot h_B-1)}
 \right]^{1/2}.
 \label{eq:J2-HHL-C}
\end{equation}

\begin{figure}[t!]
\centering
\begin{subfigure}[b]{0.45\linewidth}
\centering
\includegraphics[width=\linewidth]{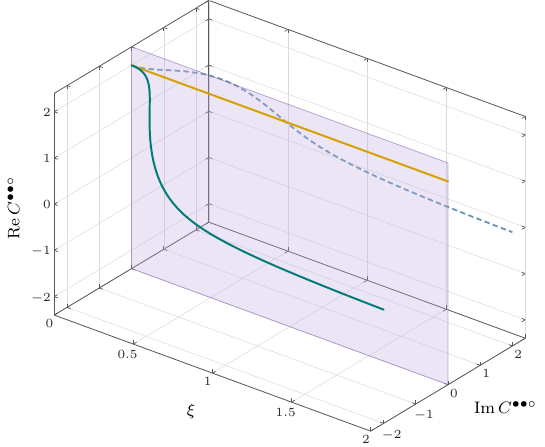}
\caption{$(S_A,S_B)=(0,0)$}
\label{fig:J2-HHL-S0S0}
\end{subfigure}\hfill
\begin{subfigure}[b]{0.45\linewidth}
\centering
\includegraphics[width=\linewidth]{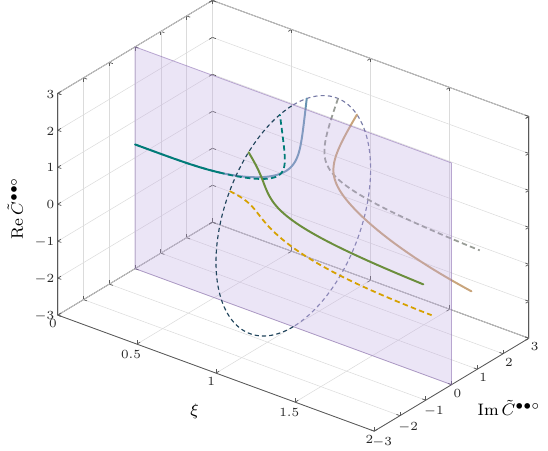}
\caption{$(S_A,S_B)=(0,2)$}
\label{fig:J2-HHL-S0S2}
\end{subfigure}
\caption{Real and imaginary parts of the $J=2$ HHL structure constant as functions of the common coupling: (a)~$C_{J=2}^{\bullet\bullet\circ}$ for two scalar operators and (b)~$\tilde C_{J=2}^{\bullet\bullet\circ}$ for one scalar and one spin-two operator, using the map~\eqref{eq:Ctilde}. Solid and dashed lines follow the scalar's physical and shadow branches. In (b), the two teal curves for $\xi<1$ are the physical and shadow branches of the scalar with the still-real spin-two operator. Beyond $\xi=1$, the spin-two becomes complex and olive curves represent physical--physical and shadow--shadow, while gold curves represent physical--shadow and shadow--physical.}
\label{plot:J2-HHL}
\end{figure}

When the two heavy operators are scalars, the structure constant remains finite with constant norm $|C_{J=2}^{\bullet\bullet\circ}|=2$, while its phase changes with the coupling (see Figure~\ref{plot:J2-HHL}).

A more interesting behaviour is displayed by the structure constant between the $S=2$ and the $S=0$ operator depicted also in Figure~\ref{plot:J2-HHL}. Since the scalar is complex at any $\xi > 0$, the structure constant is complex as well.
 However, when the spin-two operator meets its shadow at $\xi=1$, both branches of $C_{J=2}^{\bullet\bullet\circ}$ diverge and re-emerge as four trajectories in the complex plane, depending on the branch chosen for each operator.

Besides the shadow collision, the HLL~\eqref{eq:J2-Cpt} and HHL~\eqref{eq:J2-HHL-C} structure constants have poles at $\gamma=2,4,\dots$; however, these poles are never reached at real values of the coupling, since all the dimensions~\eqref{eq:J2-spectrum} decrease with the coupling. This is not the case for the length-three operators considered below, whose dimensions can grow with the coupling, generating very non-trivial trajectories that cross and branch as we now explain.

\subsection{Length three -- lowest family}
\label{sec:plots-J3}

We consider the lowest parity-even $J=3$ operators, schematically
$\operatorname{tr}[\bar\partial^S\phi^3]$, with even spin
$S=\dot h-h$ and bare dimension $\Delta = \tfrac{3}{2} +S$. In integrability language, these states are defined by the property $a_0=\dot a_0=0$
in~\eqref{eq:transfer-matrix-expansion}.

Due to the simplicity of the lowest family, the coordinate wave function can be computed explicitly
in terms of hypergeometric functions.
Substituting this wave function into the graph-building
operator~\eqref{eq:graph-building-operator} and imposing
stationarity~\eqref{eq:graph-building-stationarity} fixes the coupling to be
\begin{equation}
 \xi^6=\frac{2\Gamma(\tfrac34)^4}{\pi^2}
 \frac{
 \Gamma\!\left(\tfrac78-\tfrac h2\right)
 \Gamma\!\left(\tfrac98-\tfrac h2\right)
 \Gamma\!\left(\tfrac38+\tfrac{\dot h}2\right)
 \Gamma\!\left(\tfrac58+\tfrac{\dot h}2\right)}
 {
 \Gamma\!\left(\tfrac38-\tfrac h2\right)
 \Gamma\!\left(\tfrac58-\tfrac h2\right)
 \Gamma\!\left(-\tfrac18+\tfrac{\dot h}2\right)
 \Gamma\!\left(\tfrac18+\tfrac{\dot h}2\right)}.
 \label{eq:J3-leading-twist-spectrum}
\end{equation}
Expanding this in the coupling gives
\begin{equation}
 \Delta_S=\frac32+S+\gamma_S\,\xi^6+O(\xi^{12})\,,
 \qquad
 \gamma_S=-\frac{2\sqrt2\,\pi^{5/2}}{\Gamma\!\left(\tfrac34\right)^{6}}\,
  \frac{\Gamma\!\left(\tfrac S2+\tfrac14\right)\Gamma\!\left(\tfrac S2+\tfrac12\right)}
       {\Gamma\!\left(\tfrac S2+\tfrac34\right)\Gamma\!\left(\tfrac S2+1\right)}\,,
 \label{eq:J3-leading-twist-gamma}
\end{equation}
which is the perturbative correction conjectured in~\cite{Ekhammar:2025fvn}.

\begin{figure}[t]
  \centering
  \includegraphics[width=0.6\textwidth]{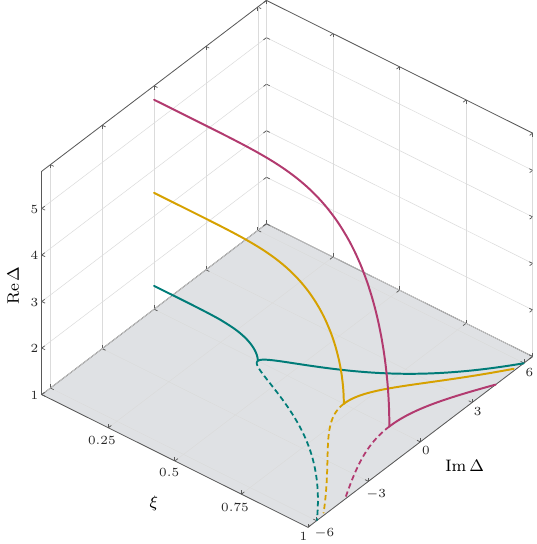}
  \caption{Real and imaginary parts of the scaling dimension $\Delta$ of the $J=3$ lowest-family states with spin $S=0$ (teal), $S=2$ (gold) and $S=4$ (raspberry), as functions of the coupling $\xi$. The grey plane $\Re \Delta=1$ marks where each operator meets its shadow, at $\xi_\ast$ given by~\eqref{eq:J3-shadow-collision}; beyond this point the two dimensions form a complex-conjugate pair. Solid lines follow the physical operator and dashed lines its shadow.}
  \label{plot:J3-Spins-Dimensions}
\end{figure}

The physical branch starts at $\Delta=\tfrac32+S$, and as the coupling increases, the dimensions decrease until they meet their shadows at $\Delta=1$, at the coupling $\xi_\ast$ given by
\begin{equation}
  \xi_*^3=\frac{\sqrt2\,\Gamma(\tfrac34)^2}{\pi}\,
  \frac{\Gamma(\tfrac58+\tfrac S4)\,\Gamma(\tfrac78+\tfrac S4)}
       {\Gamma(\tfrac18+\tfrac S4)\,\Gamma(\tfrac38+\tfrac S4)}\,.
  \label{eq:J3-shadow-collision}
\end{equation}
After this, both the physical and shadow trajectories branch into the complex plane as conjugates of each other, as depicted in Figure~\ref{plot:J3-Spins-Dimensions}.

\paragraph{HLL structure constants.} Even though the wave functions of these states are known explicitly, computing their structure constants in coordinate space is difficult. The two-point normalisation, which in coordinate space is the coefficient of the logarithmic divergence of the norm~\eqref{eq:coordinate-logarithmic-overlap}, involves multiple integrals of products of hypergeometric functions, which are hard to evaluate even numerically.

\begin{figure}[t]
  \centering
  \includegraphics[width=0.6\textwidth]{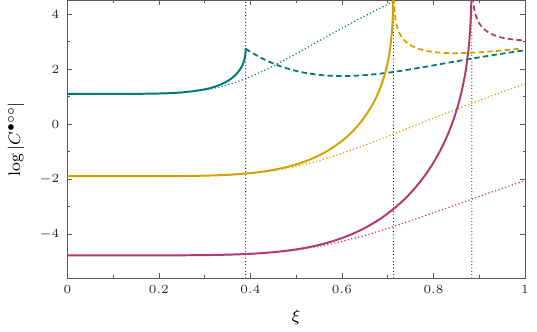}
  \caption{Logarithm of the absolute value of the structure constant $C_{J=3,\ell=1}^{\bullet\circ\circ}$ as a function of the coupling for the same $J=3$ operators of Figure~\ref{plot:J3-Spins-Dimensions}, with the same colour code. Solid lines indicate a real structure constant and dashed lines a complex one. The coloured dotted lines are the first perturbative correction computed in~\eqref{eq:C3-leading-twist-analytic}. The black dotted vertical lines mark the collisions with the shadow at $\xi_\ast$ given by~\eqref{eq:J3-shadow-collision}, where the structure constants of $S=2,4$ diverge before becoming complex. For $S=0$ the structure constant has an infinite slope at the shadow collision, but it remains finite.}
  \label{plot:J3-Spins-LogAbsC}
\end{figure}

\begin{figure}[t!]
\centering
\begin{subfigure}[b]{0.45\linewidth}
\centering
\includegraphics[width=\linewidth]{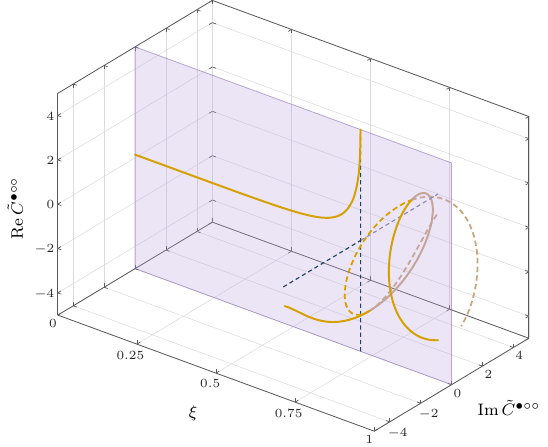}
\caption{$S=2$}
\label{fig:J3-S2}
\end{subfigure}\hfill
\begin{subfigure}[b]{0.45\linewidth}
\centering
\includegraphics[width=\linewidth]{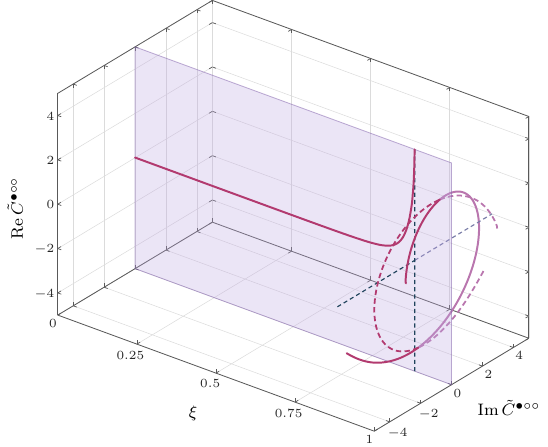}
\caption{$S=4$}
\label{fig:J3-S4}
\end{subfigure}
\caption{Real and imaginary parts of the normalised structure constant $\tilde C_{J=3,\ell=1}^{\bullet\circ\circ}$ of the $J=3$ operators with (a)~$S=2$ and (b)~$S=4$, as functions of the coupling. Before the collision with the shadow, $\tilde C_{J=3,\ell=1}^{\bullet\circ\circ}$ is real and grows logarithmically as the structure constant diverges at $\xi_\ast$ given by~\eqref{eq:J3-shadow-collision}. Beyond the collision, the physical operator (solid line) and its shadow (dashed line) have complex-conjugate structure constants, which return from infinity along $\mp\ii$. The shaded plane $\Im \tilde C=0$ is drawn to guide the eye.}
\label{plot:J3-S24-ComplexC}
\end{figure}

This is where the SoV formalism comes in handy. As for $J=2$ in~\eqref{eq:J2-Cpt}, we present the point-split block coefficient, which for $J=3$ and $\ell=1$ is
\begin{equation}
  C_{J=3,\ell=1}^{\bullet\circ\circ}
  =\tfrac13\bigl(\mathcal C_1^{\bullet\circ\circ}\bigr)^2 ,
  \label{eq:J3-Cpt}
\end{equation}
where $\mathcal C_1^{\bullet\circ\circ}$ is the HLL coefficient~\eqref{eq:untwisting-normalised-limits}. Using the package \texttt{BSolver2D}, we computed $C_{J=3,\ell=1}^{\bullet\circ\circ}$ numerically for spins $S=0,2,4$ as a function of the coupling, shown in Figure~\ref{plot:J3-Spins-LogAbsC}. At weak coupling the structure constants start real and grow with the coupling until the operator meets its shadow at $\xi_\ast$ in~\eqref{eq:J3-shadow-collision}. For $S=2$ and $S=4$ the structure constant diverges as a simple pole, and beyond the collision the physical and shadow states acquire complex-conjugate structure constants. To follow them through the divergence, we plot the normalised structure constant~\eqref{eq:Ctilde} in Figure~\ref{plot:J3-S24-ComplexC}, whose physical and shadow branches return from infinity along $\mp\ii$ beyond the collision, as was the case for the length-two operators.

\begin{figure}[t]
  \centering
  \includegraphics[width=0.6\textwidth]{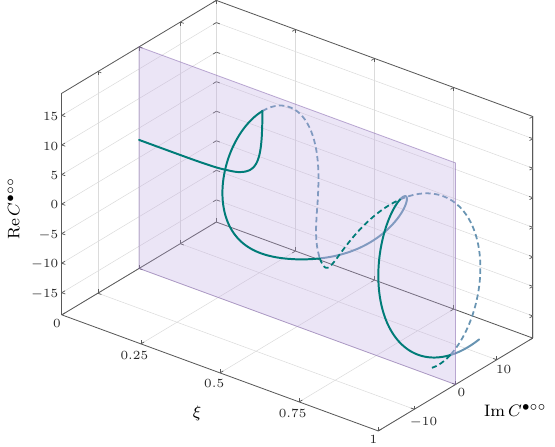}
  \caption{Real and imaginary parts of the structure constant $C_{J=3,\ell=1}^{\bullet\circ\circ}$ of the $J=3$ scalar operator as functions of the coupling. The solid line (physical operator) and the dashed line (shadow) are complex conjugates of each other. The shaded plane at zero imaginary value is drawn to guide the eye.}
  \label{plot:J3-S0-Raw}
\end{figure}

The exception is again the scalar, whose structure constant remains finite at the shadow collision and evaluates to $C_{J=3,\ell=1,S=0}^{\bullet\circ\circ}\simeq15.530$, see Figure~\ref{plot:J3-S0-Raw}. Unlike the length-two case, this collision happens at the finite coupling $\xi_\ast\simeq0.390$.

\paragraph{Weak coupling checks.} The structure constants of Figure~\ref{plot:J3-Spins-LogAbsC} are the first new data presented so far: unlike the spectrum~\eqref{eq:J3-leading-twist-spectrum}, they are not known in closed form for finite values of the coupling. Fitting the numerical data at small values of the coupling, we find
\begin{subequations}
\label{eq:C3-leading-twist-weak-coupling}
\begin{align}
  C_{J=3,\ell=1,S=0}^{\bullet\circ\circ} &= 3\left[1+212.458724873\,\xi^6
  +\mathcal{O}\left(\xi^{12}\right)\right]\,,
  \\
  C_{J=3,\ell=1,S=2}^{\bullet\circ\circ} &= \frac{3}{20}\left[1+27.7469587596\,\xi^6
  +\mathcal{O}\left(\xi^{12}\right)\right]\,,
  \\
  C_{J=3,\ell=1,S=4}^{\bullet\circ\circ} &= \frac{7}{832}\left[1+14.1685656530\,\xi^6
  +\mathcal{O}\left(\xi^{12}\right)\right]\,.
\end{align}
\end{subequations}
As a sanity check, we compare them with the analytic structure constants at weak coupling, denoted by $C_{J=3,\ell=1}^{\bullet\circ\circ}|_{\text{analytic}}$. One way to obtain these analytic structure constants is by extracting them from the OPE of the four-point correlator
\begin{equation}
  \langle \operatorname{tr} \big[\phi_1^{2}(x_1)\phi_1(x_2) \big] \operatorname{tr} \big[\phi^{\dagger\,2}_1(x_3)\phi^{\dagger}_1(x_4)\big] \rangle\,.
  \label{eq:tree-J3-fourpoint}
\end{equation}
At tree level this correlator is the sum of the three planar Wick contractions, whose block expansion reproduces the tree-level values of~\eqref{eq:C3-leading-twist-weak-coupling}. Its first correction, of order $\xi^6$, comes from a single wheel. By expanding the correlator in conformal blocks one can read off the anomalous dimensions and OPE coefficients\footnote{For $S=0,2,4$ there are no degeneracies and each anomalous dimension and structure constant can be directly read off the block expansion.} obtaining
\begin{subequations}
\label{eq:C3-leading-twist-analytic}
\begin{align}
  C_{J=3,\ell=1,S=0}^{\bullet\circ\circ}\big|_{\text{analytic}} &= 3\left(1-\gamma_{0}\log 16 \,\xi^{6} + \mathcal{O}\left(\xi^{12}\right)\right)\,,
  \\
  C_{J=3,\ell=1,S=2}^{\bullet\circ\circ}\big|_{\text{analytic}} &= \frac{3}{20}\left(1-\gamma_2\left(\log 16-\frac{3}{5}\right) \,\xi^{6} + \mathcal{O}\left(\xi^{12}\right)\right)\,,
  \\
  C_{J=3,\ell=1,S=4}^{\bullet\circ\circ}\big|_{\text{analytic}} &= \frac{7}{832}\left(1-\gamma_4\left(\log 16 -\frac{821}{1170}\right)\,\xi^{6} + \mathcal{O}\left(\xi^{12}\right)\right)\,,
\end{align}
\end{subequations}
which agree with the numerical fits~\eqref{eq:C3-leading-twist-weak-coupling}, within our numerical precision, with relative accuracy of order $10^{-12}$.

\begin{figure}[t!]
\centering
\begin{subfigure}[b]{0.45\linewidth}
\centering
\includegraphics[width=\linewidth]{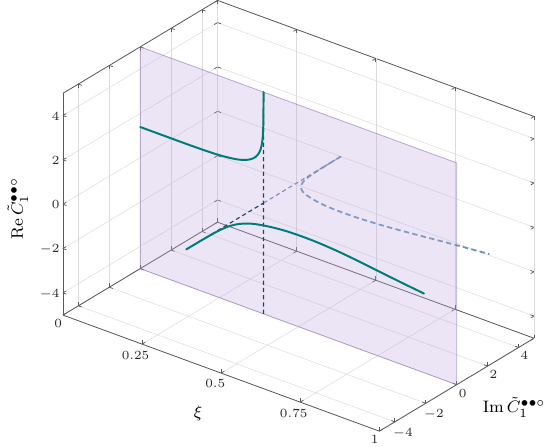}
\caption{$\ell=1$}
\label{fig:J3-HHL-l1}
\end{subfigure}\hfill
\begin{subfigure}[b]{0.45\linewidth}
\centering
\includegraphics[width=\linewidth]{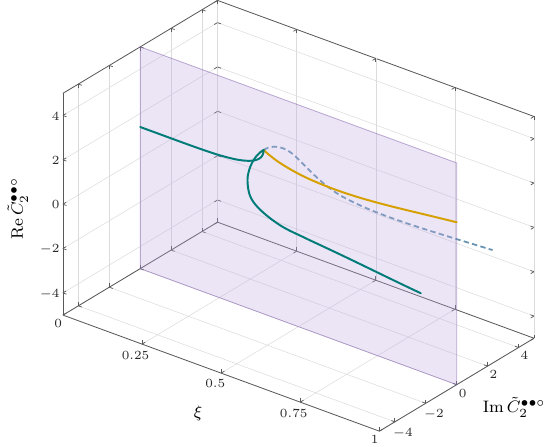}
\caption{$\ell=2$}
\label{fig:J3-HHL-l2}
\end{subfigure}
\caption{Real and imaginary parts of the normalised HHL structure constant $\tilde C_{J=3,\ell}^{\bullet\bullet\circ}$ of two $J=3$ scalars of the lowest family at a common coupling, for bridge lengths (a)~$\ell=1$ and (b)~$\ell=2$, using the map~\eqref{eq:Ctilde}. Below the shadow collision $\xi_\ast$ the two operators coincide and the structure constant is real. Beyond it, the solid teal line has both operators physical and the dashed teal line both shadows. In (a) the structure constant diverges at $\xi_\ast$, marked by the dashed diameters, and the mixed physical--shadow structure constants are infinite. In (b) all branches meet at a finite value at $\xi_\ast$, and the gold line shows the physical--shadow and shadow--physical structure constants, which coincide and are real. The shaded plane $\Im \tilde C=0$ is drawn to guide the eye.}
\label{plot:J3-HHL}
\end{figure}

\paragraph{HHL structure constants.} We can also evaluate the structure constants between two coupling-dependent operators of length three and one light operator; these structure constants are obtained here for the first time. We consider the two heavy operators to be the scalars of the lowest family, which can share either $\ell=1$ or $\ell=2$ legs. Both HHL structure constants are shown in Figure~\ref{plot:J3-HHL}. In the HLL case, the configurations with $\ell$ and $J-\ell$ contractions are equal by symmetry, $C^{\bullet\circ\circ}_{J,\ell}=C^{\bullet\circ\circ}_{J,J-\ell}$, and checking this equality between these very distinct SoV formulations is another non-trivial sanity check we performed for length-three states.

At zero coupling both $C_{J=3,\ell=1,S_A=S_B=0}^{\bullet\bullet\circ}$ and $C_{J=3,\ell=2,S_A=S_B=0}^{\bullet\bullet\circ}$ are real and grow with the coupling until they hit their shadows at $\xi_\ast$ in~\eqref{eq:J3-shadow-collision}. After the collision, the two cases $\ell=1,2$ behave differently. For $\ell=1$ the structure constant diverges at $\xi_\ast$, and beyond the collision, the physical--physical and shadow--shadow structure constants are purely imaginary, starting from $\mp\ii\infty$, respectively. The mixed ones, instead, remain at the pole, since their dimensions are such that the overlap grows logarithmically as $\phi\to0$ instead of as a power of the twist.

For $\ell=2$ the structure constant stays finite at the collision $C_{J=3,\ell=2,S_A=S_B=0}^{\bullet\bullet\circ}\simeq9.916$. Beyond the collision, the physical--physical and shadow--shadow structure constants are complex conjugates of each other, while the other two mixed structure constants coincide and are real.

\subsection{Length three -- scalars}
\label{sec:plots-J3-scalars}

We now turn to the scalar operators of length $J=3$ studied in Section~5.1 of~\cite{Ekhammar:2025fvn}, where their spectrum was computed numerically from weak to strong coupling. We label the trajectories that start at $\xi=0$ with $\Delta=\tfrac32$, $\tfrac52$ and $\tfrac72$ as trajectories~1, 2 and~3, respectively. The first two lie in the $a_0=\dot a_0=0$ sector and follow distinct branches of~\eqref{eq:J3-leading-twist-spectrum}, starting at $h=\dot h=\tfrac34$ and $\tfrac54$, respectively. Trajectory~3, instead, starts from $a_0=\mp\tfrac{\sqrt3}{8}$, so its Baxter equation is not fixed by the conformal weights and its dimension is only known numerically. For none of these three states was the structure constant known before.

\begin{figure}[!htbp]
\centering
\begin{subfigure}[b]{0.52\linewidth}
\centering
\includegraphics[width=\linewidth]{plots/J3-Dimensions.pdf}
\caption{}
\label{fig:J3-Scalars-Dims-3D}
\end{subfigure}\hfill
\begin{subfigure}[b]{0.45\linewidth}
\centering
\includegraphics[width=\linewidth]{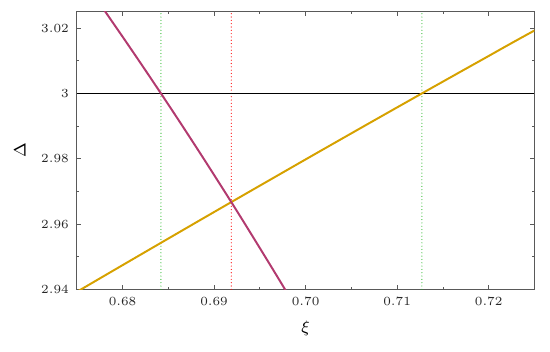}
\caption{}
\label{fig:J3-Scalars-Dims-Zoom}
\end{subfigure}

\medskip

\begin{subfigure}[b]{0.52\linewidth}
\centering
\includegraphics[width=\linewidth]{plots/J3-LogAbsC.pdf}
\caption{}
\label{fig:J3-Scalars-LogAbsC-Global}
\end{subfigure}\hfill
\begin{subfigure}[b]{0.45\linewidth}
\centering
\includegraphics[width=\linewidth]{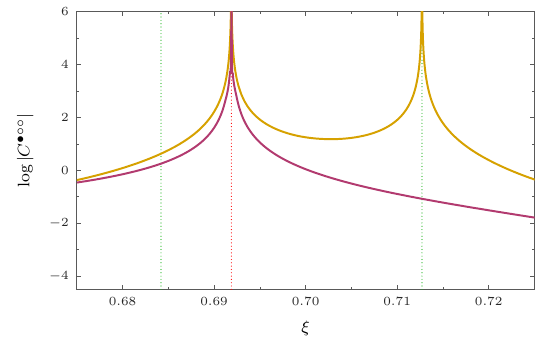}
\caption{}
\label{fig:J3-Scalars-LogAbsC-Zoom}
\end{subfigure}
\caption{The first three scalar $J=3$ trajectories, starting at $\Delta=\tfrac32$ (teal), $\tfrac52$ (gold) and $\tfrac72$ (raspberry), as functions of the coupling $\xi$: dimensions in the upper panels and structure constants in the lower panels. (a)~Real and imaginary parts of the scaling dimension $\Delta$. The black lines are the protected states with $\Delta=2$ and $\Delta=3$, and the grey plane marks the shadow collisions, beyond which the dimensions become complex. (b)~Zoom of the real dimensions of trajectories~2 and~3 for $0.675\leq\xi\leq0.725$, showing their unprotected and protected collisions. (c)~Logarithm of the absolute value of the structure constant. Solid lines indicate a real structure constant and dashed lines a complex one. The dotted vertical lines mark the shadow collisions (black), the protected collisions with $\Delta=2$ (blue) and $\Delta=3$ (green), and the unprotected collision of trajectories~2 and~3 (red), while the grey line marks a zero for trajectory~3. (d)~Zoom on the region $0.65\leq\xi\leq0.75$, where trajectories~2 and~3 collide. \vspace{2em}}
\label{plot:J3-Scalars}
\end{figure}

Their dimensions and structure constants are shown in Figure~\ref{plot:J3-Scalars}. The spectrum is already considerably richer than for the lowest family. As the coupling grows, each trajectory goes through a sequence of events: a \emph{protected collision}, when its dimension crosses one of the protected states with $\Delta=2$ or $\Delta=3$; an \emph{unprotected collision}, when it crosses another unprotected trajectory; a \emph{shadow collision}, when it meets its shadow at $\Delta=1$ and becomes complex; and a \emph{decoupling}, when its structure constant vanishes, all depicted in Figure~\ref{plot:J3-Scalars}. Trajectory~1 is the scalar of the lowest family of Section~\ref{sec:plots-J3}, and we comment no further on it. The other trajectories are summarised below.

The trajectories cross each other and the protected states. Such level crossings are possible because we work in the planar limit, where the mixing between operators that would otherwise repel each other is suppressed. Away from the planar limit, $1/N$ corrections are expected to mix them and to resolve the level crossings, as was worked out for Konishi and the lowest double-trace operator in $\mathcal N=4$ SYM~\cite{Korchemsky:2015cyx,Chester:2023ehi}.

\definecolor{trajGold}{RGB}{214,161,0}
\definecolor{trajRaspberry}{RGB}{178,58,111}
\definecolor{guideRed}{rgb}{1,0,0}
\definecolor{guideBlue}{rgb}{0,0.3,1}
\definecolor{guideGreen}{rgb}{0.1,0.7,0.1}
\definecolor{decouplingGray}{gray}{0.55}
\newcommand{\guideline}[1]{\tikz[baseline=-0.6ex]\draw[#1,line width=1.4pt,dash pattern=on 1.4pt off 1.6pt] (0,-0.5em)--(0,1.05em);}

\begin{table}[h]
\centering
\renewcommand{\arraystretch}{1.25}
\begin{tabular}{c l c c c}
\multicolumn{5}{c}{\color{trajGold}\bfseries Trajectory 2} \\
\toprule
 & Event & $\xi$ & $\Delta$ & $C^{\bullet\circ\circ}$ \\
\midrule
\guideline{guideRed} & Unprotected collision & $0.692$ & $2.967$ & pole \\
\guideline{guideGreen} & Protected collision & $0.713$ & $3$ & pole \\
\bottomrule
\end{tabular}
\label{tab:J3-traj2}
\end{table}
\vspace{-1em}
\paragraph{Trajectory~2.} In the free theory the structure constant of trajectory~2 vanishes, and its first correction is $O(\xi^{12})$. Its dimension grows with the coupling and stays real. It first has an unprotected collision with trajectory~3, at $\Delta\simeq2.967$, and then a protected collision with the state with $\Delta=3$. In both, the structure constant diverges as a simple pole.

\begin{table}[h]
\centering
\renewcommand{\arraystretch}{1.25}
\begin{tabular}{c l c c c}
\multicolumn{5}{c}{\color{trajRaspberry}\bfseries Trajectory 3} \\
\toprule
 & Event & $\xi$ & $\Delta$ & $C^{\bullet\circ\circ}$ \\
\midrule
\guideline{guideGreen} & Protected collision & $0.684$ & $3$ & $1.289$ \\
\guideline{guideRed} & Unprotected collision & $0.692$ & $2.967$ & pole \\
\guideline{decouplingGray} & Decoupling & $0.768$ & $\tfrac52$ & $0$ \\
\guideline{guideBlue} & Protected collision & $0.813$ & $2$ & pole \\
\guideline{black} & Shadow collision & $0.844$ & $1$ & $24.31$ \\
\bottomrule
\end{tabular}
\end{table}
\vspace{-1em}
\paragraph{Trajectory~3.} In the free theory the structure constant of trajectory~3 is finite, $C_{J=3}^{\bullet\circ\circ}=\tfrac1{12}$. As the coupling increases, the dimension of the operator decreases. It first has a protected collision with the state with $\Delta=3$, where its structure constant stays finite, and then the unprotected collision with trajectory~2, where it has a simple pole. At $\xi\simeq0.768$, where $\Delta=\tfrac52$, the structure constant vanishes and the operator decouples from the three-point function. The trajectory then has a protected collision with the state with $\Delta=2$, where the structure constant has again a simple pole, and finally the shadow collision at $\xi\simeq0.844$, where it remains finite, $C_{J=3}^{\bullet\circ\circ}\simeq24.31$, before becoming complex.

The divergence of the structure constants at a collision is naturally interpreted as a sign of mixing between the colliding operators, which become degenerate. At the shadow and unprotected collisions, the divergence comes from a zero of the two-point normalisation~\eqref{eq:norms-two-point-normalisation}. Which of its factors vanishes, however, depends on the collision. At the shadow collision, the dimension has a square-root branch point, so $\mathcal D_\xi$ diverges, and the structure constant can diverge. At the unprotected collisions, instead, $\mathcal D_\xi$ stays finite and the zero comes from the SoV determinant itself. In this case, the two states are always orthogonal via the SoV inner product and their coincident point is no exception, where the state becomes orthogonal to itself and has zero norm. Thus, close to the collision the two eigenstates are no longer independent and one expects that the operators should be redefined there by a mixing matrix that resolves this divergence.

Not every protected collision produces a pole. The structure constant of trajectory~3 remains finite at the protected collision with $\Delta=3$. This suggests some extra symmetry that forbids trajectory~3 from mixing with this state. However, at these finite-coupling points the only tools available to us are the Q-functions, which obscure such selection rules, and we could not identify a simple mechanism that forbids the mixing. Similar crossings without mixing also appear in the BFKL regime of $\mathcal N=4$ SYM~\cite{Ekhammar:2024neh,Ekhammar:2025vig}.

Finally, the decoupling of trajectory~3 does not seem accidental either. It happens at $\Delta = \tfrac{5}{2}$, precisely the free dimension of trajectory~2, for which its structure constant is also zero. From the field theory side it suggests that the four-point function~\eqref{eq:tree-J3-fourpoint} does not exchange a scalar with dimension $\Delta=\tfrac52$ at any value of the coupling.

\section{Conclusions and outlook}
\label{sec:conclusions}
In this paper we show that, in the two-dimensional bi-scalar fishnet theory, the same Q-functions controlling the spectrum also determine structure constants at finite coupling. We identify the separated wave function with a product of glued Q-functions and express overlaps, norms and three-point functions as finite determinants of one-dimensional integrals of q-bilinears. We treat three-point functions with one or two unprotected operators. In planar fishnet theory, this seems to be the most general configuration, and we were unable to construct three-point correlators where all three operators have non-trivial coupling dependence. The integral over q-functions can be recast as a sum over residues, a task for which we provide a highly efficient bespoke C++ code, shared on GitHub.

Furthermore, we find that the two-point normalisation takes a form analogous to the Gaudin formula for Bethe-state norms. The determinant entries are finite sums of residues that pair derivatives of the gluing conditions with respect to the conserved charges.

In the limit of vanishing twist, we obtain closed-form OPE coefficients at $J=2$ and recover known results. At $J=3$, we check the weak-coupling OPE coefficients against an independent perturbative calculation and test the finite-twist overlaps by direct numerical evaluation in coordinate space. The cost of the SoV calculation grows polynomially with $J$, compared with exponential growth for direct coordinate-space integration.

Several questions remain open for future and ongoing investigation. It would be interesting to extend the analysis to operators with magnons and to more general light operators, and to understand the selection rules behind the crossings of states that do not produce poles in their OPE coefficients.

The ultimate goal of this programme is to develop a framework for computing correlation functions from Q-functions in any theory that admits a quantum spectral curve. The same steps, a twisted SoV basis, determinants of Q-function moments and a norm given by variations of the gluing data, should carry over to the four-dimensional fishnet~\cite{Cavaglia:2021mft}, and the AdS$_3$ QSC appears structurally particularly close to the present case. In planar $\mathcal{N}=4$ SYM, our results suggest that the normalisation of~\cite{Ekhammar:2026structure}, currently given by TBA Fredholm determinants, should admit a finite representation in terms of the QSC gluing data, and that structure constants with two unprotected operators, as well as correlators with integrable topological defects, should be within reach. We hope that this will help to uncover the still enigmatic structure of non-perturbative planar OPE coefficients in $\mathcal{N}=4$ SYM.

\acknowledgments
We are grateful to A.~Cavagli\`a, F. Levkovich-Maslyuk and A.~Sever for collaboration at the initial stages of this project or on closely related projects.
The work of N.G. and C.B. were supported by the Science and Technology Facilities
Council (STFC) under the grant ST/X000753/1. The work of N.G. and C.B.\ were
partially supported by the European Research Council (ERC) under the European
Union's Horizon 2020 research and innovation programme (grant agreement
No.~865075) EXACTC. The work of C.B.\ is funded by UK Research and Innovation (UKRI) under the UK government's Horizon
Europe funding Guarantee [grant number EP/X042618/1]. N.G. is grateful to the Perimeter Institute and Yale
University for their hospitality while part of this work was done. N.G. is
grateful to the organizers of the workshop ``Exact techniques and their
applications -- in Tropea 2026'' for creating a productive environment. The work of S.E. was partially supported by the Swedish Research Council, Vetenskapsr{\aa}det, grant
VR 2024-00598. S.E. thanks DESY for its hospitality during part of this work. The work of P.R. was funded by the Deutsche
Forschungsgemeinschaft (DFG, German Research Foundation) – 460391856. P.R. acknowledges support from DESY (Hamburg, Germany), a member of the
Helmholtz Association HGF, and the Deutsche Forschungsgemeinschaft (DFG, German
Research Foundation) under Germany’s Excellence Strategy – EXC 2121 “Quantum Universe”
– 390833306. P.R. is grateful to King's College London for warm hospitality during part of this work.

\section*{Use of AI tools}

AI tools were used in the preparation of this work. The derivations of the core examples were carried out independently by the authors and with the assistance of an LLM, and the results were cross-checked. Symbolic and numerical verifications were performed in Mathematica through the Wolfbook interface~\cite{Wolfbook}, which records the computations of both the authors and the AI in shared notebooks and allows LaTeX notes to be revised collaboratively, so that each step can be inspected and validated. The authors take full responsibility for the content of the paper.

\appendix
\section{The BSolver2D package}
\label{app:bsolver}
In this appendix, we introduce
\texttt{BSolver2D}, a Wolfram Language package with C++ backends for solving
the Baxter equation and evaluating Q-function bilinears.
The source and build instructions are in the
\href{https://github.com/vanbaalon/BSolver2D}{BSolver2D repository}. We refer the reader to the repository for a complete user manual and provide a brief summary here.

Run this setup cell from the package directory before evaluating the
examples in this appendix.
\begin{lstlisting}[style=bsolver]
Get["BSolver2D.wl"];
\end{lstlisting}

The inputs are the chain length \texttt{J}, the inhomogeneities
$\{\theta_1,\ldots,\theta_J\}$, the site spins $\{s_1,\ldots,s_J\}$,
the twist \texttt{phi} and the weights \texttt{\{h, hDot\}}.
The undotted and dotted seeds \texttt{aSeed} and \texttt{aDotSeed}
each contain one initial guess per free transfer-matrix charge.
\begin{lstlisting}[style=bsolver]
(* Define a chain of length J with inhomogeneities and spins *)
chain = BaxterChain[J, thetas, spins];

(* Asymptotic order, recursion shift and working precision in digits *)
sol = BaxterSolution[chain, phi, {h, hDot}, aSeed, maxN, nShift, wp];
\end{lstlisting}
To quantize the solution, we also need the dotted spin chain. It is enough to supplement the dotted seed \texttt{aDotSeed}, and then run
\begin{lstlisting}[style=bsolver]
(* Solve the gluing conditions and attach the result to sol *)
on = sol["SolveOnshell"][aDotSeed];
sol["SetOnshell"][on];
\end{lstlisting}
which finds the on-shell solution and attaches it to the original solution object. Finally, to compute any overlap, we need to compute sums over residues of Q-functions; this is implemented in \texttt{BSolver2D} as
\begin{lstlisting}[style=bsolver]
(* Lower-half-plane residues with the down-arrow measure *)
undotted = bilinearProduct[solA, solB, alpha, "poles" -> "down",
  "measure" -> "down", "qs" -> "downup", "insertion" -> r];
dotted = bilinearProduct[solA["Dotted"], solB["Dotted"], alpha,
  "poles" -> "down", "measure" -> "down", "qs" -> "updown",
  "insertion" -> r];
\end{lstlisting}
Each call returns a $2\times 2$ matrix. The $ij$ entry pairs $q_i^{A\downarrow}$ with $q_j^{B\uparrow}$ in the undotted matrix and $\dot q_i^{A\uparrow}$ with $\dot q_j^{B\downarrow}$ in the dotted matrix, according to the specified measure and insertion.

In the remainder of this appendix, we will give examples of how to use \texttt{BSolver2D} to numerically compute overlaps, providing checks of the formulas presented in the main text.

\subsection{\texorpdfstring{$J=1$}{J=1} overlap}

As a first example, we use \texttt{BSolver2D} to numerically compute the overlap $\LRpair{\psi_A^{\rm C}}{\psi_B^{\rm C}}$
of two one-site states in the companion frame. The states have the
same site spin $s=\dot s$ and zero inhomogeneity, with twists
$\lambda_X=e^{\ii\phi_X}$.

We recall that the overlap is
\begin{equation}
 \LRpair{\psi_A^{\rm C}}{\psi_B^{\rm C}}
 =-2\pi\ii\,\mathcal N_1^{\rm SoV}\mathcal A_A\mathcal A_B\,
 \Gamma_A^{ik}\Gamma_B^{jl}\,
 \towerbracket{q_i^{A\downarrow}q_j^{B\uparrow}}
              {\dot q_k^{A\uparrow}\dot q_l^{B\downarrow}} .
\end{equation}
Here $\Gamma_X=\operatorname{diag}(1,c_X)$ is the gluing matrix and
$\mathcal A_X=\psi_X^{\rm C}(0)/
\mathbb Q_{+,X}(-\ii s,-\ii s)$ fixes the wave-function normalisation. To explicitly rewrite this in terms of residues we introduce the notation
\begin{equation*}
 R_p^\sigma[H]=
 \sum_{r\geq0}\operatorname*{Res}_{w=p-\ii r}
 \bigl[\mu^\sigma(w)H(w)\bigr],
 \qquad \sigma=\downarrow,\uparrow\,,
\end{equation*}
so that
\begin{equation}
 \begin{aligned}
 \towerbracket{q_i^{A\downarrow}q_j^{B\uparrow}}
              {\dot q_k^{A\uparrow}\dot q_l^{B\downarrow}}
 &= k(s)\,R_{-\ii s}^\downarrow\bigl[q_i^{A\downarrow}q_j^{B\uparrow}\bigr]\,
          R_{-\ii s}^\downarrow\bigl[\dot q_k^{A\uparrow}\dot q_l^{B\downarrow}\bigr]\\
 &\quad+ k(1-s)\,
          R_{-\ii(1-s)}^\uparrow\bigl[q_i^{A\downarrow}q_j^{B\uparrow}\bigr]\,
          R_{-\ii(1-s)}^\uparrow\bigl[\dot q_k^{A\uparrow}\dot q_l^{B\downarrow}\bigr] .
 \end{aligned}
 \label{eq:bsolver-A1-one-site-bracket}
\end{equation}
Here $k(s)=-\ii\Gamma(2s)\Gamma(1-2s)=-k(1-s)$ by
\eqref{eq:J1-local-measure-conversions}.
We evaluate the bracket with \texttt{BSolver2D} and compare the resulting
overlap with the hypergeometric expression~\eqref{eq:J1CompanionOverlapClosed}.

\begin{lstlisting}[style=bsolver]
wp = 40;
{s, hA, hB} = N[{1/4, 1/5, 3/20}, wp];
{phiA, phiB} = {7/10, 11/10};
{lambdaA, lambdaB} = N[Exp[I {phiA, phiB}], wp];

(* One site with zero inhomogeneity and scalar weights. *)
chain = BaxterChain[1, {0}, {s}];
solA = BaxterSolution[chain, phiA, {hA, hA}, {}, 24, 80, wp];
solB = BaxterSolution[chain, phiB, {hB, hB}, {}, 24, 80, wp];
gA = {1, solA["SetOnshell"][{}]["c"]};
gB = {1, solB["SetOnshell"][{}]["c"]};
{dotA, dotB} = {solA["Dotted"], solB["Dotted"]};

norm[sol_, lambda_, h_] :=
  (lambda^2 - 1)^s (Conjugate[lambda]^2 - 1)^s *
  (-lambda)^(h - s) (-Conjugate[lambda])^(h - s) /
  sol["QPlus"][-I s, -I s];

(* Ordinary and shadow chiral sums over the lower towers. *)
chiral[a_, b_, measure_, arrows_] := bilinearProduct[a, b, 1,
  "poles" -> "down", "measure" -> measure, "qs" -> arrows,
  "TowerLength" -> 4000];
tower[mu_] := gA . (chiral[solA, solB, mu, "downup"] *
  chiral[dotA, dotB, mu, "updown"]) . gB;
bilinear = Gamma[1 - 2 s]^2 norm[solA, lambdaA, hA] norm[solB, lambdaB, hB] *
  (tower["up"] - tower["down"]);

(* Independent hypergeometric overlap for these scalar states. *)
chi = N[(Sin[(phiA - phiB)/2]/Sin[(phiA + phiB)/2])^2, wp];
zeta = 1 - chi;
block[a_] := zeta^a Hypergeometric2F1[a - hA, a + hB, 2 a, zeta];
betaFactor[h_] := Gamma[1 - s + h] Gamma[1 - s - h] /
  (Gamma[s + h] Gamma[s - h]);
beta = Gamma[2 s] Gamma[2 s - 1] betaFactor[hA] betaFactor[hB] /
  (Gamma[1 - 2 s] Gamma[2 - 2 s]);
closed = chi^(hB - hA) (block[s]^2 + beta block[1 - s]^2);
Grid[{{"BSolver2D overlap", bilinear}, {"Hypergeometric overlap", closed},
      {"Absolute difference", Abs[bilinear - closed]}}, Frame -> All]
\end{lstlisting}
Both methods give $1.2980233488368664856283394879\ldots$,
with an absolute difference of about $6.2\times10^{-29}$.

\subsection{\texorpdfstring{$J=2$}{J=2} overlap}

We consider the twisted $J=2$ overlaps
$\LRpair{\psi_A^{\rm C}}{\Omega_B^{\rm C}}$ and
$\LRpair{\psi_A^{\rm C}}{\psi_C^{\rm C}}$ with $s=1/4$ and
$\boldsymbol\theta=(0,1/5)$ for all three states.
We take $A$ and $C$ to have weights
$h_A=\dot h_A=3/10$ and $h_C=\dot h_C=7/20$ while $B$ is the reference state $h_B=\dot h_B=1/2$. Their twists are $\phi_A=7/10$ and
$\phi_B=\phi_C=11/20$.

We start by initializing the interacting states $A$ and $C$.
We use precomputed values as initial guesses for the undotted and dotted
transfer-matrix coefficients and refine them with \texttt{SolveOnshell}.

\begin{lstlisting}[style=bsolver]
theta = {0, 1/5}; chain = BaxterChain[2, theta, {1/4, 1/4}];
{phiA, phiB} = {7/10, 11/20};
hA = 3/10;
state[phi_, h_, a_, ad_] := Module[{sol, on},
  sol = BaxterSolution[chain, phi, {h, h}, {a}, 100, 100, 60];
  on = sol["SetOnshell"][sol["SolveOnshell"][{ad},
    "TargetPrecision" -> 30, "MaxIterations" -> 15]];
  {sol, {1, on["c"]}}];
{solA, gA} = state[phiA, hA,
  -11302943297724170/10^17, -3572331050871877/10^17];
{solC, gC} = state[phiB, 7/20,
  -12508404755354270/10^17, -5190783550325042/10^17];
\end{lstlisting}

\paragraph{Reference-state pairing.}
Equating the SoV expression with the reference-state identity~\eqref{eq:HLL-overlap-wavefunction} we find
\begin{equation}
 \LRpair{\psi_A^{\rm C}}{\Omega_B^{\rm C}}
 =4\sin\phi_A\sin\phi_B\,
 \frac{\lvert 1-\lambda_B/\lambda_A\rvert^{2h_A-1}}
      {\lvert 1-\lambda_A\lambda_B\rvert^{2h_A+1}} = \mathfrak n_A\mathfrak n_B\,\frac{\det\mathcal M^{A,B}}{\det \mathcal M^{A,B,(0)}}\,,
 \label{eq:bsolver-tree-coordinate-overlap}
\end{equation}
where the entries of the moment matrix are given by
\begin{equation}
 \frac{\mathcal M^{A,B}_{nm}}{-2\pi\ii}=\sum_{\alpha,i=1}^{2} k_\alpha(s)\Gamma_A^{ii}\,R_{\svx_{\alpha,0}}^{\downarrow}\bigl[u^{n-1}q_i^{A\downarrow}e^{-\phi_Bu}\bigr]\,R_{\svx_{\alpha,0}}^{\downarrow}\bigl[\dot u^{m-1}\dot q_i^{A\uparrow}e^{\phi_B\dot u}\bigr]\,.
 \label{eq:bsolver-tree-residue-moments}
\end{equation}
The matrix $\mathcal M^{A,B,(0)}$ is obtained by keeping only the $r=0$
term in each residue sum in~\eqref{eq:bsolver-tree-residue-moments}.
The lack of shadow poles comes from the fact that $\Gamma_B=\operatorname{diag}(1,0)$. To check this equality using \texttt{BSolver2D}, we run
\begin{lstlisting}[style=bsolver]
originProduct = 4 Sin[phiA] Sin[phiB];
treeSoV = originProduct solA["EllSplitterRatio"][phiB, 2];
treeCoordinate = N[originProduct
  Abs[1 - Exp[I (phiB - phiA)]]^(2 hA - 1)/
  Abs[1 - Exp[I (phiA + phiB)]]^(2 hA + 1), 55];
Grid[{{"SoV overlap", N[treeSoV, 16]},
      {"Closed coordinate overlap", N[treeCoordinate, 16]},
      {"Relative difference", Abs[(treeSoV - treeCoordinate)/treeCoordinate]}},
  Frame -> All]
\end{lstlisting}
Both methods give $2.23793774219599455688\ldots$, with a relative
difference of about $7.2\times10^{-35}$.

\paragraph{Interacting pair $A,C$.}
For $n,m=1,2$, write the chiral products as
\begin{equation*}
 F_{n,ij}=u^{n-1}q_i^{A\downarrow}q_j^{C\uparrow},
 \qquad
 G_{m,ij}=\dot u^{m-1}\dot q_i^{A\uparrow}\dot q_j^{C\downarrow}.
\end{equation*}
Expanding the site bracket~\eqref{eq:paired-tower-bracket} in terms
of $R$ gives
\begin{equation}
 \begin{aligned}
 \frac{\mathcal M^{A,C}_{nm}}{-2\pi\ii}
 &=\sum_{\alpha,i,j=1}^{2}\Gamma_A^{ii}\Gamma_C^{jj}\,
     \towerbracket{F_{n,ij}}{G_{m,ij}}_\alpha\\
 &=\sum_{\alpha,i,j=1}^{2}\Gamma_A^{ii}\Gamma_C^{jj}
 \Bigl[
 k_\alpha(s)\,
 R_{\svx_{\alpha,0}}^\downarrow[F_{n,ij}]\,
 R_{\svx_{\alpha,0}}^\downarrow[G_{m,ij}]\\
 &\qquad\qquad\qquad\quad
 +k_\alpha(1-s)\,
 R_{\widetilde\svx_{\alpha,0}}^\uparrow[F_{n,ij}]\,
 R_{\widetilde\svx_{\alpha,0}}^\uparrow[G_{m,ij}]
 \Bigr].
 \end{aligned}
 \label{eq:bsolver-AC-residue-moments}
\end{equation}
For this chain, $k_\alpha(s)=k_\alpha(1-s)
=-\pi\tanh\!\bigl(\pi(\theta_\alpha-\theta_{3-\alpha})\bigr)$.
The code forms the determinant of $\mathcal M^{A,C}/(-2\pi\ii)$
and applies the prefactor in~\eqref{eq:overlaps-bilinear-determinant}.
\begin{lstlisting}[style=bsolver]
dotA = solA["Dotted"]; dotC = solC["Dotted"]; gluing = Outer[Times, gA, gC];

chiral[a_, b_, site_, mu_, arrows_, r_] := bilinearProduct[a, b, site,
  "poles" -> "down", "measure" -> mu, "qs" -> arrows, "insertion" -> r,
  "TowerLength" -> 8000];
tower[site_, mu_, n_, m_] := Total[Flatten[gluing
  chiral[solA, solC, site, mu, "downup", n - 1]
  chiral[dotA, dotC, site, mu, "updown", m - 1]]];
kSite[site_] := -Pi Tanh[Pi (theta[[site]] - theta[[3 - site]])];
sites = Table[kSite[site] (tower[site, "down", n, m] + tower[site, "up", n, m]),
  {site, 2}, {n, 2}, {m, 2}];
norm[sol_] := sol["SoVNormalizationDataJ2"]["Normalization"];
overlap = -norm[solA] norm[solC] Det[Total[sites]];
\end{lstlisting}

BSolver2D also computes overlaps directly from coordinate-space wave
functions. The next cell calls \texttt{LRPairOverlap} on the same
solutions and compares the result with the SoV overlap. The coordinate
quadrature took about 32 minutes with eight threads in this run.
\begin{lstlisting}[style=bsolver]
SetEnvironment["WAVEFUN_FAR_RADIUS" -> "12"];
SetEnvironment["WAVEFUN_THREADS" -> "8"];
coordinate = solA["LRPairOverlap", "SingularOrder" -> 6,
  "BulkOrder" -> 8, "OuterOrder" -> 4, "GradeLevels" -> 1,
  "TRange" -> 5, "MaxTerms" -> 260, "WorkingPrecision" -> 40][solC]["Overlap"];
Grid[{{"SoV overlap", N[overlap, 12]}, {"Coordinate overlap", coordinate},
      {"Relative difference", Abs[(coordinate - overlap)/overlap]}}, Frame -> All]
\end{lstlisting}
The SoV overlap is $1.797316724272$. The independent coordinate-space
quadrature from \texttt{LRPairOverlap} gives
$1.799050494496+0.000398901111\ii$, a relative difference of
$9.90\times10^{-4}$.

\subsection{Projected wave functions}
\label{sec:endpoints-tests}
\label{app:projected-wavefunction-checks}

For $J_B=2,3$ and $1\leq\ell\leq J_B$, we compare the ratios
\begin{equation*}
 \frac{\psi_B^{\rm C}(z^\ell,0^{J_B-\ell})}
      {\psi_B^{\rm C}(\boldsymbol0)}
\end{equation*}
using \texttt{EllSplitterRatio} and direct coordinate-space evaluations
with \texttt{Psi}. 

We take $z=e^{\ii\phi_A}$ with $\phi_A=3/32$ and use $800$ terms in
each residue tower for $J_B=2$. The coordinate evaluation uses the
companion-frame wave function at $(0,0)$, $(z,0)$ and $(z,z)$.
The helper in the first cell evaluates both ratios and their relative
differences for Table~\ref{tab:bsolver-projected-j2}.

\begin{lstlisting}[style=bsolver]
projectedCheck[sol_, opts___Rule] := Module[
 {j = sol["chain"]["J"], phi = 3/32, z, sov, values, coordinate},
 z = Exp[I phi];
 sov = Internal`InheritedBlock[{bilinearProduct},
   SetOptions[bilinearProduct, "TowerLength" -> 800,
     "FitOrder" -> 8, "SBPDepth" -> 6];
   Table[sol["EllSplitterRatio"][phi, ell], {ell, j}]];
 values = sol["Psi", "Frame" -> "Companion", opts][
   Table[Join[ConstantArray[z, ell], ConstantArray[0, j - ell]],
     {ell, 0, j}]];
 coordinate = Rest[values]/First[values];
 Grid[Prepend[
   Transpose[{Range[j], N[sov, 15], N[coordinate, 15],
     Abs[(sov - coordinate)/coordinate]}],
   {"Cut", "SoV ratio", "Coordinate ratio", "Relative difference"}],
  Frame -> All]];

Module[{sol},
 sol = BaxterSolution[BaxterChain[2, {0, 1/17}, {1/4, 1/4}],
   7/10, {3/5, 3/5}, {-0.13497064997914111449`30}, 24, 60, 32];
 sol["SetOnshell"][sol["SolveOnshell"][{-0.089496460291774687172`30},
   "TargetPrecision" -> 20]];
 projectedCheck[sol]]
\end{lstlisting}

\begin{table}[H]
 \centering
 \small
 \begin{tabular}{@{}cllc@{}}
 $\ell$ & SoV ratio & Coordinate ratio & Relative difference\\
 \hline
 $1$ & $1.424291900596-0.034063409120\ii$
     & $1.424291900596-0.034063409120\ii$ & $1.2\times10^{-16}$\\
 $2$ & $1.588937968920$
     & $1.588937968920$ & $5.9\times10^{-16}$
 \end{tabular}
 \caption{Normalised projected wave functions for $J_B=2$.}
 \label{tab:bsolver-projected-j2}
\end{table}

The second cell reuses \texttt{projectedCheck} for $J_B=3$. We keep
the weights, twists and site spin and take
$\boldsymbol\theta=(0,1/17,-2/19)$ and again use $800$ terms per residue
tower. The coordinate evaluation uses
an extrapolation from ten points near each coincident configuration.
Table~\ref{tab:bsolver-projected-j3} gives the three cuts.

\begin{lstlisting}[style=bsolver]
Module[{sol, seed},
 seed = {-0.01487413210922453873520627654`30,
         -0.23533858419544265727342571748`30};
 sol = BaxterSolution[BaxterChain[3, {0, 1/17, -2/19}, {1/4, 1/4, 1/4}],
   7/10, {3/5, 3/5}, seed, 30, 60, 35];
 sol["SetOnshell"][sol["SolveOnshell"][{-First[seed], Last[seed]},
   "TargetPrecision" -> 20]];
 projectedCheck[sol, "CoincidentLimit" -> {10, 1/20, 1/2}]]
\end{lstlisting}

\begin{table}[H]
 \centering
 \small
 \begin{tabular}{@{}cllc@{}}
 $\ell$ & SoV ratio & Coordinate ratio & Relative difference\\
 \hline
 $1$ & $1.093048459611-0.014045587318\ii$
     & $1.093045949100-0.014049673686\ii$ & $4.4\times10^{-6}$\\
 $2$ & $1.455501778438-0.109011823948\ii$
     & $1.455499523781-0.109017961193\ii$ & $4.5\times10^{-6}$\\
 $3$ & $2.338867204682051$
     & $2.338867204682042$ & $3.7\times10^{-15}$
 \end{tabular}
 \caption{Normalised projected wave functions for $J_B=3$.}
 \label{tab:bsolver-projected-j3}
\end{table}

\subsection{Twisted HHL overlaps}
\label{app:HHL-residues}
\label{app:HHL-tests}
\label{sec:HHL-tests}

The partial shadow transform~\eqref{eq:endpoints-partial-shadow-q}
puts the glued ordinary towers and both spectator families on poles of
$\hat\mu^\downarrow$. The glued shadow towers use $\hat\mu^\uparrow$.
The measure conversion in~\eqref{eq:khatAndKShHat} gives the corresponding
coefficients $\hat k_\alpha$ and $\hat k_\alpha^{\rm sh}$.

For the mixed-chain measures $\hat\mu^\sigma$ of
\eqref{eq:HHL-partial-shadow-measure}, define
\begin{equation*}
 R_p^{\sigma,\hat\mu}[H]
 =\sum_{r\geq0}\operatorname*{Res}_{w=p-\ii r}
   \bigl[\hat\mu^\sigma(w)H(w)\bigr],
 \qquad \sigma=\downarrow,\uparrow.
\end{equation*}
For $n,m=1,\ldots,\ell$, set
$F=u^{n-1}\hat q_i^{A\downarrow}\hat q_j^{B\uparrow}$ and
$G=\dot u^{m-1}\dot{\hat q}_k^{A\uparrow}\dot{\hat q}_l^{B\downarrow}$.
Expanding the site
bracket in~\eqref{eq:HHL-moment-towers} over the four families in
Table~\ref{tab:poleFamiliesAB} gives
\begin{equation}
 \begin{aligned}
 (\mathcal M_\ell^{A,B})_{nm}
 &=-2\pi\ii\Gamma_A^{ik}\Gamma_B^{jl}\Biggl[
 \sum_{\alpha=1}^{\ell}\hat k_\alpha
 R_{\svx_{\alpha,0}}^{\downarrow,\hat\mu}[F]
 R_{\dot\svx_{\alpha,0}}^{\downarrow,\hat\mu}[G] \\
 &\qquad+\sum_{\alpha=1}^{\ell}\hat k_\alpha^{\rm sh}
 R_{\widetilde\svx_{\alpha,0}}^{\uparrow,\hat\mu}[F]
 R_{\dot{\widetilde\svx}_{\alpha,0}}^{\uparrow,\hat\mu}[G] \\
 &\qquad+\sum_{\alpha=\ell+1}^{J_A}\hat k_{\alpha,A}
 R_{\widetilde\svx_{\alpha,0}^A}^{\downarrow,\hat\mu}[F]
 R_{\dot{\widetilde\svx}_{\alpha,0}^A}^{\downarrow,\hat\mu}[G] \\
 &\qquad+\sum_{\alpha=\ell+1}^{J_B}\hat k_{\alpha,B}
 R_{\widetilde\svx_{\alpha,0}^B}^{\downarrow,\hat\mu}[F]
 R_{\dot{\widetilde\svx}_{\alpha,0}^B}^{\downarrow,\hat\mu}[G]\Biggr].
 \end{aligned}
 \label{eq:HHL-four-residue-products}
\end{equation}

\paragraph{$J_A=J_B=2$, $\ell=1$.}
The moment matrix has one entry, $(\mathcal M_1^{A,B})_{11}$. The code
evaluates the four terms of~\eqref{eq:HHL-four-residue-products} in order,
with one glued site and one spectator in each chain.

\begin{lstlisting}[style=bsolver]
Module[{state, b, ad, bd, g, chiral, term, terms},
 state[theta_, phi_, h_, seed_] := Module[{sol, on},
   sol = BaxterSolution[BaxterChain[2, {0, theta}, {1/4, 1/4}],
     phi, {h, h}, {seed}, 24, 40, 30];
   on = sol["SolveOnshell"][{seed}, "TargetPrecision" -> 18];
   sol["SetOnshell"][on]; sol];
 solAHHL = state[1/5, 7/10, 1/10, -7/100];
 b = state[-1/7, 1/2, 3/20, -6/100];
 ad = solAHHL["Dotted"]; bd = b["Dotted"];
 g = Outer[Times, {1, solAHHL["Onshell"]["c"]}, {1, b["Onshell"]["c"]}];
 chiral[x_, y_, k_, chain_, qs_, {mu_, poles_}] := bilinearProduct[x, y, k,
   "measureAt" -> {1}, "poleChain" -> chain, "qs" -> qs,
   "measure" -> mu, "poles" -> poles, "TowerLength" -> 500,
   "FitOrder" -> 8, "SBPDepth" -> 6];
 term[k_, chain_, u_, d_] := -2 Pi I Total[Flatten[g
   chiral[solAHHL, b, k, chain, "downup", u] chiral[ad, bd, k, chain, "updown", d]]];
 terms = {term[1, "A", {"down", "down"}, {"up", "valuesOrdinary"}],
   I Pi term[1, "A", {"up", "down"}, {"up", "down"}],
   term[2, "A", {"up", "down"}, {"down", "valuesShadow"}],
   term[2, "B", {"down", "valuesShadow"}, {"up", "down"}]};
 Grid[{{"Glued ordinary", N[terms[[1]], 12]},
  {"Glued shadow", N[terms[[2]], 12]},
  {"Spectator of A", N[terms[[3]], 12]},
  {"Spectator of B", N[terms[[4]], 12]},
  {"Total", N[Chop[Total[terms], 10^-11], 12]}}, Frame -> All]]
\end{lstlisting}

\paragraph{Reference-ket check.}
For a reference ket $B$ of length $\ell$, the reproducing property
\eqref{eq:shadow-inverse-propagator} evaluates the glued coordinates of
$\psi_A^{\rm C}$ at $\lambda_B$. The spectator coordinates
$\boldsymbol R$ of $A$ are sent to infinity, giving
\begin{equation}
 \frac{\LRpair{\psi_A^{\rm C}}{\Omega_B^{\rm C}}_\ell}
      {\psi_A^{\rm C}(\boldsymbol0)\,\mathfrak n^{(\ell)}_{\lambda_B}}
 =\lim_{\boldsymbol R\to\infty}
 \prod_{a=1}^{J_A-\ell}[R_a]^{2s}\,
 \frac{\psi_A^{\rm C}(\boldsymbol R,\lambda_B^\ell)}
      {\psi_A^{\rm C}(\boldsymbol0)} .
 \label{eq:HHL-reference-ket}
\end{equation}
On the SoV side, $\Gamma_B=\operatorname{diag}(1,0)$ leaves the
glued ordinary and $A$-spectator towers of~\eqref{eq:HHL-four-residue-products}.

\begin{lstlisting}[style=bsolver]
Module[{sov, coordinate, a3, seed, phiB = 1/2},
 sov[sol_, ell_] /; ell == sol["chain"]["J"] - 1 := Module[
   {theta = sol["chain"]["thetas"], s = 1/4, shortTheta, chain,
    reference, ref, refDot, dot, g, chiral, term, moment, q, kappa},
   shortTheta = Take[theta, ell];
   chain = BaxterChain[ell, shortTheta, ConstantArray[s, ell]];
   (* Exact reference charges for one- and two-site chains. *)
   reference[phi_] := BaxterSolution[chain, phi, {ell s, ell s},
     If[ell == 1, {}, {2 Re[Exp[-I phi] (Times @@ (-shortTheta + I s))]}],
     sol["maxN"], sol["nshift"], sol["Ndig"]];
   {ref, refDot} = reference /@ {phiB, -phiB};
   dot = sol["Dotted"]; g = {1, sol["Onshell"]["c"]};
   chiral[a_, b_, k_, qs_, r_, {mu_, poles_}] := bilinearProduct[a, b, k,
     "measureAt" -> Range[ell], "qs" -> qs, "insertion" -> r,
     "measure" -> mu, "poles" -> poles, "TowerLength" -> 800,
     "FitOrder" -> 8, "SBPDepth" -> 6][[All, 1]];
   term[k_, n_, m_, u_, d_] := g . (
     chiral[sol, ref, k, "downup", n, u] * chiral[dot, refDot, k, "updown", m, d]);
   moment = Table[-2 Pi I (
     Sum[term[k, n, m, {"down", "down"}, {"up", "valuesOrdinary"}],
       {k, ell}] +
     term[ell + 1, n, m, {"up", "down"}, {"down", "valuesShadow"}]),
     {n, 0, ell - 1}, {m, 0, ell - 1}];
   q[x_] := sol["QPlus"][x, x];
   kappa = q[Last[theta] + I s]/q[Last[theta] - I s];
   kappa Det[moment]/((2 Pi)^ell (Times @@ (q /@ (shortTheta - I s))))];
 coordinate[sol_, ell_, opts___Rule] := Module[{chain = sol["chain"], j, spins, lam, z, rev, psi},
   j = chain["J"]; spins = Reverse[chain["spins"]];
   lam = Exp[I sol["phi"]]; z = Exp[I phiB];
   rev = BaxterSolution[BaxterChain[j, Reverse[chain["thetas"]], spins],
     sol["phi"], {sol["h"], sol["hDot"]}, sol["aFree"],
     sol["maxN"], sol["nshift"], sol["Ndig"]];
   rev["SetOnshell"][sol["Onshell"]];
   psi = rev["Psi", "Frame" -> "Central", opts][{ConstantArray[-1/lam, j],
     Join[ConstantArray[-lam, j - ell], ConstantArray[(1 - lam z)/(z - lam), ell]]}];
   (Times @@ (Abs[z - lam]^(-4 #) & /@ Take[spins, -ell])) Last[psi]/First[psi]];
 seed = {-0.01487413210922453873520627654`30,
         -0.23533858419544265727342571748`30};
 a3 = BaxterSolution[BaxterChain[3, {0, 1/17, -2/19},
   ConstantArray[1/4, 3]], 7/10, {3/5, 3/5}, seed, 30, 60, 35];
 a3["SetOnshell"][a3["SolveOnshell"][{-First[seed], Last[seed]},
   "TargetPrecision" -> 20]];
 Grid[{{"J_A", "ell", "SoV overlap", "Coordinate limit"},
   {2, 1, N[sov[solAHHL, 1], 13], N[coordinate[solAHHL, 1], 13]},
   {3, 2, N[sov[a3, 2], 13],
     N[coordinate[a3, 2, "CoincidentLimit" -> {22, 1/20, 1/2}], 13]}},
  Frame -> All]]
\end{lstlisting}
Table~\ref{tab:bsolver-reference-ket} gives both outputs of the cell.
\begin{table}[H]
 \centering
 \small
 \begin{tabular}{@{}cc ll@{}}
 $J_A$ & $\ell$ & SoV overlap & Coordinate limit\\
 \hline
 $2$ & $1$ & $1.214794938710-0.120975113070\ii$
           & $1.214794938710-0.120975113070\ii$\\
 $3$ & $2$ & $0.797962758993+0.059764527319\ii$
           & $0.797962760298+0.059764526620\ii$
 \end{tabular}
 \caption{Normalised reference-ket overlaps in~\eqref{eq:HHL-reference-ket}.}
 \label{tab:bsolver-reference-ket}
\end{table}

\subsection{Check of the untwisted norm matrix}
\label{sec:untwisting-checks}

For a scalar three-site state with $h=\dot h=7/10$ and
$\boldsymbol\theta=(0,1/5,2/5)$, the regular norm matrix has one entry.
We evaluate it from the first-pole boundary Jacobians in
\eqref{eq:untwisting-norm-boundary-matrix} and independently by summing
the finite-part q-bilinears.

Both methods give $\boldsymbol{\mathcal M}_{11}=26.01793305445533$,
with relative agreement within $10^{-24}$.

\subsection{Untwisted norms and HHL coefficients}
\label{app:untwisting-evaluation}

We give the reduced HHL moment and the origin-normalised numerator for
two three-site scalar states at zero twist, and then the HHL OPE
coefficient of a two-site state.

\paragraph{A three-site HHL example.}
Take scalar states with $h_A=h_B=3/5$ and inhomogeneities
$\boldsymbol\theta^A=(0,1/5,2/5)$ and
$\boldsymbol\theta^B=(0,1/5,-2/5)$.
Gluing the first two sites gives $\eta=\dot\eta=7/10$ and a
$1\times1$ reduced matrix.
The reduced moment is $10.0581663364$ and the origin-normalised
numerator~\eqref{eq:untwisting-HHL-unnormalised} is
$-9.7370299241-5.2847297494\ii$.

\paragraph{From the numerator to the OPE coefficient.}
Take $A=B$ to be the homogeneous two-site scalar with $h=\dot h=7/10$,
so $\xi^4=-1/25$. For the
one-site bridge, the cyclic factor is $2$ and
$\mathcal N_{2,C}=1$. The spectral Jacobian is
$\boldsymbol{\mathcal D}_\xi=2h-1$.
This gives $\mathcal C_1^{\bullet\bullet\circ}=0.377063202688$.

\section{The one-site chain in closed form}
\label{app:J1}

This appendix collects and details the one-site results used in Section~\ref{sec:J1}.

\paragraph{The chain relation.}
With the complex-power convention~\eqref{eq:complex-propagator}, the
chain relation reads~\cite{Derkachov:2001yn}
\begin{equation}
 \begin{aligned}
 &\int\dd^2w\,
 \frac{1}{[z-w]^{\{a,\dot a\}}[w-y]^{\{b,\dot b\}}}
 \\[-2pt]
 &\qquad=\pi\,
 \frac{\Gamma(1-a)\Gamma(1-b)\Gamma(\dot a+\dot b-1)}
      {\Gamma(\dot a)\Gamma(\dot b)\Gamma(2-a-b)}\,
 \frac{1}{[z-y]^{\{a+b-1,\dot a+\dot b-1\}}},
 \qquad a-\dot a,\ b-\dot b\in\mathbb Z\,,
 \end{aligned}
 \label{eq:chain-relation}
\end{equation}
understood by
meromorphic continuation, with distributional limits where needed.

\Needspace{7\baselineskip}
\subsection{Baxter functions and the glued Q-function}
\label{app:J1-Q-functions}
The one-site example of~\cite{Ekhammar:2025fvn} gives the coordinate wave function and the spectral condition at $s=\tfrac14$. Here we give the Q-functions for general $s$ and a non-degenerate twist, $\lambda^2\neq1$. At $\theta=0$, the Baxter equation~\eqref{eq:baxter-equation} reads
\begin{equation}
 (u+\ii s)q(u+\ii)-t(u)q(u)+(u-\ii s)q(u-\ii)=0,
 \qquad
 t(u)=(\lambda+\lambda^{-1})u-\ii h(\lambda-\lambda^{-1}).
 \label{eq:J1-baxter-closed}
\end{equation}
Its solutions with the normalised asymptotics~\eqref{eq:baxter-asymptotics} are
\begin{align}
 q_1^\downarrow
 &=(-\ii)^{s-h}(1-\lambda^{-2})^{1-s-h}\lambda^{\ii u}
 \frac{\Gamma(1-s-\ii u)}{\Gamma(1-h-\ii u)}
 {}_2F_1(1-s-h,1-s-\ii u;1-h-\ii u;\lambda^{-2}),
 \notag\\
 q_1^\uparrow
 &=\ii^{s-h}(1-\lambda^2)^{1-s-h}\lambda^{\ii u}
 \frac{\Gamma(1-s+\ii u)}{\Gamma(1-h+\ii u)}
 {}_2F_1(1-s-h,1-s+\ii u;1-h+\ii u;\lambda^2),
  \label{eq:J1-q-closed}\\
 q_2^\downarrow
 &=(-\ii)^{s+h}(1-\lambda^2)^{1-s+h}\lambda^{-\ii u}
 \frac{\Gamma(1-s-\ii u)}{\Gamma(1+h-\ii u)}
 {}_2F_1(1-s+h,1-s-\ii u;1+h-\ii u;\lambda^2),
 \notag\\
 q_2^\uparrow
 &=\ii^{s+h}(1-\lambda^{-2})^{1-s+h}\lambda^{-\ii u}
 \frac{\Gamma(1-s+\ii u)}{\Gamma(1+h+\ii u)}
 {}_2F_1(1-s+h,1-s+\ii u;1+h+\ii u;\lambda^{-2}).
\nonumber
\end{align}

These solutions can be found by applying a Mellin transform to the Baxter equation~\eqref{eq:J1-baxter-closed}, turning it into a first-order differential equation, which can be readily solved.

The connection formula of the hypergeometric function under $\lambda\to \lambda^{-1}$ relates the two bases $q_a^{\uparrow}$ and $q_a^{\downarrow}$, as in~\eqref{eq:baxter-connection-matrix}. Imposing the gluing condition~\eqref{eq:baxter-gluing-condition} then gives
\begin{equation}
 \mathbb Q_+(u,\dot u)
 =q_1^\downarrow(u)\dot q_1^\uparrow(\dot u)
  +c\,q_2^\downarrow(u)\dot q_2^\uparrow(\dot u),
 \quad
 c=\frac{1}{[\lambda-\lambda^{-1}]^{\{2h,2\dot h\}}}\frac{\Gamma(1-s+h)\,\Gamma(s+\dot h)}
      {\Gamma(1-s-h)\,\Gamma(s-\dot h)}\,,
 \label{eq:J1-Qplus-closed}
\end{equation}
where the dotted factors have twist $\dot\lambda$.

\Needspace{7\baselineskip}
\subsection{The Mellin relation}
\label{app:J1-Q-mellin}
We now derive~\eqref{eq:J1-Q-mellin}. Acting with the one-site kernel~\eqref{eq:J1-integral-Q-kernel} on the diagonal-frame wave function~\eqref{eq:reference-J1-diagonal} gives, up to a factor independent of $u$ and $\dot u$,
\begin{equation}
 \mathbb Q_{+}(u,\dot u)\propto
 [\lambda]^{\ii\{u,\dot u\}}
 \int\frac{\dd^2t}
 {[t-1]^{-\ii\{u,\dot u\}+s}
  [\lambda^2t-1]^{\ii\{u,\dot u\}+s}
  [t]^{\{h,\dot h\}+1-s}}.
 \label{eq:J1-Q-kernel-integral}
\end{equation}
Replacing $t=(\lambda z-1)/(\lambda(z-\lambda))$ above gives
\begin{equation}
 \mathbb Q_{+}(u,\dot u)\propto
 \int\dd^2z\,[z]^{-\ii\{u,\dot u\}-s}\,
 \frac{1}{[z-\lambda]^{\{1-s-h,1-s-\dot h\}}[1-\lambda z]^{\{1-s+h,1-s+\dot h\}}}.
\end{equation}

Up to a constant, the second factor is the companion wave function~\eqref{eq:reference-J1-companion} with $s$ replaced by $1-s$, i.e.\ the shadow transform $\bfS\psi^{\rm C}$ of the wave function, see~\eqref{eq:shadow-operator}. Since the kernel of $\bfS$ is symmetric, the shadow transform can be moved from $\psi^{\rm C}$ onto the power. For pure powers, inversion is trivial, $[z]^{-\ii\{u,\dot u\}-s}=\cD[z]^{\ii\{u,\dot u\}-s}$, and the integral above becomes
\begin{equation}
 \mathbb Q_{+}(u,\dot u)\propto
 \int\dd^2z\,\psi^{\rm C}(z,\bar z)\,\bigl[\bfS\cD[z]^{\ii\{u,\dot u\}-s}\bigr](z).
 \label{eq:J1-Q-shadow-power}
\end{equation}
Evaluating the shadow transform of the power with the chain relation gives a ratio of Gamma functions of $u$ and $\dot u$, which is precisely the inverse of the measure~\eqref{eq:overlaps-one-pair-measure},
\begin{equation}
 \begin{aligned}
 \bfS\cD[z]^{\ii\{u,\dot u\}-s}=\frac{\Gamma(2s)}{\pi\Gamma(1-2s)\mu_s(u,\dot u)}\,
 [z]^{-\ii\{u,\dot u\}+s-1}.
 \end{aligned}
 \label{eq:J1-shadow-Mellin-mode}
\end{equation}

Inserting~\eqref{eq:J1-shadow-Mellin-mode} into~\eqref{eq:J1-Q-shadow-power} leaves the Mellin transform of the wave function,
\begin{equation}
 \mathbb Q_{+}(u,\dot u)\propto
 \frac{1}{\mu_s(u,\dot u)}\int\dd^2z\,[z]^{-\ii\{u,\dot u\}+s-1}\psi^{\rm C}(z,\bar z)
 =\frac{M_{\psi^{\rm C}}(u,\dot u)}{\mu_s(u,\dot u)},
\end{equation}
which is~\eqref{eq:J1-Q-mellin} up to a $u$-independent constant.

To fix the constant, we evaluate~\eqref{eq:J1-Q-mellin} at the lowest point of the ordinary tower~\eqref{eq:J1-overlap-lower-pole-families}, $u=\dot u=-\ii s$. There both $M_{\psi^{\rm C}}$ and $\mu_s$ have a simple pole, and the ratio of their residues fixes $\mathcal A$. Set $u=\dot u=-\ii(s+\epsilon)$. The Mellin integrand becomes $[z]^{-1-\epsilon}\psi^{\rm C}(z,\bar z)$, so the pole comes from the region $z\to0$, where $\psi^{\rm C}$ is regular and
\begin{equation}
 M_{\psi^{\rm C}}=-\frac{\pi}{\epsilon}\,\psi^{\rm C}(0)+O(1),
 \qquad
 2\pi^2\mathcal N_1^{\rm SoV}\mu_s=-\frac{\pi}{\epsilon}+O(1),
\end{equation}
so~\eqref{eq:J1-Q-mellin} gives
\begin{equation}
 \mathcal A=\frac{\psi^{\rm C}(0)}{\mathbb Q_{+}(-\ii s,-\ii s)}
 =\frac{[\lambda^2-1]^s[-\lambda]^{\{h-s,\dot h-s\}}}
 {\mathbb Q_{+}(-\ii s,-\ii s)},
 \label{eq:J1-mellin-normalisation}
\end{equation}
in agreement with~\eqref{eq:sov-state-normalisation}.

\Needspace{7\baselineskip}
\subsection{The residue conversion}
\label{app:J1-overlap-residues}
To pass from~\eqref{eq:J1_Overlap_SoV} to~\eqref{eq:J1-overlap-q-bilinears-skeleton}, each residue picked up in Section~\ref{sec:J1-overlap-q-bilinears} must factorise into an undotted and a dotted residue. At each pole only one chiral factor of~\eqref{eq:J1-overlap-mixed-q-pairs} is singular, and the value of the regular one is converted into a residue with the coefficients~\eqref{eq:J1-measure-conversion-coefficients}. At one site they follow from the Gamma reflection formula and are the same at every level $n$,
\begin{equation}
 k(s)=-\ii\,\Gamma(2s)\Gamma(1-2s)=-k(1-s).
 \label{eq:J1-local-measure-conversions}
\end{equation}
Since the shadow-map phases~\eqref{eq:shadow-q-map} of the two chiral factors cancel, the local residues factorise as
\begin{equation}
 \begin{aligned}
 \operatorname*{Res}_{u=\svx_n}
 [\mathbb Q_{+,A}\widetilde{\mathbb Q}_{+,B}]
 &=k(s)\,\Gamma_A^{ik}\Gamma_B^{jl}
 \operatorname*{Res}_{u=\svx_n}
 [\mu_s^\downarrow q_i^{A\downarrow}q_j^{B\uparrow}]
 \operatorname*{Res}_{\dot u=\dot\svx_{\dot n}}
 [\mu_s^\downarrow\dot q_k^{A\uparrow}\dot q_l^{B\downarrow}],\\
 \operatorname*{Res}_{u=\widetilde\svx_n}
 [\mathbb Q_{+,A}\widetilde{\mathbb Q}_{+,B}]
 &=k(1-s)\,\Gamma_A^{ik}\Gamma_B^{jl}
 \operatorname*{Res}_{u=\widetilde\svx_n}
 [\mu_s^\uparrow q_i^{A\downarrow}q_j^{B\uparrow}]
 \operatorname*{Res}_{\dot u=\dot{\widetilde\svx}_{\dot n}}
 [\mu_s^\uparrow\dot q_k^{A\uparrow}\dot q_l^{B\downarrow}],
 \end{aligned}
 \label{eq:J1-local-Q-shadow-residues}
\end{equation}
with $\dot n=n+m$. Summed over $n$ and $\dot n$, these chiral residues give the contour brackets~\eqref{eq:lower-tower-contour} of the measure-weighted products. The clockwise contour factor $-2\pi\ii$ and the normalisation~\eqref{eq:J1-sov-resolution-normalisation} then give the prefactors $-2\pi\ii\,\mathcal N_1^{\rm SoV}k(s)=-\Gamma(1-2s)^2$ of the ordinary tower and $-2\pi\ii\,\mathcal N_1^{\rm SoV}k(1-s)=\Gamma(1-2s)^2$ of the shadow tower. This is the tower bracket~\eqref{eq:paired-tower-bracket}, and is the form evaluated in Appendix~\ref{app:bsolver}. Equivalently, the measures can be absorbed into the canonical shadow Q-functions with the shadow map~\eqref{eq:shadow-q-map}, $\mu^\downarrow q^\downarrow=e^{-\frac{\ii\pi}{2}(2s-1)}\widetilde q^\downarrow$ and $\mu^\uparrow q^\uparrow=e^{+\frac{\ii\pi}{2}(2s-1)}\widetilde q^\uparrow$. Each tower then carries two such phases, which turn $k(s)$ and $k(1-s)$ into the coefficients $k^{\rm c}(s)$ and $k^{\rm c}(1-s)$ of~\eqref{eq:contour-conversion-coefficients}. This gives the brackets~\eqref{eq:J1-ordinary-residue-bilinear} and the form~\eqref{eq:J1-overlap-q-bilinears-skeleton}.

\Needspace{7\baselineskip}
\subsection{Evaluating the conformal integral}
\label{app:J1-coordinate-overlap}
We now evaluate the overlap~\eqref{eq:J1-overlap-shadow-bracket} directly, which gives~\eqref{eq:J1CompanionOverlapClosed} including its normalisation. The companion pairing~\eqref{eq:companion-pairing} is invariant under a common M\"obius transformation, $\langle g\,f|\cD{g\,f'}\rangle=\langle f|\cD{f'}\rangle$, where, displaying only the undotted part,
\begin{equation}
  g\cdot f(z)=(cz+d)^{-2s}f\left(\frac{az+b}{cz+d}\right),
  \qquad
  g=\begin{pmatrix}a&b\\c&d\end{pmatrix}\in\mathrm{SL}(2,\mathbb C).
  \label{eq:shadow-bracket-invariance}
\end{equation}
The map
\begin{equation}
  z\mapsto\kappa\,\frac{z-\lambda_B^{-1}}{z-\lambda_B},
  \qquad
  \kappa=\frac{\lambda_A-\lambda_B}{\lambda_A-\lambda_B^{-1}},
  \label{eq:J1-overlap-mobius-map}
\end{equation}
sends $\{\lambda_B^{-1},\lambda_B,\lambda_A,\lambda_A^{-1}\}$ to $\{0,\infty,1,\chi\}$, and likewise in the dotted sector. One coordinate integration is then done with the chain relation, leaving the conformal three-point integral
\begin{equation}
  \mathcal{I}_{h_A,-h_B}(\chi,\dot\chi)
  \equiv \int \dd^{2}z\, \frac{1}{[z-\chi]^{\{s+h_{A},s+\dot h_A\}}[z-1]^{\{s-h_{A},s-\dot h_A\}}[z]^{\{1-s-h_{B},1-s-\dot h_B\}}},
  \label{eq:J1-overlap-reduced-integral}
\end{equation}
where $-h_B$ appears because the inversion $\boldsymbol\lambda_B\to1/\boldsymbol\lambda_B$ in~\eqref{eq:J1-overlap-shadow-bracket} acts on the three-point structure of the second state as $h_B\to-h_B$. The overlap is
\begin{equation}
 \begin{aligned}
  \LRpair{\psi_A^{\rm C}}{\psi_B^{\rm C}}
  &=K\,
  \frac{[\lambda_A^2-1]^s[\lambda_B^2-1]^s}
       {[\lambda_A-\lambda_B]^{2s}}
  \chi^{s+\frac{h_A-h_B}{2}}
  \dot\chi^{s+\frac{\dot h_A-\dot h_B}{2}}
  \mathcal I_{h_A,-h_B}(\chi,\dot\chi),\\
  K&=-(-1)^{S_A+S_B}\,\frac{(2s-1)^2}{\pi}
  \frac{\Gamma(2s-1)}{\Gamma(2-2s)}
  \frac{\Gamma(1-s+\dot h_B)\Gamma(1-s-\dot h_B)}
       {\Gamma(s+h_B)\Gamma(s-h_B)},
 \end{aligned}
\label{eq:J1-overlap-reduction-prefactor}
\end{equation}
where the cross-ratios are defined in~\eqref{eq:cross-ratios-J1} and the constant $K$ comes from the normalisation of the shadow transform, the M\"obius transformation and the integration already performed with the chain relation.

The integral~\eqref{eq:J1-overlap-reduced-integral} is fixed up to two constants by a differential equation. Its undotted factor obeys the hypergeometric equation
\begin{equation}
 \Bigl[\chi(1-\chi)\partial_\chi^2+\bigl[1+h_A-h_B-(1+2s+h_A-h_B)\chi\bigr]\partial_\chi-(s+h_A)(s-h_B)\Bigr]\,\mathcal I_{h_A,-h_B}=0,
 \label{eq:J1-overlap-hypergeometric-ODE}
\end{equation}
and its dotted factor the same equation with dotted weights. Indeed, acting on the undotted part of the integrand, the operator in~\eqref{eq:J1-overlap-hypergeometric-ODE} gives a total derivative,
\begin{equation}
 \begin{aligned}
 &\Bigl[\chi(1-\chi)\partial_\chi^2+\bigl[1+h_A-h_B-(1+2s+h_A-h_B)\chi\bigr]\partial_\chi\\
 &\qquad-(s+h_A)(s-h_B)\Bigr]
 \frac{1}{(z-\chi)^{s+h_A}(z-1)^{s-h_A}z^{1-s-h_B}}\\
 &\qquad=(s+h_A)\,\partial_z\,\frac{(z-1)^{1-s+h_A}\,z^{s+h_B}}{(z-\chi)^{1+s+h_A}},
 \end{aligned}
\end{equation}
whose integral vanishes. The solutions of~\eqref{eq:J1-overlap-hypergeometric-ODE} are $F_1(\chi)$ and $\chi^{h_B-h_A}F_2(\chi)$, with
\begin{align}
 F_1(\chi)&={}_2F_1(s+h_A,s-h_B;1+h_A-h_B;\chi),
 \\
 F_2(\chi)&={}_2F_1(s+h_B,s-h_A;1-h_A+h_B;\chi).
\end{align}
Since $\mathcal I_{h_A,-h_B}$ solves both equations, it is a combination of products of an undotted and a dotted solution. Around $\chi=0$ the products $F_1\dot F_1$ and $[\chi]^{\{h_B-h_A,\dot h_B-\dot h_A\}}F_2\dot F_2$ are single-valued, while the mixed ones are not, so
\begin{equation}
 \mathcal I_{h_A,-h_B}(\chi,\dot\chi)
 =I_1\,F_1(\chi)\dot F_1(\dot\chi)
 +I_2\,[\chi]^{\{h_B-h_A,\dot h_B-\dot h_A\}}F_2(\chi)\dot F_2(\dot\chi),
 \label{eq:J1-overlap-regions}
\end{equation}
where $\dot F_{1,2}$ carry the dotted weights and $I_1$, $I_2$ are constants. These can be explicitly computed by considering the limit $\chi\to0$, in which two singular points of~\eqref{eq:J1-overlap-reduced-integral} collide and the integral receives contributions from two regions. Region one is when $|z|\gg|\chi|$, where one can set $[z-\chi]\to[z]$. Region two is when $z\sim\chi$, where one can set $[z-1]\to[-1]$. In this limit, each region reduces the integral to the following form, which is~\eqref{eq:chain-relation} at $z=1$, $y=0$, $a=1-B$, $b=1-A$,
\begin{equation}
 \int\dd^2z\,[z]^{\{A-1,\dot A-1\}}[1-z]^{\{B-1,\dot B-1\}}
 =\pi\,\frac{\Gamma(A)\Gamma(B)\Gamma(1-\dot A-\dot B)}{\Gamma(1-\dot A)\Gamma(1-\dot B)\Gamma(A+B)},
 \qquad A-\dot A,\ B-\dot B\in\mathbb Z,
 \label{eq:J1-chain-integral}
\end{equation}
which can be evaluated directly. In region one the integral does not depend on $\chi$, while in region two it scales as $[\chi]^{\{h_B-h_A,\dot h_B-\dot h_A\}}$. These are the leading behaviours of the two terms in~\eqref{eq:J1-overlap-regions}, and matching them gives
\begin{equation}
 \begin{aligned}
 I_1&=(-1)^{S_A}\pi\,
 \frac{\Gamma(h_B-h_A)\Gamma(1-s+h_A)\Gamma(s-\dot h_B)}
      {\Gamma(1+\dot h_A-\dot h_B)\Gamma(s-\dot h_A)\Gamma(1-s+h_B)},\\
 I_2&=\pi\,
 \frac{\Gamma(s+h_B)\Gamma(1-s-h_A)\Gamma(\dot h_A-\dot h_B)}
      {\Gamma(s+\dot h_A)\Gamma(1-s-\dot h_B)\Gamma(1+h_B-h_A)}.
 \end{aligned}
 \label{eq:J1-overlap-region-coefficients}
\end{equation}
Inserting these constants into~\eqref{eq:J1-overlap-regions}, and the result into~\eqref{eq:J1-overlap-reduction-prefactor}, gives the overlap~\eqref{eq:J1CompanionOverlapClosed}.

\section{Connection data for the norm}
\label{app:omega}
{We derive the connection representation of the finite norm
minor used in Section~\ref{sec:untwisted}.\par}

The analytic properties of the connection matrices and the logarithmic
cancellations are derived in Sections~\ref{sec:setup-omega}
and~\ref{sec:norms-onshell}. Here we collect the explicit first-pole
formulas, check the cutoff finite part, and relate the connection
formula to the Q-function Jacobians of the norm minor. We use equal site spins
$s_\alpha=\dot s_\alpha=s$, as in Section~\ref{sec:norms}.

\paragraph{Explicit Wronskians.}
\label{app:omega-bases}
Evaluating the normalisation~\eqref{eq:setup-wronskian-normalisation}
with the Gamma-function factors~\eqref{eq:shadow-chain-mu-factors} gives
\begin{equation}
 \begin{aligned}
 \mathcal W[q_1^\downarrow,q_2^\downarrow](u)
 &=(\lambda-\lambda^{-1})e^{-\ii\pi Js}
 \prod_{\alpha=1}^{J}
 \frac{\Gamma(1-s-\ii(u-\theta_\alpha))}
      {\Gamma(1+s-\ii(u-\theta_\alpha))},\\
 \mathcal W[q_1^\uparrow,q_2^\uparrow](u)
 &=(\lambda-\lambda^{-1})e^{+\ii\pi Js}
 \prod_{\alpha=1}^{J}
 \frac{\Gamma(-s+\ii(u-\theta_\alpha))}
      {\Gamma(s+\ii(u-\theta_\alpha))}.
 \end{aligned}
 \label{eq:omega-casoratians}
\end{equation}

\Needspace{7\baselineskip}
\subsection{Tower sums from the Baxter equation}
\label{app:omega-tower-sums}
We check~\eqref{eq:norms-first-pole-boundary} directly in Q-functions
and derive the endpoint forms of~\eqref{eq:norms-log-coefficient}
and~\eqref{eq:norms-finite-moment-connection}.

We first use Cramer's rule to express single residues of $q^\downarrow$ and
$q^\uparrow$ through the connection matrix. Use~\eqref{eq:omega-cramer}
with $\bar j\equiv3-j$. Its numerator is $(-1)^j$ times
\[
 \mathcal W[q_i^\downarrow,q_j^\uparrow](u)
 =q_i^\downarrow(u)q_j^\uparrow(u+\ii)
  -q_i^\downarrow(u+\ii)q_j^\uparrow(u).
\]
The second term is regular at both first-pole points, so its residue
vanishes. In the first term, only $q_j^\uparrow(u+\ii)$ has a pole
at $\svx_{\alpha,0}$, whereas only $q_i^\downarrow(u)$
has a pole at $\widetilde\svx_{\alpha,0}$.
Dividing by the up-arrow Wronskian therefore gives
\begin{equation}
 \begin{aligned}
 (\Omega^{-1})_{i\bar j}(\svx_{\alpha,0})
 &=(-1)^j\,
 \frac{q_i^\downarrow(\svx_{\alpha,0})
 \operatorname{Res}_{u=\svx_{\alpha,0}+\ii}q_j^\uparrow(u)}
 {\operatorname{Res}_{u=\svx_{\alpha,0}}
 \mathcal W[q_1^\uparrow,q_2^\uparrow](u)},\\
 \operatorname{Res}_{u=\widetilde\svx_{\alpha,0}}
 (\Omega^{-1})_{i\bar j}(u)
 &=(-1)^j\,
 \frac{[\operatorname{Res}_{u=\widetilde\svx_{\alpha,0}}q_i^\downarrow(u)]
 q_j^\uparrow(\widetilde\svx_{\alpha,0}+\ii)}
 {\mathcal W[q_1^\uparrow,q_2^\uparrow](\widetilde\svx_{\alpha,0})}.
 \end{aligned}
 \label{eq:norms-connection-endpoint-factors}
\end{equation}
The same formulas hold for the mixed matrix built from $q^{A\downarrow}$ and
$q^{B\uparrow}$, provided the two states share the twist and site data:
the canonical Wronskian is independent of the charges, including $h$.
The normalisation~\eqref{eq:setup-wronskian-normalisation} fixes the
remaining Wronskian. At $x\equiv\svx_{\alpha,0}$, $P_+$ has a
simple zero, so it fixes the pole coefficient of the Wronskian. Together with
\eqref{eq:J1-measure-conversion-coefficients}, it gives
\begin{equation}
 \begin{aligned}
 P_+'(x)\operatorname{Res}_{u=x}\mathcal W[q_1^\uparrow,q_2^\uparrow]
 &=\frac{(\lambda-\lambda^{-1})
 e^{\frac{\ii\pi}{2}(2Js-J)}}{\mu^\uparrow(x)},\\
 \frac{\operatorname{Res}_{u=x}\mu^\downarrow}
      {\mu^\uparrow(x)}
 &=\frac1{k_\alpha(s)},\qquad
 \operatorname*{Res}_{u=x}\det\Omega(u)
 =\frac{e^{\ii\pi(2Js-J)}}{k_\alpha(s)}.
 \end{aligned}
 \label{eq:app-omega-wronskian-residues}
\end{equation}
The last relation follows from~\eqref{eq:setup-omega-determinant}.

\paragraph{First-pole boundary term.}
In $D_{ij}$ of~\eqref{eq:norms-D-bilinear}, the second term is
regular at $x$. In the first, the zero of $P_+$ meets two simple poles, so
\[
 \operatorname*{Res}_{u=x}D_{ij}
 =P_+'(x)\left[\operatorname*{Res}_{u=x}\widetilde q_i^{A\downarrow}(u)\right]
 \left[\operatorname*{Res}_{u=x+\ii}q_j^{B\uparrow}(u)\right].
\]
The down-arrow shadow map gives
$\operatorname{Res}_x\widetilde q_i^{\,\downarrow}
=e^{\frac{\ii\pi}{2}(2Js-J)}(\operatorname{Res}_x\mu^\downarrow)
q_i^\downarrow(x)$. Inserting the first line
of~\eqref{eq:norms-connection-endpoint-factors}
and~\eqref{eq:app-omega-wronskian-residues} gives
\[
 \operatorname*{Res}_{u=x}D_{ij}
 =(-1)^j(\lambda-\lambda^{-1})\,
 \bigl(\operatorname*{Res}_{u=x}\det\Omega\bigr)\,
 (\Omega^{-1})_{i\bar j}(x),
\]
which is~\eqref{eq:norms-first-pole-boundary}, since
$\widetilde{\mathsf R}_\alpha=\ii e^{-\frac{\ii\pi}{2}(2Js-J)}
(\operatorname{Res}_x\det\Omega)\,\Omega^{-1}(x)$.

\paragraph{The cutoff term and the logarithm.}
At equal states, the left-hand side of~\eqref{eq:norms-baxter-tower-sum}
vanishes, so $\operatorname{Res}D_{ij}$ is the same at every
$\svx_{\alpha,N}$. Consider its resonant $i2$ part as
$N\to\infty$. Equations~\eqref{eq:baxter-asymptotics}
and~\eqref{eq:shadow-q-map} give
\begin{equation}
 \operatorname{Res}_{u=\svx_{\alpha,N}}
 [\widetilde q_i^\downarrow(u)q_2^\uparrow(u)]_{\mathrm{res}}=O(N^{-J}),
 \qquad P_\pm(\svx_{\alpha,N})\simeq(-\ii N)^J.
 \label{eq:norms-cutoff-asymptotics}
\end{equation}
For the resonant component, the shift ratios tend to $\lambda$ for
$\widetilde q_{i,\mathrm{res}}^{\,\downarrow}\propto\widetilde q_1^{\,\uparrow}$ and to
$\lambda^{-1}$ for $q_2^\uparrow$. Hence the cutoff term
$D_{ij}(\svx_{\alpha,N}-\ii)$ gives
\begin{equation}
 \lim_{N\to\infty}
 (\svx_{\alpha,N})^J
 \operatorname{Res}_{u=\svx_{\alpha,N}}
 [\widetilde q_i^\downarrow(u)q_2^\uparrow(u)]_{\mathrm{res}}
 =\frac{P_+'(\svx_{\alpha,0})}{\lambda-\lambda^{-1}}
 \left[\operatorname{Res}_{u=\svx_{\alpha,0}}\widetilde q_i^\downarrow(u)\right]
\left[ \operatorname{Res}_{u=\svx_{\alpha,0}}q_2^\uparrow(u+\ii)\right].
 \label{eq:norms-cutoff-limit}
\end{equation}
For the resonant $i1$ part, replace $q_2^\uparrow$ by
$q_1^\uparrow$ and $\lambda-\lambda^{-1}$ by $\lambda^{-1}-\lambda$.
By the shadow map~\eqref{eq:shadow-q-map}, the bracket
$(\widetilde q_i^\downarrow u^{J-1}q_j^\uparrow)_\alpha$ of~\eqref{eq:J1-ordinary-residue-bilinear}
equals $e^{\frac{\ii\pi}{2}(2Js-J)}\bigl(\mu^\downarrow u^{J-1}q_i^\downarrow q_j^\uparrow\bigr)_\alpha$.
Removing this phase, the coefficient of $\mathbb L$ in
$\bigl(\mu^\downarrow u^{J-1}q_i^\downarrow q_j^\uparrow\bigr)_\alpha$
is therefore
\begin{equation}
 (\widetilde{\mathsf R}_\alpha\sigma_1)_{ij}
 =\frac{\ii e^{-\frac{\ii\pi}{2}(2Js-J)}
       (-1)^j}{\lambda-\lambda^{-1}}\,
 P_+'(\svx_{\alpha,0})
 \left[\operatorname{Res}_{u=\svx_{\alpha,0}}
 \widetilde q_i^\downarrow(u)\right]
 \left[\operatorname{Res}_{u=\svx_{\alpha,0}+\ii}
 q_j^\uparrow(u)\right].
 \label{eq:norms-bilinear-log-coefficient}
\end{equation}
This reproduces~\eqref{eq:norms-log-coefficient}, which the main text obtained
directly from the tails of~\eqref{eq:norms-tower-residues}. The two routes
are independent checks of each other.

\paragraph{Convergent moments.}
With the deformation $t_B=t+\eta_bu^{b-1}$, $1\le b<J$, and
$\partial^B_{\eta_b}$ acting on the up-arrow factor only,
differentiating~\eqref{eq:norms-baxter-tower-sum} at $\eta=0$ gives
\begin{equation}
 \begin{aligned}
 \sum_{n=0}^{\infty}\operatorname{Res}_{u=\svx_{\alpha,n}}
 [u^{b-1}\widetilde q_i^\downarrow(u)q_j^\uparrow(u)]
 &=P_+'(\svx_{\alpha,0})
 \left[\operatorname{Res}_{u=\svx_{\alpha,0}}
 \widetilde q_i^\downarrow(u)\right]
 \left.\left[{\partial_{\eta_b}}
 \operatorname{Res}_{u=\svx_{\alpha,0}+\ii}
 q_{j,\eta}^\uparrow(u)\right]\right|_{\eta=0}\\
 &=-\ii(\lambda-\lambda^{-1})e^{\frac{\ii\pi}{2}(2Js-J)}
 (\partial_{\eta_b}^{B}\widetilde{\mathsf R}_\alpha\,\epsilon)_{ij}.
 \end{aligned}
 \label{eq:norms-finite-moment-endpoint}
\end{equation}
The derivative of the cutoff residue vanishes as $N\to\infty$,
since this deformation changes only subleading asymptotic terms.
Removing the shadow-map phase as above gives~\eqref{eq:norms-finite-moment-connection}.
For $b=J$, the same deformation changes $h$ and the cutoff endpoint
contributes $(\widetilde{\mathsf R}_\alpha\sigma_1)_{ij}\log(-\ii N)$.
Subtracting only $(\widetilde{\mathsf R}_\alpha\sigma_1)_{ij}\log N$ defines the
finite part, denoted $\mathrm{fin}$, and leaves
\begin{equation}
 \bigl(\mu^\downarrow u^{J-1}q_i^\downarrow q_j^\uparrow\bigr)_{\alpha,\mathrm{fin}}
 =-\ii(\lambda-\lambda^{-1})
 (\partial_{\eta_J}^{B}\widetilde{\mathsf R}_\alpha\,\epsilon)_{ij}
 +\log(-\ii)(\widetilde{\mathsf R}_\alpha\sigma_1)_{ij}.
 \label{eq:norms-highest-moment-endpoint}
\end{equation}
The cutoff contribution is derived in
Appendix~\ref{app:omega-highest-moment}.
The shadow version follows by $s\to1-s$. For the dotted version, dot all
chain data: $\partial^B_{\eta_b}$ varies the dotted up-arrow basis through
$\dot t(u)+\eta_bu^{b-1}$.

\paragraph{Both pole families as residues of one function.}
\label{app:omega-mixed-log}
In~\eqref{eq:norms-moment-connection-form} the ordinary and shadow towers
appear as residues of one function. We verify this here.
On the ordinary tower, the products of the coefficients in
\eqref{eq:overlaps-moment-residues} are, by the general
conversion~\eqref{eq:setup-residue-value-conversion},
\begin{equation}
 k_\alpha(s)(\widetilde{\mathsf R}_\alpha\sigma_1)_{ij}(\dot{\widetilde{\mathsf R}}_\alpha\sigma_1)_{ji}
 =-\operatorname{Res}_{u=\svx_{\alpha,0}}
 \left[\det\Omega(u)(\Omega^{-1}\sigma_1)_{ij}(u)
                   (\dot\Omega^{-1}\sigma_1)_{ji}(u)\right],
\label{eq:norms-ordinary-tail-product}
\end{equation}
because $\Omega^{-1}$ is regular there.
The expression in brackets is invariant under the shadow
map~\eqref{eq:setup-omega-shadow}: the determinant is inverted and each
inverse connection matrix acquires a factor $\det\Omega$.
Replacing $s\to1-s$ in the ordinary residue identity therefore gives
\begin{equation}
 k_\alpha(1-s)(\mathsf R_\alpha\sigma_1)_{ij}
                 (\dot{\mathsf R}_\alpha\sigma_1)_{ji}
 =-\operatorname{Res}_{u=\widetilde\svx_{\alpha,0}}
 \left[\det\Omega(u)(\Omega^{-1}\sigma_1)_{ij}(u)
                   (\dot\Omega^{-1}\sigma_1)_{ji}(u)\right].
 \label{eq:norms-tail-connection-data}
\end{equation}
The products with one or two charge derivatives, built
from~\eqref{eq:norms-finite-moment-connection}, behave in the same way,
because $\det\Omega$ depends only on the site data. Collecting the index
signs into $\sigma_1$ and $\epsilon$ gives~\eqref{eq:norms-moment-connection-form}.

\paragraph{The highest moment and the finite part.}
\label{app:omega-highest-moment}
We check the cutoff finite part~\eqref{eq:norms-highest-moment-endpoint}
and its independence of the common subtraction constant in the norm.
For $b=J$ we retain the cutoff term in
\eqref{eq:norms-baxter-tower-sum}.
For fixed $\alpha,j,i$, denote its phase-normalised residue by
\[
 \begin{aligned}
 \mathcal F_N(\eta)
 &\equiv e^{-\frac{\ii\pi}{2}(2Js-J)}
 \operatorname{Res}_{u=\svx_{\alpha,N}}
 \Bigl[P_+(u-\ii)\dot{\widetilde q}_j^\downarrow(u-\ii)
                   \dot q_{i,\eta}^\uparrow(u)\\
 &\hspace{42mm}
       -P_-(u)\dot{\widetilde q}_j^\downarrow(u)
                   \dot q_{i,\eta}^\uparrow(u-\ii)\Bigr].
 \end{aligned}
\]
Differentiating the tower-sum identity~\eqref{eq:norms-baxter-tower-sum} with respect to $\eta_J$ gives the finite-cutoff moment
\[
 \bigl(\mu^\downarrow(\dot u)\,\dot u^{J-1}
       \dot q_j^\downarrow\dot q_i^\uparrow\bigr)_{\alpha,N}
 =-\ii(\dot\lambda-\dot\lambda^{-1})
    (\partial_{\eta_J}^{B}\dot{\widetilde{\mathsf R}}_\alpha\,\epsilon)_{ji}
   -\left.\partial_{\eta_J}\mathcal F_N(\eta)\right|_{\eta=0}.
\]
The first term is the first-pole contribution already used above;
the second is the endpoint that vanished for $b<J$.

For $b=J$, differentiating the powers in
\eqref{eq:baxter-asymptotics} produces $\log u$.
Using the resonant endpoint limit~\eqref{eq:norms-cutoff-limit} and
$\partial_{\eta_J}\dot h=\ii/(\dot\lambda-\dot\lambda^{-1})$ gives
\[
 -\left.\partial_{\eta_J}\mathcal F_N(\eta)\right|_{\eta=0}
 =(\dot{\widetilde{\mathsf R}}_\alpha\sigma_1)_{ji}\log(-\ii N)+o(1).
\]
Subtracting the $\log N$ term gives the
dotted version of~\eqref{eq:norms-highest-moment-endpoint}.
The remaining $\log(-\ii)$ term contributes the mixed-log coefficient
after contraction, and hence vanishes
by~\eqref{eq:norms-double-log-zero}.
The same holds for any common subtraction constant.
Thus the boundary identity also applies at $b=J$.

\subsection{The single logarithm}
\label{app:omega-single-log}
Here we give the two steps of Section~\ref{sec:norms-onshell} that lead
from~\eqref{eq:norms-B-matrices} to~\eqref{eq:norms-graded-trace}. As there,
$\partial\equiv\partial_{\dot a_{m-1}}$ at fixed twists and site data.

\paragraph{Details of the variation.}
At an ordinary first pole $p=\svx_{\alpha,0}$, consider the
Wronskian expression~\eqref{eq:omega-cramer} for the inverse connection
matrix, together with the shadow map~\eqref{eq:shadow-q-map}. In the
shifted Wronskian, the term $q_i^{A\downarrow}(u+\ii)q_j^{B\uparrow}(u)$ is
regular at $u=p$, so the pole comes only from $q_j^{B\uparrow}(u+\ii)$ and
the residue factorises,
\[
 (\widetilde{\mathsf R}_{\alpha,(AB)}\epsilon)_{ij}
 \propto q_i^{A\downarrow}(p)\,
 \operatorname*{Res}_{u=p}q_j^{B\uparrow}(u+\ii),
\]
and similarly in the dotted sector. The proportionality constant depends
only on the twist, since the Wronskian denominator is charge independent.
The variation $\partial^B$ acts on the up-arrow factor, while
$\partial^A\equiv\partial-\partial^B$ acts on the down-arrow factor. We now
show that the $\partial^A$ term vanishes by gluing, so that $\partial^B$
can be completed to the full variation.
Using $\sigma_1=\sigma_3\epsilon$, the extra term in the contraction factorises as
\begin{equation}
 \begin{aligned}
 &\sum_{i,j}\Gamma^{ii}\Gamma^{jj}
 (\widetilde{\mathsf R}_\alpha\sigma_1)_{ij}
 \bigl[\partial^A
 \dot{\widetilde{\mathsf R}}_\alpha\epsilon\bigr]_{ji}\\
 &\quad\propto
 \underbrace{\operatorname*{Res}_{u=p}\Qp(u,u+\ii)}_{=0}
 \sum_j\Gamma^{jj}(-1)^j
 \bigl[\operatorname*{Res}_{u=p}q_j^\uparrow(u+\ii)\bigr]
 \partial^A\dot q_j^\downarrow(p)=0.
 \end{aligned}
 \label{eq:norms-lower-moment-rank-one}
\end{equation}
The zero is precisely the analyticity of ${\mathbb Q}_+$ on shell~\eqref{eq:setup-Qplus-first-pole-regularity}.
The shadow family follows by $s\to1-s$.
We may therefore replace $\partial^B$ by $\partial$, which
gives~\eqref{eq:norms-single-log-connection}.

\paragraph{Details of the trace reduction.}
We now simplify the trace under this residue.
Gluing~\eqref{eq:setup-omega-gluing}, used only at the undeformed state,
moves the left $\Gamma$ next to $\sigma_1$. The two $\Gamma$ matrices
can then be replaced by the scalar factor $c$:
\[
 \Gamma\Omega^{-1}\sigma_1\Gamma
 =(\dot\Omega^{-1})^{\mathsf T}
 \underbrace{\Gamma\sigma_1\Gamma}_{c\sigma_1}
 =c(\dot\Omega^{-1})^{\mathsf T}\sigma_1.
\]
Applying~\eqref{eq:omega-two-by-two-inverse} to $\dot\Omega$ gives
$\epsilon(\dot\Omega^{-1})^{\mathsf T}
=\dot\Omega\epsilon/\det\dot\Omega$, so the determinant cancels due to
$\det\Omega=\det\dot\Omega$ from~\eqref{eq:setup-omega-determinant}.
Using $\epsilon\sigma_1=\sigma_3$ and cycling $\dot\Omega$ to the right gives
\[
 \begin{aligned}
 \det\dot\Omega\,
 \operatorname{tr}\!\bigl[(\dot\Omega^{-1})^{\mathsf T}\sigma_1
 (\partial\dot\Omega^{-1})\epsilon\bigr]
 &=\operatorname{tr}\!\bigl[\dot\Omega\epsilon\sigma_1
 (\partial\dot\Omega^{-1})\bigr]=\operatorname{tr}\!\bigl[\sigma_3(\partial\dot\Omega^{-1})\dot\Omega\bigr].
 \end{aligned}
\]
Finally, differentiating $\dot\Omega^{-1}\dot\Omega=1$ gives
$(\partial\dot\Omega^{-1})\dot\Omega=-\dot\Omega^{-1}\partial\dot\Omega$.
Substituting these steps into~\eqref{eq:norms-single-log-connection}
gives~\eqref{eq:norms-graded-trace}.

\subsection{The norm minor in connection form}
\label{sec:norms-boundary-variation}
We have isolated the logarithm; it remains to evaluate the finite $(J-1)\times(J-1)$ matrix in~\eqref{eq:norms-two-point-normalisation}.
For $b,d<J$, \eqref{eq:norms-moment-connection-form} reads
\begin{equation}
 \mathcal M_{bd}
 =-2\pi\ii(\lambda-\lambda^{-1})(\dot\lambda-\dot\lambda^{-1})
 \sum_p\operatorname*{Res}_{u=p}\,\det\Omega\,
 \operatorname{tr}\!\bigl[\Gamma(\partial^B_{a_{b-1}}\Omega^{-1})\epsilon\,
 \Gamma(\partial^B_{\dot a_{d-1}}\dot\Omega^{-1})\epsilon\bigr],
 \qquad b,d<J,
 \label{eq:norms-minor-up-variation}
\end{equation}
where $p$ runs over the $2J$ ordinary and shadow poles in one period strip, as defined in Section~\ref{sec:norms-finite}.
We would like a form involving full charge variations, so that we can
later take the untwisting limit directly on the connection matrices in Section~\ref{sec:untwisted}.

As in the single-log calculation, we will show that the complementary
down-arrow variation $\partial^A$ contributes no residue. The full charge-space
connection will then reproduce the original one-sided moments.
Here the derivatives with respect to the charges $a_{b-1}$ in~\eqref{eq:transfer-matrix-expansion} are the same as the derivatives $\partial_{\eta_b}$ used for the deformation $t(u)\to t(u)+\eta_bu^{b-1}$ above, since this deformation just shifts $a_{b-1}\to a_{b-1}+\eta_b$; that is, $\partial_{\eta_b}=\partial_{a_{b-1}}$.

\paragraph{Full charge variations at a first pole.}
Introduce the connection in charge space,
\begin{equation}
 \mathsf E_b(u)\equiv\Omega^{-1}(u)\partial_{a_{b-1}}\Omega(u)
 =-(\partial_{a_{b-1}}\Omega^{-1}(u))\Omega(u).
 \label{eq:norms-connection-data}
\end{equation}
The second equality follows by differentiating $\Omega^{-1}\Omega=1$.
This is a natural object because it does not distinguish the two pole
families: the scalar factor in the shadow
map~\eqref{eq:setup-omega-shadow} is charge independent, so
$\widetilde{\mathsf E}_b=\mathsf E_b$.
Split the variation by defining
$\mathsf E_b^A\equiv-(\partial^A_{a_{b-1}}\Omega^{-1})\Omega$
and $\mathsf E_b^B$ with $\partial^B$ in place of $\partial^A$. Then
\[
 \mathsf E_b=\mathsf E_b^A+\mathsf E_b^B
 =\widetilde{\mathsf E}_b^A+\widetilde{\mathsf E}_b^B.
\]
The same scalar rescaling commutes with either one-sided variation.
Each part is therefore shadow invariant:
$\mathsf E_b^A=\widetilde{\mathsf E}_b^A$, and similarly for $B$.

We now show that $\mathsf E_b^A$ has zero residue at both pole types.

This plays the role of the $\partial^B\to\partial$ step
in~\eqref{eq:norms-lower-moment-rank-one}, but requires no quantisation
condition here. The inverse-matrix identity~\eqref{eq:omega-two-by-two-inverse}
gives
\[
 \mathsf E_b^A=(\det\Omega)(\partial^A_{a_{b-1}}\Omega^{-1})
   \epsilon(\Omega^{-1})^{\mathsf T}\epsilon.
\]
First take $p=\svx_{\alpha,0}=\theta_\alpha-\ii s$.
Only $\det\Omega$ has a pole there. In the mixed version of the
Wronskian ratio~\eqref{eq:omega-cramer}, only the product
$q_i^{A\downarrow}(u)q_k^{B\uparrow}(u+\ii)$ survives: the other product
is regular and vanishes upon division by the pole in the denominator.
Thus $\partial^A$ varies only the down-arrow factor. Setting $A=B$
after the variation leaves two identical up-arrow factors paired
through $\epsilon$, so
\[
 \begin{aligned}
 &\bigl[(\operatorname{Res}_p\mathsf E_b^A)\epsilon\bigr]_{ij}\\
 &\quad\propto
 (\partial_{a_{b-1}}q_i^\downarrow(p))q_j^\downarrow(p)
 \underbrace{\sum_{k,l=1}^2\epsilon_{kl}
 \bigl[\operatorname*{Res}_{u=p}q_k^\uparrow(u+\ii)\bigr]
 \bigl[\operatorname*{Res}_{u=p}q_l^\uparrow(u+\ii)\bigr]}_{=0}=0.
 \end{aligned}
\]
Since $\mathsf E_b^A=\widetilde{\mathsf E}_b^A$, the same result with
$s\to1-s$ immediately gives zero residue at
$p=\widetilde\svx_{\alpha,0}$ as well. Therefore only
$\mathsf E_b^B$ contributes at either pole family.
At an ordinary point this gives explicitly
\begin{equation}
 \begin{aligned}
 \operatorname{Res}_p\mathsf E_b
 &=\operatorname{Res}_p\mathsf E_b^B\\
 &=(\operatorname{Res}_p\det\Omega)\,
   (\partial^B_{a_{b-1}}\Omega^{-1})(p)\,
   \epsilon\Omega^{-1}(p)^{\mathsf T}\epsilon,
 \qquad p=\svx_{\alpha,0}.
 \end{aligned}
 \label{eq:norms-local-connection-residue}
\end{equation}
\paragraph{The norm from the connection residues.}
The same epsilon contraction that killed $\mathsf E_b^A$ gives the
surviving $12$ entry explicitly. At $p=\svx_{\alpha,0}$,
\[
 (\operatorname{Res}_p\mathsf E_b)_{12}
 =(\operatorname{Res}_p\det\Omega)
 \bigl[(\Omega^{-1})_{12}(\partial^B_{a_{b-1}}\Omega^{-1})_{11}
       -(\Omega^{-1})_{11}(\partial^B_{a_{b-1}}\Omega^{-1})_{12}\bigr]_{u=p}.
\]
Multiply this by its dotted counterpart. Since the gluing matrix is
$\Gamma=\operatorname{diag}(1,c)$, the gluing condition~\eqref{eq:setup-omega-gluing}
gives $(\dot\Omega^{-1})_{11}=(\Omega^{-1})_{11}$ and
$(\dot\Omega^{-1})_{12}=c(\Omega^{-1})_{21}$.

The fixed column factors of the one-sided variations then combine
the four products into the trace appearing in the moment matrix:
\[
 \begin{aligned}
 &(\operatorname{Res}_p\mathsf E_b)_{12}
  (\operatorname{Res}_p\dot{\mathsf E}_d)_{12}\\
 &\quad=(\operatorname{Res}_p\det\Omega)^2
       [(\Omega^{-1})_{11}(p)]^2\\
 &\qquad\quad\times\operatorname{tr}\!\bigl[
 \Gamma(\partial^B_{a_{b-1}}\Omega^{-1})\epsilon
 \Gamma(\partial^B_{\dot a_{d-1}}\dot\Omega^{-1})\epsilon\bigr]_{u=p},
 \quad p=\svx_{\alpha,0}.
 \end{aligned}
\]
The component multiplication is written out in
Appendix~\ref{app:omega-norm-component}.
To recover the moment integrand, divide by the residue of
\[
 \rho(u)\equiv\frac{\Omega_{22}(u)^2}{\det\Omega(u)},\qquad
 \operatorname{Res}_p\rho
 =(\operatorname{Res}_p\det\Omega)[(\Omega^{-1})_{11}(p)]^2,
 \quad p=\svx_{\alpha,0},
\]
where $\Omega_{22}=(\det\Omega)(\Omega^{-1})_{11}$.
This removes one determinant residue and the squared matrix entry,
leaving exactly the weighted trace in~\eqref{eq:norms-minor-up-variation}.

The weight $\rho$ is itself a connection response. The unit basis
variation $\delta q_1^\uparrow=q_2^\uparrow$, $\delta q_2^\uparrow=0$,
with the down-arrow basis fixed, gives
\[
 [\Omega^{-1}\delta\Omega]_{12}
 =(\Omega^{-1})_{11}\Omega_{22}=\rho.
\]
Thus $\rho$ measures the response to mixing the two up-arrow solutions,
just as $\mathsf E_b$ measures the response to varying a charge.

The moment integrand is shadow invariant: the two variation factors
$\det\Omega$ combine with $\det\widetilde\Omega=1/\det\Omega$ to give
the original weight. Since $\mathsf E_b$ and $\rho$ are also shadow
invariant, the same quotient supplies the shadow poles. At generic data
these poles are simple, so the three factors can be combined under one
residue, giving~\eqref{eq:norms-connection-matrix}.

Here $p$ runs over both pole families in one period strip.
This resembles the Gaudin formula for the norm~\cite{Gaudin:1981,Korepin:1982gg}, which uses the Jacobian
of the Bethe equations. Here $\mathsf E_b$ measures charge variations
of the connection data entering the quantisation condition, and
$\mathcal M$ pairs the dotted and undotted variations.
This expression will be useful for taking the untwisting limit in
Section~\ref{sec:untwisted}.

{Further details of the derivation and a
covariant form of~\eqref{eq:norms-connection-matrix} are given below.\par}

\Needspace{7\baselineskip}
\subsection{The norm minor as a pairing of boundary Jacobians}
\label{app:omega-boundary-jacobians}
Substituting the endpoint forms~\eqref{eq:norms-finite-moment-endpoint}
into~\eqref{eq:overlaps-moment-residues} gives, for the convergent moments,
\begin{equation}
 \begin{aligned}
 \mathcal M_{bc}
 ={}&2\pi\ii(\lambda-\lambda^{-1})(\dot\lambda-\dot\lambda^{-1})
 \sum_{\alpha=1}^{J}\sum_{i,j=1}^{2}\Gamma^{ii}\Gamma^{jj}\\
 &\quad\times\Bigl[
 k_\alpha(s)
 (\partial_{\eta_b}^{B}\widetilde{\mathsf R}_\alpha\,\epsilon)_{ij}
 (\partial_{\dot\eta_c}^{B}\dot{\widetilde{\mathsf R}}_\alpha\,\epsilon)_{ji}
 +(s\to1-s)\Bigr],\qquad b,c<J,
 \end{aligned}
 \label{eq:norms-convergent-moment-derivatives}
\end{equation}
with the deformations
$t_\eta(u)=t(u)+\sum_{b=1}^{J-1}\eta_bu^{b-1}$ and
$\dot t_{\dot\eta}(u)=\dot t(u)+\sum_{c=1}^{J-1}\dot\eta_cu^{c-1}$.
All derivatives are evaluated at $\eta=\dot\eta=0$.

Here $s\to1-s$ replaces $\widetilde{\mathsf R}_\alpha$ by
$\mathsf R_\alpha$ and likewise in the dotted factor.
Using~\eqref{eq:setup-omega-residue-matrices} and
\eqref{eq:app-omega-wronskian-residues} gives~\eqref{eq:norms-minor-up-variation}.
To expose the relation to the on-shell condition, define
\begin{equation}
 \begin{aligned}
 \mathcal F_\alpha(s;\eta)
 &\equiv\sum_j\Gamma^{jj}
 \left[\operatorname{Res}_{u=\svx_{\alpha,0}}
 \dot{\widetilde q}_j^\downarrow(u)\right]
 \left[\operatorname{Res}_{u=\svx_{\alpha,0}+\ii}
 q_{j,\eta}^\uparrow(u)\right],\\
 \dot{\mathcal F}_\alpha(s;\dot\eta)
 &\equiv\sum_i\Gamma^{ii}
 \left[\operatorname{Res}_{u=\svx_{\alpha,0}}
 \widetilde q_i^\downarrow(u)\right]
 \left[\operatorname{Res}_{u=\svx_{\alpha,0}+\ii}
 \dot q_{i,\dot\eta}^\uparrow(u)\right].
 \end{aligned}
 \label{eq:norms-boundary-conditions}
\end{equation}
By~\eqref{eq:shadow-q-map}, the down-arrow shadow residues are
proportional to the ordinary down-arrow values, with a common factor
for both indices. Thus, at the physical state, $\mathcal F_\alpha(s;0)$ and
$\dot{\mathcal F}_\alpha(s;0)$ are proportional to the two vanishing
residues in~\eqref{eq:setup-Qplus-first-pole-regularity}. These are the
same exit-pole conditions used in~\eqref{eq:sov-exit-pole-cancellation}.
The shadow conditions follow by $s\to1-s$.
Substituting the endpoint
formula~\eqref{eq:norms-bilinear-log-coefficient} into
\eqref{eq:norms-convergent-moment-derivatives} separates the $i$ and
$j$ sums. The sign factors cancel those in the two coefficients,
giving
\begin{equation}
 \begin{aligned}
 \mathcal M_{bc}
 ={}&-2\pi\ii\sum_{\alpha=1}^{J}\Bigl[
 k_\alpha(s)e^{-\ii\pi(2Js-J)}
 [P_+'(\svx_{\alpha,0})]^2\\
 &\qquad\times
 (\partial_{\eta_b}\mathcal F_\alpha(s;\eta))
 (\partial_{\dot\eta_c}\dot{\mathcal F}_\alpha(s;\dot\eta))
 +(s\to1-s)\Bigr],\qquad b,c<J.
 \end{aligned}
 \label{eq:norms-boundary-variation-matrix}
\end{equation}
The ordinary derivatives here act only on the up-arrow solutions,
as specified in the definitions of $\mathcal F$ and $\dot{\mathcal F}$.
The dotted and undotted $P_+$ coincide for our common site data.
The norm minor is therefore a weighted pairing of the dotted and
undotted Jacobians of these endpoint conditions.

At zero twist, taking $u\to x_\alpha=\theta_\alpha-\ii s$ in the
Baxter equation, where $P_+(x_\alpha)=0$, gives
\begin{equation}
 P_+'(x_\alpha)\operatorname{Res}_{u=x_\alpha+\ii}q_j^{0\uparrow}(u)
 =t^0(x_\alpha)q_j^{0\uparrow}(x_\alpha)
  -P_-(x_\alpha)q_j^{0\uparrow}(x_\alpha-\ii).
 \label{eq:untwisting-exit-residue}
\end{equation}
Substituting this identity and its dotted counterpart into
\eqref{eq:norms-boundary-conditions}, and using the shadow
map~\eqref{eq:shadow-q-map} for the opposite-arrow residues, gives
the regular norm block~\eqref{eq:untwisting-norm-boundary-matrix}
as a pairing of derivatives of regular Q-values. The charge
derivative acts on both $t^0$ and $q^{0\uparrow}$ in the right-hand
side, with the analogous variation in the dotted sector.

\paragraph{The norm matrix in connection variables.}
\label{app:omega-norm-matrix}
We relate the connection formula~\eqref{eq:norms-connection-matrix}
to the boundary-Jacobian formula~\eqref{eq:norms-boundary-variation-matrix}.
This conversion is used in the numerical checks and the BSolver2D appendix.
Write $x\equiv\svx_{\alpha,0}$ and
$R\equiv\operatorname{Res}_{u=x}\mu^\downarrow(u)$.
In this paragraph
$\mathcal W^\downarrow(u)\equiv
\mathcal W[q_1^\downarrow,q_2^\downarrow](u)$,
with the analogous dotted definition.
All charge derivatives preserve the twist and weights, so the
canonical Wronskians~\eqref{eq:omega-casoratians} are fixed.

First, Cramer's rule~\eqref{eq:omega-cramer} gives
\begin{equation}
 (\operatorname{Res}_{u=x}\mathsf E_b(u))_{12}
 =\frac{q_1^\downarrow(x)}{\mathcal W^\downarrow(x)}
 \sum_i(\Omega^{-1})_{1i}(x)
 \partial_{a_{b-1}}^B
 \left[\operatorname{Res}_{u=x+\ii}q_i^\uparrow(u)\right].
 \label{eq:omega-norm-variation-residue}
\end{equation}
Only the up-arrow variation contributes to this residue.
The down-arrow contribution to $(\mathsf E_b)_{12}$ is
$-\mathcal W[q_1^\downarrow,\partial_{a_{b-1}}q_1^\downarrow]
/\mathcal W^\downarrow$.
This ratio is regular at the ordinary first pole.
At the shadow first pole its numerator has at most a simple pole,
while its denominator has a simple pole, so it is regular there too.

At $x$, the inverse connection matrices have rank one.
The endpoint identity~\eqref{eq:norms-connection-endpoint-factors}
and the on-shell condition~\eqref{eq:setup-omega-gluing} give
\[
 (\Omega^{-1})_{1i}(x)
 =(\Omega^{-1})_{11}(x)
   \frac{\Gamma^{ii}\dot q_i^\downarrow(x)}
        {\dot q_1^\downarrow(x)}.
\]
The sum in~\eqref{eq:omega-norm-variation-residue} is therefore
proportional to the boundary Jacobian
$\partial_{a_{b-1}}^B\mathcal F_\alpha$.
Its proportionality factor follows directly from the down-arrow
shadow map~\eqref{eq:shadow-q-map}.
Multiplying by the dotted identity cancels
$q_1^\downarrow(x)/\dot q_1^\downarrow(x)$ against its inverse.
Since $(\Omega^{-1})_{11}=(\dot\Omega^{-1})_{11}$, this gives
\[
 \begin{aligned}
 &(\operatorname{Res}_{u=x}\mathsf E_b)_{12}
  (\operatorname{Res}_{u=x}\dot{\mathsf E}_c)_{12}\\
 &\quad=
 \frac{e^{-\ii\pi(2Js-J)}[(\Omega^{-1})_{11}(x)]^2}
      {R^2\mathcal W^\downarrow(x)\dot{\mathcal W}^\downarrow(x)}
 (\partial_{a_{b-1}}^B\mathcal F_\alpha)
 (\partial_{\dot a_{c-1}}^B\dot{\mathcal F}_\alpha).
 \end{aligned}
\]
The Wronskian normalisation~\eqref{eq:omega-casoratians} implies
\[
 P_+'(x)R\,\mathcal W^\downarrow(x)
 =(\lambda-\lambda^{-1})e^{-\frac{\ii\pi}{2}(2Js-J)},
\]
and likewise for dots.
Hence the local conversion is
\begin{equation}
 \begin{aligned}
 &(\operatorname{Res}_{u=x}\mathsf E_b)_{12}
  (\operatorname{Res}_{u=x}\dot{\mathsf E}_c)_{12}\\
 &\quad=
 \frac{[P_+'(x)]^2[(\Omega^{-1})_{11}(x)]^2}
 {(\lambda-\lambda^{-1})(\dot\lambda-\dot\lambda^{-1})}
 (\partial_{a_{b-1}}^B\mathcal F_\alpha)
 (\partial_{\dot a_{c-1}}^B\dot{\mathcal F}_\alpha).
 \end{aligned}
 \label{eq:omega-norm-local-conversion}
\end{equation}
Finally, $\Omega^{-1}$ is regular at $x$, so
\eqref{eq:setup-omega-determinant} gives
\[
 \operatorname{Res}_{u=x}\rho(u)
 =[(\Omega^{-1})_{11}(x)]^2
   \operatorname{Res}_{u=x}\det\Omega(u)
 =\frac{e^{\ii\pi(2Js-J)}}{k_\alpha(s)}
   [(\Omega^{-1})_{11}(x)]^2.
\]
Substitution into~\eqref{eq:norms-boundary-variation-matrix}
proves the ordinary-pole term of
\eqref{eq:norms-connection-matrix}.
The shadow term follows from~\eqref{eq:setup-omega-shadow}:
both $\rho$ and $\mathsf E_b$ are unchanged by the shadow
transformation, because $\det\Omega$ depends only on site data.
It supplies the shadow first poles of the same functions.

\paragraph{The component contraction for the norm.}
\label{app:omega-norm-component}
{We spell out the component multiplication used after
\eqref{eq:norms-local-connection-residue} above.
Throughout this paragraph $p=\svx_{\alpha,0}$, and
$\rho$ is the weight defined there.\par}
Taking the $12$ entry of~\eqref{eq:norms-local-connection-residue} gives
\[
 \begin{aligned}
 (\operatorname{Res}_p\mathsf E_b)_{12}
 &=(\operatorname{Res}_p\det\Omega)
 \bigl[(\Omega^{-1})_{12}(\partial^B_{a_{b-1}}\Omega^{-1})_{11}
       -(\Omega^{-1})_{11}(\partial^B_{a_{b-1}}\Omega^{-1})_{12}\bigr]_{u=p}\\
 &=-\operatorname{Res}_p\rho\,
 \left.\partial^B_{a_{b-1}}
 \frac{(\Omega^{-1})_{12}}{(\Omega^{-1})_{11}}\right|_{u=p}.
 \end{aligned}
\]
The dotted relation has the same weight, since gluing gives
$(\dot\Omega^{-1})_{11}=(\Omega^{-1})_{11}$.

To expand the trace in~\eqref{eq:norms-minor-up-variation}, recall that
the one-sided variation changes only the row factor. Its two rows are
therefore proportional:
\[
 (\partial^B_{a_{b-1}}\Omega^{-1})_{2j}(p)
 =\frac{(\Omega^{-1})_{21}(p)}{(\Omega^{-1})_{11}(p)}
   (\partial^B_{a_{b-1}}\Omega^{-1})_{1j}(p),
\]
and similarly in the dotted sector. With
$\Gamma=\operatorname{diag}(1,c)$, gluing~\eqref{eq:setup-omega-gluing}
gives $(\dot\Omega^{-1})_{21}=(\Omega^{-1})_{12}/c$ and
$(\dot\Omega^{-1})_{12}=c(\Omega^{-1})_{21}$.
Substituting these row relations into the two-by-two trace factorises it:
\[
 \begin{aligned}
 &\operatorname{tr}\!\bigl[
 \Gamma(\partial^B_{a_{b-1}}\Omega^{-1})\epsilon
 \Gamma(\partial^B_{\dot a_{c-1}}\dot\Omega^{-1})\epsilon\bigr]_{u=p}\\
 &\quad=[(\Omega^{-1})_{11}(p)]^2
 \left.\partial^B_{a_{b-1}}
 \frac{(\Omega^{-1})_{12}}{(\Omega^{-1})_{11}}\right|_{u=p}
 \left.\partial^B_{\dot a_{c-1}}
 \frac{(\dot\Omega^{-1})_{12}}{(\dot\Omega^{-1})_{11}}\right|_{u=p}.
 \end{aligned}
\]

\Needspace{7\baselineskip}
\subsection{A covariant form of the norm minor}
\label{app:omega-norm-covariance}
{Equation~\eqref{eq:norms-connection-matrix} was derived above in
the canonical basis. Here we give the equivalent covariant contraction.\par}
At an ordinary point $p=\svx_{\alpha,0}$, all columns of
$\partial^B\Omega^{-1}(p)$ lie in the one-dimensional image of
$\Omega^{-1}(p)$. The latter matrix acts on this image by its nonzero
eigenvalue, $\operatorname{tr}\Omega^{-1}(p)$, so
\begin{equation}
 \Omega^{-1}(p)(\partial^B_{a_{b-1}}\Omega^{-1})(p)
 =\operatorname{tr}\Omega^{-1}(p)\,
   (\partial^B_{a_{b-1}}\Omega^{-1})(p).
 \label{eq:norms-local-rank-one-variation}
\end{equation}
Charge independence of $\det\Omega$ also gives
\[
 \operatorname{tr}\mathsf E_b
 =\operatorname{tr}\!\left[\Omega^{-1}\partial_{a_{b-1}}\Omega\right]
 =\partial_{a_{b-1}}\log\det\Omega=0.
\]
Hence $\mathsf E_b\epsilon$ is symmetric, and the gluing matrix pairs
the two symmetric residue tensors.
Substitute~\eqref{eq:norms-local-connection-residue} and its dotted
counterpart into this pairing, and use
gluing~\eqref{eq:setup-omega-gluing} to relate their regular matrices.
Each regular rank-one matrix acts on its one-sided variation through
\eqref{eq:norms-local-rank-one-variation}, giving one factor
$\operatorname{tr}\Omega^{-1}(p)$ in each sector. Explicitly,
\begin{equation}
 \begin{aligned}
 &\operatorname{tr}\!\left[
 \Gamma(\operatorname{Res}_p\mathsf E_b)\epsilon
 \Gamma(\operatorname{Res}_p\dot{\mathsf E}_c)\epsilon\right]\\
 &\quad=(\operatorname{Res}_p\det\Omega)^2
       [\operatorname{tr}\Omega^{-1}(p)]^2\\
 &\qquad\quad\times\operatorname{tr}\!\left[
 \Gamma(\partial^B_{a_{b-1}}\Omega^{-1})(p)\epsilon
 \Gamma(\partial^B_{\dot a_{c-1}}\dot\Omega^{-1})(p)\epsilon\right].
 \end{aligned}
 \label{eq:norms-local-tensor-contraction}
\end{equation}
The same scalar factors occur in the denominator. Indeed,
$\operatorname{tr}\Omega=(\det\Omega)\operatorname{tr}\Omega^{-1}$ gives
\[
 \operatorname{Res}_p\!\left[
 \frac{(\operatorname{tr}\Omega)^2}{\det\Omega}\right]
 =(\operatorname{Res}_p\det\Omega)
   [\operatorname{tr}\Omega^{-1}(p)]^2.
\]
Dividing cancels the squared trace and one determinant residue.
What remains is exactly the ordinary-pole contribution to
\eqref{eq:norms-minor-up-variation}.

{The shadow invariance established above extends this
ordinary-pole identity to the shadow family. Thus\par}
\begin{equation}
 \mathcal M_{bc}
 =-2\pi\ii(\lambda-\lambda^{-1})(\dot\lambda-\dot\lambda^{-1})
 \sum_p\frac{\operatorname{tr}\!\left[
 \Gamma(\operatorname{Res}_p \mathsf E_b)\epsilon
 \Gamma(\operatorname{Res}_p\dot{\mathsf E}_c)\epsilon\right]}
 {\operatorname{Res}_p\!\left[(\operatorname{tr}\Omega)^2/\det\Omega\right]}.
 \label{eq:norms-connection-covariant}
\end{equation}
Here $p$ runs over the $2J$ ordinary and shadow points in one period strip.
This form is invariant under a common, charge-independent $SL(2)$ change
of both arrow bases in each sector: each $\mathsf E_b\epsilon$ carries two basis factors,
which are cancelled by the inverse factors in $\Gamma$, while the
denominator contains only traces and determinants.

\paragraph{Evaluating the norm from connection data.}
The functions \texttt{Omega}, \texttt{OmegaInv} and
\texttt{OmegaResidues} supply every quantity in
\eqref{eq:norms-connection-covariant} and its component form
\eqref{eq:norms-connection-matrix}.
The two arrays returned by \texttt{OmegaResidues} contain the
residues of $\Omega$ and $\Omega^{-1}$, respectively.
Charge differences use off-shell displaced solutions at the same
twist and weights. Since the pole positions stay fixed,
\begin{equation}
 \operatorname{Res}_{u=p}\mathsf E_b(u)=
 \begin{cases}
 \Omega^{-1}(p)\,
 \partial_{a_{b-1}}(\operatorname{Res}_{u=p}\Omega(u)),
       &p\ \text{ordinary},\\
 (\operatorname{Res}_{u=p}\Omega^{-1}(u))\,
 \partial_{a_{b-1}}\Omega(p),
       &p\ \text{shadow}.
 \end{cases}
 \label{eq:untwisting-omega-residue-evaluation}
\end{equation}
The covariant denominator uses only unvaried connection data:
\[
 \operatorname{Res}_{u=p}\!\left[
 \frac{(\operatorname{tr}\Omega(u))^2}{\det\Omega(u)}\right]
 =\begin{cases}
 \operatorname{tr}(\operatorname{Res}_{u=p}\Omega(u))\,
 \operatorname{tr}\Omega^{-1}(p),&p\ \text{ordinary},\\
 \operatorname{tr}\Omega(p)\,
 \operatorname{tr}(\operatorname{Res}_{u=p}\Omega^{-1}(u)),
       &p\ \text{shadow}.
 \end{cases}
\]
For the component form, take the $12$ entries of the residue
matrices and replace the two traces by $\Omega_{22}$ and
$(\Omega^{-1})_{11}$, respectively.
The same expressions apply in the unit-leading zero-twist basis
for the regular charge variations.

\section{Exact results for the untwisted \texorpdfstring{$J=2$}{J=2} chain}
\label{sec:J2-untwisted}

For $J=2$ without twist, conformal symmetry again fixes the wave function up to an overall constant.
 A primary state of weights $(h,\dot h)$ obeys
\begin{equation}
 (\widehat{\mathbb S}^{z}-h)\psi^{\rm D}=0,
 \qquad
 \widehat{\mathbb S}^{-}\psi^{\rm D}=0,
\end{equation}
and the corresponding dotted equations, with the global generators defined in \eqref{eq:global-sl2-generators}. Their solution is
\begin{equation}
 \psi^{\rm D}(z_1,z_2) = \mathcal A\,\frac{[z_1-z_2]^{\{h,\dot h\}-2s}}
 {[z_1]^{\{h,\dot h\}}[z_2]^{\{h,\dot h\}}},
 \label{eq:J2-untwisted-wavefunction}
\end{equation}
where the constant $\mathcal A$ depends only on the coupling and cancels in all normalised quantities.
This is the standard three-point form for a primary of dimension $\Delta=h+\dot h$ and spin $S=\dot h-h$ at the origin, coupled to two scalar probes of dimension $2s$ at points $z_1$ and $z_2$, as in~\eqref{eq:CFT-wavefunction}. Thus no integrability input is needed to determine its coordinate dependence.

The three-point coefficients are the normalised planar brackets of these wave functions. Repeating~\eqref{eq:untwisting-normalised-limits} for $J_A=J_B=2$ and bridge length $\ell=1$, so that $J_C=2$, we have
\begin{equation}
 \begin{aligned}
 \mathcal C_1^{\bullet\circ\circ}
 &=\lim_{\theta_\alpha^A,\theta_\alpha^B\to0}\lim_{\phi\to0^+}
 \frac{2\LRpair{\Omega_{2}^{\rm C}}{\psi_B^{\rm C}}_1}
 {2\sin\phi\,\sqrt{\mathcal N_{2,B}\mathcal N_{2,C}}},\\[5pt]
 \mathcal C_1^{\bullet\bullet\circ}
 &=\lim_{\theta_\alpha^A,\theta_\alpha^B\to0}\lim_{\phi\to0^+}
 \frac{2\LRpair{\psi_A^{\rm C}}{\psi_B^{\rm C}}_1}
 {2\sin\phi\,\sqrt{\mathcal N_{2,A}\mathcal N_{2,B}\mathcal N_{2,C}}},
 \end{aligned}
 \label{eq:J2-untwisted-C-definition}
\end{equation}
with the planar bracket $\LRpair{\psi_A^{\rm C}}{\psi_B^{\rm C}}_1$ of~\eqref{eq:HHL-definition}. Here, one line of $A$ is glued to $B$, and the remaining leg of each ends on $\mathcal O_C=\operatorname{tr}[\phi_1\phi_1^\dagger]$ at the origin. As a free operator, $\mathcal O_C$ has unit norm, $\mathcal N_{2,C}=1$.

At zero twist these brackets can be evaluated directly. We place the three operators at finite points $x_A$, $x_B$ and $0$, and glue the wave functions~\eqref{eq:J2-untwisted-wavefunction} along their shared lines. Each glued line is amputated with the shadow transform~\eqref{eq:shadow-operator}.

Every integral that appears is a two-dimensional star-triangle relation. It reads
\begin{equation}
 \int\dd^2w\,\prod_{i=1}^{3}[w-z_i]^{-\{a_i,\dot a_i\}}
 =\pi\prod_{i=1}^{3}\frac{\Gamma(1-\dot a_i)}{\Gamma(a_i)}\;
 [z_{12}]^{\{a_3,\dot a_3\}-1}[z_{13}]^{\{a_2,\dot a_2\}-1}[z_{23}]^{\{a_1,\dot a_1\}-1},
 \label{eq:J2-untwisted-star-triangle}
\end{equation}
valid whenever $\sum_ia_i=\sum_i\dot a_i=2$ and $a_i-\dot a_i\in\mathbb Z$.

\paragraph{HHL.}
State $A$ sits at $x_A$ and state $B$ at $x_B$. They share one line, attached at $w$, and each sends its other leg to $\mathcal O_C$ at the origin. Let $\psi_X$ denote the wave function~\eqref{eq:J2-untwisted-wavefunction} with the operator moved to $x_X$. The glued diagram is
\begin{equation}
 \int\dd^2w\;\psi_{A}(w,0)\,\bigl[\bfS^{(w)}\psi_{B}(\cdot,0)\bigr](w).
\end{equation}
The shadow transform and the $w$ integral are each a single application of~\eqref{eq:J2-untwisted-star-triangle}. The result is the standard three-point structure,
\begin{equation}
 \frac{[x_{AB}]^{\{\frac12-h_A-h_B,\,\frac12-\dot h_A-\dot h_B\}}}
 {[x_A]^{\{\frac12+h_A-h_B,\,\frac12+\dot h_A-\dot h_B\}}\,[x_B]^{\{\frac12-h_A+h_B,\,\frac12-\dot h_A+\dot h_B\}}},
\end{equation}
multiplied by the coefficient
\begin{equation}
 (-1)^{S_B}\mathcal A_A\mathcal A_B\,
 \frac{\Gamma(1-\dot h_A)\,\Gamma(1-\dot h_B)\,\Gamma(\dot h_A+\dot h_B-\tfrac12)}
 {\Gamma(h_A)\,\Gamma(h_B)\,\Gamma(\tfrac32-h_A-h_B)}.
 \label{eq:J2-untwisted-HHL-glue}
\end{equation}
This is the zero-twist counterpart of~\eqref{eq:untwisting-J2-HHL-homogeneous}.

\paragraph{Norm.}
For the norm we glue both legs. After the two shadow transforms, the $z_2$ integral is again a star-triangle, and we are left with
\begin{equation}
 \LRpair{\psi}{\psi}
 =(-1)^S\mathcal A^2\,
 \frac{\Gamma(2\dot h)\,\Gamma(1-\dot h)^2}{4\pi\,\Gamma(h)^2\,\Gamma(1-2h)}\;
 [x_{AB}]^{\{1-2h,1-2\dot h\}}\int\frac{\dd^2z_1}{|z_1-x_A|^2\,|z_1-x_B|^2}.
\end{equation}
In the frame $x_A=0$, $x_B\to\infty$, the prefactor $[x_{AB}]^{\{1-2h,1-2\dot h\}}$ and $|z_1-x_B|^{-2}$ combine into the two-point coordinate factor $[x_{AB}]^{-\{2h,2\dot h\}}$, and the remaining integral over $\delta<|z_1|<R$ equals $2\pi\log(R/\delta)$. Comparing with~\eqref{eq:coordinate-logarithmic-overlap} gives $\mathcal C_{\log}$.

To convert this into the norm we need $\mathcal D_\xi$. From the spectrum~\eqref{eq:J2-spectrum} at fixed spin,
\begin{equation}
 \mathcal D_\xi=\frac{(1-2h)(2\dot h-1)}{1-\Delta}.
\end{equation}
The relation~\eqref{eq:coordinate-general-norm-relation} at $J=2$ then gives
\begin{equation}
 \mathcal N_2
 =(-1)^S\mathcal A^2\,
 \frac{(\Delta-1)\,\Gamma(2\dot h-1)\,\Gamma(1-\dot h)^2}
 {\Gamma(h)^2\,\Gamma(2-2h)}.
 \label{eq:J2-untwisted-direct-norm}
\end{equation}
This is the single-ordering norm. At the free point $h=\dot h=\tfrac12$
it equals $\mathcal A^2$, and the physical norm
$\mathcal N_2^{\theta=0}=2\mathcal N_2=2\mathcal A^2$ is the planar Wick
count of $\mathcal A\operatorname{tr}\phi_1^2$.

\paragraph{Structure constants.}
The light operators are free, with $\mathcal N^{\theta=0}_{2,\Omega_2}=2$ and $\mathcal N_{2,C}=1$. For HLL, the numerator is just $\psi_B$ with its two legs at the positions of $\Omega_2$ and $\mathcal O_C$, so that, including the cyclic factor $\sqrt{J_AJ_B}=2$ of~\eqref{eq:untwisting-normalised-limits}, $\mathcal C_1^{\bullet\circ\circ}=2\,\mathcal A_B/\sqrt{\mathcal N_{2,B}}$. Using~\eqref{eq:J2-untwisted-direct-norm} and Euler's reflection formula,
\begin{equation}
 \bigl(\mathcal C_1^{\bullet\circ\circ}\bigr)^2
 =(-1)^S\,\frac{4\,\Gamma(2-2h)\,\Gamma(\dot h)^2}
 {(\Delta-1)\,\Gamma(2\dot h-1)\,\Gamma(1-h)^2}.
 \label{eq:J2-untwisted-direct-HLL}
\end{equation}

For HHL, we multiply~\eqref{eq:J2-untwisted-HHL-glue} by the cyclic factor $2$ and divide by $\sqrt{\mathcal N_{2,A}\mathcal N_{2,B}}$. The factors $\Gamma(1-\dot h)/\Gamma(h)$ cancel against the norms, and up to the phase convention for their square roots we find
\begin{equation}
 \mathcal C_1^{\bullet\bullet\circ}
 =2\,\frac{\Gamma(h_A+h_B-\tfrac12)}{\Gamma(\tfrac32-\dot h_A-\dot h_B)}
 \left[\frac{\Gamma(2-2\dot h_A)\,\Gamma(2-2\dot h_B)}
 {(\Delta_A-1)(\Delta_B-1)\,\Gamma(2h_A-1)\,\Gamma(2h_B-1)}\right]^{1/2}.
 \label{eq:J2-untwisted-direct-C}
\end{equation}
Both results agree with the SoV computation, \eqref{eq:untwisting-J2-HLL-OPE} and~\eqref{eq:untwisting-J2-HHL-OPE}.

\section{Computing the \texorpdfstring{$J=2$}{J=2} wave function}
\label{sec:coordinate-J2-ode}

Unlike the untwisted example in Appendix~\ref{sec:J2-untwisted}, the
twisted two-site wave function is not fixed by conformal symmetry alone.
We therefore use the additional conserved charges encoded in the
fundamental transfer matrices \eqref{eq:transfer-matrix-expansion}.
Their coefficients are commuting differential operators, whose
eigenvalue equations determine the coordinate wave function.
Requiring it to be single-valued selects the allowed transfer matrix
eigenvalues, which we compare with the QSC quantisation condition.

\paragraph{Kinematics.}
We take equal local spins $s_1=s_2=s$ and a generic non-zero complex
twist $\lambda=e^{\varrho+\ii\phi}$. A shift of $u$ lets us choose
$\theta_1=-\theta_2=\theta$, with the same inhomogeneities in the dotted sector.

The two probes at $z_1$ and $z_2$
and the two twist fixed points $0,\;\infty$ give $\psi^{\rm D}$ the
kinematics of a four-point function. After removing a kinematic factor, its coordinate dependence is through
a single cross-ratio, which is
$\chi=z_1/z_2$. Let us denote
\begin{equation}\label{eqn:centerrelcoords}
Z=\frac{z_1+z_2}{2},\qquad
w=\frac{z_1-z_2}{z_1+z_2},\qquad
z_1=Z(1+w),\qquad z_2=Z(1-w),
\end{equation}
with the same for their dotted counterparts. The Cartan weights
in~\eqref{eq:weights-dimension-spin} fix the overall scaling, so we write
\begin{equation}
\psi^{\text{D}}(z_1,\dot z_1,z_2,\dot z_2)
= \frac{G(w,\dot w)}{Z^{h+2s}\dot Z^{\dot h+2\dot s}} .
\label{eqn:wavecentrerel}
\end{equation}
Just like usual four-point functions, conformal symmetry fixes the prefactor, but leaves $G(w,\dot w)$ to be determined
by the dynamics. For the physical fishnet chain, this information is, in principle,
contained in the graph-building equation~\eqref{eq:graph-building-stationarity}. However, solving that integral equation directly is cumbersome. A more practical
route is to use the commuting differential charges encoded in the
fundamental transfer matrix~\eqref{eq:transfer-matrix-expansion}. This is what we turn to next.

\paragraph{The additional conserved charge.}
This is where integrability enters. The two-site transfer matrix is
\begin{equation}
 \hat t(u)= (\lambda+\lambda^{-1})u^2 -\ii (\lambda-\lambda^{-1})\widehat{\mathbb S}^{z}u +\hat{\mathsf I}_0.
\end{equation}

The Cartan charge has already been taken care of by the prefactor.
The remaining charge $\hat{\mathsf I}_0$ is a second-order differential
operator~\eqref{eq:transfer-matrix-expansion}. Requiring $\psi$ to be its eigenfunction with eigenvalue $a_0$, and similarly in the dotted sector, supplies the extra information needed to determine~\eqref{eqn:wavecentrerel}.

\Needspace{7\baselineskip}
\subsection{Reduced ODE}
Substituting~\eqref{eqn:wavecentrerel} into the eigenvalue equation for
$\hat{\mathsf I}_0$ leaves a second-order equation for the undotted
factors of $G$:
\begin{equation}\label{eqn:reducedode}
B_2(w)\partial^2_{w}G(w,\dot{w})+B_1(w)\partial_{w}G(w,\dot{w})+B_0(w,a_0)G(w,\dot{w})=0\,,
\end{equation}
and similarly for the dotted sector. The coefficients $B_2(w)$, $B_1(w)$ and $B_0(w,a_0)$ follow by substituting~\eqref{eqn:wavecentrerel} into the transfer matrix eigenvalue equation.

\begin{figure}[t]
\centering
\includegraphics[width=0.55\linewidth]{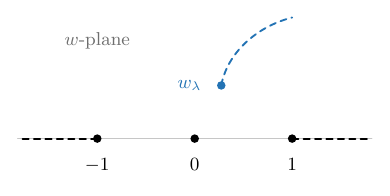}
\caption{Singular points in the $w$-plane. The points $-1$ and $1$
correspond to $z_1=0$ and $z_2=0$, respectively. The bilinear
$G=f_+\dot f_+ + r f_-\dot f_-$ is locally single-valued at ordinary
coincidence $w=0$, so no cut is shown there. The blue cut starts at
$w_\lambda=(\lambda^2-1)/(\lambda^2+1)$, where the twisted propagator
is singular, $z_1=\lambda^2z_2$. For generic charges its discontinuity
remains. Requiring cancellation there and at $w=\pm1$ fixes $r$ and
quantises the charges, as discussed below. The cut paths are schematic;
on shell the full bilinear is single-valued, although the individual
chiral solutions remain multivalued.}
\label{fig:J2-wave-coordinate-and-w-planes}
\end{figure}

The only information we require about them is that they have regular singularities at $w=0,\pm 1, w_\lambda \equiv (\lambda^2-1)/(\lambda^2+1)$ and $w=\infty$, see Figure~\ref{fig:J2-wave-coordinate-and-w-planes}. Hence, two independent solutions $f_\pm(w)$ of the ODE \eqref{eqn:reducedode} expanded around one of the singular points $w_0$ can generally be classified according to their Frobenius exponents $\rho_\pm$:
\begin{equation}
  f_\pm(w) = (w-w_0)^{\rho_\pm}(1+\mathcal{O}(w-w_0))\,.
\end{equation}

The singular points describe collisions where a probe meets the operator ($z_\alpha = 0$), where the probes meet each other ($z_1 = z_2$), or where the probes meet the twisted image of each other ($z_1 = \lambda^2 z_2$). In Table~\ref{tab:J2-frobenius}, we list these
configurations and their local Frobenius exponents $\rho_\pm$. The additional point
$w=\infty$ corresponds to $Z=0$ and reflects our choice of kinematic factor. Equivalently, written in terms of the standard cross-ratio
$
\chi=\frac{z_1}{z_2}=\frac{1+w}{1-w},
$
the four singular points are located at
$
\chi=\{0,1,\infty,\lambda^2\}
$.
\begin{table}[H]
\centering
\renewcommand{\arraystretch}{1.2}
\begin{tabular}{c|c|c}
\toprule
$w$ & Configuration & Frobenius exponents $\{\rho_+,\rho_-\}$ \\
\midrule
$-1$ & $z_1=0$ & $\{0,\,-h\}$ \\
$0$ & $z_1=z_2$ & $\{0,\,1-2s-2\ii\theta\}$ \\
$1$ & $z_2=0$ & $\{0,\,-h\}$ \\
$w_\lambda$ & $z_1=\lambda^2 z_2$ & $\{0,\,1-2s+2\ii\theta\}$ \\
$\infty$ & $z_1=-z_2$ & $\{h+2s,\,h+2s+1\}$ \\
\bottomrule
\end{tabular}
\caption{Singular points of the reduced ODE~\eqref{eqn:reducedode} with their corresponding coordinate configurations and their Frobenius exponents.}
\label{tab:J2-frobenius}
\end{table}

\Needspace{7\baselineskip}
\subsection{Single-valuedness and quantisation}
\label{sec:coordinate-J2-quantisation}

The transfer equations determine the local wave function, but do not yet
select a physical state. The missing condition has a simple coordinate
meaning: the full two-dimensional wave function must be single-valued.
This is the coordinate counterpart of the QSC gluing condition in
Section~\ref{sec:setup-baxter}.

Let $f(w)=(f_+(w),f_-(w))$ and
$\dot f(\dot w)=(\dot f_+(\dot w),\dot f_-(\dot w))$ be bases of the two
equations~\eqref{eqn:reducedode}. Their general joint solution is
\begin{equation}
 G(w,\dot w)=f(w)C\dot f(\dot w)^{\mathsf T},
 \label{eqn:reducedwave}
\end{equation}
where $C$ is a constant $2\times2$ matrix. The individual solutions $f_\pm$ are multivalued around the singular points of Table~\ref{tab:J2-frobenius}, so a generic $G$ is not a well-defined function on the plane. On the Euclidean slice $\dot w=\bar w$, the transport along a closed path $\gamma$ acts as $f \to f M_\gamma$, and along the conjugate path as $\dot{f} \to \dot{f} \dot{M}_\gamma$. Hence, $G(w,\dot{w})$ is single-valued if and only if
\begin{equation}
 M_\gamma C\dot M_\gamma^{\mathsf T}=C
 \qquad\text{for every independent loop }\gamma.
 \label{eq:coordinate-J2-monodromy}
\end{equation}
We do not assume that the dotted equation is the complex conjugate of the undotted one, nor that $C$ is Hermitian.

\paragraph{Local pairing and global quantisation.}
At $z_1=z_2$, or $w=0$, the unit-leading Frobenius solutions have
exponents $0$ and $1-2s-2\ii\theta$. For generic, non-resonant
data, compatible dotted exponents pair diagonally: their differences
must be integers, whereas the crossed differences must not.
Fixing the regular coefficient to one gives
\begin{equation}
 G(w,\dot w)=f_+(w)\dot f_+(\dot w)
            +r f_-(w)\dot f_-(\dot w).
 \label{eqn:w0expansion}
\end{equation}
This is single-valued around $w=0$ for any $r$, $a_0$, $\dot a_0$, but it is generally not single-valued around $w=\pm1$ and $w_\lambda$.
Requiring this very same pairing to hold at the remaining three points $w=\{-1,w_\lambda,1\}$ fixes the constant $r$ and quantises the eigenvalues $(a_0,\dot a_0)$. For example, at $s=\dot s=\tfrac14$, $\theta=\tfrac17$, $\lambda=e^{7\ii/10}$ and
$h=\dot h=3/5$, coordinate monodromy gives
\footnote{For the real scalar weights and pure-rotation twist used here,
$r$ is real if $\theta\in\ii\mathbb R$ (including zero) and
$\dot a_0=a_0^*$: the two unit-leading Frobenius bases are then complex
conjugates on the Euclidean slice.}
\begin{equation}
  a_0 = \dot{a}_0 \simeq 0.26889681817897, \quad r \simeq -1.57419510485850 +0.92234420983643\,\ii.
\end{equation}
Following the solutions between collision points is already a demanding
numerical task for $J=2$, and becomes more involved as $J$ grows.
Fortunately, QSC gluing offers a more practical route to the same
spectrum. We find agreement between the two approaches for $J=2$ and $J=3$, where the latter can be treated in a similar way to $J=2$.

The algorithm for computing these coordinate wave functions is also included in the \texttt{BSolver2D} package.

\paragraph{Single-valued does not mean analytic.}
Equation~\eqref{eqn:w0expansion} contains both an analytic sector and a non-analytic sector
proportional to
$w^{1-2s-2\ii\theta}\dot w^{1-2\dot s-2\ii\theta}$.
At zero inhomogeneities and $s=\dot s$, the latter is $|w|^{2-4s}$
on the Euclidean slice, reducing to $|w|$ at the fishnet point.
For generic $s$, this term is single-valued but cannot be represented
by a Taylor series at $z_1=z_2$. This is why a basis built only from polynomial
wave functions needs an extension at finite coupling.

\section{Four-point functions and perturbative calculations}
\label{app:4pt}
In this appendix we consider the four-point function
\begin{equation}\label{eq:G34App}
    G^{(3)}_{4}\left(x_1,x_2,x_3,x_4\right) = \langle \operatorname{tr}\left[\phi^{2}_1(x_1) \phi_1(x_2)\right] \operatorname{tr}\left[\phi^{\dagger,2}_1(x_3)\phi^{\dagger}_1(x_4)\right] \rangle
\end{equation}
in the planar limit and use it to perturbatively extract some of the OPE coefficients controlling $\operatorname{tr}\left[ \phi^{2}_1(x)\phi_1(y)\right]$ as $x\rightarrow y$. This will provide a check of the data found from our integrability approach and presented in Section~\ref{sec:CFTData}.

We start by recasting \eqref{eq:G34App} as
\begin{equation}
    G^{(3)}_{4}(x_1,x_2,x_3,x_4) = \frac{1}{\left(2\pi\right)^{3}}\frac{1}{|x_{12}|^{3/2}|x_{34}|^{3/2}}\left(\frac{|x_{24}|}{|x_{13}|}\right)^{1/2}\mathcal{G}^{(3)}\left(z,\bar{z}\right)\,,
\end{equation}
where we have defined the conformal cross-ratios
\begin{equation}
    z \bar{z} = \frac{x^{2}_{12}x^{2}_{34}}{x^{2}_{13}x^{2}_{24}}\,,
    \quad
    (1-z)(1-\bar{z}) = \frac{x^{2}_{14}x^{2}_{23}}{x^{2}_{13}x^{2}_{24}}\,.
\end{equation}
The reduced correlator $\mathcal{G}^{(3)}$ has the following expansion in global conformal blocks:
\begin{equation}\label{eq:ConformalBlockApp}
    \mathcal{G}^{(3)}(z,\bar{z}) = \sum_a c_{a}\,k_{2h_{a}}(z)\, k_{2\dot{h}_a}(\bar{z})\,,
    \quad
    k_{2h} = z^{h}{}_{2}F_{1}\left(h-\frac{1}{4},h+\frac{1}{4};2h;z\right)\,.
\end{equation}
In an orthogonal basis of exchanged primaries, $c_a$ is the product of the two three-point couplings divided by the two-point normalisation of the exchanged primary, with the external normalisation fixed by the definition of $\mathcal{G}^{(3)}$.

\Needspace{7\baselineskip}
\subsection{Tree-level structure constants} At $\xi=0$, the three planar Wick contractions of~\eqref{eq:G34App} give the reduced correlator
\begin{equation}
    \mathcal{G}^{(3)}_{0} = |z|^{3/2}\left(1+\frac{2}{|1-z|}\right)\,.
\end{equation}
To recast this into the language of conformal blocks we use the identities
\begin{equation}
    \frac{z^{3/4}}{(1-z)^{1/2}} = \sum_{n=0}^{\infty}\frac{(\frac{1}{2})^{3}_{n}}{4^n n!(\frac{1}{4})_n (\frac{3}{4})_n}k_{3/2+2n}(z)\,,
    \quad
    z^{3/4} = \sum_{n=0}^{\infty}\frac{(-1)^{n}(\frac{1}{2})^{2}_{n}}{4^n(\frac{1}{4})_n (\frac{3}{4})_n}k_{3/2+2n}(z)\,.
\end{equation}
Applying these identities to $\mathcal{G}^{(3)}_0$ and matching to~\eqref{eq:ConformalBlockApp} gives the tree-level block coefficients.
We focus on leading-twist primaries with $\dot h-h=S$ and $h+\dot h=\tfrac32+S$, built by distributing $S$ antiholomorphic derivatives among the three fields in $\operatorname{tr}\phi_1^3$.
The degeneracy $d_S$ is the dimension of the space of these cyclic operators modulo total derivatives.
The block expansion determines the sum $c_S^{(0)}=\sum_{a=1}^{d_S}c_{S,a}\big|_{\xi=0}$ over primaries with these quantum numbers.
The degeneracies and summed coefficients are
\begin{center}
\begin{tabular}{c|ccccccc}
 $S$ & $0$ & $1$ & $2$ & $3$ & $4$ & $5$ & $6$ \\
 \hline
 $d_S$ & $1$ & $0$ & $1$ & $2$ & $1$ & $2$ & $3$ \\
 $c_S^{(0)}$ & $3$ & $0$ & $\frac{3}{20}$ & $-\frac{5}{616}$
 & $\frac{7}{832}$ & $-\frac{315}{454784}$
 & $\frac{24519}{49447424}$
\end{tabular}
\end{center}

\subsection{Perturbative corrections}
The first correction to the reduced four-point function appears at order $\xi^6$. The expansion is
\begin{equation}
    \mathcal{G}^{(3)}(z,\bar{z}) = \mathcal{G}^{(3)}_{0}(z,\bar{z}) + \xi^{6} \,\mathcal{G}^{(3)}_{1}(z,\bar{z}) + \mathcal{O}\left(\xi^{12}\right)\,.
\end{equation}
The block expansion~\eqref{eq:ConformalBlockApp} gives the corrections to the block coefficients and conformal dimensions.
At fixed spin, write $c_a=c_a^{(0)}+\xi^6\delta c_a+O(\xi^{12})$ and $\Delta_a=\Delta_a^{(0)}+\xi^6\gamma_a+O(\xi^{12})$. The order-$\xi^6$ contribution is then
\begin{equation}
 \mathcal G_1^{(3)}(z,\bar z)
 =\sum_a\left[\delta c_a
 +\frac{c_a^{(0)}\gamma_a}{2}\left(\partial_h+\partial_{\dot h}\right)
 \right]k_{2h}(z)\,k_{2\dot h}(\bar z)\Big|_{h=h_a^{(0)},\,\dot h=\dot h_a^{(0)}},
 \label{eq:J3-paper-one-wheel-blocks}
\end{equation}
with $h^{(0)}_{a},\dot{h}_a^{(0)}$ the tree-level weights of state $a$. In particular, we note that since $k_{2h}(z) \sim z^{h}$ it is enough to expand \eqref{eq:J3-paper-one-wheel-blocks} for small $z,\bar{z}$ up to orders $h^{(0)}_{a}, \dot{h}^{(0)}_{a}$ to read the correction. To find this small $z,\bar{z}$ expansion we need to compute $\mathcal{G}$ perturbatively.

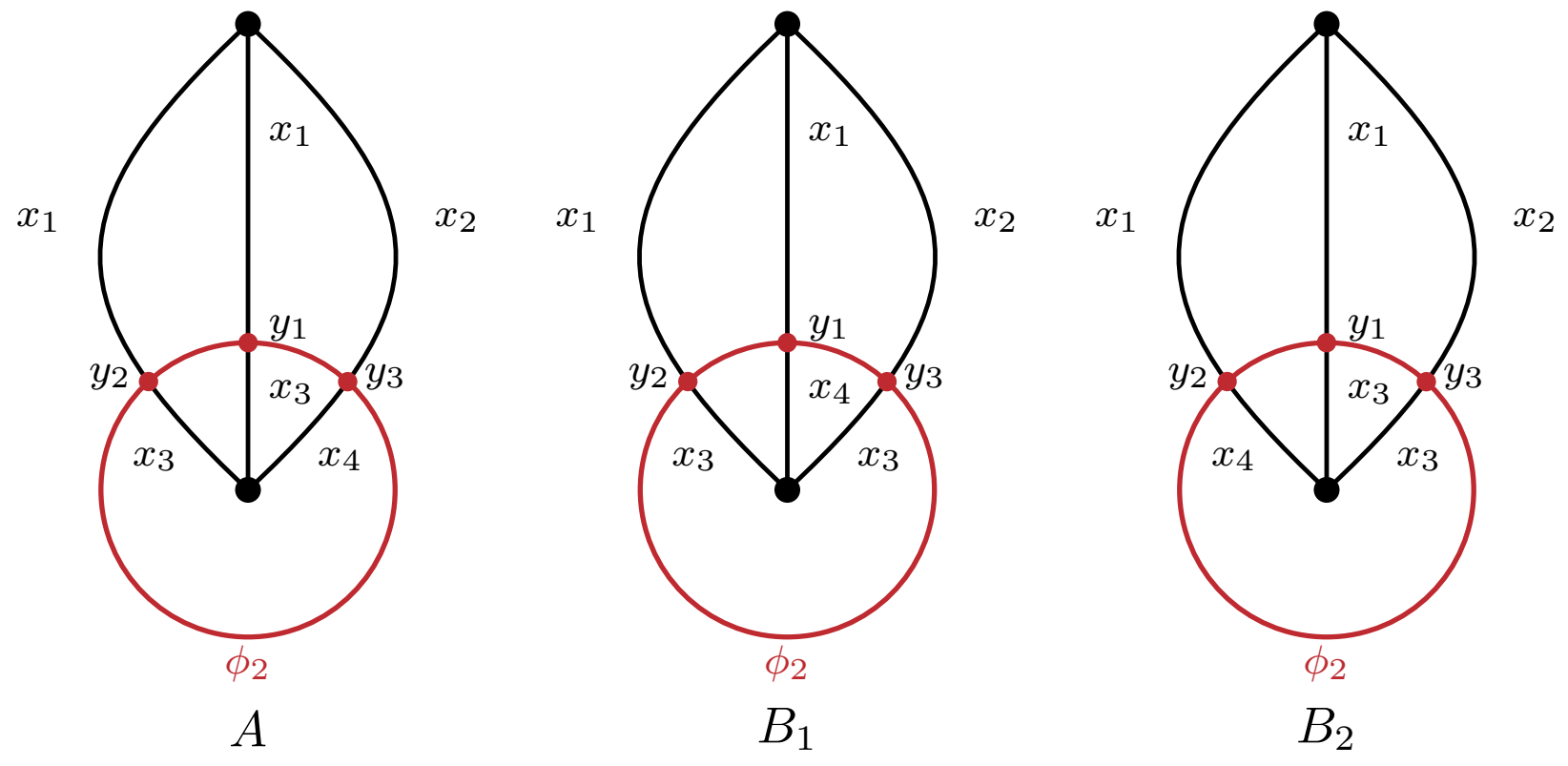
\begin{figure}[htbp]
\centering
\definecolor{wheelred}{RGB}{190,42,48}
\begin{tikzpicture}[
  scale=0.85,
  line cap=round,
  line join=round,
  font=\small,
  propagator/.style={black,line width=.9pt},
  wheel/.style={wheelred,line width=1pt},
  vertex/.style={circle,fill=wheelred,inner sep=1.35pt},
  trace/.style={circle,fill=black,inner sep=1.8pt},
  lab/.style={font=\scriptsize,inner sep=1.5pt}
]
\def\R{1.2}

\foreach \shift/\name/\iTop/\iLeft/\iRight in {
  0/A/3/3/4,
  4.4/B_1/4/3/3,
  8.8/B_2/3/4/3
}{
  \begin{scope}[xshift=\shift cm]
    \coordinate (I) at (0,0);
    \coordinate (O) at (0,3.8);

    \draw[propagator,name path=top]
      (I) -- (O);
    \draw[propagator,name path=left]
      (I) .. controls +(137:2.2) and +(-137:2.2) .. (O);
    \draw[propagator,name path=right]
      (I) .. controls +(43:2.2) and +(-43:2.2) .. (O);

    \draw[wheel,name path=ring]
      (I) circle[radius=\R];

    \foreach \strand in {top,left,right}{
      \path[
        name intersections={of=ring and \strand,by=crossing}
      ];
      \node[vertex] at (crossing) {};
    }

    \node[lab,right] at (0.1,\R+0.12) {$y_1$};
    \node[lab,anchor=east] at (-0.88,0.92) {$y_2$};
    \node[lab,anchor=west] at (0.88,0.92) {$y_3$};
    \node[lab,wheelred,below] at (270:\R) {$\phi_2$};

    \node[trace] at (I) {};
    \node[trace] at (O) {};

    \node[lab,right] at (0.1,2.9) {$x_1$};
    \node[lab,left] at (-1.45,2.2) {$x_1$};
    \node[lab,right] at (1.45,2.2) {$x_2$};

    \node[lab,right] at (0.1,0.8) {$x_{\iTop}$};
    \node[lab,anchor=east] at (-0.5,0.25) {$x_{\iLeft}$};
    \node[lab,anchor=west] at (0.5,0.25) {$x_{\iRight}$};

    \node at (0,-\R-0.75) {$\name$};
  \end{scope}
}
\end{tikzpicture}
\caption{The three planar one-wheel pairings $A$, $B_1$ and $B_2$.
Black lines are $\phi_1$ propagators, the red circle is the
$\phi_2$ wheel, and the red dots are integrated interaction vertices.
The black dots represent the two traces; the labels on the
propagator ends specify their spacetime insertions.}
\label{fig:J3-one-wheel-pairings}
\end{figure}

In the conformal frame $(x_1,x_2,x_3,x_4)=(0,z,1,\infty)$, the computation of $\mathcal{G}^{(3)}_1$ reduces to the following integral
\begin{align}
 \mathcal G_1^{(3)}(z,\bar z)
 &=\frac{|z|^{3/2}}{\pi^3}
 \int
 \frac{\dd^2y_1\,\dd^2y_2\,\dd^2y_3}
 {|y_{12}|\,|y_{23}|\,|y_{31}|\,|y_1|\,|y_2|\,|y_3-z|}
 \nonumber\\[-2pt]
 &\qquad\times\left[
 \frac{1}{|1-y_1|\,|1-y_2|}
 +\frac{1}{|1-y_2|\,|1-y_3|}
 +\frac{1}{|1-y_1|\,|1-y_3|}
 \right].
 \label{eq:J3-paper-one-wheel-integral}
\end{align}
The three terms in brackets in~\eqref{eq:J3-paper-one-wheel-integral} correspond, in order, to the pairings $A$, $B_1$ and $B_2$ in Figure~\ref{fig:J3-one-wheel-pairings}. Exchanging $y_1$ and $y_2$ maps $B_1$ to $B_2$, so their integrals are equal.

Let $W_A$ be the integral for $A$ and $W_B$ the common integral for $B_1$ and $B_2$. Then $\mathcal{G}^{(3)}_1=|z|^{3/2}(W_A+2W_B)/\pi^3$, with
\begin{align}
W_A(z,\bar z)
&=
\int_{\mathbb C^3}
\frac{\dd^2y_1\,\dd^2y_2\,\dd^2y_3}
{|y_{12}|\,|y_{23}|\,|y_{31}|\,
 |y_1|\,|y_2|\,|y_3-z|\,
 |1-y_1|\,|1-y_2|}, \label{eq:WAApp}
\\
W_B(z,\bar z)
&=
\int_{\mathbb C^3}
\frac{\dd^2y_1\,\dd^2y_2\,\dd^2y_3}
{|y_{12}|\,|y_{23}|\,|y_{31}|\,
 |y_1|\,|y_2|\,|y_3-z|\,
 |1-y_2|\,|1-y_3|}.
\end{align}
Let us first consider $W_A$. Using conformal symmetry we fix
$$
    \int \frac{\dd^2y_1\, \dd^2y_2}{|y_1||y_1-1||y_{13}||y_2||y_2-1||y_{23}||y_{12}|} = \frac{K_A}{|y_3||y_3-1|}\,,
$$
where $K_A$ is a constant. To determine $K_A$, we apply a M\"obius transformation that fixes $0$ and $1$ and sends $y_3$ to infinity. Using~\cite{DOTSENKO1985691}, we obtain
\begin{equation}
\begin{split}
    K_A &= \int \frac{\dd^2u\, \dd^2v}{|u||u-1||u-v||v||v-1|} = \frac{2\pi^6}{\Gamma(\tfrac34)^{8}}\,.
\end{split}
\end{equation}
Substituting this result into \eqref{eq:WAApp} reduces $W_A$ to $K_A$ times the isotropic two-dimensional box integral. Evaluating the box using the Yangian Ward identities~\cite{Corcoran:2021gda} gives
\begin{equation}
\begin{split}
W_A(z,\bar{z})
&=K_A\left(
\vcenter{\hbox{%
\begin{tikzpicture}[scale=.65, every path/.style={line width=.7pt}]
  \coordinate (y) at (0,0);
  \draw (y) -- (-1,0) node[left] {$0$};
  \draw (y) -- (1,0) node[right] {$1$};
  \draw (y) -- (0,-1) node[below] {$z$};
  \draw[line width=1pt, dash pattern=on 3pt off 2pt]
    (y) -- (0,1) node[above] {$\infty$};
  \fill (y) circle (3.5pt);
  \node[above right, inner sep=4pt] at (y) {$y_3$};
\end{tikzpicture}%
}}
\right)=K_A\,T(z,\bar{z}),\\
T(z,\bar{z})
&=\int_{\mathbb C}\frac{\dd^2y_3}{|y_3-z|\,|y_3|\,|y_3-1|}=4\left[\mathrm K(z)\mathrm K(1-\bar{z})+\mathrm K(\bar{z})\mathrm K(1-z)\right].
\end{split}
\end{equation}
Here $\mathrm K$ is the complete elliptic integral of the first kind
in the parameter convention,
\begin{equation*}
 \mathrm K(m)=\int_0^{\pi/2}\frac{\dd\vartheta}
                         {\sqrt{1-m\sin^2\vartheta}}.
\end{equation*}

Integrating first over $y_1$ and $y_2$ and setting $v=y_3$ gives
\begin{equation*}
 W_B(z,\bar z)=\int_{\mathbb C}
 \frac{\dd^2v}{|v-z|\,|1-v|}\,B(v,\bar v).
\end{equation*}
The function $B$ is the two-loop Basso--Dixon ladder~\cite{Basso:2017jwq}
given by
\begin{equation}
B(v,\bar v)
=
\left(
\vcenter{\hbox{%
\begin{tikzpicture}[scale=.7, every path/.style={line width=.7pt}]
  \coordinate (y1) at (-.7,0);
  \coordinate (y2) at (.7,0);
  \coordinate (zero) at (0,1);
  \coordinate (v) at (0,-1);

  \draw (y1) -- (y2);
  \draw (y1) -- (zero) -- (y2);
  \draw (y1) -- (v) -- (y2);
  \draw (y2) -- (1.7,0) node[right] {$1$};
  \draw[line width=1pt, dash pattern=on 3pt off 2pt]
    (y1) -- (-1.7,0) node[left] {$\infty$};

  \fill (y1) circle (3.5pt);
  \fill (y2) circle (3.5pt);
  \node[above left, inner sep=4pt] at (y1) {$y_1$};
  \node[above right, inner sep=4pt] at (y2) {$y_2$};
  \node[above] at (zero) {$0$};
  \node[below] at (v) {$v$};
\end{tikzpicture}%
}}
\right)
=
\int_{\mathbb C^2}
\frac{\dd^2y_1\,\dd^2y_2}
{|y_{12}|\,|y_1|\,|y_2|\,
 |y_1-v|\,|y_2-v|\,|1-y_2|}.
\end{equation}
These integrals were treated in detail in~\cite{Derkachov:2018rot}. The elliptic representation of~\cite{Duhr:2022pch} gives
\begin{equation}
B(v,\bar v)
=
\frac{1}{2|v|}\,
T\!\left(
\frac{1-\sqrt v}{2},
\frac{1-\sqrt{\bar v}}{2}
\right)^2.
\end{equation}
\paragraph{Scalar contribution.}
The elliptic expression for $W_A$ gives its small-$z$ behaviour.
To obtain the corresponding term in $W_B$, subtract the two wheel
integrands before taking $z\to0$. The subtracted integrand is integrable
in this limit and is antisymmetric under $y_1\leftrightarrow y_3$, so
$W_A-W_B=o(1)$. Thus
\begin{equation}
 W_A,\,W_B=2\pi K_A\bigl(\log16-\log|z|\bigr)+o(1).
\end{equation}
Combining the three pairings and matching to the scalar block in
\eqref{eq:J3-paper-one-wheel-blocks} gives
\begin{equation}
 \frac{\mathcal G^{(3)}_1}{|z|^{3/2}}
 =\frac{6K_A}{\pi^2}\bigl(\log16-\log|z|\bigr)+o(1)
 =\delta c_0+c_0^{(0)}\gamma_0\log|z|+o(1).
\end{equation}
Since $c_0^{(0)}=3$, we find $\gamma_0=-2K_A/\pi^2$ and
$\delta c_0=-3\gamma_0\log16$.

\paragraph{Higher spins.}
We expand $W_A$ directly and obtain the leading-twist coefficients of
$W_B$ by passing the ladder differential equation through the remaining
integral. The scalar limit and the absence of a spin-one primary fix
the boundary data. Matching order by order to
\eqref{eq:J3-paper-one-wheel-blocks}, including the dimension derivatives
of the lower-spin descendant contributions, gives the results below.

\begin{table}[htbp]
 \centering
 \renewcommand{\arraystretch}{1.7}
 \begin{tabular}{c|ccc}
  $S$ & $0$ & $2$ & $4$ \\
  \hline
  $\gamma_S/\gamma_0$ & $1$ & $\dfrac16$ & $\dfrac5{56}$ \\
  $\delta c_S$ & $-3\gamma_0\log16$ &
   $-\dfrac{\gamma_0}{40}\left(\log16-\dfrac35\right)$ &
   $-\dfrac{5\gamma_0}{6656}
    \left(\log16-\dfrac{821}{1170}\right)$
 \end{tabular}
 \caption{Order-$\xi^6$ corrections to the dimensions and block
 coefficients of the leading-twist $J=3$ primaries with $S=0,2,4$.}
 \label{tab:J3-one-wheel-data}
\end{table}

The anomalous dimensions in Table~\ref{tab:J3-one-wheel-data} agree
with the spectrum in~\eqref{eq:J3-leading-twist-gamma}. The relative
shifts $\delta c_S/c_S^{(0)}$ agree with the weak-coupling coefficients
fitted to the BSolver2D data in~\eqref{eq:C3-leading-twist-weak-coupling}
to a relative accuracy of order $10^{-12}$.

\section{Further details on untwisting}
\label{app:untwisting-descendants}

\subsection{Parent weights and crossover matching}
\label{app:untwisting-crossover-matching}

\paragraph{The parent weight.}
In general, the twisted spectrum has only degeneracies due to discrete parity symmetries. When the full 2D conformal symmetry is restored at zero twist, states that are independent in the twisted case have to join conformal families. Most states approach descendants of a conformal primary in the untwisting limit.

For each twisted branch, its \emph{parent} is the zero-twist
highest-weight primary whose global conformal family contains the
limiting state. We denote the parent's weights by
$(\boldsymbol{h},\dot{\boldsymbol{h}})$.
A branch approaching a descendant of levels
$(n_{\mathrm d},\dot n_{\mathrm d})$ thus has
\begin{equation}
 h(\phi)\underset{\phi\to0}{\longrightarrow}\boldsymbol{h}+n_{\mathrm d},\qquad
 \dot h(\phi)\underset{\phi\to0}{\longrightarrow}\dot{\boldsymbol{h}}+\dot n_{\mathrm d}.
 \label{eq:untwisting-parent-weight}
\end{equation}
The nonnegative integers count the holomorphic and antiholomorphic
derivatives acting on this primary. The zero-twist Casimir
is $\boldsymbol{h}(\boldsymbol{h}-1)$, and the corresponding power-law
basis and Wronskians are given in Section~\ref{sec:untwisting-whittaker}.
We now retain arbitrary descendant levels in the crossover region
$x=\phi u$, with the same unit-leading twisted normalisation.

Expanding~\eqref{eq:baxter-equation} at fixed $x$, the first two
orders cancel by~\eqref{eq:leading-transfer-eigenvalues}.
The leading order $\phi^{2-J}$ gives
\begin{equation}
 x^2\mathtt{q}''+2Js\,x\mathtt{q}'
 +\left[-x^2+2(\boldsymbol{h}+n_{\mathrm d})x
        -(\boldsymbol{h}-\tfrac12)^2+(Js-\tfrac12)^2\right]\mathtt{q}=0.
 \label{eq:untwisting-whittaker-equation}
\end{equation}
Here the linear coefficient in the last term contains the descendant weight,
whereas the constant term contains the Casimir.

The equation is solved explicitly by the two solutions below, which carry the unit-leading twisted normalisation matching the canonical normalisation of $q_i$:
\begin{equation}
 \begin{aligned}
 \mathtt{q}_1(x)&=(2\phi)^{-n_{\mathrm d}}
 \left(\frac{x}{\phi}\right)^{\boldsymbol{h}-Js}
 e^{-x}U(-n_{\mathrm d},2\boldsymbol{h},2x),\\
 \mathtt{q}_2(x)&=(-2\phi)^{2\boldsymbol{h}+n_{\mathrm d}}
 \left(\frac{x}{\phi}\right)^{\boldsymbol{h}-Js}
 e^{x}U(2\boldsymbol{h}+n_{\mathrm d},2\boldsymbol{h},-2x).
 \end{aligned}
 \label{eq:untwisting-crossover-solutions}
\end{equation}
Here $U$ is Tricomi's function, with
$U(a,b,z)\simeq z^{-a}$ at large $z$.
For $n_{\mathrm d}\in\mathbb Z_{\ge0}$, the factor in the first solution
$U(-n_{\mathrm d},2\boldsymbol{h},z)$ is a degree-$n_{\mathrm d}$ polynomial
with leading coefficient one and value
$(-1)^{n_{\mathrm d}}\Gamma(2\boldsymbol{h}+n_{\mathrm d})/\Gamma(2\boldsymbol{h})$ at zero.
For generic $\boldsymbol{h}$, the two small-$x$ terms of $\mathtt{q}_2$ give
\begin{equation}
 \begin{aligned}
 \mathtt{q}_2(x)={}&(-2\phi)^{2\boldsymbol{h}+n_{\mathrm d}}
 \left(\frac{x}{\phi}\right)^{\boldsymbol{h}-Js}
 \Bigl[\frac{\Gamma(1-2\boldsymbol{h})}{\Gamma(1+n_{\mathrm d})}
 (1+O(x))\\
 &\qquad\qquad+
 \frac{\Gamma(2\boldsymbol{h}-1)}{\Gamma(2\boldsymbol{h}+n_{\mathrm d})}
 (-2x)^{1-2\boldsymbol{h}}(1+O(x))\Bigr].
 \end{aligned}
 \label{eq:untwisting-g2-small-x}
\end{equation}
Since $x=u\phi$, at fixed $u$ as $\phi\to0$ the second line dominates for our working domain $\Re \boldsymbol{h}>1/2$.
 The leading terms at small and large $x$ thus are
\begin{equation}
 \begin{array}{c|c|c}
 &1\ll|u|\ll\phi^{-1}&|u|\gg\phi^{-1}\\ \hline
 \mathtt{q}_1(\phi u)&
 \dfrac{\Gamma(2\boldsymbol{h}+n_{\mathrm d})}
 {\Gamma(2\boldsymbol{h})(-2\phi)^{n_{\mathrm d}}}
 u^{\boldsymbol{h}-Js}&
 e^{-\phi u}u^{\boldsymbol{h}+n_{\mathrm d}-Js}\\[6pt]
 \mathtt{q}_2(\phi u)&
 \dfrac{(-2\phi)^{n_{\mathrm d}+1}\Gamma(2\boldsymbol{h}-1)}
 {\Gamma(2\boldsymbol{h}+n_{\mathrm d})}
 u^{1-\boldsymbol{h}-Js}&
 e^{\phi u}u^{-\boldsymbol{h}-n_{\mathrm d}-Js}
 \end{array}
 \label{eq:untwisting-crossover-powers}
\end{equation}
The right column reproduces~\eqref{eq:baxter-asymptotics}; the left
column fixes the coefficients of the zero-twist powers
\eqref{eq:untwisting-zero-twist-asymptotics}, providing non-trivial matching coefficients for the two regimes, as
the small-$x$ regime here still means $|u|\gg1$. So we see that the $u\sim 1/\phi$ regime explains the $n_{\mathrm d}$ mismatch in the asymptotics between twisted and untwisted Q-functions.

{
For $n_{\mathrm d}=\dot n_{\mathrm d}=0$ the main text writes $h\equiv\boldsymbol h$ and $\dot h\equiv\dot{\boldsymbol h}$. The two coefficients in the left column of~\eqref{eq:untwisting-crossover-powers} are then $1$ and $2\phi/(1-2h)$; repeating with $(h,\phi)\mapsto(\dot h,-\phi)$ gives~\eqref{eq:untwisting-basis-limit}. Keeping the levels arbitrary gives the descendant factor~\eqref{eq:untwisting-descendant-normalised-factor} derived below.\par}

\paragraph{Mixing of the two solutions.}
The main text needs to know which zero-twist solutions each $q_i$
contains in the matching region. For a primary, $n_{\mathrm d}=0$,
the crossover solutions~\eqref{eq:untwisting-crossover-solutions} read
\begin{equation}
 \mathtt q_1(x)=u^{h-Js}e^{-x}U(0,2h,2x),\qquad
 \mathtt q_2(x)=(-2\phi)^{2h}u^{h-Js}e^{x}U(2h,2h,-2x).
\end{equation}
The Tricomi factor in the first solution is trivial, $U(0,2h,2x)=1$.
At small $x$, $\mathtt q_1$ is therefore the single power $u^{h-Js}$ and
matches onto pure $\boldsymbol q_1$.
 The second solution has two
small-$x$ terms by~\eqref{eq:untwisting-g2-small-x},
\begin{equation}
 \mathtt q_2(\phi u)\simeq(-2\phi)^{2h}\,\Gamma(1-2h)\,u^{h-Js}
 +\frac{2\phi}{1-2h}\,u^{1-h-Js},
\end{equation}
so it contains both zero-twist solutions. For $\Re h>1/2$ the first
term is suppressed relative to the second by $\phi^{2h-1}$.

The simple form of $\mathtt q_1$ relies on
$U(-n_{\mathrm d},2\boldsymbol h,2x)$ being a polynomial, i.e.\ on
$n_{\mathrm d}=h(\phi)-\boldsymbol h$ being a nonnegative integer.
Varying the Casimir weight to $\boldsymbol h_\epsilon$ at fixed
$h(\phi)$, as in Section~\ref{sec:untwisting-norm}, makes
$n_{\mathrm d}$ non-integer, so $\mathtt q_1$ acquires a $\boldsymbol q_2$
component.

We obtain this component from the same two-power expansion as for $\mathtt q_2$. In the
first line of~\eqref{eq:untwisting-crossover-solutions} we replace
\begin{equation}
 \boldsymbol h\to\boldsymbol h_\epsilon,
 \qquad
 n_{\mathrm d}\to h(\phi)-\boldsymbol h_\epsilon.
 \label{eq:untwisting-norm-crossover-replacement}
\end{equation}
The coefficient of the additional power $u^{1-\boldsymbol h_\epsilon-Js}$
is then
\begin{equation}
 (2\phi)^{1-\boldsymbol{h}_\epsilon-h(\phi)}
 \frac{\Gamma(2\boldsymbol{h}_\epsilon-1)}
      {\Gamma(\boldsymbol{h}_\epsilon-h(\phi))}.
 \label{eq:untwisting-norm-extra-coefficient}
\end{equation}
On the unperturbed primary branch this vanishes, since $1/\Gamma(0)=0$.
The variation switches it on. Using $1/\Gamma(z)=z+O(z^2)$, its
$\epsilon$ derivative is
\begin{equation}
 -\frac{\Gamma(2h-1)}{2h-1}\,(2\phi)^{1-2h}.
 \label{eq:untwisting-norm-extra-derivative}
\end{equation}
Unlike the suppressed admixture in $q_2$, this term dominates at small
$\phi$ for $\Re h>1/2$. It therefore supplies the leading variation of
$q_1$. Matching this contribution to $\boldsymbol q_2^\uparrow$
gives~\eqref{eq:untwisting-norm-weight-variation}.

\subsection{Descendant matrix elements and normalisation}

The main text extracts the highest-weight OPE coefficient. Branches
that approach global descendants contain the same dynamical data:
their three-point matrix elements follow by the conformal Ward
identities. We derive the resulting ratio of descendant to primary
matrix elements,~\eqref{eq:untwisting-descendant-normalised-factor},
and show that it gives exactly the finite factors in the untwisting limit.
To lighten notation, in this subsection we drop the bold font and write
$h\equiv\boldsymbol h$ and $\dot h\equiv\dot{\boldsymbol h}$ for the
parent weights, and $n,\dot n$ for the levels
$n_{\mathrm d},\dot n_{\mathrm d}$. In the notation
of~\eqref{eq:untwisting-parent-weight}, the twisted weight $h(\phi)$
then tends to $h+n$, with $n\in\mathbb Z_{\ge0}$, and the dotted
level $\dot n$ is independent.

The general crossover equation is already
\eqref{eq:untwisting-whittaker-equation}.
This separation of the asymptotic weight from the Casimir also occurs
in the defect-fusion limit of~\cite{Chernikov:2026lcv},
Section~4.1.2, equations~(4.10)--(4.13).

\paragraph{The polynomial crossover.}
Substituting $\mathtt{q}=(x/\phi)^{h-Js}e^{-x}P(2x)$ and putting $z=2x$ gives
$zP''+(2h-z)P'+nP=0$.
 Its polynomial solution with $P(0)=1$ is
\begin{equation}
 P_n(z;h)\equiv\sum_{r=0}^{n}
 \frac{(-n)_r}{(2h)_r}\frac{z^r}{r!},\qquad
 \mathtt{q}(x)\propto (x/\phi)^{h-Js}e^{-x}P_n(2x;h),
 \label{eq:untwisting-descendant-polynomial}
\end{equation}
where $(a)_r\equiv\Gamma(a+r)/\Gamma(a)$ is the rising factorial.
The coefficient recurrence from the differential equation terminates
at $r=n$; equivalently, $P_n$ is an associated Laguerre polynomial
divided by its value at zero. Its highest coefficient is
$(-1)^n/(2h)_n$, so matching the unit-leading twisted asymptotics gives
\begin{equation}
 q_1^{\rm tw}(u)\sim(-1)^n(2h)_n(2\phi)^{-n}q_1^0(u)
 \quad\text{at fixed }u.
 \label{eq:untwisting-descendant-matching}
\end{equation}
The second-solution scale follows from the same Wronskian ratio as
in the main text. The dotted matching uses
$(h,n,\phi)\mapsto(\dot h,\dot n,-\phi)$.
For $n=0$ the polynomial is one and the primary matching is recovered.
Already $n=1$ gives $P_1(z;h)=1-z/(2h)$, so replacing the twisted
weight by the parent weight without this factor is insufficient.

\paragraph{Finite factors in the moment tails.}
For a common twist, the two crossovers multiply the undotted tail by
$P_{n_A}(z;h_A)P_{n_B}(z;h_B)$. Expanding this product under the
Laplace integral gives the finite factor
\begin{equation}
 R_{n_A n_B}(a;h_A,h_B)\equiv
 \sum_{r=0}^{n_A}\sum_{s=0}^{n_B}
 \frac{(-n_A)_r(-n_B)_s(a)_{r+s}}
      {(2h_A)_r(2h_B)_s\,r!\,s!}.
 \label{eq:untwisting-descendant-tail-factor}
\end{equation}
Indeed, its product with $\Gamma(a)$ is
$\int_0^\infty dz\,z^{a-1}e^{-z}
P_{n_A}(z;h_A)P_{n_B}(z;h_B)$, initially for $\Re a>0$ and then by
analytic continuation. Thus each undotted Abel factor
$\Gamma(a)$ is replaced by $\Gamma(a)R_{n_A n_B}(a;h_A,h_B)$;
the dotted factor changes in the same way.

Remove the compact-set matching factors
of~\eqref{eq:untwisting-descendant-matching} from each glued
Q-function and denote the resulting moment matrix by
$\widehat{\mathcal M}$. Its regular block is the same
$\mathcal M^0$ as for the parent primaries. The joint tail still
factorises into an undotted row factor and a dotted column factor.
The contour cancellations of Section~\ref{sec:untwisting-moment-reduction}
therefore apply unchanged: the polynomials change the coefficients
at fixed $\phi u$, but not the powers of $\phi$. With
$\eta=h_A+h_B-h_C$ and $\dot\eta=\dot h_A+\dot h_B-\dot h_C$ as in the
main text, where $h_C=\dot h_C=J_C/4$, now computed from the parent
weights, the determinant reduction becomes
\begin{equation}
 \begin{aligned}
 \lim_{\phi\to0}(2\phi)^{\eta+\dot\eta}
 \det\widehat{\mathcal M}
 ={}&\frac{2\pi\Gamma(\eta)}{\Gamma(1-\dot\eta)}
 R_{n_A n_B}(\eta;h_A,h_B)\\
 &\times R_{\dot n_A\dot n_B}(\dot\eta;\dot h_A,\dot h_B)
 \det\mathcal M^0.
 \end{aligned}
 \label{eq:untwisting-descendant-minor}
\end{equation}
The parent-weight convergence conditions of the main text still apply.
This gives the leading coefficient at generic points where the displayed
factors are nonzero.

\paragraph{Norm and numerator coefficients.}
Two special cases of the finite sum are needed. For HLL the reference
polynomial is one, which gives the first identity below, while for the
norm the two states coincide, $A=B$, which gives the second,
\begin{equation}
 R_{0n}(a;h_A,h_B)=\frac{(2h_B-a)_n}{(2h_B)_n},\qquad
 R_{nn}(2h-1;h,h)=\frac{n!}{(2h)_n}.
 \label{eq:untwisting-descendant-special-factors}
\end{equation}
The first equality is the terminating Vandermonde identity.
 For the
second, write the associated Laguerre polynomial as a sum of the
polynomials with index $2h-2$ and use their orthogonality with weight
$z^{2h-2}e^{-z}$. The resulting sum is
$\sum_{k=0}^n(2h-1)_k/k!=(2h)_n/n!$, giving the stated factor after
normalising $P_n(0)=1$.

The matching factors cancel between the state normalisations and the
moment determinants. In the norm the remaining product of first- and
second-solution scales is fixed by the Wronskians and is independent
of the levels. Hence the norm of a branch approaching level
$(n,\dot n)$ obeys
\begin{equation}
 \lim_{\phi\to0}(2\phi)^{2\Delta-J}\mathcal N_2^{(n,\dot n)}
 =\frac{n!\,\dot n!}{(2h)_n(2\dot h)_{\dot n}}
 \mathcal N_2^0,
 \label{eq:untwisting-descendant-norm}
\end{equation}
where $\mathcal N_2^0$ is the parent coefficient
in~\eqref{eq:untwisting-norm-limit}.
 The origin-normalised HLL and
HHL numerator coefficients acquire the two $R$ factors
in~\eqref{eq:untwisting-descendant-minor}; their zero-twist Q-values
and reduced q-bilinears are unchanged.

\paragraph{The standard descendant relation.}
To identify the normalised factor, use radial quantisation with $B$
at $0$, the light primary $C$ at $1$ and the dual of $A$ at infinity.
The primary matrix element $\langle A|\widehat{\mathcal O}_C(1)|B\rangle$
equals $(-1)^{S_A+S_B}\mathcal C_\ell^{\bullet\bullet\circ}$: the map $z\mapsto1-z$
exchanges the positions of $B$ and $C$, with the integer-spin phase
fixed by the dotted and undotted branches of the main text.
This fixes the frame also for the descendants.
In one $SL(2)$ factor, the Ward identities
$[L_j,\mathcal O_C(z)]=(z^{j+1}\partial_z+(j+1)h_Cz^j)
\mathcal O_C(z)$, $j=-1,0,1$, imply
\begin{equation}
 \frac{\langle A|e^{uL_1}\widehat{\mathcal O}_C(1)
                   e^{vL_{-1}}|B\rangle}
      {\langle A|\widehat{\mathcal O}_C(1)|B\rangle}
 =(1-u)^{\eta-2h_A}(1-v)^{\eta-2h_B}(1-uv)^{-\eta}.
 \label{eq:untwisting-descendant-Ward-kernel}
\end{equation}
Indeed, the bra generating insertion is $u^{-2h_A}\mathcal O_A(1/u)$
and the ket insertion is $\mathcal O_B(v)$; substituting these points
in the primary three-point function gives the right-hand side.

The same kernel generates the finite tail factors. The polynomial
in~\eqref{eq:untwisting-descendant-polynomial} obeys
\[
 \sum_{n\ge0}\frac{(2h)_n}{n!}P_n(z;h)u^n
 =(1-u)^{-2h}\exp\!\left[-\frac{zu}{1-u}\right].
\]
Inserting this identity twice in the Laplace integral for $R$ shows
that the coefficient of $u^nv^m$ in
\eqref{eq:untwisting-descendant-Ward-kernel} is
$(2h_A)_n(2h_B)_mR_{nm}(\eta;h_A,h_B)/(n!m!)$.
Coefficient comparison proves that the unnormalised descendant
matrix element is multiplied by $(2h_A)_n(2h_B)_mR_{nm}$.
The algebra $[L_1,L_{-1}]=2L_0$ gives
$\langle h|L_1^nL_{-1}^n|h\rangle=n!(2h)_n$ by induction.
Thus the factor $n!/(2h)_n$ in
\eqref{eq:untwisting-descendant-norm} is the standard norm of
$L_{-1}^n|h\rangle/(2h)_n$, the normalisation selected by the
compact-set matching.

Let $|A;n,\dot n\rangle$ denote
$L_{-1}^n\dot L_{-1}^{\dot n}|A\rangle$ divided by
$\sqrt{n!(2h_A)_n\,\dot n!(2\dot h_A)_{\dot n}}$, and define
$|B;n,\dot n\rangle$ in the same way, with the dual
normalised by the same pairing.
 The standard relation is therefore
\begin{equation}
 \boxed{\begin{aligned}
 &\frac{\langle A;n_A,\dot n_A|\widehat{\mathcal O}_C(1)
                          |B;n_B,\dot n_B\rangle}
      {\langle A|\widehat{\mathcal O}_C(1)|B\rangle}\\
 &\qquad=\frac{R_{n_A n_B}(\eta;h_A,h_B)
        R_{\dot n_A\dot n_B}(\dot\eta;\dot h_A,\dot h_B)}
 {\displaystyle\sqrt{
   \frac{n_A!\,\dot n_A!\,n_B!\,\dot n_B!}
        {(2h_A)_{n_A}(2\dot h_A)_{\dot n_A}
         (2h_B)_{n_B}(2\dot h_B)_{\dot n_B}}}}.
 \end{aligned}}
 \label{eq:untwisting-descendant-normalised-factor}
\end{equation}
The ratio of numerator and norm factors in the untwisting limit
is precisely this expression. Thus the crossover calculation
reproduces the conformal Ward identities.
For example, if only $n_A=1$ is nonzero, the ratio is
$(h_A-h_B+h_C)/\sqrt{2h_A}=(2h_A-\eta)/\sqrt{2h_A}$.
The dotted calculation is independent, so the result holds at
nonzero spin as well.

For evaluation, obtain the parent coefficient from the BSolver2D calls
in Appendix~\ref{app:untwisting-evaluation} and multiply by the
finite factor above in this radial frame. Conversely, divide a
normalised descendant-branch overlap by this factor to recover the
highest-weight coefficient, with the position factors and permutation
phase treated in the same frame. No additional q-bilinears are needed.

\paragraph{Checks of the descendant relation.}
{Both solutions in~\eqref{eq:untwisting-crossover-solutions} solve
\eqref{eq:untwisting-whittaker-equation} symbolically for $J=2,4$,
$n=0,1,2,3$ and weights $h=7/10,17/10$. The norm and HLL identities
in~\eqref{eq:untwisting-descendant-special-factors} were checked
symbolically through $n=5$. For $J_A=J_B=4$, $\ell=2$, $h_A=7/10$ and $h_B=4/5$, direct Laplace
quadrature and the
finite double sum agree to a relative accuracy of $10^{-45}$ for levels
$(0,0),(1,0),(1,1),(2,1),(3,2)$ and spins $(S_A,S_B)=(0,0),(1,0),(1,1)$.

The tail factors also agree symbolically with the global descendant
three-point matrix elements through $n_A,n_B=4$ at arbitrary parent
weights. These checks test the crossover and tail factors, rather than a
finite-twist continuation of descendant states.\par}

\Needspace{7\baselineskip}
\subsection{Determinant reduction at nonzero spin}
\label{app:untwisting-spinning-reduction}

For $S_A+S_B>0$, $\dot\eta=\eta+S_A+S_B$ can exceed one, so
additional dotted brackets can have divergent tails. The joint
singular terms factorise into an undotted row factor and a dotted
column factor. Row and column operations collect these terms into
the corner of the moment matrix.

The product of the two nonanalytic terms
in~\eqref{eq:untwisting-Abel-tail} gives
\begin{equation}
 \begin{aligned}
 (\mathcal M_{bc})_{\rm joint}\simeq{}&
 2(-1)^{\ell-c}e^{\pi\ii(\eta-\dot\eta)}\sin(\pi\eta)
 \Gamma(\eta+b-\ell)\Gamma(\dot\eta+c-\ell)\\
 &{}\times(2\phi)^{2\ell-b-c-\eta-\dot\eta}.
 \end{aligned}
 \label{eq:untwisting-spinning-tail}
\end{equation}
This factorisation holds for the full singular series.

Indeed, the large-$u$ Baxter expansions are common to all towers
in each arrow basis.
When expressed in the same spectral coordinate, their singular
Abel terms are independent of the tower origin; only the
$\boldsymbol{\rho}_{AB}$-residue coefficients differ.
Summing those coefficients therefore leaves one product
of an undotted and a dotted series.

From each row $b<\ell$, subtract the last row multiplied by the
first ratio below, and do the analogous subtraction in each
column $c<\ell$ with the second:
\[
 \frac{(\mathcal M_{b\ell})_{\rm joint}}
      {(\mathcal M_{\ell\ell})_{\rm joint}},
 \qquad
 \frac{(\mathcal M_{\ell c})_{\rm joint}}
      {(\mathcal M_{\ell\ell})_{\rm joint}}.
\]
Using the full singular series removes the joint tail from every
entry except the corner.
These operations preserve the determinant, and their coefficients
vanish as $(2\phi)^{\ell-b}$ and $(2\phi)^{\ell-c}$.
They therefore leave the regular constant block unchanged.

The mixed terms still cancel by the contour argument of
Section~\ref{sec:untwisting-moment-reduction}.
If a dotted bracket needs a finite part, its powers differ from
noninteger $\dot\eta$ by integers.
There is no $u^{-1}$ term, so the large semicircle supplies no
constant term.
The resulting $\boldsymbol{\mathcal M}$ is thus obtained by
finite-part summing each q-bilinear before contracting Baxter
indices and adding the towers.

For completeness, the remaining diagonal channels must also leave
this constant block unchanged.
Work first with $0<\eta<1$ and, for $\ell>1$,
$\eta+\dot\eta<2\min(\Delta_A,\Delta_B)$.
Replacing both first solutions of state $A$ by its second
solutions changes the large-$n$ power by $2-2\Delta_A$.
Its largest reduced-block contribution is consequently
$(2\phi)^{2\Delta_A-\eta-\dot\eta}$, which vanishes;
the same holds for $B$, and replacing both states improves the bound.
The undotted mixed brackets decay faster in these channels,
while any undotted tail in a reduced row carries at least
$(2\phi)^{\ell-b-\eta}$, a positive power.
Thus these channels give no additional constant, also after
the row and column operations.
For $\ell=1$ there is no reduced block and no additional bound.
This establishes~\eqref{eq:untwisting-numerator-minor}
at nonzero spin; the remaining weights follow by meromorphic
continuation on each fixed-spin branch.

\bibliographystyle{JHEP}
\bibliography{references}

\end{document}